\documentclass[12pt]{article}
\usepackage{amsmath}
\usepackage{amsfonts}
\usepackage{graphicx}
\usepackage{enumerate}
\usepackage{natbib}
\setcitestyle{authoryear,open={(},close={)}} %

\usepackage{url} %

\usepackage{moreverb}
\usepackage{booktabs}
\usepackage{eqparbox} %

\usepackage{caption}
\usepackage{subcaption}

\usepackage{float}

\usepackage{makecell}
\usepackage{multirow}
\usepackage{setspace}%

\usepackage{pifont} %
\newcommand{\cmark}{\ding{51}}%

\usepackage{amsthm}
\newtheorem{prop}{Proposition}
\newcommand{\bmath}{\boldsymbol}

\usepackage{algorithm}
\usepackage{algpseudocode}

\def\bSig\mathbf{\Sigma}

\usepackage{authblk}

\usepackage{titlesec}
\titlespacing*{\section}{0pt}{3em}{1em}

\usepackage{minitoc}

\makeatletter
\renewcommand{\l@section}[2]{\vspace{0.4em}\@dottedtocline{1}{0em}{10em}{\textbf{#1}}{\textbf{#2}}}
\renewcommand{\l@subsection}{\@dottedtocline{2}{2.5em}{10em}} 
\renewcommand{\l@subsubsection}{\@dottedtocline{3}{5em}{10.5em}}\makeatother

\begin{document}

\title{
Estimating Associations Between Cumulative Exposure and Health via Generalized Distributed Lag Non-Linear Models using Penalized Splines
}

\author[1]{Tianyi Pan}
\author[2,3]{Hwashin Hyun Shin}
\author[1]{Glen McGee}
\author[1]{Alex Stringer}

\affil[1]{Department of Statistics and Actuarial Science, University of Waterloo}
\affil[2]{Environmental Health Science and Research Bureau, Health Canada}
\affil[3]{Department of Mathematics and Statistics, Queen's University}

\date{}
\maketitle

\begin{abstract}
Quantifying associations between short-term exposure to ambient air pollution and health outcomes is an important public health priority. 
Many studies have investigated the association considering delayed effects within the past few days. 
Adaptive cumulative exposure distributed lag non-linear models (ACE-DLNMs) quantify associations between health outcomes and cumulative exposure that is specified in a data-adaptive way. 
While the ACE-DLNM framework is highly interpretable, it is limited to continuous outcomes and does not scale well to large datasets. Motivated by a large analysis of daily pollution and respiratory hospitalization counts in Canada between 2001 and 2018, we propose a generalized ACE-DLNM incorporating penalized splines, improving upon existing ACE-DLNM methods to accommodate general response types. 
We then develop a computationally efficient estimation strategy based on profile likelihood and Laplace approximate marginal likelihood with Newton-type methods. 
We demonstrate the performance and practical advantages of the proposed method through simulations. In application to the motivating analysis, the proposed method yields more stable inferences compared to generalized additive models with fixed exposures, while retaining interpretability.
\end{abstract}

\noindent%
\vfill

\newpage

\doparttoc 
\faketableofcontents 
\part{}  
\vspace{-1.1cm}
\section{Introduction}
\label{s:intro}

Quantifying associations between short-term exposure to air pollution and health outcomes remains a public health priority in Canada and around the world. 
For example,  Health Canada and Environment and Climate Change Canada are tasked to monitor air quality and their associated risks to public health \citep{AHTI}. 
The motivation for this work is to characterize associations between daily hospitalization counts and $\text{PM}_\text{2.5}$ measurements from 2001 to 2018 aggregated by census division across Canada. 
Many studies have recognized that the association between air pollution and health is potentially non-linear and subject to delayed effects \citep{chen2017fine,liu2019ambient}. 
To appropriately account for the delayed effect in a non-linear framework, two distinct classes of distributed lag non-linear models have been proposed. 
However, one class (e.g. \citealp{gasparrini2014modeling}) can be challenging to interpret, while the other (e.g. \citealp{wilson2022kernel}) requires continuous outcomes and therefore cannot be applied to our motivating analysis of health outcome counts. 
In this paper, we present a fast, scalable, and flexible framework to estimate associations between exposure and daily health outcome counts, accounting for the lagged effects and retaining interpretability. 

A common approach %
for estimating non-linear associations between health outcomes and short-term exposure to air pollution is to fit a generalized additive model (GAM). 
This approach typically considers either
(1) a single-day exposure \citep{powell2015ambient, shin2021comparison}, %
or (2) fixed cumulative exposure for lagged effects \citep{chen2017fine,liu2019ambient}, which 
takes a simple average of exposures over several days. 
While the former approach ignores exposures on any other days, the latter can be sensitive to the choice of time window (e.g. \citealp{shin2021comparison,shin2023circulatory}). %
In our motivating application, 
fitting GAMs for daily $\text{PM}_\text{2.5}$ with different definitions of exposure led to different inferences and wide intervals (Section \ref{s:example}).  
These results highlight the need for a more sophisticated modelling approach, %
allowing cumulative exposure to be specified adaptively rather than fixed in advance.

Distributed lag models (DLMs) 
address this by modelling the cumulative exposure as a weighted average with unknown weights estimated from the data  \citep{schwartz2000distributed, zanobetti2000generalized}. 
We refer to the cumulative exposure under unknown weights as \textit{adaptive cumulative exposure} (ACE), and the association function between ACE and response as \textit{adaptive cumulative exposure-response function} (ACERF). The ACE generalizes the 
simple average exposure, and the ACERF is analogous to the exposure-response function in a GAM. 
However, the DLM restricts associations to be linear, which 
may not be appropriate for modelling air pollution and health.
Two distinct classes of distributed lag non-linear model (DLNM) have been proposed to extend DLMs to handle non-linear associations.

The first class---which we call distributed response function DLNMs (DRF-DLNMs)---fits distinct non-linear associations at each lag via a bivariate exposure-lag-response function
\citep{gasparrini2010distributed, gasparrini2014modeling, gasparrini2017penalized, mork2022treed}.
Fitting DRF-DLNMs is computationally convenient (e.g. \citealp{gasparrini2011distributed}), but interpretation is 
not straightforward, 
relying on post-hoc summary statistics from the estimated bivariate surface, such as the exposure-response curve at a specific lag. 
These quantities are not directly comparable to the associations between cumulative exposure and response from the broadly used GAMs.

The second class of model instead specifies a non-linear function (i.e. ACERF) of the ACE itself \citep{wilson2022kernel, wang2023semiparametric}. 
We refer to this model as an \textit{adaptive cumulative exposure distributed lag non-linear model} (ACE-DLNM).
This is more interpretable, yielding inference similar to that from simpler GAMs: the lag weights indicate at which times exposure effects are strongest, and the ACERF characterizes the effect of cumulative exposure. 

While the ACE-DLNM framework is highly interpretable,
all existing ACE-DLNMs require continuous responses, preventing their 
applications to daily hospitalization counts and $\text{PM}_\text{2.5}$.
In addition, they face difficulties in inference and computation.
\citet{wilson2022kernel} proposed the Bayesian kernel machine regression and distributed lag model for Gaussian responses, but it relies on an MCMC approach for a number of kernel function coefficients equal to the sample size.
\citet{wang2023semiparametric} introduced a distributed lag quantile regression model for continuous responses, but its estimation is based on a derivative-free algorithm, 
and it uses unpenalized splines requiring manual smoothness selection. 
Some functional regression models share similar forms to ACE-DLNMs \citep{kong2010statistical,ma2016estimation};
these too require continuous responses, which simplifies the objective functions into least-square problems that do not extend naturally to count responses. 
These approaches fit unpenalized splines and avoid over-fitting by grid-search for numbers of knots. 
The discrete number of knots cannot properly control smoothness \citep{eilers1996flexible} and grid-search is infeasible with many smooth functions.
Our motivating application considers five non-linear functions---the ACERF, lag weights, secular time trend, seasonality, temperature associations---and hence it requires a more efficient approach.

In this paper, we propose an inferential and computational framework for ACE-DLNMs, improving upon existing ACE-DLNM methods to accommodate general response types and scale to large datasets.
We introduce a novel estimation approach, using profile likelihood and Laplace approximate marginal likelihood implemented by fast and accurate Newton-type methods. 
This framework: (1) is designed for possibly over-dispersed count responses but applies to more general types of data (including the exponential family),
(2) allows for flexible non-linearity in weight function and ACERF (as well as association functions of covariates), (3) imposes smoothness penalties to avoid over-fitting, 
and (4) scales well to large datasets. 
We successfully apply the proposed ACE-DLNM to the motivating analysis of daily hospitalization counts and PM$_{2.5}$ in five cities across Canada between 2001 and 2008, 
finding that it yields more stable inferences than simpler GAM approaches. 
We compare the proposed approach with the DRF-DLNM in this analysis and illustrate the advantage of the ACE-DLNM in interpretability.

\section{Adaptive Cumulative Exposure Distributed Lag Non-Linear Model}
\label{s:model}

Let $Y_t$ be a response, which may be count, continuous or otherwise, 
observed at a set of $N$ discrete time points, $t \in \mathcal{T} =  \left\{1, 2, \cdots, N\right\}$.
Let $X: \mathbb{R} \rightarrow \mathbb{R}$ be a continuous exposure process. 
Denote $x_s$ as the value of $X$ at discrete time point $s \in \mathcal{T}^x$, where 
$\mathcal{T}^x$ is not necessarily equal to or a subset of $\mathcal{T}$ but its range is bounded by the range of $\mathcal{T}$. 
Let $\bmath{z}_t \in \mathbb{R}^p$ represent the vector of covariates. 
We assume that $Y_t | X, \bmath{z}_t$ follows a distribution specified in \citet{wood2016smoothing} 
with a finite mean $\mu_t \in \mathcal{M} \subseteq \mathbb{R}$ and unknown parameters $\bmath{\theta}$.

An adaptive cumulative exposure distributed lag non-linear model (ACE-DLNM) is 
\begin{equation}
    g(\mu_t) = f \left\{\int_{0}^{L} w(l) X(t-l) dl\right\} + \sum_{j=1}^p h_j (z_{tj}),
\label{eq:model}
\end{equation}
where $g: \mathcal{M} \rightarrow \mathbb{R}$ is a smooth, monotone link function. Define $E(t) = \int_{0}^{L} w(l) X(t-l) dl$ as the adaptive cumulative exposure (ACE) at time $t$ with respect to the unknown smooth weight function $w(\cdot)$ and the maximum lag $L$. The $f(\cdot)$ is the unknown smooth adaptive cumulative exposure-response function (ACERF). The $h_j(\cdot)$ are $p$ unknown smooth functions of covariates $z_{tj}$, $j = 1, \cdots, p$, accommodating non-linear smooth functions (e.g., non-linear time trend and temperature association in the motivating application), linear components, and random effects (see e.g. \citealp{wood2016smoothing}).

This model extends existing ACE-DLNMs %
from continuous responses to more general types of data. This extension introduces inferential and computational challenges. 
In Section \ref{s:est}, we propose a fast, scalable, and flexible framework for fitting ACE-DLNMs.

Motivated by the analysis of daily $\text{PM}_{2.5}$ and hospitalization counts in Canada, in the remainder of the paper, we adopt a negative binomial distribution with mean $\mu_t$ and dispersion parameter $\theta > 0$ such that $\mathrm{Var}\left(Y_t | X, \bmath{z}_t\right) = \mu_t+ \mu_t^2/\theta$ and hence $\mathrm{Var}\left(Y_t | X, \bmath{z}_t\right) > \mathrm{E}\left(Y_t | X, \bmath{z}_t\right)$ allowing for potential over-dispersion; and we use the log-link, $g(\mu_t) = \log (\mu_t)$.

\subsection{Relation to GAM, DLM and DRF-DLNM}
\label{ss:othermethod}
The ACE-DLNM
includes the GAM and DLM as special cases by restricting either $w$ or $f$. 
Assuming $w$ is a fixed \textit{known} function,
the cumulative exposure $E(t)$ is therefore \textit{known} and the ACE-DLNM reduces to a GAM, 
where the \textit{known} $E(t)$ can be, e.g., the simple average over some set of days \citep{chen2017fine, liu2019ambient}.
On the other side, assuming $f$ to be \textit{linear}, the ACE-DLNM reduces to a DLM with
$g(\mu_t) = \beta \int_{0}^{L} w(l) X(t-l) dl + \sum_{j=1}^p h_j (z_{tj})$.

The DRF-DLNM generalizes the DLM in another way, fitting potentially distinct non-linear associations for $x_{t-l}$ at each lag $l$: 
$$
 g(\mu_t) = \sum_{l=0}^{L-1} \psi(x_{t-l}, l) + \sum_{j=1}^p h_j (z_{tj}),
$$
where $\psi(\cdot, \cdot)$ is a bivariate exposure-lag-response function (See Web Appendix A for details). 
Interpretation of the DRF-DLNM is based on the bivariate function $\psi(\cdot, \cdot)$.

\subsection{Continuous Exposure Process}
In ACE-DLNMs, we treat the exposure continuously, 
replacing $X(t)$ with an interpolating spline fit to the observed $x_s, s \in \mathcal{T}^x$; see Web Appendix B for details. 
The continuous exposure leads to a fast and exact evaluation of the ACEs and an identifiability proposition.

Under the spline representation, we propose a fast and exact approach to evaluate 
$E(t) = \int_{0}^{L} w(l) X(t-l) dl$ for each $t$, overcoming the need to approximate them by discretization that is 
adopted in all existing ACE-DLNMs. %
This approach uses de Boor’s algorithm \citep{de1978practical} and a modification of the method in \citet{wood2017p}; see Web Appendix C.

\subsection{Identifiability}
\label{ss:iden}
As discussed in \citet{wilson2022kernel} and \citet{wang2023semiparametric}, the model is not identifiable without constraints: we could replace $w(l)$ with $\widetilde{w}(l) = c \cdot w(l)$ and $f(E)$ with $\widetilde{f}(E) = f(E/c)$ but the likelihood holds the same. 
We impose the continuous constraints on $w$ following \citet{wilson2022kernel}: 
(1) scale constraint 
$\int_0^L w(l)^2 dl = 1$,
and (2) sign constraint 
$\int_{0}^L w(l) dl > 0$.
We then introduce Proposition \ref{prop:iden}. The proof is in Web Appendix D.2. 
\begin{prop} %
\label{prop:iden}
    Under Conditions in Web Appendix D.2, there exists a knot sequence for spline $X(t)$, such that the ACE-DNLM is identifiable under the scale and sign constraints. 
\end{prop}

\section{Estimation}
\label{s:est}
We represent $w(\cdot)$ and $f(\cdot)$ by cubic B-spline basis expansions, specifically: 
$f(E) = \sum_{q=1}^{d^f} b^f_{q}(E)\alpha^f_{q}$ and $w(l) = \sum_{q=1}^{d^w} b^w_{q}(l)\alpha^w_{q}$, 
where $b^f_{q}(\cdot)$ and $b^w_{q}(\cdot)$ are known B-spline basis functions, and $\bmath{\alpha}^f = [\alpha^f_{1}, \cdots, \alpha^f_{d^f}]^{\top}$ and $\bmath{\alpha}^w = [\alpha^w_{1},\cdots,\alpha^w_{d^w}]^{\top}$ are unknown spline coefficients. 
We represent $h_j(\cdot)$ by basis expansions: $h_j(z_{tj}) = \sum_{q = 1}^{d^{h_j}} b_{jq}^{h}(z_{tj}) \beta_{jq}$, where $b^h_{jq}(\cdot)$ are known basis functions, and let $\bmath{\beta} = [\bmath{\beta}_1^{\top},\cdots,\bmath{\beta}_p^{\top}]^{\top}$ where $\bmath{\beta}_{j} = [\beta_{j1}, \cdots, \beta_{jd^{j}}]^{\top}$ for $j = 1, \cdots, p$.

\subsection{Reparameterization}
\label{ss:repa}
For model identifiability, the sign and scale constraints (Section \ref{ss:iden}) are imposed on function $w$ and hence its spline coefficients $\bmath{\alpha}^w$. 
We propose the following reparametrization to map the constrained $\bmath{\alpha}^w$ to an unconstrained $\boldsymbol{\phi}^w \in \mathbb{R}^{d^w-1}$. 

We first reparametrize $w(l) = \sum_{q=1}^{d^w} b^w_{q}(l)\alpha^w_{q}$ via the sum-to-zero reparametrization (details in \citealp{wood2017generalized}):  
$w(l) = \alpha^{w+}_{1} + w^+(l)$,
where $w^+(l) = \sum_{q=2}^{d_w} b^{w+}_{q}(l)\alpha^{w+}_{q}$ and $\sum_{j=1}^J w^+(l_j) = 0$, 
with $\left\{l_j\right\}_{j = 1}^J$ being an evenly spaced partition of $[0, L]$. 
We replace $\int_{0}^L w^+(l)$ by $(L/J)\sum_{j=1}^J w^+(l_j)$ with a large $J$, 
and then the sign of $\int_{0}^L w(l) dl$ is determined by the sign of $\alpha^{w+}_{1}$.
Define $\bmath{\alpha}^{w+} = \left[\alpha^{w+}_{1}, \cdots, \alpha^{w+}_{d^w}\right]^\top$ and $\mathbf{C}$ as a $d^w \times d^w$ matrix with $C_{i,j} = \int_{0}^L b^{w+}_i(l) b^{w+}_j(l) dl$.
We then reparametrize $\bmath{\alpha}^{w+}$ by the unconstrained $\bmath{\phi}^w \in \mathbb{R}^{d^w-1}$: 
$$
    \bmath{\alpha}^{w+} = \frac{[1,\bmath{\phi}^{w\top}]^\top}{\left([1,\bmath{\phi}^{w\top}] \mathbf{C}[1,\bmath{\phi}^{w\top}]^\top\right)^{1/2}}.
$$

Through the reparameterization above, %
the function $w$ satisfies the sign and scale constraints, ensuring that the ACE-DLNM is identifiable under the Conditions in Proposition \ref{prop:iden}; see details in Web Appendix D.3. The computational complexity of the reparameterization is $O(J(d^w)^2)$, being dominated by the QR decomposition of a $J\times d^w$-dimensional matrix.

Under the same identifiability constraints, 
\citet{wilson2022kernel} sampled parameters on a constrained parameter space. 
\citet{ma2016estimation} imposed alternative constraints and minimized an objective function using a constrained optimization. 
The constrained sampling/optimization is challenging.
In this proposed framework, the reparameterization absorbs the constraints, 
facilitating a simple unconstrained estimation. %

\subsection{Penalized Log-Likelihood}
To control the roughness of the functions and avoid over-fitting, 
quadratic penalties, denoted by $\mathcal{P}$, are imposed on $\bmath{\alpha}^{w}$, $\bmath{\alpha}^{f}$ and $\bmath{\beta}$: 
$\mathcal{P}(\bmath{\alpha}^{w}; \lambda^w) = (1/2)\lambda^w \bmath{\alpha}^{w\top} \bmath{S}^w \bmath{\alpha}^{w}$,
$\mathcal{P}(\bmath{\alpha}^{f}; \lambda^f) = (1/2)\lambda^f \bmath{\alpha}^{f\top} \bmath{S}^f \bmath{\alpha}^{f}$, and 
$\mathcal{P}(\bmath{\beta}; \bmath{\lambda}^{h}) = \sum_{j=1}^p (1/2) \lambda^{h}_j \bmath{\beta}_j^{\top} \bmath{S}_j^{h} \bmath{\beta}_j$, 
where $\bmath{\lambda} = \left[ \lambda^w, \lambda^f, \bmath{\lambda}^{h\top} \right]^\top$, $\bmath{\lambda}^{h} = [\lambda^{h}_1, \cdots, \lambda^{h}_p]^{\top}$ are unknown smoothing parameters, 
and $\bmath{S}^w$, $\bmath{S}^f$ and $\bmath{S}^{h}_j, j = 1,\cdots,p$ are \textit{fixed}, known penalty matrices. 
We set $\bmath{S}^w$ and $\bmath{S}^f$ as integrated squared second-order derivative penalty matrices \citep{wood2017p}. 
Specifically, $\bmath{S}^f = (S^f_{i, j})_{1\leq i,j \leq d^f}$ where $S^f_{i,j} = \int b^{f[2]}_{i}(x) b^{f[2]}_{j}(x) dx$ and $b^{f[2]}_i(x)$ denotes the second-order derivative of $b^{f}_i(x)$.
We specify $h_j$ using any basis expansion with the corresponding default penalty (see \citealp{wood2016smoothing} and \citealp{wood2015package}).

The basis functions $b_1^f(\cdot),\cdots,b_{d^f}^f(\cdot)$, and thus $\bmath{S}^f$, rely on the range of $E(t) = \int_0^L w(l)X(t-l)dl$,
which varies by $w(\cdot)$ and can change at each optimization iteration. 
These changes make the derivatives of $\bmath{\alpha}^w$ and $\bmath{\alpha}^f$ intractable and require recomputing the expensive penalty matrix $\bmath{S}^f$ at each iteration.
To address this, for a generic exposure $X$ and any given $t$, we derive the predetermined bounds for $E(t)$ using the Cauchy-Schwarz inequality: 
$$
- \sqrt{\int_{0}^L X (t-l)^2 dl} \le E(t) \le \sqrt{\int_{0}^L X (t-l)^2 dl}.
$$
It follows that the range of $E(t)$ over all $t \in \mathcal{T}$ is contained within $[-\overline{E}, \overline{E}]$ where $\overline{E} = \max_{t \in \mathcal{T}}\sqrt{\int_{0}^L X (t-l)^2 dl}$. 
We then define the basis functions and compute $\bmath{S}^f$ only once at the beginning. %

For identifiability, we replace $\bmath{\alpha}^w$ with the unconstrained parameter $\bmath{\phi}^w \in \mathbb{R}^{d^w-1}$ as described in Section \ref{ss:repa}. The penalized log-likelihood is 
$$
\begin{aligned}
    \mathcal{L}(\bmath{\phi}^w, \bmath{\alpha}^{f}, \bmath{\beta}; \bmath{\lambda}, \bmath{\theta}) = \bmath{l}(\bmath{\phi}^w, \bmath{\alpha}^{f}, \bmath{\beta}; \bmath{\lambda}, \bmath{\theta}) - 
    \mathcal{P}(\bmath{\phi}^w; \lambda^w) - 
    \mathcal{P}(\bmath{\alpha}^{f}; \lambda^f) - 
    \mathcal{P}(\bmath{\beta}; \bmath{\lambda}^{h}), 
\end{aligned}
$$
where  
$\bmath{l}(\bmath{\phi}^{w}, \bmath{\alpha}^{f}, \bmath{\beta}; \bmath{\lambda}, \bmath{\theta}) = \sum_{t = 1}^{N} \log p(y_t; \mu_t, \bmath{\theta})$
is the log-likelihood.

\subsection{Optimization: Nested Profiling Approach}
There are two challenges in maximizing $\mathcal{L}$ for estimations.
The first challenge is that estimating the coefficients for $w$ and $f$ simultaneously is infeasible \citep{ma2016estimation}, due to the nested, non-linear, and high-dimensional structure. 
\citet{kong2010statistical} and \citet{ma2016estimation} used iterative approaches
for continuous responses, which cannot be applied to general response types. 
\citet{wang2023semiparametric} used a profile likelihood approach for continuous responses
but relied on a computationally inefficient derivative-free optimization, stating that implicit functions in the profile likelihood are not analytically tractable. 
The second challenge is the need to estimate both $\bmath{\theta}$ and $\bmath{\lambda}$, which has been described by \citet{wood2017generalized} as the biggest computational challenge encountered when fitting penalized spline models.

To address both challenges, we propose a nested profiling approach. 
In the inner stage, given $\bmath{\lambda}$ and $\bmath{\theta}$, the $\bmath{\phi}^{w}$, $\bmath{\alpha}^{f}$ and $\bmath{\beta}$ are estimated by a profile likelihood method. 
Rather than employing a derivative-free optimization as in \citet{wang2023semiparametric}, we derive the analytical gradient and Hessian for fast and accurate Newton's methods. 
In the outer stage, the mode of $[\bmath{\phi}^{w},\bmath{\alpha}^{f}, \bmath{\beta}]$ from the inner stage is plugged into the Laplace approximate marginal likelihood for estimating $\bmath{\lambda}$ and $\bmath{\theta}$ via Broyden–Fletcher–Goldfarb–Shanno (BFGS) method. See Web Appendix E for complete algorithms.

\subsubsection{Inner Stage}
Given $\bmath{\lambda}$ and $\bmath{\theta}$, we estimate 
$\bmath{\phi}^{w}$ (for $w$), $\bmath{\alpha}^{f}$ (for $f$) and $\bmath{\beta}$ (for $h_j$)
using a profile likelihood method. 
For a given $\bmath{\phi}^w$, the estimation is reduced to an unpenalized estimation for a GAM. We denote the estimator as
\begin{equation}
\begin{aligned}
    \left[\widehat{\bmath{\alpha}}^{f}(\bmath{\phi}^w; \bmath{\lambda}, \bmath{\theta}), \widehat{\bmath{\beta}}(\bmath{\phi}^w; \bmath{\lambda}, \bmath{\theta})\right] =& \arg\max_{\bmath{\alpha}^{f}, \bmath{\beta}} \mathcal{L}(\bmath{\phi}^w, \bmath{\alpha}^{f}, \bmath{\beta}; \bmath{\lambda}, \bmath{\theta}).
\end{aligned}
\label{eq:est_fh}
\end{equation}
Then, we define the profile log-likelihood for $\bmath{\phi}^w$ as
$$
    \mathbf{Q}(\bmath{\phi}^w; \bmath{\lambda}, \bmath{\theta}) = \mathcal{L}\left(\bmath{\phi}^w,\widehat{\bmath{\alpha}}^{f}(\bmath{\phi}^w; \bmath{\lambda}, \bmath{\theta}), \widehat{\bmath{\beta}}(\bmath{\phi}^w; \bmath{\lambda}, \bmath{\theta}); \  \bmath{\lambda}, \bmath{\theta}\right). 
$$
The profile likelihood estimators are obtained by
\begin{equation}
\widehat{\bmath{\phi}}^{w}(\bmath{\lambda}, \bmath{\theta}) = \arg\max_{\bmath{\phi}^w} \mathbf{Q}(\bmath{\phi}^w; \bmath{\lambda}, \bmath{\theta}).
\label{eq:est_w}
\end{equation}
Plugging $\widehat{\bmath{\phi}}^{w}(\bmath{\lambda}, \bmath{\theta})$ into Equation \ref{eq:est_fh}, we obtain the estimators 
$\widehat{\bmath{\alpha}}^{f}(\bmath{\lambda}, \bmath{\theta}) = \widehat{\bmath{\alpha}}^{f}(\widehat{\bmath{\phi}}^w(\bmath{\lambda}, \bmath{\theta}); \bmath{\lambda}, \bmath{\theta})$ 
and $\widehat{\bmath{\beta}}(\bmath{\lambda}, \bmath{\theta}) = \widehat{\bmath{\beta}}(\widehat{\bmath{\phi}}^w(\bmath{\lambda}, \bmath{\theta}); \bmath{\lambda}, \bmath{\theta})$.

The solution of Equation \ref{eq:est_w} relies on Equation \ref{eq:est_fh} and contributes to the outer stage; thus, the stability and speed of the optimization are important. 
We employ a Newton's method with two modifications: 
(1) the Q-Newton method of \citet{truong2023fast} ensures that the Newton step is in an ascent direction and of appropriate length, and (2) step-halving ensures that the objective function increases after each Newton update. %

Newton's methods need the gradient and Hessian of the objective functions in Equations \ref{eq:est_fh} and \ref{eq:est_w}. 
For Equation \ref{eq:est_fh}, we describe the analytic forms of the gradient and Hessian of $\mathcal{L}$ in Web Appendix F.1. 
For Equation \ref{eq:est_w}, it is challenging to derive the gradient and Hessian of the profile likelihood $\mathbf{Q}$ because of the implicit functions $\widehat{\bmath{\alpha}}^{f}(\bmath{\phi}^w; \bmath{\lambda}, \bmath{\theta})$ and $\widehat{\bmath{\beta}}(\bmath{\phi}^w; \bmath{\lambda}, \bmath{\theta})$ (see e.g. \citealp{wang2023semiparametric}). 
In this approach, we analytically evaluate these derivatives using the implicit differentiation technique; see details in Web Appendix F.2.

\subsubsection{Outer Stage}
\label{sss:outer}
Denote $\bmath{u} = \left[\bmath{\phi}^w, \bmath{\alpha}^{f}, \bmath{\beta}\right]^\top$ and the estimator from the inner stage as $\widehat{\bmath{u}}\left(\bmath{\lambda}, \bmath{\theta}\right)$. 
Inference for $\bmath{\lambda}$ and $\bmath{\theta}$ is based on the Laplace approximate marginal likelihood (LAML; \citealp{wood2011fast}) that depends on $\widehat{\bmath{u}}\left(\bmath{\lambda}, \bmath{\theta}\right)$. 
LAML is preferred over prediction error criteria such as Akaike’s information criterion (AIC) and generalized cross-validation (GCV), due to its greater resistance to over-fitting as well as other advantages \citep{wood2011fast, wood2017generalized}. 

We estimate $\bmath{\lambda}$ and $\bmath{\theta}$ by maximizing the log-LAML: 
$$
\begin{aligned}
\mathcal{L}^*_{\text{LA}}(\bmath{\lambda}, \bmath{\theta}) =&
\mathcal{L}(\widehat{\bmath{u}}\left(\bmath{\lambda}, \bmath{\theta}\right); \bmath{\lambda}, \bmath{\theta}) - 
\frac{1}{2} \log \left\{\mathrm{det} \bmath{\mathcal{H}}\left(\widehat{\bmath{u}}\left(\bmath{\lambda}, \bmath{\theta}\right); \bmath{\lambda}, \bmath{\theta}\right)\right\} \\ &+ 
\frac{1}{2}\log|\lambda^w \bmath{S}^w|_{+} + \frac{1}{2}\log|\lambda^f \bmath{S}^f|_{+} + \frac{1}{2}\sum_{j=1}^p\log|\lambda^{h}_j \bmath{S}^{h}_j|_{+} + C,
\end{aligned}
$$
where $\bmath{\mathcal{H}}\left(\widehat{\bmath{u}}\left(\bmath{\lambda}, \bmath{\theta}\right); \bmath{\lambda}, \bmath{\theta}\right) = -\partial^2\mathcal{L}(\widehat{\bmath{u}}\left(\bmath{\lambda}, \bmath{\theta}\right); \bmath{\lambda}, \bmath{\theta})/\partial \bmath{u}^2$, $\log |\bmath{A}|_+$ is the product of positive eigenvalues of matrix $\bmath{A}$, 
and $C$ is a constant. 
We write the estimators as 
$(\log\widehat{\bmath{\lambda}}, \log\widehat{\bmath{\theta}}) = \arg\max
_{\log\bmath{\lambda}, \log\bmath{\theta}}\mathcal{L}^*_{\text{LA}}$. 
Finally, plugging $\widehat{\bmath{\lambda}}, \widehat{\bmath{\theta}}$ into $\widehat{\bmath{u}}\left(\bmath{\lambda}, \bmath{\theta}\right)$ yields $\widehat{\bmath{u}} = \widehat{\bmath{u}}(\widehat{\bmath{\lambda}}, \widehat{\bmath{\theta}})$.

To maximize $\mathcal{L}^*_{\text{LA}}$, the derivative-free optimization algorithms are inefficient. We propose to use a Newton-type method, BFGS. 
Unlike the inner stage, we do not adopt Newton's method here, since computing the Hessian of $\mathcal{L}^*_{\text{LA}}$ is both analytically and algorithmically complicated. 
The BFGS needs the gradient of $\mathcal{L}^*_{\text{LA}}$, 
involving that of $\mathcal{L}(\widehat{\bmath{u}}\left(\bmath{\lambda}, \bmath{\theta}\right); \bmath{\lambda}, \bmath{\theta})$, $\log |\bmath{A}|_+$ terms and $\log \left\{\mathrm{det}\bmath{\mathcal{H}}\left(\widehat{\bmath{u}}\left(\bmath{\lambda}, \bmath{\theta}\right); \bmath{\lambda}, \bmath{\theta}\right)\right\}$. 
We find the gradient of $\mathcal{L}(\widehat{\bmath{u}}\left(\bmath{\lambda}, \bmath{\theta}\right); \bmath{\lambda}, \bmath{\theta})$ analytically using the implicit differentiation; see Web Appendix F.3. 
The derivatives of $\log |\bmath{A}|_+$ terms follow directly from \citet{wood2016smoothing}. %
It is difficult to obtain analytical gradient of $\log \left\{\mathrm{det}\bmath{\mathcal{H}}\left(\widehat{\bmath{u}}\left(\bmath{\lambda}, \bmath{\theta}\right); \bmath{\lambda}, \bmath{\theta}\right)\right\}$, and 
we compute it via the automatic differentiation of the Cholesky algorithm, achieved using the {\tt autodiff} 
\citep{autodiff} library in {\tt C++}.

\subsection{Confidence Intervals}
To estimate the confidence intervals for $w$, $f$ and $h_j$, we need to estimate those for the coefficients $\bmath{\alpha}^w$, $\bmath{\alpha}^{f}$ and  $\bmath{\beta}$. 
In this framework, given $\bmath{\lambda}$ and $\bmath{\theta}$, the unconstrained $\bmath{u} = [\bmath{\phi}^w, \bmath{\alpha}^{f}, \bmath{\beta}]^\top$ is estimated by maximizing the log-likelihood, where $\bmath{\phi}^w$ is from a non-linear transformation of $\bmath{\alpha}^w$. 
We propose to estimate the confidence intervals via a sampling strategy similar to one presented in \citet{stringer2024semi}. 
First, we sample $\bmath{u}^{(i)} = [\bmath{\phi}^{w(i)}, \bmath{\alpha}^{f(i)}, \bmath{\beta}^{(i)}]^\top$ from $\mathcal{N}(\widehat{\bmath{u}}, [\bmath{\mathcal{H}}(\widehat{\bmath{u}}; \widehat{\bmath{\lambda}}, \widehat{\bmath{\theta}})]^{-1})$
for $i = 1, 2, \cdots, R$, where $R$ is the number of samples chosen by the user. Second, for each $\bmath{\phi}^{w(i)}$, we compute $\bmath{\alpha}^{w(i)}$ by the reparameterization in Section \ref{ss:repa}. Finally, the approximated $(1-\alpha)$ confidence intervals are obtained by the $(\alpha/2, 1-\alpha/2)$ sample quantiles of $\bmath{\alpha}^{w(i)}$, $\bmath{\alpha}^{f(i)}$ and $\bmath{\beta}^{(i)}$. %
In this procedure, we only use 
the point estimates of $\bmath{\lambda}$ and $\bmath{\theta}$, ignoring their uncertainty, as is typical in a GAM (see e.g. \citealp{wood2016smoothing}). 

The delta method can also be used to quantify the uncertainty in transformed maximum likelihood estimators (see Web Appendix G for details).
In simulations, however, we found that it yielded poor finite sample performance (see Web Table 5). Instead we recommend the sampling method described above.

\section{Simulations}
\label{s:simulation}
In this section, we conduct two simulation studies to investigate the performance of the proposed framework, implemented in the R package {\tt aceDLNM}.  
Simulation A demonstrates the absolute performance, showing low RMSE and appropriate coverage. 
Simulation B compares the ACE-DLNM with the GAM and DRF-DLNM, illustrating its practical advantages.

\subsection{Simulation A: Absolute Performance}
\label{ss:sim1}
\begin{figure}
    \centering
        \centering
        \includegraphics[width=0.9\textwidth]{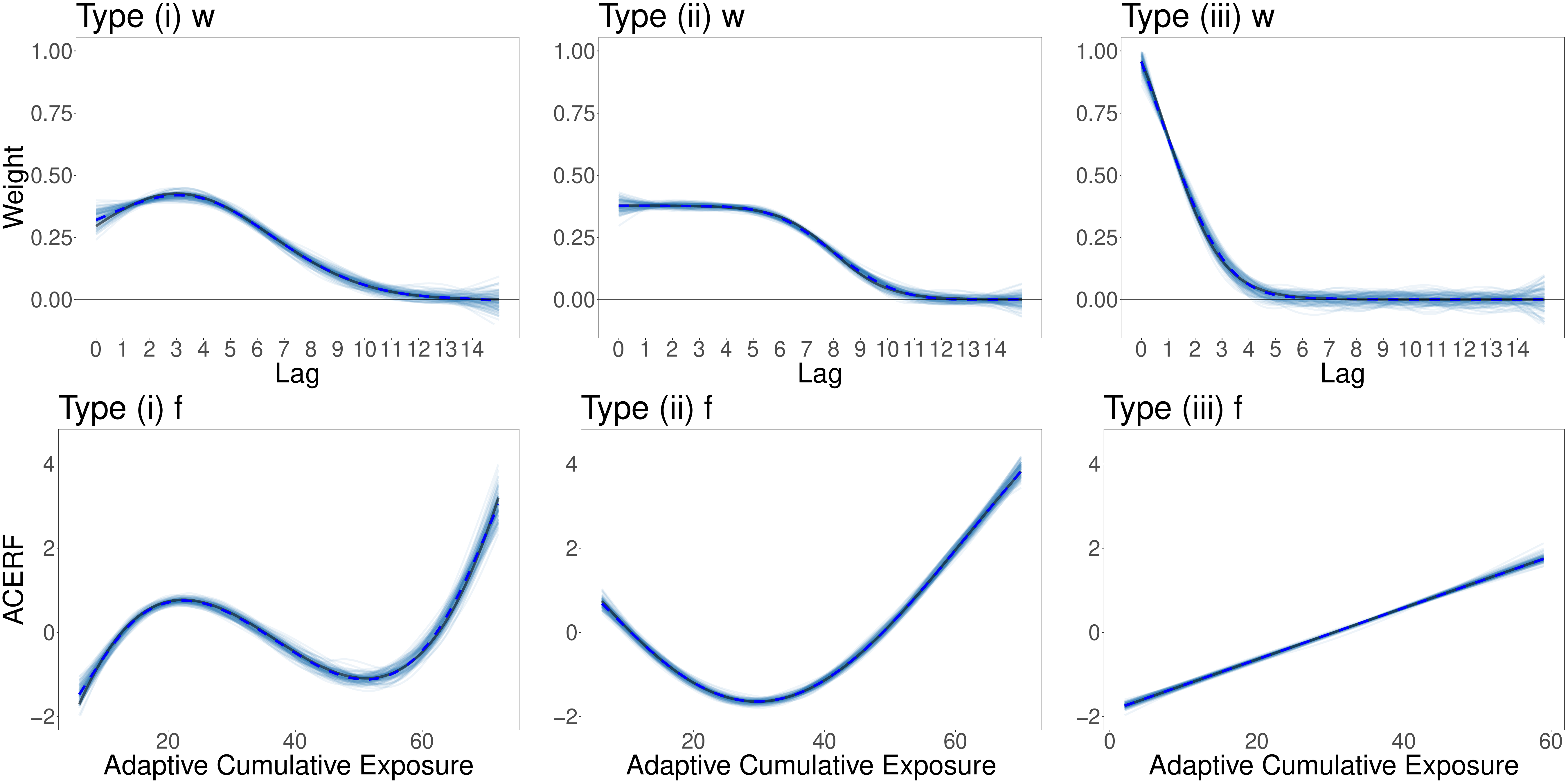}
    \caption{Visualization of Simulation A Results. The data are generated with 
    (1) the first column: Type (i) weight function and Type (i) ACERF; 
    (2) the second column: Type (ii) weight function and Type (ii) ACERF;
    and (3) the third column: Type (iii) weight function and Type (iii) ACERF. 
    The sample size is 1,000. The estimates from the first 100 simulated datasets for the weight function $w$ (first row) and the ACEFR $f$ (second row) are shown as blue curves. 
    The dashed blue curves show the mean of the estimates over all of the 10,000 simulated datasets. 
    The true functions are the solid black curves.}
    \label{fig:sim1-main}
\end{figure}

In this simulation, we demonstrate the performance of the proposed ACE-DLNM. 
The continuous exposure is obtained by interpolating the true daily $\text{PM}_{2.5}$ in Waterloo (visualized in Web Appendix H.1), 
starting from January 1, 2001, with a sample size of $1,000$ or $2,000$. 
We set a time trend $h(t) = 0.5\sin(t/150)$. 
We consider three weight functions $w$ with a two-week maximum lag (Type (i) $w(l)$ peaks at $l=3$ and then gradually decreases, Type (ii) $w(l)$ is flat up to around $l=5$ and decreases, and Type (iii) $w(l)$ decays exponentially from $l=0$), 
and three ACERFs $f$ (Type (i) $f$ is cubic, Type (ii) $f$ is from the standard normal density function, and Type (iii) $f$ is linear). These functions are shown as solid black curves in Figure \ref{fig:sim1-main}.
The true dispersion parameter in the negative binomial distribution is $\theta = 8$, yielding large over-dispersion. 
We generate the count outcome according to the model: 
$$
\log(\mu_t) = f \left\{\int_0^{15} w(l) X(t-l) dl\right\} + h(t). 
$$

Table \ref{tab:sim1} reports root mean square error (RMSE), 95\% confidence intervals coverage (Cvg) and average widths (Width) for estimates of $w$ and $f$ across grids of evenly spaced values over 10,000 replicates.  
In Figure \ref{fig:sim1-main}, we plot the estimates from the first 100 simulated datasets for three cases as examples; remaining results are provided in Web Appendix H.2.1.
In all settings, we observe low RMSE and nominal interval coverage in estimating $w$ and $f$, indicating the validity of the proposed approach. 
We observe that the RMSE for $f$ is higher than that for $w$. A possible explanation is that $f$ measures the association between the ACE and the outcome, while the ACE itself is computed using the estimated $w$. 
Estimation error in $w$ introduces additional variability into the estimation of $f$, resulting in a higher RMSE. 
When the true $f$ is linear (Type (iii)), the model simplifies to a DLM. 
We compare the ACE-DLNM and the correctly specified DLM under this setting in Web Appendix H.3, and find that the ACE-DLNM achieves similar performance.
Web Table 4 shows that the proposed method performs well in estimating the time trend $h(t)$ and the dispersion parameter $\theta$.

\begin{table}
    \centering
    \caption{Results of Simulation A under different weight functions $w$ and ACERFs $f$. The RMSE, 95\% confidence intervals coverage (Cvg) and average widths (Width) for $w$ and $f$ with the sample size $N = 1000$ and $2000$ are reported.}
    \label{tab:sim1}
    \begin{tabular}{@{}cccccccc@{}}
\toprule
      & \multicolumn{3}{c}{$w$}                 &  & \multicolumn{3}{c}{$f$}                 \\ \cmidrule(lr){2-4} \cmidrule(l){6-8} 
$N$ & RMSE        & Cvg        & Width        &  & RMSE        & Cvg        & Width        \\ \midrule
      & \multicolumn{3}{c}{\textbf{Type (i) $w$}} &  & \multicolumn{3}{c}{\textbf{Type (i) $f$}} \\
1000  & 0.015       & 0.970      & (0.064)      &  & 0.126       & 0.943      & (0.467)      \\
2000  & 0.011       & 0.979      & (0.049)      &  & 0.102       & 0.951      & (0.355)      \\
      & \multicolumn{3}{c}{}                    &  & \multicolumn{3}{c}{\textbf{Type (ii) $f$}} \\
1000  & 0.012       & 0.977      & (0.055)      &  & 0.085       & 0.964      & (0.352)      \\
2000  & 0.009       & 0.980      & (0.042)      &  & 0.062       & 0.965      & (0.263)      \\
      & \multicolumn{3}{c}{}                    &  & \multicolumn{3}{c}{\textbf{Type (iii) $f$}} \\
1000  & 0.018       & 0.961      & (0.075)      &  & 0.027       & 0.945      & (0.104)      \\
2000  & 0.014       & 0.973      & (0.060)      &  & 0.019       & 0.954      & (0.078)      \\
      &             &            &              &  &             &            &              \\
      & \multicolumn{3}{c}{\textbf{Type (ii) $w$}} &  & \multicolumn{3}{c}{\textbf{Type (i) $f$}} \\
1000  & 0.014       & 0.965      & (0.060)      &  & 0.121       & 0.949      & (0.449)      \\
2000  & 0.010       & 0.966      & (0.044)      &  & 0.092       & 0.948      & (0.318)      \\
      & \multicolumn{3}{c}{}                    &  & \multicolumn{3}{c}{\textbf{Type (ii) $f$}} \\
1000  & 0.012       & 0.968      & (0.051)      &  & 0.079       & 0.962      & (0.337)      \\
2000  & 0.008       & 0.973      & (0.038)      &  & 0.056       & 0.964      & (0.235)      \\
      & \multicolumn{3}{c}{}                    &  & \multicolumn{3}{c}{\textbf{Type (iii) $f$}} \\
1000  & 0.017       & 0.954      & (0.071)      &  & 0.026       & 0.948      & (0.102)      \\
2000  & 0.013       & 0.964      & (0.055)      &  & 0.018       & 0.951      & (0.073)      \\
      &             &            &              &  &             &            &              \\
      & \multicolumn{3}{c}{\textbf{Type (iii) $w$}} &  & \multicolumn{3}{c}{\textbf{Type (i) $f$}} \\
1000  & 0.017       & 0.964      & (0.063)      &  & 0.148       & 0.950      & (0.550)      \\
2000  & 0.011       & 0.975      & (0.050)      &  & 0.117       & 0.950      & (0.427)      \\
      & \multicolumn{3}{c}{}                    &  & \multicolumn{3}{c}{\textbf{Type (ii) $f$}} \\
1000  & 0.013       & 0.975      & (0.060)      &  & 0.100       & 0.961      & (0.410)      \\
2000  & 0.010       & 0.979      & (0.047)      &  & 0.078       & 0.955      & (0.316)      \\
      & \multicolumn{3}{c}{}                    &  & \multicolumn{3}{c}{\textbf{Type (iii) $f$}} \\
1000  & 0.021       & 0.969      & (0.093)      &  & 0.030       & 0.943      & (0.114)      \\
2000  & 0.017       & 0.976      & (0.078)      &  & 0.023       & 0.946      & (0.090)      \\ \bottomrule
\end{tabular}
\end{table}

\subsection{Simulation B: Relative Performance}
\label{ss:sim2}
This simulation compares the ACE-DLNM with the GAM, which pre-specifies the cumulative exposure as if the lag weight was known \textit{a priori}, and the DRF-DLNM, which fits a bivariate exposure-lag-response surface, to illustrate the advantages of the ACE-DLNM in practice. 

The data are generated from the GAMs: $\log(\mu_t) = f (E_t) + h(t)$, 
where the cumulative exposure $E_t$ is specified as one of the following: 
(i) same-day exposure $E_t = x_t$; 
(ii) average of 0- and 1-day lags $E_t = (x_t + x_{t-1})/2$; 
(iii) average from 0- to 7-day lags $E_t = (x_t + x_{t-1} + \cdots +  x_{t-7})/8$; 
(iv) average from 0- to 14-day lags $E_t = (x_t + x_{t-1} + \cdots +  x_{t-14})/15$; 
or (v) weighted average $E_t = w_0 x_t + w_1 x_{t-1} + \cdots + w_{14} x_{t-14}$ where the non-constant weights are from Type (i) $w(l)$ in Simulation A. %
The true $f$ is Type (i) from Simulation A. The sample size is 2,000, and all other settings remain the same as in Simulation A.

We compare the ACE-DLNM with: 
(a) GAM using the same-day exposure (GAM lag0), 
(b) GAM using the average of 0- and 1-day lags (GAM avg lag0-1), 
(c) GAM using the average from 0- to 7-day lags (GAM avg lag0-7),
(d) GAM using the average from 0- to 14-day lags (GAM avg lag0-14), and (e) DRF-DLNM \citep{gasparrini2017penalized}. 
Models (a)---(d) are correctly specified under scenarios (i)---(iv) respectively. 
We evaluate performance using RMSE and 95\% confidence intervals coverage (Cvg) for $f(E_t)$ evaluated over $t\in \mathcal{T}$ across 10,000 replicates, as reported in Table \ref{tab:sim2}. 

\begin{table}
    \centering
    \caption{Results of Simulation B under different data generations. The RMSE and 95\% confidence intervals coverage (Cvg) for $f(E_t)$ are reported. }
    \label{tab:sim2}
    \begin{tabular}{@{}ccccccc@{}}
    \toprule
          & \begin{tabular}[c]{@{}c@{}}ACE-DLNM\end{tabular}  & \begin{tabular}[c]{@{}c@{}}GAM\\ lag 0\end{tabular} & \begin{tabular}[c]{@{}c@{}}GAM\\ avg lag 0-1\end{tabular} & \begin{tabular}[c]{@{}c@{}}GAM \\ avg lag 0-7\end{tabular} & \begin{tabular}[c]{@{}c@{}}GAM\\ avg lag 0-14\end{tabular}  & \begin{tabular}[c]{@{}c@{}}DRF-DLNM\end{tabular}   \\ \midrule
          \multicolumn{7}{c}{(i) $E_t = x_t$}                                                                                                                                                                                                                                                                   \\[6pt] 
     RMSE &  \textbf{0.069}                                                                                  &  \textbf{0.040}                                     &    0.427                                                  &   0.600                                                    &   0.612                                                  &   \textbf{0.116}    \\
     Cvg  &  \textbf{0.932}                                                                                  &  \textbf{0.959}                                     &    0.165                                                  &   0.094                                                    &   0.084                                                  &   \textbf{0.944}     \\ \midrule          
          \multicolumn{7}{c}{(ii) $E_t = (x_t + x_{t-1})/2$}                                                                                                                                                                                                                                                                                                              \\[6pt] 
     RMSE &  \textbf{0.062}                                                                                   &   0.405                                             &    \textbf{0.040}                                         &   0.518                                                   &   0.537                                                  &   0.267     \\
     Cvg  &  \textbf{0.939}                                                                                  &   0.184                                             &    \textbf{0.959}                                         &    0.115                                                   &   0.096                                                    &   0.750    \\ \midrule
         \multicolumn{7}{c}{(iii) $E_t = (x_t + x_{t-1} + \cdots +  x_{t-7})/8$}                                                                                                                                                                                                                                                                                             \\[6pt] 
     RMSE &  \textbf{0.060}                                                                                    &  0.520                                              &    0.504                                                  &   \textbf{0.041}                                           &   0.442                                                   &   0.359   \\
     Cvg  &  \textbf{0.922}                                                                         &  0.083                                              &    0.101                                                  &   \textbf{0.964}                                           &   0.188                                                    &   0.554   \\ \midrule         
          \multicolumn{7}{c}{(iv) $E_t = (x_t + x_{t-1} + \cdots +  x_{t-14})/15$}                                                                                                                                                                                                                                                                                                                \\[6pt] 
     RMSE &  \textbf{0.048}                                                                                   &   0.597                                             &    0.586                                                  &    0.457                                                   &   \textbf{0.042}                                           &   0.340     \\
     Cvg  &  \textbf{0.954}                                                                                     &   0.063                                             &    0.064                                                  &    0.133                                                   &   \textbf{0.967}                                           &   0.355   \\ \midrule         
          \multicolumn{7}{c}{(v) $E_t = w_0 x_t + w_1 x_{t-1} + \cdots + w_{14} x_{t-14}$}                                                                                                                                                                                                                                                                                                                            \\[6pt] 
     RMSE &  \textbf{0.051}                                                                                   &  0.529                                              &    0.512                                                  &     0.160                                                  &   0.411                                               &   0.357         \\
     Cvg  &  \textbf{0.955}                                                                                   &  0.080                                              &    0.091                                                  &     0.481                                                  &   0.167                                                &   0.484        \\\bottomrule
    \end{tabular}
\end{table}

Under scenarios (i)---(iv) where the data are generated from the GAM using a simple average, 
the GAM with the correctly specified lag window achieves the lowest RMSE and near nominal coverage. 
In contrast, GAMs with misspecified lag windows perform poorly. 
The ACE-DLNM yields the second-best performance,
close to the correctly specified GAM, and substantially outperforms the misspecified GAMs. 
Under scenario (v) where the lag weights are non-constant, 
the ACE-DLNM is the only model that achieves low RMSE and appropriate coverage, while all other models perform poorly with high RMSE and substantial undercoverage.
In practice, the true relevant lag window for the cumulative exposure is unknown, and the weights are also unknown and are not expected to remain constant across the lag window. 
The GAM, on its own, fails to offer a practical way to define the appropriate cumulative exposure, whereas the ACE-DLNM addresses this by treating cumulative exposure adaptively. 

Mild under-coverage is observed for the ACE-DLNM (0.932, 0.939 and 0.922), due to the difficulty of making inferences in these scenarios, where the true weight holds constant over the relevant lag window and then drops to zero, 
causing a discontinuity that is challenging to capture using a spline.

The DRF-DLNM models a separate function of exposure at each lag, rather than a single function applied to the cumulative exposure as in ACE-DLNM and GAM. 
It is correctly specified only under scenario (i), 
where the cumulative exposure is the same-day exposure, and performs better than the misspecified GAMs in this case. 
However, the RMSE is higher than that of the ACE-DLNM and GAM lag0, 
potentially because of the difficulty of estimating 15 separate functions when 14 of them are truly null. 
Under scenarios (ii)---(v), %
the DRF-DLNM is misspecified and behaves poorly as expected. 
Note that ACE-DLNM and DRF-DLNM rely on different model assumptions, 
and their performance depends on how well these align with the true data-generating mechanism; see also Web Appendix H.4 for two supplementary simulations where the DRF-DLNM is correctly specified.
In practice, 
selection may be based on model fit or interpretability considerations, 
which we further illustrate in Section \ref{s:example}.

\section{Application: Adverse Health Effect of Air Pollution in Canada}
\label{s:example}

\subsection{Background and Aim}

Health Canada monitors adverse health effect of ambient air pollution in collaboration with Environment and Climate Change Canada, including the public health risks attributable to short-term exposure to fine particulate matter smaller than 2.5 micrometers in diameter ($\text{PM}_{2.5}$). $\text{PM}_{2.5}$ is emitted directly into the air from vehicles, residential wood-burning, industry, wildfires, and burning waste. Numerous epidemiological studies have established significant associations between exposure to airborne $\text{PM}_{2.5}$ and adverse health effects, including non-accidental all-cause hospitalizations 
\citep{dominici2006fine} and %
mortality \citep{pope1995particulate}. 
The goal of this analysis is to quantify \textit{non-linear} associations between \textit{cumulative} exposure to $\text{PM}_\text{2.5}$ and hospitalizations (respiratory morbidity) in five cities across Canada from January 1, 2001 to December 31, 2018. 

\subsection{Data sources}
Daily counts of hospitalization (morbidity) were collected from the Canadian Institute for Health Information (CIHI) for 2001-2018. Based on the International Classification of Diseases 10th version 
\citep{WHO_ICD10}, 
we extracted the respiratory (J00-J99) morbidity and aggregated them by city (census division). 
Where there were multiple causes (up to 10) diagnosed by a medical doctor, we used the first cause in this analysis. 

Hourly $\text{PM}_{2.5}$ concentrations were obtained from the National Air Pollution Surveillance (NAPS) program \citep{NAPS}. 
For each NAPS monitoring station, daily average $\text{PM}_{2.5}$ concentrations were calculated if at least 18 hourly concentrations were available for the given day. 
In the case of short gaps (up to 10 consecutive hours), the interpolation method of
\citet{burr2015bias} %
was employed. 
For cities with multiple NAPS stations, average concentrations of available stations were taken to represent city level air pollution concentrations. More details can be found in \citet{shin2024PM2.5}.
 
Daily mean temperatures were also obtained from the National Climate Data and Information Archive 
\citep{ECCC}. 
Daily temperatures for each city were calculated using the same approach as for daily $\text{PM}_{2.5}$ concentrations.

\subsection{Analysis}
We analyzed five cities---Waterloo, Peel, Hamilton, Calgary and Vancouver---that represent a diverse range of geographies, climates, industries, and populations. (See Web Table 9 for summary data.)
The outcomes were counts, and their sample variances were relatively large (e.g., Peel had an average respiratory morbidity of 16.4 with a variance of 37.2), indicating potential over-dispersion. Moreover, the datasets were quite large: outcomes and exposure were measured daily between 2001 and 2018 for a total of 6,574 days for each city.

We therefore fit the generalized ACE-DLNM with the negative binomial distribution separately for each of the five cities. The mean model is
$$
\begin{aligned}
    \log(\mu_t) = f \left\{\int_0^{15} w(l) X(t-l) dl\right\} + h_1(t) + h_2(\text{Month}_t)
    + h_3(\text{Temp}_t) + \sum_{p=1}^6 \beta^{\text{DOW}}_p\text{DOW}_{pt},
\end{aligned}
$$
where $Y_t$ is the respiratory morbidity count, $X$ is the $\text{PM}_\text{2.5}$, 
$h_1(t)$ and $h_3(\text{Temp}_t)$ are cubic B-splines for long-term trend and temperature association, and $h_2(\text{Month}_t)$ is a cyclic cubic spline of month to capture seasonality. %
Indicator variables for the day of week (DOW) are included.  %
We quantify the associations without making causal claims in this paper, and leave open the possibility of the presence of confounders.

We compare the ACE-DLNM to simpler GAMs fitted under different pre-specified cumulative exposures (via R package {\tt mgcv}; \citealp{wood2015package}): 
(a) same-day exposure (lag0), (b) average of 0- and 1-day lags (avg lag0-1), (c) average from 0- to 7-day lags (avg lag0-7), and (d) average from 0- to 14-day lags (avg lag0-14),  
while keeping all other specifications same with those of the ACE-DLNM.
The cumulative exposures are scaled by constants, 
to align with the constraints in ACE-DLNM, making the association functions comparable. 
The estimates for all quantities are almost identical among these models (see Web Appendix I.2), except for the association function $f$.

We also fit the DRF-DLNM \citep{gasparrini2017penalized} via the {\tt dlnm} package \citep{gasparrini2011distributed}. 
Its model specification is similar to the ACE-DLNM, except that the non-linear association with the ACE is replaced by 15 distinct lag-specific functions, i.e. $\sum_{l=0}^{14} \psi (x_{t-l},l)$.

\subsection{Results}
\label{ss:app-results}
\begin{figure}
    \centering
    \begin{subfigure}{\textwidth}
        \centering
        \includegraphics[width=\linewidth]{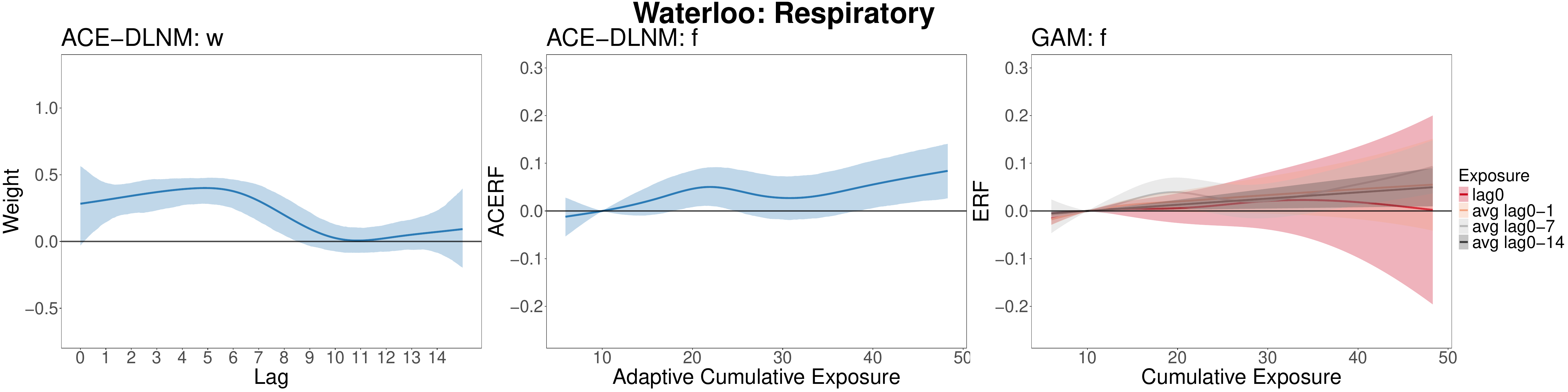}
    \end{subfigure}%
    
    \vspace{0.1cm}
    \centering
    \begin{subfigure}{\textwidth}
        \centering
        \includegraphics[width=\linewidth]{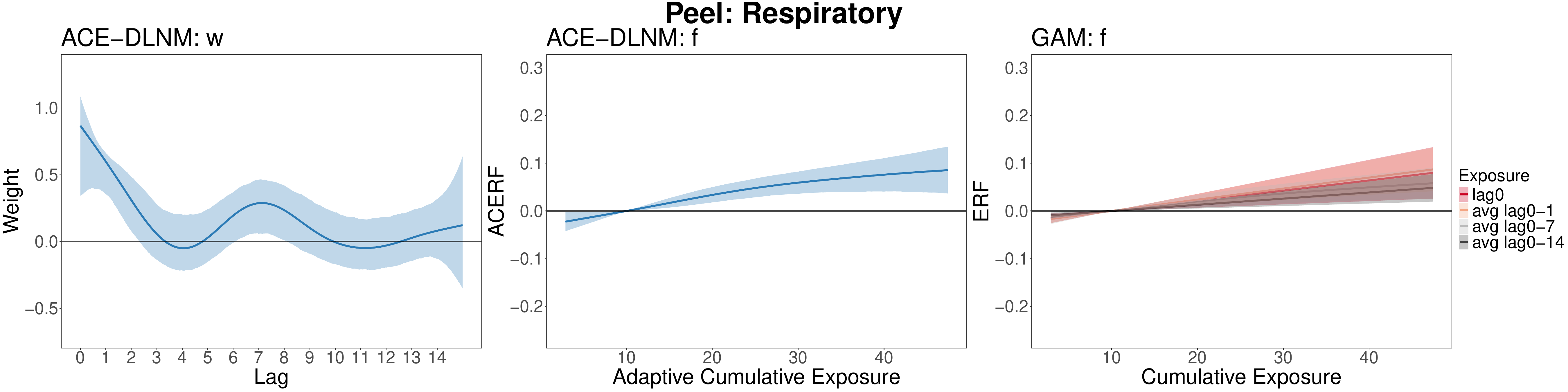}
    \end{subfigure}%
    
    \vspace{0.1cm}
    \centering
    \begin{subfigure}{\textwidth}
        \centering
        \includegraphics[width=\linewidth]{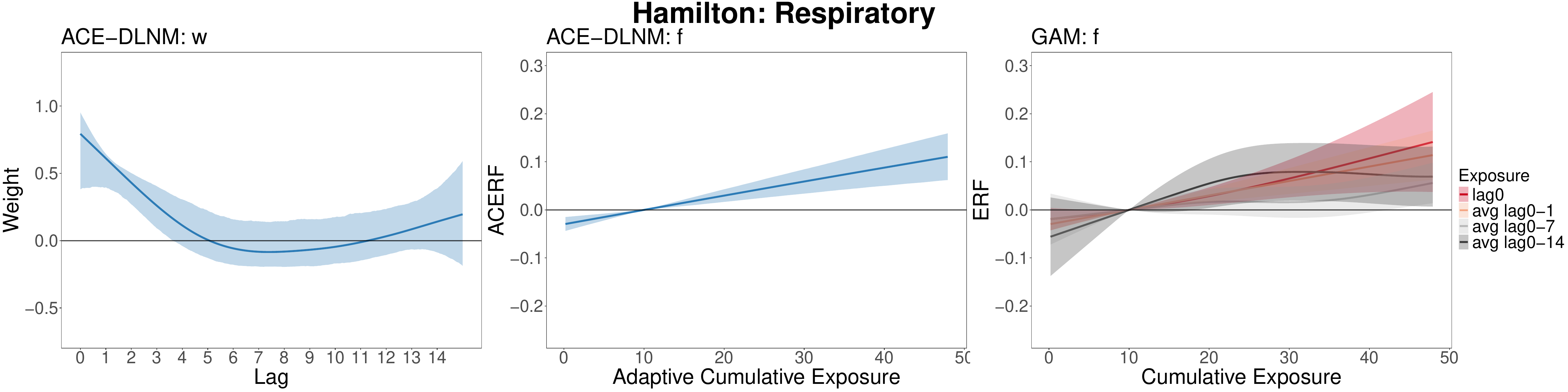}
    \end{subfigure}%

    \vspace{0.1cm}
    \centering
    \begin{subfigure}{\textwidth}
        \centering
        \includegraphics[width=\linewidth]{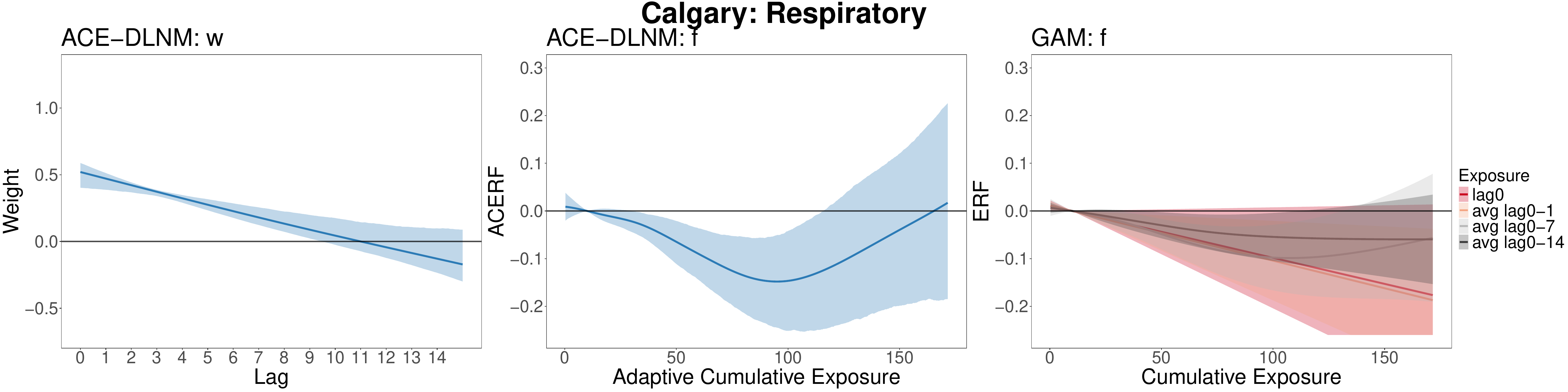}
    \end{subfigure}%

    \vspace{0.1cm}
    \centering
    \begin{subfigure}{\textwidth}
        \centering
        \includegraphics[width=\linewidth]{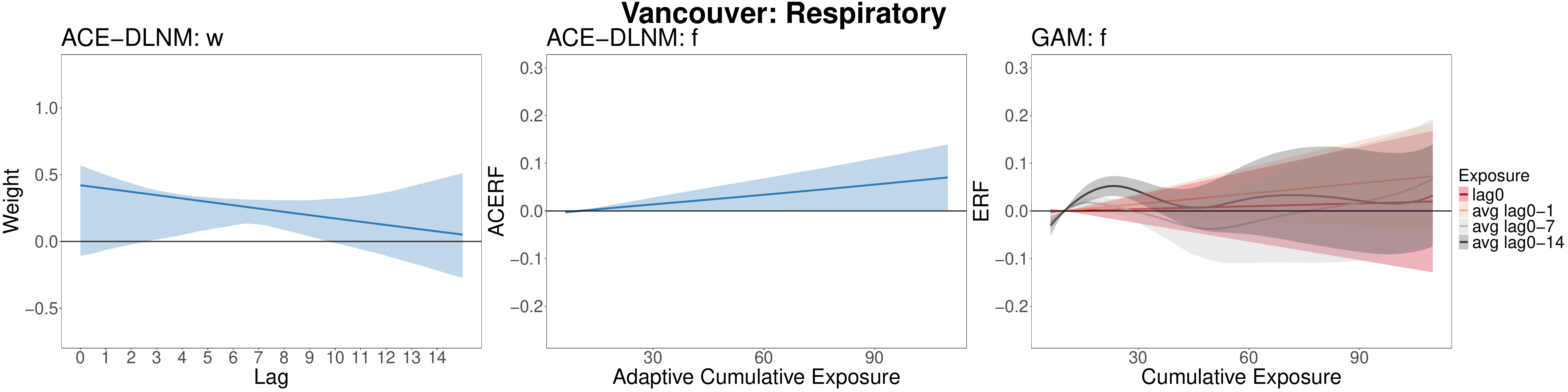}
    \end{subfigure}%
    \caption{Estimated weight (left column) and ACERF (middle column) from ACE-DLNM and the estimated ERF (right column) from GAM for respiratory morbidity in five cities: Waterloo (top row), Peel, Hamilton, Calgary and Vancouver (bottom row). The point estimates are in solid curves with 95\% confidence intervals in shade.}
    \label{fig:application_P}
\end{figure}

Figure \ref{fig:application_P} demonstrates the results for respiratory morbidity in the five selected cities.
Figure \ref{fig:application_P} demonstrates the results for respiratory morbidity in the five selected cities. 
The estimated weight function and ACERF from the ACE-DLNM, as well as the exposure-response function (ERF) from the GAMs, are plotted.
The ACE-DLNM estimates non-linear associations in Waterloo and Calgary, a slight non-linear association in Peel, and a linear association in Hamilton. 
In contrast, the GAMs, with pre-specified cumulative exposures, fail to capture these non-null associations in some cases. 
In Peel and Hamilton, where the relevant lag windows cover about two or three days, the GAM with same-day exposure (lag0) yields a similar association to the ACE-DLNM. However, in Waterloo and Calgary, where the relevant lag windows extend to about one week, the estimated associations from the GAM lag0 are nonsignificant with wide confidence intervals. 
This highlights the advantage of ACE-DLNMs in making stable inferences by adaptively estimating cumulative exposure without the need for pre-specification.

ACE-DLNMs require non-constant (non-null) ACERFs for identifiability (see Web Appendix D). 
In Vancouver, the association is close to zero, making the weight function uninformative. Even in this nonidentifiable case, the ACE-DLNM provides a stable and narrower confidence interval for the association than GAMs. 

For cities with non-constant ACERFs, estimated weight functions behave as expected, starting positive at zero lag and nearing zero at large lags. But the shape varies by city. 
In Waterloo, for example, 
the weight function maintains a large value for up to one week, suggesting long-lasting exposure associations, whereas in Peel, it declines sharply after two days. 
The variation in weight functions emphasizes the flexibility of ACE-DLNM in capturing city-specific lag patterns.

Model diagnostics based on randomized quantile residuals \citep{dunn1996randomized} show no evidence of a lack of model fit (Web Appendix I.2).
We compare the fitted negative binomial models to more restrictive Poisson models using AIC; the results in Web Table 15 support our choice of the negative binomial distribution.

\begin{figure}[H]
    \centering
     \begin{subfigure}[t]{0.5\textwidth}
        \centering
        \includegraphics[width=\linewidth]{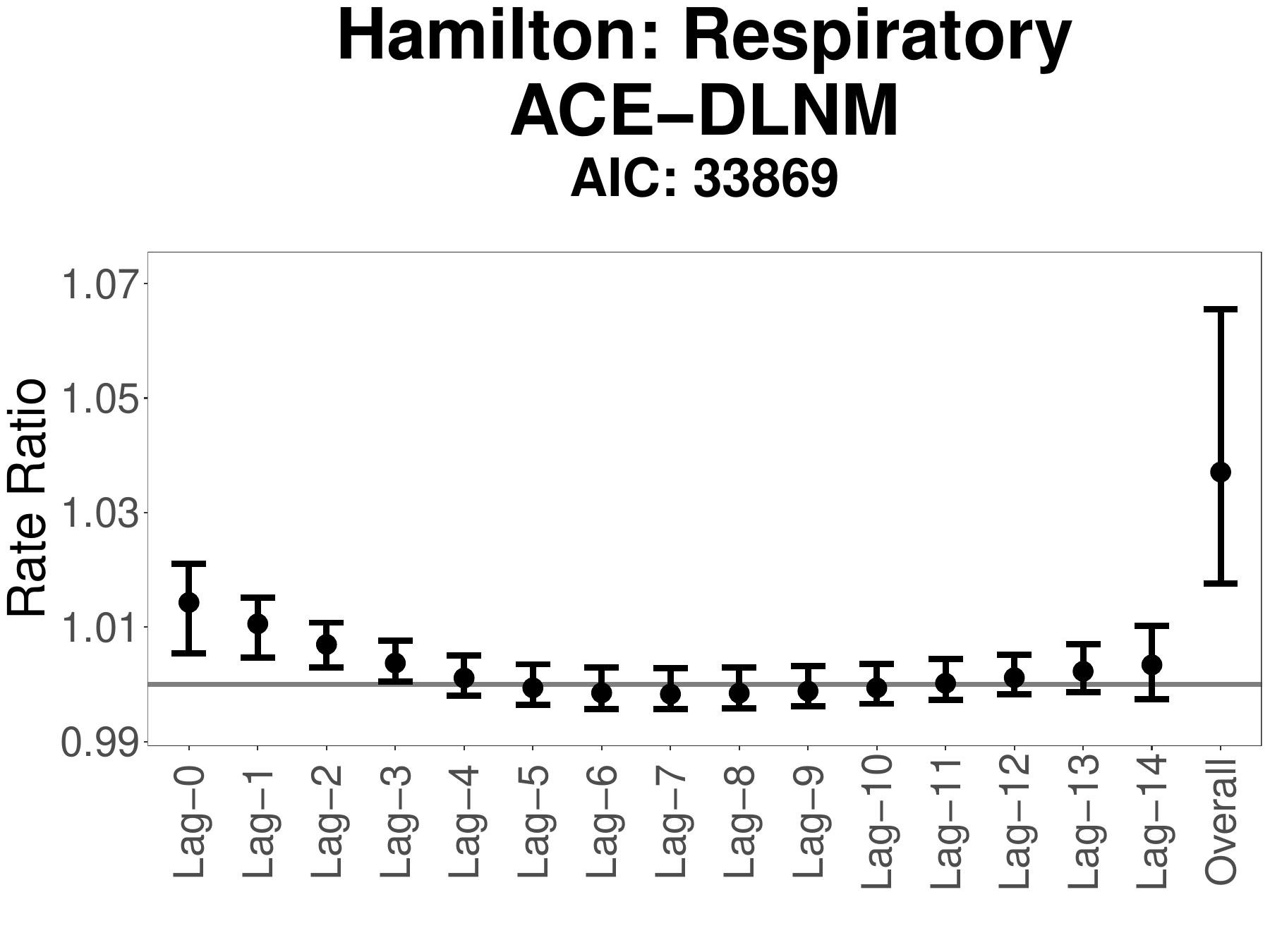}
        \caption{Rate ratios from ACE-DLNM}
        \label{fig:app-compare-RR-ACE-DLNM}
    \end{subfigure}%
    \hfill
    \begin{subfigure}[t]{0.5\textwidth}
        \centering
        \includegraphics[width=\linewidth]{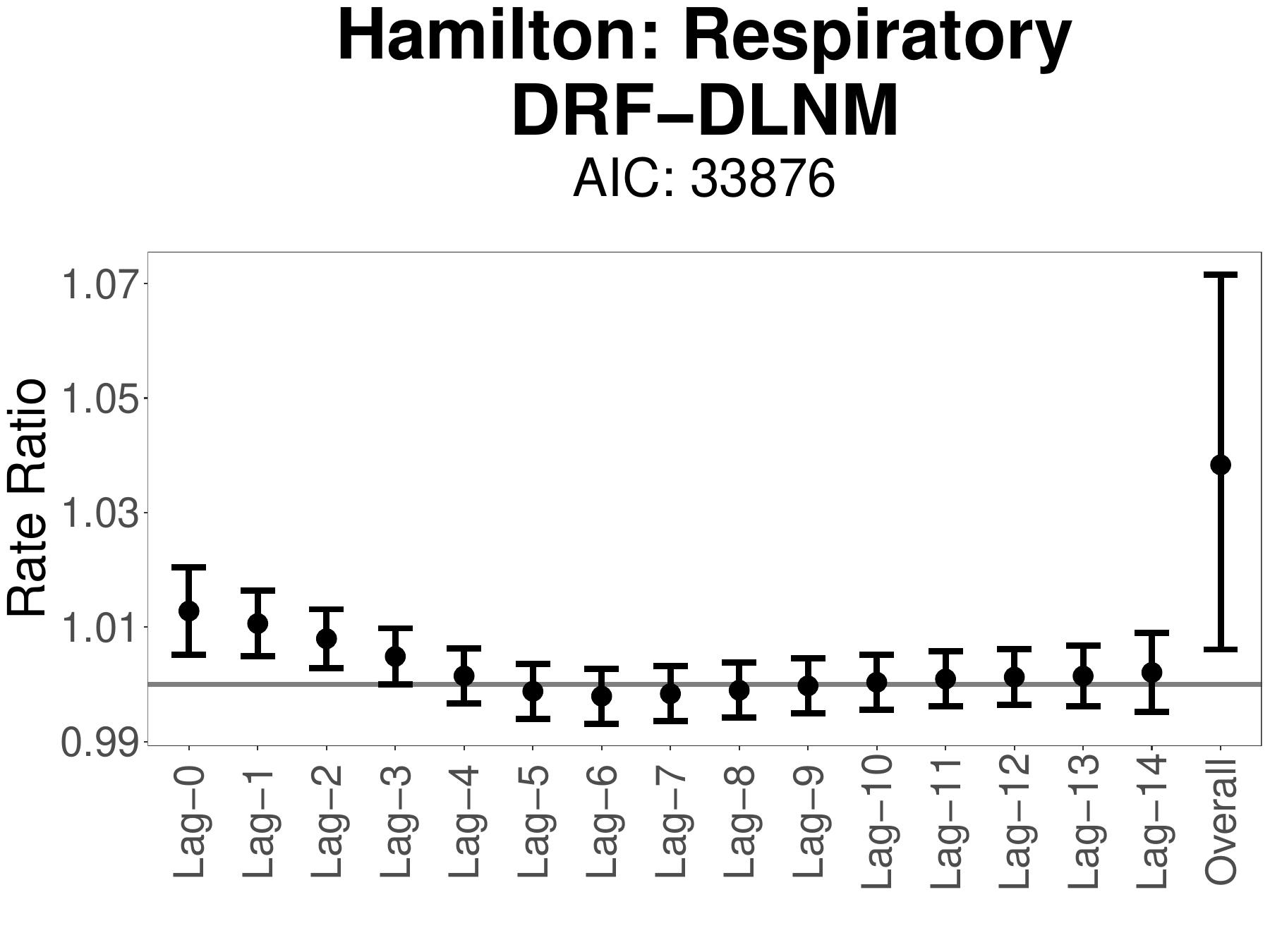}
        \caption{Rate ratios from DRF-DLNM}
        \label{fig:app-compare-RR-DRF-DLNM}
    \end{subfigure}

    \vspace{0.5cm}
    \centering
    
    \begin{subfigure}[c]{0.48\textwidth}
    \centering
    \includegraphics[width=0.9\linewidth]{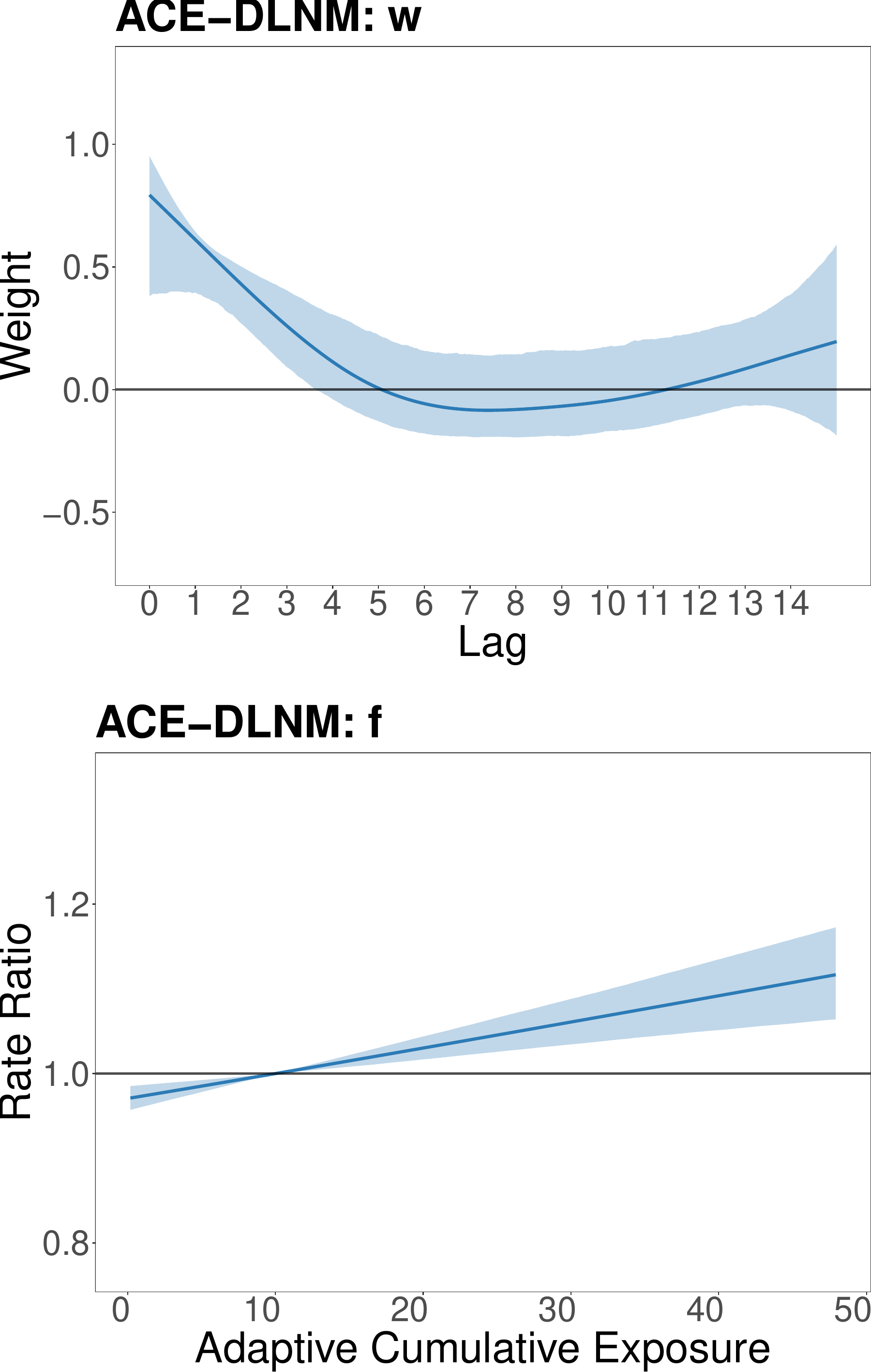}
    \caption{Estimated curves from ACE-DLNM}
    \label{fig:app-compare-ind-ACE-DLNM}
    \end{subfigure}
    \hfill
    \begin{subfigure}[c]{0.48\textwidth}
    \centering
    \begin{subfigure}[b]{\textwidth}
      \centering
      \includegraphics[width=\linewidth]{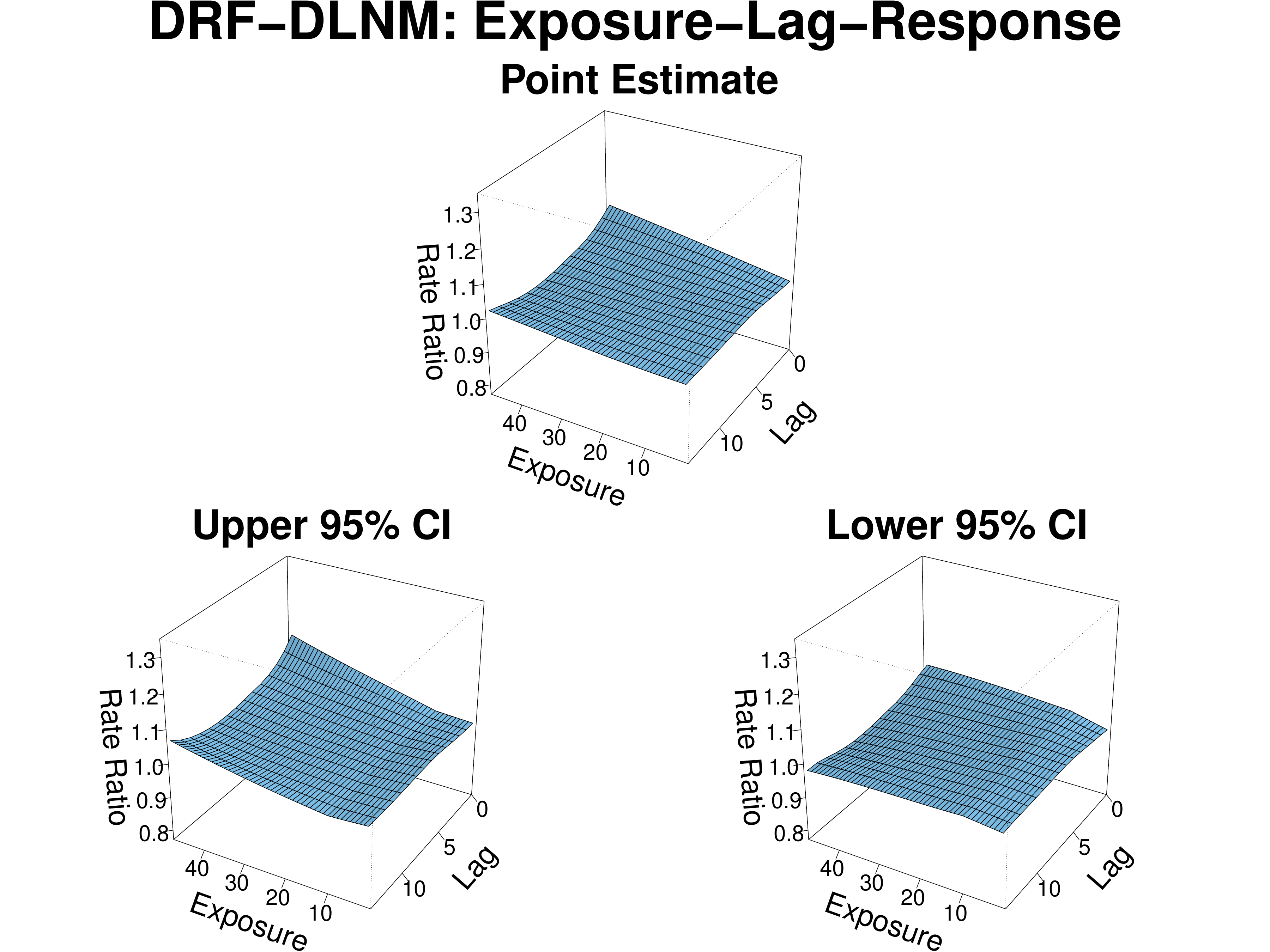}
      \caption{Estimated surfaces from DRF-DLNM}
      \label{fig:app-compare-DRF-DLNM}
    \end{subfigure}
    \begin{subfigure}[b]{\textwidth}
      \centering
      \includegraphics[width=0.9\linewidth]{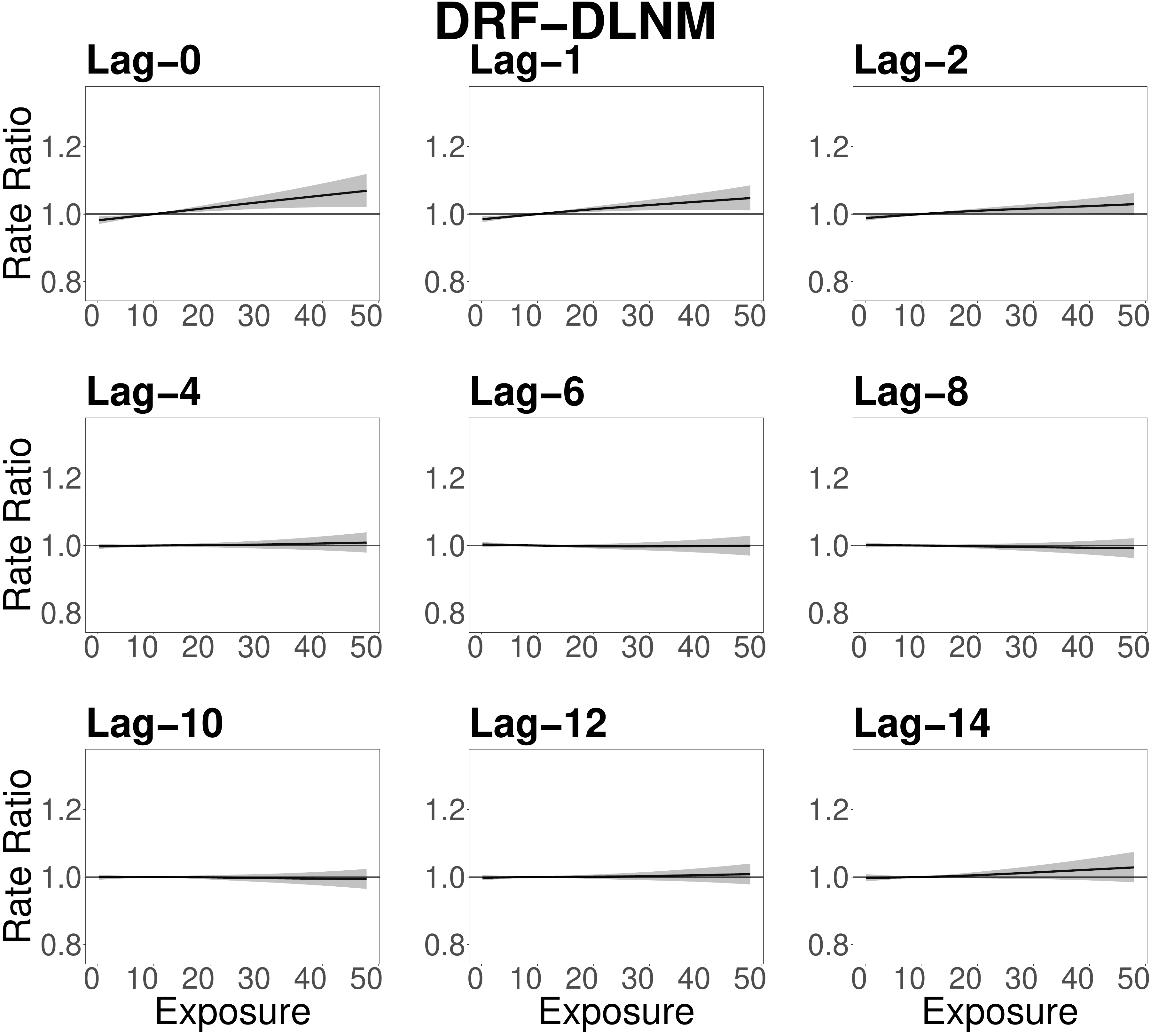}
      \caption{Estimated curves from DRF-DLNM}
      \label{fig:app-compare-ind-DRF-DLNM}
    \end{subfigure}
  \end{subfigure}
    \caption{Comparisons between the ACE-DLNM and DRF-DLNM in the Hamilton dataset. The lag-specific and overall rate ratios from the ACE-DLNM (a) and DRF-DLNM (b) are plotted. Interpretation of the ACE-DLNM is based on the two curves shown in (c). Interpretation of the DRF-DLNM relies on the bivariate surfaces in (d), and we plot the estimated curves at selected lags in (e) (the full results are provided in Web Figure 14). The AICs for the two fitted models are reported. }
    \label{fig:app-compare}
\end{figure}

To compare ACE-DLNM and DRF-DLNM, we report two types of rate ratios: 
(1) the lag-specific rate ratio, defined as the ratio of morbidity rates when 
the exposure at a specific lag increases from the lower quartile ($Q_1$) to upper quartile ($Q_3$) while all other lags are held at $Q_1$; and (2) the overall rate ratio, defined as the ratio of morbidity rates when the entire exposure profile shifts from $Q_1$ to $Q_3$ simultaneously. 
We use the Hamilton data as an example where $Q_1 = 4.33$ and $Q_3 = 11.11$. 
Figures \ref{fig:app-compare-RR-ACE-DLNM} and \ref{fig:app-compare-RR-DRF-DLNM} show the rate ratios from ACE-DLNM and DRF-DLNM. 
The two models yield similar lag-specific rate ratios---significant from lag-0 to lag-2 and then gradually declining to null. The ACE-DLNM provides a more efficient estimate of the overall rate ratio with a narrower confidence interval. 
AIC comparisons also favour the ACE-DLNM. The results for the remaining four cities are provided in Web Appendix I.2. The patterns of estimated rate ratio are broadly similar between the two models. 

However, the rate ratio comparing $Q_3$ and $Q_1$ does not reflect nonlinearity, 
and the underlying interpretations of the models differ a lot. 
The DRF-DLNM estimates a bivariate exposure-lag-response surface---estimating a separate non-linear function at each lag. 
We plot the estimated surface with its 95\% confidence interval in Figure \ref{fig:app-compare-DRF-DLNM}. 
It can be difficult to interpret the bivariate surfaces directly, so we plot the separate non-linear function evaluated at each lag in Figure \ref{fig:app-compare-ind-DRF-DLNM} and Web Figure 14.
At a specific lag, the estimated function, 
among the 15 functions, describes how the exposure at that lag is associated with the response. 
Although these functions vary in scale, we cannot observe much variation in their shape---which is also observed in the other four cities (Web Appendix I.2)---suggesting that modelling separate functions at each lag may be unnecessary but complicate the interpretations. 
In contrast, the ACE-DLNM uses only two univariate functions: the lag weight and ACERF, as shown in Figure \ref{fig:app-compare-ind-ACE-DLNM}. 
The ACERF describe the association between the cumulative exposure and response, while the lag weight indicates the relative contribution of each lag. Both estimated functions are interpretable and informative in this analysis.

\section{Discussion}
\label{s:discuss}

The ACE-DLNM quantifies the associations between cumulative exposure and health outcomes. 
However, the existing ACE-DLNM methods are limited to continuous outcomes and do not scale well to large datasets. 
We present an inferential and computational framework for ACE-DLNMs that addresses these limitations, allowing for general response types and large datasets. 
We successfully apply the proposed ACE-DLNM to our motivating analysis, and the results demonstrate that the ACE-DLNM provides more stable inferences than simpler GAMs and is more interpretable compared to the DRF-DLNM.

The distributed lag model framework is conceptually similar to the weighted cumulative exposure (WCE) models. 
\citet{breslow1983multiplicative} and \citet{thomas1988models} proposed the WCE, %
and some flexible modelling approaches were developed by, for example, \citet{hauptmann2000analysis} and \citet{sylvestre2009flexible}.
However, the existing WCE models are limited by the assumption of a linear exposure-response association. 
This limitation can be addressed by the DRF-DLNM \citep{gasparrini2014modeling} or by the ACE-DLNM as discussed in this paper.

This paper uses air pollution data without gaps over the study period and assumes no measurement error, as in \citet{shin2024PM2.5}. 
However, exposures may be subject to measurement error, which can lead to effect attenuation (see e.g. \citealp{chang2011estimating}). 
\citet{mauff2017extension} and \citet{wagner2021time} addressed exposure measurement error in linear WCE models (similar to DLMs). 
Measurement error in non-linear models, such as the DRF-DLNM and ACE-DLNM, has not been studied and remains an important direction for future work.

The model and motivating analysis in this paper focus on a single pollutant. 
A natural extension is to accommodate multiple pollutants $X_1, \cdots, X_M$ using an additive model:
$$
g(\mu_t) = f_1 \left\{\int_{0}^{L_1} w_1(l) X_1(t-l) dl\right\} + \cdots + f_M \left\{\int_{0}^{L_M} w_M(l) X_M(t-l) dl\right\} + \sum_{j=1}^p h_j (z_{tj}). 
$$
We have implemented this in the R package {\tt aceDLNMadditive}, 
available at \url{github.com/tianyi-pan/aceDLNMadditive}. 
However, this framework does not allow for interactions either among multiple exposures or between exposure and covariate $z_{tj}$. 
An important direction for future work is to extend the ACE-DLNM to allow for non-additive interactions, 
while retaining flexibility for general response types as well as smooth functions in large datasets, as in the present paper. 

\clearpage
\section*{Acknowledgements}
The authors gratefully acknowledge data stewardship and contribution: the Canadian Institute for Health Information (CIHI) for hospitalization data, and Environment and Climate Change Canada (ECCC) for air pollution and weather data. However, the analyses, conclusions, opinions, and statements expressed herein are those of the author, and not necessarily those of the data providers.
\vspace*{-25pt}

\section*{Supplementary Materials}
The R package is available at \url{https://github.com/tianyi-pan/aceDLNM}. 
Code to reproduce the results in this paper is available at \url{https://github.com/tianyi-pan/aceDLNM-paper-code}.
\vspace*{-25pt}

\section*{Funding}
This study was funded under ``Canadian Environmental Sustainability Indicators'' of Environment and Climate Change Canada and ``Addressing Air Pollution Horizontal Initiative'' of Health Canada (\#AAPHI-859104; Principal Investigator, HH Shin).
\vspace*{-25pt}

\section*{Competing Interests}
The authors also declare that they have no actual or potential competing financial interests, and that this study did not have any relationships or support that might be perceived as constituting a conflict of interest.
\vspace*{-25pt}

\section*{Data Availability}
The air pollution and temperature datasets analyzed in this study are open to the public by Environment and Climate Change Canada (\url{http://data.ec.gc.ca/data/air/monitor/national-air-pollution-surveillance-naps-program/} and \url{https://climatedata.ca/}). The health data used for this study cannot be shared freely due to the policy of the CIHI for the use and distribution of sensitive records. Our data source, the Discharge Abstract Database (DAD), captures administrative, clinical, and demographic information on hospital discharges. Details regarding the DAD can be found at: \url{https://www.cihi.ca/en/discharge-abstract-database-metadata-dad}. The CIHI is the primary custodian of the DAD. CIHI is an independent, not-for-profit organization that provides essential information on Canada’s health systems and the health of people living in Canada. More information about CIHI can be found at: \url{https://www.cihi.ca/en/about-cihi}. Restrictions apply to the availability of the data as the data was used under a data sharing agreement and is not publicly available. General data inquiries about accessing CIHI data can be made at privacy@cihi.ca at the CIHI (\url{https://www.cihi.ca/en/access-data-and-reports/data-holdings/make-a-data-request}) or the author (hwashin.shin@hc-sc.gc.ca).
\vspace*{-25pt}

\bibliography{refs.bib}

\clearpage

\appendix

\renewcommand{\thesection}{Web Appendix \Alph{section}}

\renewcommand{\thesection}{Web Appendix \Alph{section}}
\captionsetup[figure]{labelformat=empty}
\captionsetup[table]{labelformat=empty}
\newtheorem{property}{Property}[section]
\renewcommand{\theproperty}{\Alph{section}.\arabic{property}}
\renewcommand{\theequation}{S.\arabic{equation}}
\newtheorem{remark}{Remark}

\part{Appendix}
\renewcommand{\ptctitle}{}
\parttoc

\section{DRF-DLNM}
In this Web Appendix, we review the DRF-DLNMs. 
The DRF-DLNM proposed by \citet{gasparrini2010distributed} and \citet{gasparrini2014modeling}
is specified as 
$$
 g(\mu_t) = \sum_{l=0}^{L-1} \psi(x_{t-l}, l) + \sum_{j=1}^p h_j (z_{tj}),
$$
where $\psi(\cdot, \cdot)$ is a bivariate exposure-lag-response function, modelled by 
splines and expressed as a linear combination of spline coefficients through a cross-basis expansion. The fitting of the model is convenient using the packages {\tt dlnm} \citep{gasparrini2011distributed} and {\tt mgcv} \citep{wood2015package}. 
\citet{gasparrini2017penalized} proposed a penalized framework, 
where the penalty matrices $\boldsymbol{S}_x$ and $\boldsymbol{S}_l$ along with smoothing parameters $\lambda_x$ and $\lambda_l$ 
are imposed on the two dimensions of the exposure-lag-response function. 
The framework supports various types of smoothers, 
and the model referred to as the ``primary'' model in \citet{gasparrini2017penalized} is based on P-splines \citep{eilers1996flexible}, 
where difference penalties are applied to the spline coefficients. 
 In our paper, all DRF-DLNMs are fitted with the P-spline model.

A similar model based on splines is described in Chapter 7.4.2 in \citet{wood2017generalized} and can be fitted by tensor product splines via ``summation convention'' in {\tt mgcv}. 
\citet{mork2022treed} proposed a DRF-DLNM based on the Bayesian additive regression tree to relax the smoothness constraint and impose monotone constraints on weight functions. 

The interpretations of the DRF-DLNM rely on \textit{post-hoc} summary statistics, including the exposure-response curve at a specific lag, the lag-response curve at a specific exposure level, and the overall cumulative risk for a specific exposure history \citep{gasparrini2014modeling}.

\section{Interpolation using Cubic B-spline}
The continuous exposure process $X$ is obtained by interpolating 
the collected exposure $x_t$ using the cubic spline. 
\citet{green1993nonparametric} proposed an approach to specify the spline through its values and second-order derivatives at the observed points. It is computationally efficient, but has some limitations, including numerical instability, among others; see Chapter 3.6 in \citet{green1993nonparametric}. 
One popular solution is the basis expansion approach, which is appealing in stability and flexibility. 

In this Web Appendix, we represent $X$ by cubic B-spline basis expansion and present a  
least-square-based approach to obtain the spline coefficient. This approach has a computational complexity $O(N)$, the same as that of the approach in \citet{green1993nonparametric}, 
and is easier to implement and can take advantage of the B-spline basis expansion. 

We will first describe the cubic B-spline, and then introduce the least-square-based approach, along with a discussion on the computational complexity.

Let $b_{l,p}(x)$ be the $l^\mathrm{th}$ B-spline function of order $p$, where $l = 1, \cdots, K$. The knots sequence is denoted as $\mathbf{\tau} = \left[\tau_1, \tau_2, \cdots, \tau_{K+p}\right]$. Let $\xi_{l}$ and $\xi_{u}$, $\xi_{l} \le \xi_{u}$, be the boundaries defining the domain where we evaluate the B-splines. 
$\tau_1\le \tau_2 \le \cdots \le \tau_p \le \xi_l$ and $\xi_u \le \tau_{K+1}\le \tau_{K+2} \le \cdots \le \tau_{K+p}$ are the auxiliary knots with arbitrary values beyond the boundaries. The knots within the boundary $\tau_{p+1} \le \cdots \le \tau_K$ are interior knots. 
The B-spline function $b_{l,p}(x)$ is defined recursively and locally supported in $\left(\tau_l, \tau_{l+p}\right)$. 

The $s$-order derivative of $b_{l,p}(x)$ is given by 
\begin{equation}
\frac{\mathrm{d}^{(s)} b_{l, p}(x)}{\mathrm{d} x^{(s)}}=(p-1)\left(\frac{\mathrm{d}^{(s-1)} b_{l, p-1}(x) / \mathrm{d} x^{(s-1)}}{\tau_{l+p-1}-\tau_l} - \frac{\mathrm{d}^{(s-1)} b_{l+1, p-1}(x) / \mathrm{d} x^{(s-1)}}{\tau_{l+p}-\tau_{l+1}}\right). 
\label{eq:deriv_Bspline}
\end{equation}

See more details in \citet{wood2017generalized}. Using $p=4$ for cubic B-spline, we ignore the subscript $p$ to simplify the notation to $b_{l}(x)$.

A cubic spline is a \textit{natural cubic spline}, if the additional constraint is added: the function is linear beyond the boundary $\xi_{l}$ and $\xi_{u}$. The interpolating natural cubic spline $g$ is optimal, namely $g$ is the unique function minimizing the roughness $\int g''^{2}$ among all smooth functions that interpolate the data; see \citet{green1993nonparametric} for details. 

In the ACE-DLNM, we treat the exposure process $X(t)$ continuously. Without loss of generality, assume $\mathcal{T}^x = \mathcal{T}$. 
Let 
$\left\{t-0.5, x_t\right\}, t \in \mathcal{T} = \left\{1, 2, \cdots, N\right\}$ denote the data. We propose to interpolate the discrete data using the natural cubic spline. The exposure process is modelled as 
$$
X(t) = \sum_{q = 1}^{k^x} b^x_q(t) \alpha^x_q,
$$
where $b^x_q(t)$ are cubic B-splines and $\alpha^x_q$ are spline coefficients. We denote $\boldsymbol{b}^x(t) = \left[b^x_q(t), \cdots, b^x_{k^x}(t) \right]^\top$ and $\boldsymbol{\alpha^x} = \left[\alpha^x_1, \cdots, \alpha^x_{k^x}\right]^\top$.

We set $N$ interior knots $\left\{\tau_5, \tau_6, \cdots, \tau_{N + 4}\right\}$ at observed time points 
$\left\{0.5, 1.5, \cdots, N-0.5\right\}$ 
with corresponding observations $\left\{x_1, x_2, \cdots, x_{N}\right\}$, and hence $k^x = N+4$. 
For the auxiliary knots beyond the boundary, we set $\tau_1=\tau_5-c-1$, $\tau_2 = \tau_3=\tau_4 = \tau_5-c$ and $\tau_{N+5}=\tau_{N+6}=\tau_{N+7} = \tau_{N} + c$, $\tau_{N+8} = \tau_{N} + c + 1$, where $c >0$ is arbitrary. 
We place the empty points, i.e. setting values as 0, on the auxiliary knots, as shown in Web Table 1. 

\begin{table}[H]
\centering
\caption{Web Table 1. The knots and the corresponding values. The interior knots $\tau_5 \cdots \tau_{N + 4}$ are set at observed time points. The empty points (assuming observed data are zero) are added at the auxiliary knots $\tau_1, \cdots, \tau_4, \tau_{N+5}, \cdots, \tau_{N+8}$.}
\label{tab:X_knots}
\begin{tabular}{cccccccccc}
\toprule
Value  & 0  &$\cdots$ & 0  & $x_1$ & $\cdots$ & $x_{N}$   & 0     & $\cdots$  & 0     \\ \midrule 
Knot & $\tau_1$ & $\cdots$ & $\tau_4$ & $\tau_5$ & $\cdots$ & $\tau_{N + 4}$ & $\tau_{N + 5}$ & $\cdots$ & $\tau_{N + 8}$ \\ \bottomrule
\end{tabular}
\end{table}

The B-splines evaluated at the interior knots are represented as a matrix $\boldsymbol{B}^x$ with $l^{th}$ row as $\boldsymbol{b}^x(l)^\top$, $l = 1,\cdots, N$. Define $\widetilde{\boldsymbol{B}}^x$ as the B-spline evaluated at both interior and auxiliary knots, i.e., 
$$
\widetilde{\boldsymbol{B}}^x = \begin{bmatrix}
    \boldsymbol{b}^x(\tau_1) & \cdots & \boldsymbol{b}^x(\tau_4) &
    {\boldsymbol{B}^x}^\top &
    \boldsymbol{b}^x(\tau_{N + 5}) & \cdots & \boldsymbol{b}^x(\tau_{N + 8})
\end{bmatrix}^\top. 
$$

Denote the observed exposure as $\boldsymbol{X} = \left[x_1, \cdots, x_{N}\right]^\top$ and the vector with the empty points as $\widetilde{\boldsymbol{X}} = \left[0,0,0,0,\boldsymbol{X} ^\top, 0, 0,0,0\right]^\top$. We propose to obtain the spline coefficients $\boldsymbol{\alpha^x} = \left[\alpha^x_1, \cdots, \alpha^x_{N+4}\right]$ by
\begin{equation}
    \label{eq:interpolate}
    \boldsymbol{\alpha^x} = \left(\widetilde{\boldsymbol{B}}^{x\top}\widetilde{\boldsymbol{B}}^x\right)^{-1}\widetilde{\boldsymbol{B}}^{x\top}\widetilde{\boldsymbol{X}}. 
\end{equation}
Then, we have the following property: 
\begin{property}
    \label{property:interpolate}
    The cubic B-spline $X(t) = \sum_{q = 1}^{N+4} b^x_q(t) \alpha^x_q$ interpolates the data, if the spline coefficients $\alpha^x_q$ are computed from Equation \ref{eq:interpolate}. 
\end{property}

\begin{proof}[Proof of Property \ref{property:interpolate}]    
    According to Theorem 2.2 in \citet{green1993nonparametric}, for the observed data $\left\{t, x_t\right\}, t = 1, 2, \cdots, N$, there exists a \textit{unique} natural cubic spline with interior knots at $t = 1, 2, \cdots, N$ interpolating the observed data. 

    Assume that the spline coefficient of the interpolating natural cubic spline is $\boldsymbol{\alpha}^{*x}$. Then, we have,
    $\sum_{t = 1}^{N} \left(x_t - \boldsymbol{b}^x(t)^\top \boldsymbol{\alpha}^{*x}\right)^2  = 0$.
    
    By the property of natural cubic spline, the function is linear beyond the boundary. Hence, for any $z \in (\tau_1, \tau_2)$,
    $\boldsymbol{b}^{x''}(z)^\top \boldsymbol{\alpha}^{*x} = 0$,
    where $\boldsymbol{b}^{x''}$ is the second derivative of B-splines. 
    Applying the formula of derivative of B-spline (Equation \ref{eq:deriv_Bspline}) and the local support property, we have 
    $b_1^{x''}(z) > 0$ 
    and 
    $b_k^{x''}(z) = 0$ for $k \ge 2$, which leads to,
    $\alpha^{*x}_1 = 0$. Similarly, we have 
    $\alpha^{*x}_{Nt+4} = 0$. 
    
    Defining $\boldsymbol{v} = \left[v_1, \cdots,v_{N+4}\right]^\top$ and a function of $\boldsymbol{v}$ as follows
    $$
        L(\boldsymbol{v}) = \sum_{t = 1}^{N} \left(x_t - \boldsymbol{b}^x(t)^\top \boldsymbol{v}\right)^2 + 3 v_1^2 + 3 v_{N+4}^2. 
    $$
    Then, $L(\boldsymbol{\alpha}^{*x}) = 0$, i.e., $\boldsymbol{\alpha}^{*x}$ as minimizer of the function $ L(\boldsymbol{v}) \ge 0$. 
    
    From the induction definition of B-spline \citep{wood2017generalized}, we have $\boldsymbol{b}^x(\tau_1) = \left[0,0,\cdots,0\right]^\top$, $\boldsymbol{b}^x(s) = \left[1,0,\cdots,0\right]^\top \text{ for } s = \tau_2, \tau_3,\tau_4$, $\boldsymbol{b}^x(s) = \left[0,\cdots,0,1\right]^\top \text{ for } s = \tau_{N + 5},\tau_{N + 6},\tau_{N + 7}$, and $\boldsymbol{b}^x(\tau_{N + 8}) = \left[0,\cdots,0,0\right]^\top$. 
    Therefore, $L(\boldsymbol{v})$ can be rewritten as
    $$
    \begin{aligned}
    L(\boldsymbol{v}) =& \sum_{t = 1}^{N} \left(x_t - \boldsymbol{b}^x(t)^\top \boldsymbol{v}\right)^2 + \sum_{s \in \left\{\tau_1, \cdots,\tau_4, \tau_{N + 5}, \cdots, \tau_{N + 8}\right\}} \left(0 - \boldsymbol{b}^x(s)^\top \boldsymbol{v}\right)^2 \\
    =&  \left(\widetilde{\boldsymbol{X}} - \widetilde{\boldsymbol{B}}^x \boldsymbol{v}\right)^\top\left(\widetilde{\boldsymbol{X}} - \widetilde{\boldsymbol{B}}^x \boldsymbol{v}\right). 
    \end{aligned}
    $$
    Minimizing $L(\boldsymbol{v})$ turns to a least-square problem. The solution is
    $$
    \arg\min_v L(\boldsymbol{v}) = \left(\widetilde{\boldsymbol{B}}^{x\top}\widetilde{\boldsymbol{B}}^x\right)^{-1}\widetilde{\boldsymbol{B}}^{x\top}\widetilde{\boldsymbol{X}}. 
    $$
    Therefore, we have $\boldsymbol{\alpha}^{*x} = \left(\widetilde{\boldsymbol{B}}^{x\top}\widetilde{\boldsymbol{B}}^x\right)^{-1}\widetilde{\boldsymbol{B}}^{x\top}\widetilde{\boldsymbol{X}}$, which completes the proof.     
\end{proof}
By setting knots at observed points $t = 1, 2, \cdots, N$, there are other approaches to obtain the interpolating natural cubic spline; see \citet{green1993nonparametric} for an example. 
According to Theorem 2.2 in \citet{green1993nonparametric}, for the observed data $\left\{t, x_t\right\}, t = 1, 2, \cdots, N$, 
all the possible approaches yield the same interpolating natural cubic spline. Here we propose the least-square-type approach that is easier to implement and can take advantage of the basis expansion form, as discussed before. 

Solving the least-square-type equation requires $O(N^3)$ computational operations. Note that $\widetilde{\boldsymbol{B}}^{x}$ is a banded sparse matrix because of the local support property of B-splines. 
The equation is solved efficiently using the sparse Cholesky decomposition. 
With the techniques, the computational complexity reduces to $O(N)$; see also Chapter 5 in \citet{hastie2009elements}.

\section{Integration for Adaptive Cumulative Exposures}

Define $D_{q}(t) = \int_{0}^L  b^w_{q}(l) X(t-l)dl$. 
To construct the ACE, we need to evaluate the integral 
$D_{q}(t)$ for $q = 1, 2, \cdots, d^w$ and each $t$. 
We propose the following approach to fast and exactly evaluate these integrals using de Boor's algorithm \citep{de1978practical} and an appropriate modification of the method given by \citet{wood2017p}.

\citet{wood2017p} proposed an algorithm to compute the integral $\int b_i(x)b_j(x) d x$ where $b_i$ and $b_j$ are B-splines sharing the same knots sequence. The algorithm is fast and implemented in {\tt mgcv} package. In our settings, however, the knot sequences for $b^w_{q}$ and $X$ are different. Meanwhile, the intervals of integration for different $t$ are overlapped, and hence some evaluations are redundant. We propose a modification of \citet{wood2017p} algorithm. 

We highlight the key steps in the modifications. First, we merge and resort the two different knot sequences to a unified knot sequence within $[t-L, t]$. Second, in each partition based on the knots sequence, \citet{wood2017p} solves a linear system to obtain the polynomial coefficients. In our settings, some polynomial coefficients hold the same. We propose to solve the coefficients in advance to avoid redundant calculations. The details are described as follows.

Denote $\boldsymbol{\tau}^w = \left[\tau^w_1, \cdots, \tau^w_{d^w+4}\right]$ the knots for $b^w_{q}$ and $\boldsymbol{\tau}^x = \left[\tau^x_1, \cdots, \tau^x_{d^w+4}\right]$ the knots for $X$. We merge the sequences $t-\boldsymbol{\tau}^w = \left[t-\tau^w_1, \cdots, t-\tau^w_{d^w+4}\right]$ and $\boldsymbol{\tau}^x$ into one sequence $\boldsymbol{\tau}_t$, i.e. $\boldsymbol{\tau}_t = \left\{t-\boldsymbol{\tau}^w, \boldsymbol{\tau}^x\right\}$. Denote $\left[m_{ts}, n_{ts}\right]$ the intervals that partition the line $\left(t-L, t-0\right)$ based on $\boldsymbol{\tau}_t$. The spline functions $b^w_{q}$ and $X$ are cubic polynomial function within $\left[m_{ts}, n_{ts}\right]$. Following \citet{wood2017p}, we have
\begin{equation*}
    \begin{aligned}
        D_q(t) &= \sum_s \int_{m_{ts}}^{n_{ts}}  b^w_{q}(l) X(t-l)dl \\ 
        &= \sum_s \frac{m_{ts}-n_{ts}}{2}\int_{-1}^{1}  \sum_{i=0}^3 c_i x^i \sum_{j=0}^3 d_j x^j dx,
    \end{aligned}
\end{equation*}
where $c_i$ and $d_j$ are polynomial coefficients. The polynomial coefficients $\boldsymbol{c}$ are obtained by solving $\boldsymbol{P}\boldsymbol{c} = \boldsymbol{b}^w$ where $P_{ij} = \left(-1+2(i-1)/3\right)^{j-1}$ and $\boldsymbol{b}^w$ consists of $b_q^w$ evaluated at the 4 points evenly spaced from $a_{ts}$ to $b_{ts}$, that is, 
$$
\boldsymbol{b}^w = \left[b^w_{q}(a_{ts}), b^w_{q}(a_{ts}) + (b_{ts} - a_{ts})/3, b^w_{q}(a_{ts}) + 2(b_{ts} - a_{ts})/3, b^w_{q}(b_{ts})\right]^{\top}. 
$$
Similarly, $\boldsymbol{d}$ are obtained by solving $\boldsymbol{P}\boldsymbol{d} = \boldsymbol{b}^x$. Therefore, we have
$$
D_q(t) = \sum_s \frac{m_{ts}-n_{ts}}{2}  \boldsymbol{b}^{x\top} \boldsymbol{P}^{-\top}\boldsymbol{H}\boldsymbol{P}^{\top} \boldsymbol{b}^w,
$$
where $H_{ij} = (1+(-1)^{i+j-2})/(i+j-1)$.

For the second aspect of modification about solving the linear systems for each $s$, the polynomial coefficients hold the same within the adjacent knots. 
In our algorithm, we propose to solve the coefficients in advance to avoid redundant calculations.  
For two adjacent knots $\tau^x_s$ and $\tau^x_{s+1}$, for example, we calculate the matrix $\boldsymbol{K}$ where $K_{ij} = (\tau^x_s + (i-1)(\tau^x_{s+1} - \tau^x_{s})/3)^{j-1}$, and evaluate $\boldsymbol{b}^x$ at the 4 points evenly spaced from $\tau^x_{s}$ to $\tau^x_{s+1}$. Then, we can obtain the coefficient $\boldsymbol{d}$ for the two adjacent knots $\tau^x_s$ and $\tau^x_{s+1}$ by solving $\boldsymbol{K} \boldsymbol{d} = \boldsymbol{b}^x$. Then, $\boldsymbol{d}$ is the common coefficient for any $\left[m_{ts}, n_{ts}\right]$ within $\tau^x_s$ and $\tau^x_{s+1}$, and hence we do not need to solve $\boldsymbol{P}\boldsymbol{d} = \boldsymbol{b}^x$ redundantly. 

Furthermore, $X$ is modelled by interpolating cubic B-spline $X(t) = \sum_{q = 1}^{k^x} b^x_q(t) \alpha^x_q$ with a large number of knots, which will induce a further computational burden. In fact, only 4 of the basis functions are non-zero for each $t$. To avoid wasteful evaluation, we use de Boor's Algorithm \citep{de1978practical} to increase the computational efficiency.

The full details of the algorithm are in Algorithm \ref{alg:integral}. The algorithm is implemented using {\tt cpp} code available at the R package file \url{https://github.com/tianyi-pan/aceDLNM/blob/main/src/helpers.cpp}.

\begin{algorithm}[H]
    \caption{Evaluate the integral in weighted exposures}\label{alg:integral}
    \begin{algorithmic}[1]
    \State Define $I_{\boldsymbol{\tau}}(x)$ as a function returning the index of the largest knot smaller than $x$ in $\boldsymbol{\tau}$. 
    \State Calculate $\boldsymbol{P}$ where $P_{ij} = \left(-1+2(i-1)/3\right)^{j-1}$, and solve $\boldsymbol{P}^{-1}$. 
    \State Calculate $\boldsymbol{H}$ where $H_{ij} = (1+(-1)^{i+j-2})/(i+j-1)$. 
    \State Calculate $\boldsymbol{W} = \boldsymbol{P}^{-1\top} \boldsymbol{H} \boldsymbol{P}^{-1}$. 
    \For{$r = 1, \cdots, d^x+3$}
        \State $h^x_r = \tau^x_{r+1}-\tau^x_{r}$
        \State Evaluate $\boldsymbol{b}_r^x = \left[X(\tau^x_{r}), X(\tau^x_{r} + h^x_r/3), X(\tau^x_{r} + 2h^x_r/3), X(\tau^x_{r+1})\right]^{\top}$ using de Boor's algorithm. 
        \State Calculate $\boldsymbol{K}^x$ where $K^x_{ij} = \left(\tau_{r}^x + h_r^x (i-1) / 3\right)^{j-1}$. 
        \State Solve the linear system $\boldsymbol{K}^{x}\boldsymbol{d}_r = \boldsymbol{b}_r^x$ for the polynomial coefficients $\boldsymbol{d}_r$. 
    \EndFor
    \State Save $\boldsymbol{b}_1^x, \cdots, \boldsymbol{b}_{d^x+3}^x$ and $\boldsymbol{d}_1^x, \cdots, \boldsymbol{d}_{d^x+3}^x$.
    \State Similarly, obtain and save $\boldsymbol{b}_1^w, \cdots, \boldsymbol{b}_{d^w+3}^w$ and $\boldsymbol{d}_1^w, \cdots, \boldsymbol{d}_{d^w+3}^w$.
    \For{$t = 1, \cdots, N$}
        \State Set $D_q(t) = 0$.
        \For{$u = 1, \cdots, d^w+3$}
            \State $h^w_u = \tau^w_{u+1}-\tau^w_{u}$
            \If{$I_{\boldsymbol{\tau}^x}(t-\tau^w_u) =I_{\boldsymbol{\tau}^x}(t-\tau^w_{u+1})$}
                \State Calculate $\boldsymbol{K}^x$, where $K^x_{ij} =(t-\tau^w_{u+1} + (i-1)h^w_u/3))^{j-1}$. 
                \State $D_q(t) = D_q(t) + (h^w_u/2)[\boldsymbol{K}^x \boldsymbol{d}_{I_{\boldsymbol{\tau^x}}(t-\tau_u^w)}^x]^{\top} \boldsymbol{W} \boldsymbol{b}_{u}^w$. 
            \Else
                \State Define $\boldsymbol{\tau}^u$ as a sequence with the first element $t-\tau^w_{u+1}$ and the last element $t-\tau^w_{u}$. The elements in the middle are the knots of $X$ within $I_{\boldsymbol{\tau}^x}(t-\tau^w_{u+1})$ and $I_{\boldsymbol{\tau}^x}(t-\tau^w_u)$. 
                \For{$s = 1,\cdots, (\text{length of } \boldsymbol{\tau}^u-1)$}
                    \State $h_s = \boldsymbol{\tau}^u_{s+1} - \boldsymbol{\tau}^u_s$.
                    \State Calculate $\boldsymbol{K}^x$, where $K^x_{ij} = (\boldsymbol{\tau}^u_{s}+ (i-1)h_s / 3))^{j-1}$. 
                    \State Calculate $\boldsymbol{K}^w$, where $K^w_{ij} = (t-\boldsymbol{\tau}^u_{s} - (i-1)h_s / 3))^{j-1}$. 
                    \State $D_q(t) = D_q(t) + (h^w_s/2)[\boldsymbol{K}^x \boldsymbol{d}_{I_{\boldsymbol{\tau^x}}(\tau_s^u)}^x]^{\top} \boldsymbol{W} [\boldsymbol{K}^w \boldsymbol{d}^w_u]$. 
                \EndFor
            \EndIf
        \EndFor
    \EndFor
    \end{algorithmic}
\end{algorithm}

\clearpage

\section{Identifiability}
\label{s:iden}
Recall that the mean model in ACE-DLNM is
$$
g(\mu_t) = f \left\{\int_{0}^{L} w(l) X(t-l) dl\right\} + \sum_{j=1}^p h_j (z_{tj}). 
$$
Let $\boldsymbol{h}(\boldsymbol{z}_{t}) = \sum_{j=1}^p h_j (z_{tj})$
and $s(t;X) = f \left\{\int_{0}^{L} w(l) X(t-l) dl\right\}$. 

Given a known exposure process $X$, assuming that there exist two sets $\{\widetilde{s}, \widetilde{\boldsymbol{h}}\}$ and $\{s^*, \boldsymbol{h}^*\}$, 
where $\widetilde{s}(t) = \widetilde{f} \left\{\int_{l_{0}}^{L} \widetilde{w}(l) X(t-l) dl\right\}$ and $s^*(t) = f^* \left\{\int_{l_{0}}^{L} w^*(l) X(t-l) dl\right\}$, such that,
\begin{equation}
\widetilde{s}(t, X) + \widetilde{\boldsymbol{h}}(\boldsymbol{z}_{t}) = s^*(t, X) + \boldsymbol{h}^*(\boldsymbol{z}_{t})
\label{eq:iden_model}
\end{equation}
for all $t$ and $\boldsymbol{z}_t$. 
Following the discussions of identifiability in \citet{yuan2011identifiability} and \citet{chen2016generalized}, the ACE-DLNM is identifiable if Equation \ref{eq:iden_model} implies that 
$\widetilde{f} = f^*$, $\widetilde{w} = w^*$ and $\widetilde{\boldsymbol{h}} = \boldsymbol{h}^*$. 

The ACE-DLNM is reduced to a GAM if $s(t; X)$ is viewed as a function of $t$. The discussion on the identifiability of GAMs under linear constraints can be found in \citet{wood2016smoothing} and \citet{stringer2023identifiability}. Under some linear constraint, Equation \ref{eq:iden_model} implies that 
$\widetilde{s} = s^*$ and $\widetilde{\boldsymbol{h}} = \boldsymbol{h}^*$. 
To show the identifiability, we need to show that the following Proposition \ref{prop:ideNoshow} is true. 
\begin{prop}
\label{prop:ideNoshow}
$\widetilde{s} = s^*$ implies $\widetilde{f} = f^*$ and $\widetilde{w} = w^*$. 
\end{prop}

Let $\widetilde{w}(z) = c\cdot w^*(z)$ and $\widetilde{f}(z) = f^*(z/c)$, $c \in \mathbb{R} \setminus\{1\}$, and hence $\{\widetilde{f},\widetilde{w}\} \neq \{f^*,w^*\}$; but it is obvious that $\widetilde{s} = s^*$. Therefore, 
Proposition \ref{prop:ideNoshow} is false and the model is not identifiable. 

The constraints preventing the arbitrary $c \in \mathbb{R} \setminus\{1\}$ are required.

\subsection{Review of Identifiability Constraints}
We review the constraints used in existing models, including \citet{kong2010statistical}, \citet{ma2016estimation}, \citet{wilson2022kernel} and \citet{wang2023semiparametric}. 
In Web Table 2, we summarize the imposed constraints, and some information about the constraints including whether or not (1) the constraints are continuous (Continuous), (2) the bounds for ACE $E(t)$ are derived (Bounds for $E(t)$), (3) the constraints are fully absorbed by reparameterization (Absorbed by Reparameterization), and (4) the identifiability theorem/proposition was introduced (Proposition of Identifiability). 
We use the continuous constraints in \citet{wilson2022kernel} and address these issues simultaneously. 

\begin{table}[H]
\centering
\caption{Web Table 2. A review of identifiability constraints. }
\makebox[\textwidth]{ 
\resizebox{1.2\textwidth}{!}{
\begin{tabular}{@{}ccccccc@{}}
\toprule
Paper     & \multicolumn{1}{c}{Constraint 1} & \multicolumn{1}{c}{Constraint 2} & \multicolumn{1}{c}{Continous} & \multicolumn{1}{c}{\begin{tabular}[c]{@{}c@{}}Bounds\\ for $E(t)$\end{tabular}} & \multicolumn{1}{c}{\begin{tabular}[c]{@{}c@{}}Absorbed by \\ Reparame-\\-terization\end{tabular}} & \multicolumn{1}{c}{\begin{tabular}[c]{@{}c@{}}Proposition \\ of Identifi-\\-ability \\ \end{tabular}} \\ \midrule
\textbf{Our Paper} &      $\int w^2(l) dl = 1$     & $\int w(l) dl > 0$     &     \cmark        &     \cmark  &   \cmark     &  \cmark                                                                     \\ 
\citet{kong2010statistical}      &    $\int w(l) dl = 1$  &  $\mathrm{E}\left(\int X(t-l) w(l) dl\right) = 0$  & \cmark   &                                                                             & \cmark   & \cmark      \\
\citet{ma2016estimation}        &    $\sum_{m=1}^M\int w_k^2(l) dl = 1$\footnotemark[1]   &  monotone functions \footnotemark[2]&     \cmark  &  &                                                                                             &   \cmark                                      \\
\citet{wilson2022kernel}   &     $\int w^2(l) dl = 1$        &    $\int w(l) dl > 0$           &    \cmark                &                                                                              &                                                                                             &                                                                                        \\
\citet{wang2023semiparametric}      &   $\sum_j (\alpha_j^{w})^2 = 1$       &    $\alpha_1^{w} > 0$            &                               &                                                                              &              \cmark                                                          &                                                                                        \\
\bottomrule
\end{tabular}
}}
\vspace{2mm}
\footnotesize{\footnotemark[1] The model is for multiple $X$, and $E(t)$ is written as $f(\int \sum_{m=1}^M X_m(t-l)w_m(l)dl$.}
\footnotesize{\footnotemark[2] $w_1(\cdot)$ is non-constant and monotone nondecreasing, or $f(\cdot)$ is monotone nondecreasing.}
\end{table}

\subsection{Constraints and Identifiability Proposition 1}
Recall the two identifiability constraints described in Section 2.3:

\noindent \textit{Constraint 1 (Scale Constraint):} 
$\int_0^L w(l)^2 dl = 1$,

\noindent \textit{Constraint 2 (Sign Constraint):}
$\int_{0}^L w(l) dl > 0$.

We assume the following Conditions: 
\begin{enumerate}[(a)]
        \item $X$ is continuous and non-constant on its domain. $X(t)$ can be represented with a B-spline basis expansion, $X(t) = \sum_{q = 1}^{k_x} b^x_q(t) \alpha^x_q$, where $b^x_q(\cdot)$ are cubic B-spline functions and $\alpha^x_q$ are spline coefficients with some knot sequence. 
        \label{con:X}
        \item $f$ is continuously differentiable and non-constant on the range of $E(t)$.
        \label{con:f}
        \item $w$ is continuous on $[0,L]$.  %
        \label{con:w}
\end{enumerate}

Next, we prove Proposition 1 in Section 2.3.

\begin{proof}[Proof of Proposition 1]
To prove Proposition 1, we show that there exists a knot sequence such that Proposition \ref{prop:ideNoshow} is true under the scale and sign constraints and Conditions (a), (b) and (c). 

Assume there exist two sets $\{\widetilde{f},\widetilde{w}\}$ and 
$\{f^*,w^*\}$, such that,
\begin{equation}
\widetilde{f} \left\{\int_{0}^{L} \widetilde{w}(l) X(t-l) dl\right\} = f^* \left\{\int_{0}^{L} w^*(l) X(t-l) dl\right\},
\label{eq:proof-iden-s}
\end{equation}
for all $t$. Define $\widetilde{E}(t) = \int_{0}^L \widetilde{w}(l) X(t-l) d l$ and $E^*(t) = \int_{0}^L w^*(l) X(t-l) d l$. Condition (\ref{con:X}) and the range of $E(t)$ ensure the existence of the two integrals. 
To show Proposition 1 is true, we need to show that 
$\widetilde{f} = f^*$ and $\widetilde{w} = w^*$.

By Condition (\ref{con:X}), $X(t) = \sum_{q = 1}^{k_x} b^x_q(t) \alpha^x_q$. For $q \in \left[1, 2, \cdots, k_x\right]$, taking derivatives w.r.t. $\alpha_q^x$ on both sides of Equation \ref{eq:proof-iden-s} leads to
\begin{equation}
\begin{aligned}
\widetilde{f}^{\prime}\left\{\widetilde{E}(t)\right\} \int_{0}^L b_q^x (t-l) \widetilde{w}(l)d l = f^{*\prime}\left\{E^*(t)\right\} \int_{0}^L b_q^x (t-l) w^*(l)d l,
\end{aligned}
\label{eq:proof-iden-s2}
\end{equation}
where $\widetilde{f}^\prime(z)$ and $f^{*\prime}(z)$ are the derivatives w.r.t $z$, which exist by Condition (\ref{con:f}). Then, we take summation over $q = 1, 2, \cdots, N$, which yields
\begin{equation}
\begin{aligned}
\widetilde{f}^{\prime}\left\{\widetilde{E}(t)\right\} \int_{0}^L \widetilde{w}(l)d l = f^{*\prime}\left\{E^*(t)\right\} \int_{0}^L w^*(l)d l,
\end{aligned}
\label{eq:proof-iden-s3}
\end{equation}
using the property of B-spline, $\sum_{q = 1}^{k_x} b_q^x (z) = 1$. 

We have $\int_{0}^L w^*(l) dl > 0$ and $\int_{0}^L \widetilde{w}(l) dl > 0$ by sign constraint. 
There exists some $t$ such that $f^{*\prime}\{E^*(t)\} \neq 0$ and $\widetilde{f}^{\prime}\{\widetilde{E}(t)\} \neq 0$ by Condition (\ref{con:f}). 
For such $t$, we combine Equation \ref{eq:proof-iden-s2} and \ref{eq:proof-iden-s3}, which leads to
\begin{equation}
\frac{\int_{0}^L b_q^x (t-l) \widetilde{w}(l)d l}{\int_{0}^L \widetilde{w}(l)d l} = \frac{\int_{0}^L b_q^x (t-l) w^*(l)d l}{\int_{0}^L w^*(l)d l}. 
\label{eq:proof-iden-s4}
\end{equation}

Define $\widetilde{r}(l) = \widetilde{w}(l) / \int_{0}^L \widetilde{w}(l)d l$ and $r^*(l) = w^*(l) / \int_{0}^L w^*(l)d l$. Equation \ref{eq:proof-iden-s4} is rewritten as
\begin{equation}
\int_{0}^L b_q^x (t-l) \left\{\widetilde{r}(l) - r^*(l)\right\}d l = 0,
\label{eq:proof-iden-s5}
\end{equation}
for all $t$ and any $q \in \left[1,2,\cdots,k_x\right]$. %

From Equation \ref{eq:proof-iden-s5}, 
we claim that $\widetilde{r} = r^*$ almost everywhere, which will be proved later. 

Based on this claim, we have $\widetilde{w}(l) = c \cdot w^*(l)$ almost everywhere, where $c = \int_{0}^L w^*(l)d l / \int_{0}^L \widetilde{w}(l)d l$. By scale constraint and sign constraint, 
we have $c = 1$. Hence, $\widetilde{w} = w^*$. Back to Equation \ref{eq:proof-iden-s}, for any $t$, we have
$$
\widetilde{f}\left\{E^*(t)\right\} = f^*\left\{E^*(t)\right\},
$$
which leads to $\widetilde{f} = f^*$. 

Next, we prove the claim, $\widetilde{r} = r^*$ a.e., by contradiction: 

Suppose $\widetilde{r}(l_1) - r^*(l_1) = 0, \text{a.e.}$ is not true. Since $\widetilde{w}$ and $w^*$ are continuous (Condition (\ref{con:w})) and hence $\widetilde{r}, r^*$ are continuous, there exists an interval $\left[a,b\right] \in \left[0, L\right]$ such that $\widetilde{r}(l_1) - r^*(l_1), l_1 \in \left[a,b\right]$ are nonzero with the same sign. Assume $\widetilde{r}(l_1) - r^*(l_1) > 0$ without loss of generality. Then, we can find a knot sequence $\boldsymbol{\tau}$ such that there exists a $t_1$ and an index $j$, $\tau_j \ge t_1 - b$ and $\tau_{j+4} \le t_1 - a$. 
By the property of cubic B-spline, $b_j^x$ is locally supported in $\left(\tau_j, \tau_{j+4}\right)$ with positive values. Therefore, we have
$$
\int_{0}^L b_j^x (t_1-l) \left\{\widetilde{r}(l) - r^*(l)\right\}d l= \int_{a}^b b_j^x (t_1-l) \left\{\widetilde{r}(l) - r^*(l)\right\}d l > 0,
$$
which is contradicted by Equation \ref{eq:proof-iden-s5}. The claim is proved, and the proof of Proposition 1 is complete. 
\end{proof}

\begin{remark}
    \normalfont
    Conditions (\ref{con:f}) and (\ref{con:w}) are ready if $f$ and $w$ can be modelled by cubic B-splines. For Condition (\ref{con:X}), the continuous exposure process $X$ is obtained by interpolating the observed data using a cubic B-spline, which ensures continuity, and the non-constancy of $X$ can be easily verified.  
\end{remark}

\subsection{Reparameterization}

In Section 3.1, we present the reparameterization for $w$. In the reparameterization, 
the sequence $\left\{l_j\right\}_{j = 1}^J$ is an evenly spaced partition of $[0, L]$.
We replace $\int_{0}^L w^+(l)$ by $(L/J)\sum_{j=1}^J w^+(l_j)$ with a large $J$, which leads to 
$\int_{0}^L w(l) dl = \alpha^{w+}_{1}$. 
Then, we have
$$
\alpha^{w+}_1 = \frac{1}{\left([1,\boldsymbol{\phi}^{w\top}] \mathbf{C}[1,\boldsymbol{\phi}^{w\top}]^\top\right)^{1/2}} > 0. 
$$
Therefore, the sign constraint $\int_{0}^L w(l) dl > 0$ is guaranteed. By the definition of $\mathbf{C}$, we have
$$
\int_{0}^L w(l)^2 dl = \boldsymbol{\alpha}^{w+\top} \mathbf{C}\boldsymbol{\alpha}^{w+},
$$
where $\boldsymbol{\alpha}^{w+\top} \mathbf{C}\boldsymbol{\alpha}^{w+} = 1$.
Hence, the $w$ satisfies the two identifiability constraints.

\section{Optimization Algorithm}
\label{s:opt}

We summarize the algorithm as the following procedure:
\begin{enumerate}
    \item \textbf{Outer Stage} (LAML): Estimate $\boldsymbol{\lambda}$ and $\boldsymbol{\theta}$ by maximizing $\mathcal{L}^*_{\text{LA}}(\boldsymbol{\lambda}, \boldsymbol{\theta})$ using BFGS. \\ At each iteration of BFGS: 
    \begin{enumerate}
        \item \textbf{Inner Stage} (Profile Likelihood): Estimate $\boldsymbol{\phi}^w$, $\boldsymbol{\alpha}^{f}$ and $\boldsymbol{\beta}$ by the profile likelihood approach, where $\boldsymbol{\phi}^w$ is estimated by maximizing $\mathbf{Q}(\boldsymbol{\phi}^w; \boldsymbol{\lambda}, \boldsymbol{\theta})$ using Newton's method; at each iteration: 
        \begin{enumerate}
            \item Estimate $\boldsymbol{\alpha}^{f}$ and $\boldsymbol{\beta}$ by maximizing $\mathcal{L}(\boldsymbol{\phi}^w, \boldsymbol{\alpha}^{f}, \boldsymbol{\beta}; \boldsymbol{\lambda}, \boldsymbol{\theta})$ using Newton's method. 
        \end{enumerate}
    \end{enumerate}
\end{enumerate}

The details are presented in the following Algorithms \ref{alg:outer}, \ref{alg:middle} and \ref{alg:inner}. The algorithms are implemented using {\tt cpp} code available at the R package file \url{https://github.com/tianyi-pan/aceDLNM/blob/main/src/aceDLNM.cpp}. 

\begin{algorithm}[H]
    \caption{Outer Stage: Optimization for $\log \boldsymbol{\lambda}$ and $\log \boldsymbol{\theta}$}\label{alg:outer}
    \begin{algorithmic}[1]
    \While{not converged}
            \State Find $\widehat{\boldsymbol{\phi}}^w$, $\widehat{\boldsymbol{\alpha}}^{f}$ and $\widehat{\boldsymbol{\beta}}$ by Inner Stage optimization (Algorithm \ref{alg:middle}) with input $\boldsymbol{\lambda}$ and $\boldsymbol{\theta}$.
            \State Update $\log \boldsymbol{\lambda}$ and $\log \boldsymbol{\theta}$ by BFGS algorithm, where $\mathcal{L}^*_{\text{LA}}(\boldsymbol{\lambda}, \boldsymbol{\theta})$ and its gradient are evaluated as described in \ref{ss:deLAML}.
    \EndWhile
    \State \Return $\widehat{\boldsymbol{\lambda}}$,  $\widehat{\boldsymbol{\theta}}$, $\widehat{\boldsymbol{\phi}}^w$, $\widehat{\boldsymbol{\alpha}}^{f}$ and $\widehat{\boldsymbol{\beta}}$
    \end{algorithmic}
\end{algorithm}

\begin{algorithm}[H]
\caption{Inner Stage: Optimization for $\boldsymbol{\phi}^{w}$} \label{alg:middle}
\begin{algorithmic}[1]
\Require $\boldsymbol{\lambda}$ and $\boldsymbol{\theta}$
\While{not converged}
    \State Find $\widehat{\boldsymbol{\alpha}}^{f}$ and $\widehat{\boldsymbol{\beta}}$ by Algorithm \ref{alg:inner} with input $\boldsymbol{\phi}^{w}$, $\boldsymbol{\lambda}$ and $\boldsymbol{\theta}$.
    \State Evaluate Hessian $\nabla^2_{\boldsymbol{\phi}^w}\mathbf{Q}$ and gradient $\nabla_{\boldsymbol{\phi}^w}\mathbf{Q}$ (\ref{ss:deQ}). 
    \State Check the convergence: $\|\mathbf{\nabla_{\boldsymbol{\phi}^w} \mathbf{Q}}\|_2 \approx 0$.
    \State Find the Newton step $\Delta^{\boldsymbol{\phi}^w}$ (Algorithm \ref{alg:newton}).
    \While{$\mathbf{Q}(\boldsymbol{\phi}^w) > \mathbf{Q}(\boldsymbol{\phi}^w + \Delta^{\boldsymbol{\phi}^w})$}
        \State Step-halving: $\Delta^{\boldsymbol{\phi}^w} \leftarrow \Delta^{\boldsymbol{\phi}^w}/2$.
    \EndWhile
    \State Calculate $\boldsymbol{\alpha}^{w}$ according to $\boldsymbol{\phi}^w + \Delta^{\boldsymbol{\phi}^w}$ and compute $E(t) = \boldsymbol{\alpha}^w \boldsymbol{D}(t), \, t = 1, 2, \ldots, N$.\label{step:checkrange}
    \If{the range of $E(t)$ covers few (e.g. $\le$ 3) knots of the B-spline for $f$}
        \State Step-halving: $\Delta^{\boldsymbol{\phi}^w} \leftarrow \Delta^{\boldsymbol{\phi}^w}/2$.
        \State Go back to step \ref{step:checkrange}. 
    \EndIf
    \State Update: $\boldsymbol{\phi}^w \leftarrow \boldsymbol{\phi}^w + \Delta^{\boldsymbol{\phi}^w}$.
\EndWhile
\State \Return $\widehat{\boldsymbol{\phi}}^w$, $\widehat{\boldsymbol{\alpha}}^{f}$ and $\widehat{\boldsymbol{\beta}}$
\end{algorithmic}
\end{algorithm}

\begin{algorithm}[H]
\caption{Inner Stage: Optimization for $\boldsymbol{\alpha}^{f}$ and $\boldsymbol{\beta}$} \label{alg:inner}
\begin{algorithmic}[1]
\Require $\boldsymbol{\phi}^{w}$, $\boldsymbol{\lambda}$ and $\boldsymbol{\theta}$
\While{not converged}
    \State Evaluate Hessian $\nabla^2_{\boldsymbol{\alpha}^{f},\boldsymbol{\beta}} \mathcal{L}$ and gradient $\nabla_{\boldsymbol{\alpha}^{f},\boldsymbol{\beta}} \mathcal{L}$ (\ref{ss:deL}). 
    \State Check the convergence: $\|\nabla_{\boldsymbol{\alpha}^{f},\boldsymbol{\beta}} \mathcal{L}\|_2 \approx 0$. \label{step:inner_convergence}
    \State Find the Newton step $\Delta^{\boldsymbol{\alpha}^{f},\boldsymbol{\beta}}$ (Algorithm \ref{alg:newton}).
    \While{$\mathcal{L}([\boldsymbol{\alpha}^{f},\boldsymbol{\beta}]) > \mathcal{L}([\boldsymbol{\alpha}^{f},\boldsymbol{\beta}] + \Delta^{\boldsymbol{\alpha}^{f},\boldsymbol{\beta}})$}
        \State Step-halving: $\Delta^{\boldsymbol{\alpha}^{f},\boldsymbol{\beta}} \leftarrow \Delta^{\boldsymbol{\alpha}^{f},\boldsymbol{\beta}}/2$. \label{step:inner_halving}
    \EndWhile
    \State Update: $[\boldsymbol{\alpha}^{f},\boldsymbol{\beta}] \leftarrow [\boldsymbol{\alpha}^{f},\boldsymbol{\beta}] + \Delta^{\boldsymbol{\alpha}^{f},\boldsymbol{\beta}}$.
\EndWhile
\State \Return $\widehat{\boldsymbol{\alpha}}^{f}$ and $\widehat{\boldsymbol{\beta}}$
\end{algorithmic}
\end{algorithm}

We highlight the following step-having steps: (a) The step-havings in step 7, Algorithm \ref{alg:middle} and step 6, Algorithm \ref{alg:inner} are to ensure objective functions increase after Newton update. 
(b) The convergence of inner optimization is highly related to the current values of $\boldsymbol{\phi}^w$. If the range of $E(t)$ induced by the current $\boldsymbol{\phi}^w$ contains only few knots of B-spline for $f$, there is not enough variation for $f$, and hence the convergence of inner optimization is difficult. 
Therefore, we propose to check the range of $E(t)$, and halve the Newton step until the range contains enough knots, e.g. 4 knots, 
in step 11, Algorithm \ref{alg:middle}.

The choice of starting values influences the speed and convergence of optimization algorithms. In Newton's method in the inner stage, we choose the results from the previous iteration as the starting values. 
In the outer stage, there is no standard guidance for the choice of starting values for BFGS. We recommend fitting a GAM (with the 0-day lagged exposure, for example) and using the fitted values as the starting values for parameters excluding $\log \lambda^w$. The starting value for $\log \lambda^w$ is chosen as a moderate number arbitrarily. 
In our proposed R package {\tt aceDLNM}, we fit the GAM using the modern big additive model framework \citep{wood2017bam} provided by {\tt mgcv::bam}, and the starting value for $\log \lambda^w$ is fixed as $6$ by default. We also provide an option for users to choose starting values based on prior knowledge. 

\subsection{Newton Step}
\label{sss:newton}
Suppose the Hessian and gradient are $\mathbf{H}$ and $\nabla$. The Newton step is obtained by $\mathbf{H}^{-1}\Delta$. To maintain Newton Step in an ascent direction and an appropriate length, $\mathbf{H}$ should be positive definite. Let $\mathbf{H} = \boldsymbol{V}\boldsymbol{\Lambda}\boldsymbol{V}^{\top}$ by eigen-decomposition. $\mathbf{H}$ is not positive definite numerically, if some $\boldsymbol{\Lambda}_{ii}$ are close to or smaller than zero. 

To address the issue of $\mathbf{H}$, we adopt the following procedure. First, we check the minimum eigenvalue of $\mathbf{H}$ using the Lanczos algorithm \citep{demmel1997applied}. If the minimum eigenvalue is positive and not close to zero, we solve $\mathbf{H}^{-1}\nabla$ through Cholesky decomposition, otherwise, we find a Newton step by Algorithm \ref{alg:newton}. The Algorithm \ref{alg:newton} based on the eigen-decomposition of $\mathbf{H}$ follows the new Q-Newton's method \citep{truong2023fast}.
The checking for the minimum eigenvalue is to avoid the eigen-decomposition --- which is more expensive than Cholesky --- in some iterations.

\begin{algorithm}[H]
    \caption{Find Newton Step $\Delta$}\label{alg:newton}
    \begin{algorithmic}[1]
    \Require Hessian $\mathbf{H}$ and gradient $\nabla$ with dimension $k$.  
    \State Define $\mathcal{I}$ as a $k \times k$ diagonal matrix, with diagnomal element $\mathcal{I}_{i,i}$. 
    \State Decompose $\mathbf{H} = \boldsymbol{V}\boldsymbol{\Lambda}\boldsymbol{V}^\top$ by eigen-decomposition, with eigenvalues $\lambda_1, \cdots,\lambda_{k}$.
    \If{$\prod_{i=1}^{k} \lambda_i \approx 0$}
    \For{$i = 1, \cdots, k$}
        \State Update $\lambda_i \leftarrow \lambda_i + \|\nabla\|_2$.
    \EndFor
    \EndIf
    \For{$i = 1, \cdots, k$}
        \State $\mathcal{I}_{i,i} = 1/|\lambda_i|$. 
    \EndFor
    \State \Return Newton step $\Delta = \boldsymbol{V}\mathcal{I}\boldsymbol{V}^{\top}\nabla$.
    \end{algorithmic}
\end{algorithm}

\section{Derivative Computation}

The first- and second-order derivatives of $\mathcal{L}$ are discussed in \ref{ss:deL}.  
In \ref{ss:deQ}, we introduce the details of the first- and second-order derivatives computation of profile log-likelihood $\mathbf{Q}$. 
The first-order derivative of log-LAML $\mathcal{L}_{\text{LA}}^*$ w.r.t $\log \boldsymbol{\lambda}$ and $\log\boldsymbol{\theta}$ are described in \ref{ss:deLAML}. 

In this paper, we focus on the ACE-DLNM with a negative binomial distribution motivated by the Health Canada data. This framework applies generally to more general types of data, 
by taking the corresponding pdf/pmf in \ref{sss:pmf} and the others hold the same.

All the derivatives are computed using {\tt cpp} code available at the R package file \url{https://github.com/tianyi-pan/aceDLNM/blob/main/src/aceDLNM.cpp}. 

\subsection{Derivative of $\mathcal{L}$}
\label{ss:deL}

We highlight the following terms in the calculation of the gradient and Hessian of $\mathcal{L}$: 
the probability mass function of the negative binomial distribution, the mean model, and the Jacobian matrix. The remaining parts of the derivatives are straightforward. %

\subsubsection{Probability mass function}
\label{sss:pmf}
The pmf of the negative binomial distribution is
$$
\begin{aligned}
f(x| \mu, \theta) = \exp \left\{x \log \left(\frac{\mu}{\theta+\mu}\right) + \theta \log \left(\frac{\theta}{\theta+\mu}\right) + \log\left(\frac{\Gamma(x+\theta)}{\Gamma(\theta)x!}\right)\right\},
\end{aligned}
$$
where $\Gamma(\cdot)$ is the gamma function. 

Taking derivative of $\log f(x| \mu, \theta)$ w.r.t $\mu$ and $\theta$ leads to
$$
\begin{aligned}
    \frac{\partial}{\partial \mu} \log f(x|\mu, \theta) &= \frac{x}{\mu} - \frac{\theta+x}{\theta + \mu},\\
    \frac{\partial}{\partial \theta} \log f(x|\mu, \theta) &= \log \left(\frac{\theta}{\theta+\mu}\right) + \frac{\mu - x}{\theta+\mu} + \psi(\theta+x) - \psi(\theta),\\
    \frac{\partial}{\partial \log \theta} \log f(x|\mu, \theta) &= \theta \frac{\partial}{\partial \theta} \log f(x|\mu, \theta),
\end{aligned}
$$
where $\psi(s) = d \log 
\Gamma(s) / ds$. 

The 2nd-order derivatives are given by
$$
\begin{aligned}
\frac{\partial^2}{\partial \mu^2} \log f(x|\mu, \theta) &= -\frac{x}{\mu^2} + \frac{\theta + x}{(\theta + \mu)^2},\\
\frac{\partial^2}{\partial \theta^2} \log f(x|\mu, \theta) &= \frac{1}{\theta} - \frac{1}{\theta+\mu}-\frac{\mu-x}{\left(\theta+\mu\right)^2} + \psi'(\theta+x) - \psi'(\theta),\\
\frac{\partial^2}{\partial \log \theta^2} \log f(x|\mu, \theta) &= \theta^2 \frac{\partial^2}{\partial \theta^2} \log f(x|\mu, \theta) + \theta \frac{\partial}{\partial \theta} \log f(x|\mu, \theta),\\
\frac{\partial^2}{\partial \theta \partial \mu} \log f(x|\mu, \theta) &= \frac{x-\mu}{\left(\theta+\mu\right)^2}, \\
\frac{\partial^2}{\partial \log \theta \partial \mu} \log f(x|\mu, \theta) &= \theta \frac{\partial^2}{\partial \theta \partial \mu} \log f(x|\mu, \theta),
\end{aligned}
$$
where $\psi'(x+\theta) = d \psi(x)/ d x$. 
However, $\Gamma(s)$, $\psi(x)$ and $\psi'(x+\theta)$
are intractable analytically.
We adopt the Lanczos approximation \citep{lanczos1964precision} for the gamma function  $\Gamma(s)$. The derivative of $\Gamma(s)$ is approximated by the Algorithm AS103 in \citet{bernardo1976psi}. Both algorithms are differentiable using {\tt autodiff}, which is required for the derivative calculation of $\mathcal{L}^*_{\text{LA}}$ (\ref{ss:deLAML}).

\subsubsection{Mean model}
By the model specification, we have
$$
\log \mu_t = f\left\{\boldsymbol{\alpha}^{w\top} \boldsymbol{D}(t)\right\} = \sum_{q} b^f_q(\boldsymbol{\alpha}^{w\top} \boldsymbol{D}(t)) \alpha_{q}^f.
$$
Then, we have the 1st-order derivatives
$$
\begin{aligned}
    \frac{\partial \log \mu_t}{\partial \boldsymbol{\alpha}^{w}} &= \boldsymbol{D}(t) \sum_{q} b^{'f}_q(\boldsymbol{\alpha}^{w\top} \boldsymbol{D}(t))  \alpha_{q}^f,\\
    \frac{\partial \log \mu}{\partial \boldsymbol{\alpha}^{f}} &= \left\{b^f_q(\boldsymbol{\alpha}^{w\top} \boldsymbol{D}(t))\right\}_{q},\\
\end{aligned}
$$
and the 2nd-order derivatives
$$
\begin{aligned}
    \frac{\partial^2 \log \mu_t}{\partial \boldsymbol{\alpha}^{w2}} &= \boldsymbol{D}(t) \boldsymbol{D}(t) ^\top \sum_{q} b^{''f}_q(\boldsymbol{\alpha}^{w\top} \boldsymbol{D}(t)))  \alpha_{q}^f,\\
    \frac{\partial^2 \log \mu_t}{\partial \boldsymbol{\alpha}^{f2}} &= 0,\\
    \frac{\partial^2 \log \mu_t}{\partial \boldsymbol{\alpha}^{f} \boldsymbol{\alpha}^{w}} &= \left\{b^{'f}_q(\boldsymbol{\alpha}^{w\top} \boldsymbol{D}(t))\right\}_{j} \boldsymbol{D}(t)^\top.
\end{aligned}
$$
The derivatives of $\mu$ are given by
$$  
\begin{aligned}
    \frac{\partial \mu_t}{\partial \boldsymbol{\alpha}^{w}} &= \mu_t \frac{\partial \log \mu_t}{\partial \boldsymbol{\alpha}^{w}},\\
    \frac{\partial \mu_t}{\partial \boldsymbol{\alpha}^{f}} &= \mu_t \frac{\partial \log \mu_t}{\partial \boldsymbol{\alpha}^{f}},\\
    \frac{\partial^2 \mu_t}{\partial \boldsymbol{\alpha}^{w2}} &= \mu_t \left(\frac{\partial \log \mu_t}{\partial \boldsymbol{\alpha}^{w}}\right)^2 + \mu_t \frac{\partial^2 \log \mu_t}{\partial \boldsymbol{\alpha}^{w2}},\\
    \frac{\partial^2 \mu_t}{\partial \boldsymbol{\alpha}^{f2}} &= \mu_t \left(\frac{\partial \log \mu_t}{\partial \boldsymbol{\alpha}^{f}}\right)^2 + \mu_t \frac{\partial^2 \log \mu_t}{\partial \boldsymbol{\alpha}^{f2}},\\
    \frac{\partial^2 \mu_t}{\partial \boldsymbol{\alpha}^{w} \partial \boldsymbol{\alpha}^{f}} &= \mu_t \frac{\partial \log \mu_t}{\partial \boldsymbol{\alpha}^{w}} \frac{\partial \log \mu_t}{\partial \boldsymbol{\alpha}^{f}} + \mu_t \frac{\partial^2 \log \mu_t}{\partial \boldsymbol{\alpha}^{w} \partial \boldsymbol{\alpha}^{f}}.
\end{aligned}
$$
The $s$-order derivative of the B-spline function $b_{l,p}(x)$ is given by Equation \ref{eq:deriv_Bspline}; see more details in \citet{wood2017generalized}.

\subsubsection{Reparameterization}
\label{sss:repa}
We make inferences for the unconstrained parameter $\boldsymbol{\phi}^w$. 
Define $\boldsymbol{\phi}^{w L} = \left[1, \boldsymbol{\phi}^w\right]^\top$. The Jacobian is given by
$$
\frac{\partial \boldsymbol{\alpha}^{w}}{\partial \boldsymbol{\phi}^{wL}} = \left(\boldsymbol{\phi}^{{wL}^\top} \mathbf{C} \boldsymbol{\phi}^{wL}\right)^{-\frac{1}{2}} \mathbf{I} - \boldsymbol{\phi}^{wL} \boldsymbol{\phi}^{wL^\top} \mathbf{C}\left(\boldsymbol{\phi}^{wL^\top} \mathbf{C} \boldsymbol{\phi}^{wL}\right)^{-\frac{3}{2}},
$$

Then, we obtain $\partial \boldsymbol{\alpha}^{w}/\partial \boldsymbol{\phi}^w$ 
by removing the first column of $\partial \boldsymbol{\alpha}^{w}/\partial \boldsymbol{\phi}^{wL}$.

\subsection{Derivative of $\mathbf{Q}$}
\label{ss:deQ}
\subsubsection{First-Order Derivative of $\mathbf{Q}$}
The profile log-likelihood $\mathbf{Q}(\boldsymbol{\phi}^w; \boldsymbol{\lambda}, \boldsymbol{\theta})$
is related to the estimates from the inner optimization, which are implicit functions of $\boldsymbol{\phi}^w$. 
We obtain the derivatives by implicit differentiation. The first-order derivative is given by
$$
\begin{aligned}
    \frac{\partial \mathbf{Q}(\boldsymbol{\phi}^w; \boldsymbol{\lambda}, \boldsymbol{\theta})}{\partial \boldsymbol{\phi}^w} =& \frac{\partial \mathcal{L}\left(\boldsymbol{\phi}^w,\boldsymbol{\alpha}^f, \boldsymbol{\beta};  \boldsymbol{\lambda}, \boldsymbol{\theta}\right)}{\partial \boldsymbol{\phi}^w} \Bigr|_{\boldsymbol{\alpha}^f = \widehat{\boldsymbol{\alpha}}^{f}(\boldsymbol{\phi}^w), \boldsymbol{\beta} = \widehat{\boldsymbol{\beta}}(\boldsymbol{\phi}^w)}\\ 
    &+\frac{\partial \left[\widehat{\boldsymbol{\alpha}}^{f}(\boldsymbol{\phi}^w), \widehat{\boldsymbol{\beta}}(\boldsymbol{\phi}^w)\right]}{\partial \boldsymbol{\phi}^w} \frac{\partial \mathcal{L}\left(\boldsymbol{\phi}^w,\widehat{\boldsymbol{\alpha}}^{f}(\boldsymbol{\phi}^w), \widehat{\boldsymbol{\beta}}(\boldsymbol{\phi}^w); \boldsymbol{\lambda}, \boldsymbol{\theta}\right)}{\partial \left[\boldsymbol{\alpha}^f, \boldsymbol{\beta}\right]^{\top}}.
\end{aligned}
$$
By definition, we have $\partial \mathcal{L}\left(\boldsymbol{\phi}^w,\widehat{\boldsymbol{\alpha}}^{f}(\boldsymbol{\phi}^w), \widehat{\boldsymbol{\beta}}(\boldsymbol{\phi}^w); \boldsymbol{\lambda}, \boldsymbol{\theta}\right)/\partial \left[\boldsymbol{\alpha}^f, \boldsymbol{\beta}\right]^{\top} = 0$. Therefore, we have
$$
\frac{\partial \mathbf{Q}(\boldsymbol{\phi}^w; \boldsymbol{\lambda}, \boldsymbol{\theta})}{\partial \boldsymbol{\phi}^w} = \frac{\partial \mathcal{L}\left(\boldsymbol{\phi}^w,\boldsymbol{\alpha}^f, \boldsymbol{\beta};  \boldsymbol{\lambda}, \boldsymbol{\theta}\right)}{\partial \boldsymbol{\phi}^w} \Bigr|_{\boldsymbol{\alpha}^f = \widehat{\boldsymbol{\alpha}}^{f}(\boldsymbol{\phi}^w), \boldsymbol{\beta} = \widehat{\boldsymbol{\beta}}(\boldsymbol{\phi}^w)},
$$
which is a part of the first-order derivative of $\mathcal{L}$.

\subsubsection{Second-Order Derivative of $\mathbf{Q}$}
\label{sss:de2Q}

We define $\mathbf{H}^{\boldsymbol{\eta},w}$ as
$$
\mathbf{H}^{\boldsymbol{\eta},w} = \frac{\partial^2 \mathcal{L}\left(\boldsymbol{\phi}^w,\boldsymbol{\alpha}^f, \boldsymbol{\beta};  \boldsymbol{\lambda}, \boldsymbol{\theta}\right)}{\partial \left[\boldsymbol{\alpha}^f, \boldsymbol{\beta}\right] \partial \boldsymbol{\phi}^w}\Bigr|_{\boldsymbol{\alpha}^f = \widehat{\boldsymbol{\alpha}}^{f}(\boldsymbol{\phi}^w), \boldsymbol{\beta} = \widehat{\boldsymbol{\beta}}(\boldsymbol{\phi}^w)}.
$$
Then, the second-order derivative can be written as
$$
\begin{aligned}
    \frac{\partial^2 \mathbf{Q}(\boldsymbol{\phi}^w; \boldsymbol{\lambda}, \boldsymbol{\theta})}{\partial \boldsymbol{\phi}^{w 2}} =& 
    \frac{\partial^2 \mathcal{L}\left(\boldsymbol{\phi}^w,\boldsymbol{\alpha}^f, \boldsymbol{\beta};  \boldsymbol{\lambda}, \boldsymbol{\theta}\right)}{\partial \boldsymbol{\phi}^{w 2}} \Bigr|_{\boldsymbol{\alpha}^f = \widehat{\boldsymbol{\alpha}}^{f}(\boldsymbol{\phi}^w), \boldsymbol{\beta} = \widehat{\boldsymbol{\beta}}(\boldsymbol{\phi}^w)} \\
    &+\frac{\partial \left[\widehat{\boldsymbol{\alpha}}^{f}(\boldsymbol{\phi}^w), \widehat{\boldsymbol{\beta}}(\boldsymbol{\phi}^w)\right]}{\partial \boldsymbol{\phi}^w} \mathbf{H}^{\boldsymbol{\eta},w}. 
\end{aligned}
$$
The derivative $\partial \left[\widehat{\boldsymbol{\alpha}}^{f}(\boldsymbol{\phi}^w), \widehat{\boldsymbol{\beta}}(\boldsymbol{\phi}^w)\right]/\partial \boldsymbol{\phi}^w$ is obtained through the implicit differentiation. Given that $\partial \mathcal{L}\left(\boldsymbol{\phi}^w,\widehat{\boldsymbol{\alpha}}^{f}(\boldsymbol{\phi}^w), \widehat{\boldsymbol{\beta}}(\boldsymbol{\phi}^w); \boldsymbol{\lambda}, \boldsymbol{\theta}\right)/\partial \left[\boldsymbol{\alpha}^f, \boldsymbol{\beta}\right] = 0$, taking derivative with respect to $\boldsymbol{\phi}^w$ on both sides leads to
$$
\begin{aligned}
    \frac{\partial^2 \mathcal{L}\left(\boldsymbol{\phi}^w,\widehat{\boldsymbol{\alpha}}^{f}(\boldsymbol{\phi}^w), \widehat{\boldsymbol{\beta}}(\boldsymbol{\phi}^w); \boldsymbol{\lambda}, \boldsymbol{\theta}\right)}{\partial \left[\boldsymbol{\alpha}^f, \boldsymbol{\beta}\right]^2} \frac{\partial \left[\widehat{\boldsymbol{\alpha}}^{f}(\boldsymbol{\phi}^w), \widehat{\boldsymbol{\beta}}(\boldsymbol{\phi}^w)\right]}{\partial \boldsymbol{\phi}^w} = -\mathbf{H}^{\boldsymbol{\eta},w}.
\end{aligned}
$$
Solving this equation gives $\partial \left[\widehat{\boldsymbol{\alpha}}^{f}(\boldsymbol{\phi}^w), \widehat{\boldsymbol{\beta}}(\boldsymbol{\phi}^w)\right]/\partial \boldsymbol{\phi}^w$. 

\subsection{Derivative of $\mathcal{L}^*_{\text{LA}}$}
\label{ss:deLAML}
In the outer stage, we use the BFGS algorithm to maximize $\mathcal{L}^*_{\text{LA}}$ for $\log \boldsymbol{\lambda}$ and $\log \boldsymbol{\theta}$. In this Newton-type algorithm, the gradient is required.

Recall that the log-LAML is
$$
\begin{aligned}
\mathcal{L}^*_{\text{LA}}(\boldsymbol{\lambda}, \boldsymbol{\theta}) =&
\mathcal{L}(\widehat{\boldsymbol{u}}\left(\boldsymbol{\lambda}, \boldsymbol{\theta}\right); \boldsymbol{\lambda}, \boldsymbol{\theta}) - 
\frac{1}{2} \log \left\{\mathrm{det} \boldsymbol{\mathcal{H}}\left(\widehat{\boldsymbol{u}}\left(\boldsymbol{\lambda}, \boldsymbol{\theta}\right); \boldsymbol{\lambda}, \boldsymbol{\theta}\right)\right\} +
\frac{M}{2} \log(2 \pi) \\ &+ 
\frac{1}{2}\log|\lambda^w \boldsymbol{S}^w|_{+} + \frac{1}{2}\log|\lambda^f \boldsymbol{S}^f|_{+} + \frac{1}{2}\sum_{j=1}^p\log|\lambda^{h}_j \boldsymbol{S}^{h}_j|_{+}. 
\end{aligned}
$$

We will describe the first-order derivatives of $\mathcal{L}(\widehat{\boldsymbol{u}}\left(\boldsymbol{\lambda}, \boldsymbol{\theta}\right); \boldsymbol{\lambda}, \boldsymbol{\theta})$. The other derivatives are discussed in Section 3.2.3. 

The first-order derivative of $\mathcal{L}(\widehat{\boldsymbol{u}}\left(\boldsymbol{\lambda}, \boldsymbol{\theta}\right); \boldsymbol{\lambda}, \boldsymbol{\theta})$ is given by
\begin{equation*}
    \frac{\partial \mathcal{L}(\widehat{\boldsymbol{u}}\left(\boldsymbol{\lambda}, \boldsymbol{\theta}\right); \boldsymbol{\lambda}, \boldsymbol{\theta})}{\partial \left[\log\boldsymbol{\lambda}, \log\boldsymbol{\theta}\right]} =
    \frac{\partial \mathcal{L} (\boldsymbol{u}, \boldsymbol{\lambda}, \boldsymbol{\theta})}{\partial \left[\log\boldsymbol{\lambda}, \log\boldsymbol{\theta}\right]}\Bigr|_{\boldsymbol{u} = \widehat{\boldsymbol{u}}(\boldsymbol{\lambda}, \boldsymbol{\theta})}
    + \frac{\partial \widehat{\boldsymbol{u}}(\boldsymbol{\lambda}, \boldsymbol{\theta})}{\partial \left[\log\boldsymbol{\lambda}, \log\boldsymbol{\theta}\right]} \left[\frac{\partial \mathcal{L} (\boldsymbol{u}, \boldsymbol{\lambda}, \boldsymbol{\theta})}{\partial \boldsymbol{u}}\Bigr|_{\boldsymbol{u} = \widehat{\boldsymbol{u}}(\boldsymbol{\lambda}, \boldsymbol{\theta})}\right]. 
\label{eq:gr_LAMA}
\end{equation*}
The derivatives of $\mathcal{L}$ are described in \ref{ss:deL}. We address the unsolved part, the derivative of $\widehat{\boldsymbol{u}}(\boldsymbol{\lambda}, \boldsymbol{\theta})$, using the implicit differentiation. 

The estimator $\widehat{\boldsymbol{\phi}}^w(\boldsymbol{\lambda}, \boldsymbol{\theta})$ is obtained from maximizing the profile likelihood $\mathbf{Q}(\boldsymbol{\phi}^w; \boldsymbol{\lambda}, \boldsymbol{\theta})$. 
Therefore, we have,
$\partial \mathbf{Q}(\widehat{\boldsymbol{\phi}}^w; \boldsymbol{\lambda}, \boldsymbol{\theta})/ \partial \boldsymbol{\phi}^w = 0$. The derivative can be written as
$$
\begin{aligned}
0=\frac{\partial \mathbf{Q}(\widehat{\boldsymbol{\phi}}^w; \boldsymbol{\lambda}, \boldsymbol{\theta})}{\partial \boldsymbol{\phi}^w} =& \frac{\partial \mathcal{L}\left(\boldsymbol{\phi}^w,\boldsymbol{\alpha}^f, \boldsymbol{\beta};  \boldsymbol{\lambda}, \boldsymbol{\theta}\right)}{\partial \boldsymbol{\phi}^w} \Bigr|_{\boldsymbol{\phi}^w = \widehat{\boldsymbol{\phi}}^w(\boldsymbol{\lambda}, \boldsymbol{\theta}), \boldsymbol{\alpha}^f = \widehat{\boldsymbol{\alpha}}^{f}(\boldsymbol{\phi}^w), \boldsymbol{\beta} = \widehat{\boldsymbol{\beta}}(\boldsymbol{\phi}^w)}  \\ 
&+\frac{\partial \mathcal{L}\left(\boldsymbol{\phi}^w,\widehat{\boldsymbol{\alpha}}^{f}(\boldsymbol{\phi}^w), \widehat{\boldsymbol{\beta}}(\boldsymbol{\phi}^w); \boldsymbol{\lambda}, \boldsymbol{\theta}\right)}{\partial \left[\boldsymbol{\alpha}^f, \boldsymbol{\beta}\right]}\frac{\partial \left[\widehat{\boldsymbol{\alpha}}^{f}(\boldsymbol{\phi}^w), \widehat{\boldsymbol{\beta}}(\boldsymbol{\phi}^w)\right]}{\partial \boldsymbol{\phi}^w}.
\end{aligned}
$$
Recall that $\widehat{\boldsymbol{u}}(\boldsymbol{\lambda}, \boldsymbol{\theta}) = \left[\widehat{\boldsymbol{\alpha}}(\boldsymbol{\lambda}, \boldsymbol{\theta}),  \widehat{\boldsymbol{\beta}}(\boldsymbol{\lambda}, \boldsymbol{\theta}), \widehat{\boldsymbol{\phi}}^w(\boldsymbol{\lambda}, \boldsymbol{\theta})\right]^\top$. The estimates $\widehat{\boldsymbol{\alpha}}^{f}, \widehat{\boldsymbol{\beta}}$ are obtained from maximizing $\mathcal{L}$, and therefore,
\begin{equation}
\frac{\partial \mathcal{L}\left(\boldsymbol{\phi}^w,\widehat{\boldsymbol{\alpha}}^{f}(\boldsymbol{\phi}^w), \widehat{\boldsymbol{\beta}}(\boldsymbol{\phi}^w); \boldsymbol{\lambda}, \boldsymbol{\theta}\right)}{\partial \left[\boldsymbol{\alpha}^f, \boldsymbol{\beta}\right]} = 0.
\label{eq:laml1}
\end{equation}
Hence, we have
\begin{equation}
    \frac{\partial \mathcal{L}\left(\boldsymbol{\phi}^w,\boldsymbol{\alpha}^f, \boldsymbol{\beta};  \boldsymbol{\lambda}, \boldsymbol{\theta}\right)}{\partial \boldsymbol{\phi}^w} \Bigr|_{\boldsymbol{\phi}^w = \widehat{\boldsymbol{\phi}}^w(\boldsymbol{\lambda}, \boldsymbol{\theta}), \boldsymbol{\alpha}^f = \widehat{\boldsymbol{\alpha}}^{f}(\boldsymbol{\phi}^w), \boldsymbol{\beta} = \widehat{\boldsymbol{\beta}}(\boldsymbol{\phi}^w)} = 0.
    \label{eq:laml2}
\end{equation}
Combining the Equations \ref{eq:laml1} and \ref{eq:laml2} yields
$$
\frac{\partial \mathcal{L}\left(\widehat{\boldsymbol{u}}(\boldsymbol{\lambda}, \boldsymbol{\theta}), \boldsymbol{\lambda}, \boldsymbol{\theta}\right)}{\partial \boldsymbol{u}} = 0. 
$$
Taking derivative with respect to $\left[\log \boldsymbol{\lambda}, \log \boldsymbol{\theta}\right]$ on both sides of the equation leads to
$$
\frac{\partial^2 \mathcal{L}\left(\widehat{\boldsymbol{u}}(\boldsymbol{\lambda}, \boldsymbol{\theta}), \boldsymbol{\lambda}, \boldsymbol{\theta}\right)}{\partial \boldsymbol{u}^2} 
\frac{\partial \widehat{\boldsymbol{u}}(\boldsymbol{\lambda}, \boldsymbol{\theta})}{\partial \left[\log\boldsymbol{\lambda}, \log\boldsymbol{\theta}\right]}  = - \frac{\partial^2 \mathcal{L}(\boldsymbol{u}, \boldsymbol{\lambda}, \boldsymbol{\theta})}{\partial \boldsymbol{u} \partial \left[\log \boldsymbol{\lambda}, \log \boldsymbol{\theta}\right]}\Bigr|_{\boldsymbol{u} =\widehat{\boldsymbol{u}}( \boldsymbol{\lambda}, \boldsymbol{\theta})}.
$$
Solving this equation gives the derivative of $\widehat{\boldsymbol{u}}(\boldsymbol{\lambda}, \boldsymbol{\theta})$.

\section{Delta Method}
To quantify the uncertainty in maximum likelihood estimators with transformation, the delta method is a standard manner. The unconstrained parameters $\boldsymbol{u} = \left[\boldsymbol{\phi}^w, \boldsymbol{\alpha}^{f}, \boldsymbol{\beta}\right]^\top$ is estimated by maximizing (penalized) log-likelihood. We estimate the variance of $\widehat{\boldsymbol{u}}\left(\boldsymbol{\lambda}, \boldsymbol{\theta}\right)$ by the inverse of the negative Hessian matrix, i.e. 
$\widehat{\mathrm{Var}}(\widehat{\boldsymbol{u}}\left(\boldsymbol{\lambda}, \boldsymbol{\theta}\right)) = \left[\boldsymbol{\mathcal{H}}\left(\widehat{\boldsymbol{u}}\left(\boldsymbol{\lambda}, \boldsymbol{\theta}\right)\right)\right]^{-1}$. Then, the estimated variance of the constrained parameters, $\left[\boldsymbol{\alpha}^{w}, \boldsymbol{\alpha}^{f}, \boldsymbol{\beta}\right]^\top$, in the original parameter space is given by
$$
\widehat{\mathrm{Var}}\left(\left[\boldsymbol{\alpha}^{w}, \boldsymbol{\alpha}^{f}, \boldsymbol{\beta}\right]^\top\right) = \mathbf{J}\left[\widehat{\boldsymbol{u}}\left(\boldsymbol{\lambda}, \boldsymbol{\theta}\right)\right]\left[\boldsymbol{\mathcal{H}}\left(\widehat{\boldsymbol{u}}\left(\boldsymbol{\lambda}, \boldsymbol{\theta}\right)\right)\right]^{-1}\mathbf{J}\left[\widehat{\boldsymbol{u}}\left(\boldsymbol{\lambda}, \boldsymbol{\theta}\right)\right]^\top,
$$
where $\mathbf{J}\left[\widehat{\boldsymbol{u}}\left(\boldsymbol{\lambda}, \boldsymbol{\theta}\right)\right]$ is the Jacobian matrix of reparametrization from $\left[\boldsymbol{\phi}^w, \boldsymbol{\alpha}^{f}, \boldsymbol{\beta}\right]^\top$ to $\left[\boldsymbol{\alpha}^{w}, \boldsymbol{\alpha}^{f}, \boldsymbol{\beta}\right]^\top$ evaluated at $\widehat{\boldsymbol{u}}\left(\boldsymbol{\lambda}, \boldsymbol{\theta}\right)$. The details of the Jacobian matrix can be found in \ref{sss:repa}.  

The approximated $(1-\alpha)$ confidence intervals for $i^{th}$ parameters are obtained as
$$
\left[\widehat{\boldsymbol{\alpha}}^{w}, \widehat{\boldsymbol{\alpha}}^{f}, \widehat{\boldsymbol{\beta}}\right]_i^\top \pm z_{\alpha/2} \left\{\mathbf{J}\left[\widehat{\boldsymbol{u}}\left(\boldsymbol{\lambda}, \boldsymbol{\theta}\right)\right]\left[\boldsymbol{\mathcal{H}}\left(\widehat{\boldsymbol{u}}\left(\boldsymbol{\lambda}, \boldsymbol{\theta}\right)\right)\right]^{-1}\mathbf{J}\left[\widehat{\boldsymbol{u}}\left(\boldsymbol{\lambda}, \boldsymbol{\theta}\right)\right]^\top\right\}_{ii}^{1/2},
$$
where $z_{1-\alpha/2}$ is the $1-\alpha/2$ quantile of 
a standard normal distribution. We replace $\boldsymbol{\lambda}$ and $\boldsymbol{\theta}$ by their estimates, and their uncertainty is ignored as is typical in GAM contexts \citep{wood2016smoothing}. 

To our knowledge, there are no existing results that guarantee the performance of the delta method in finite samples. In Simulation A, we found that it yielded poor finite sample performance (see Web Table 5). Instead, we recommend the sampling method described in Section 3.4.

\clearpage

\section{Simulations}

\subsection{Data-Generation Details}

The true functions for data generation are visualized in Web Figure 1. 

\begin{figure}[H]
    \centering
    \begin{subfigure}{0.45\textwidth}
        \centering
        \includegraphics[width=\linewidth]{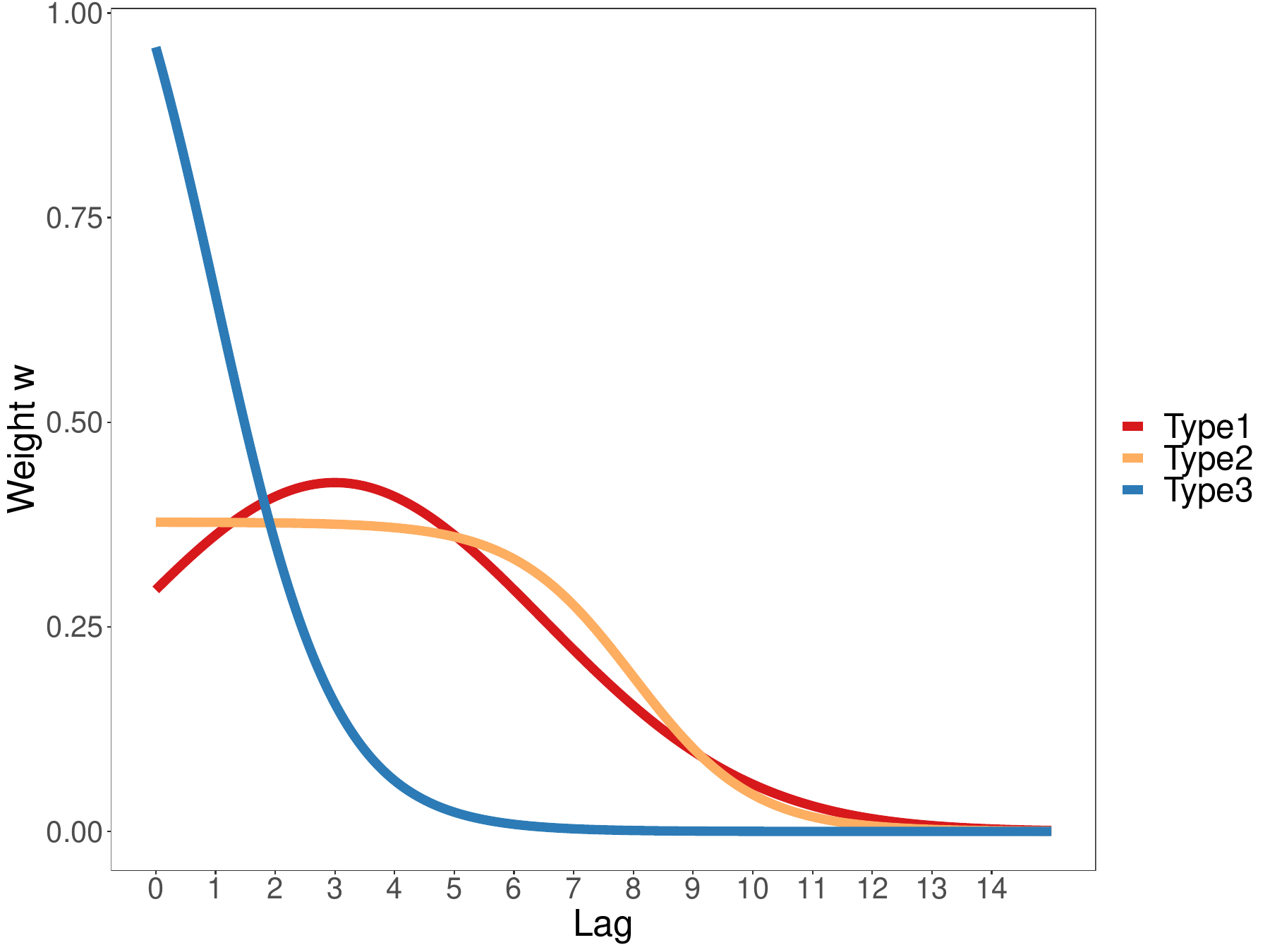}
        \caption{True Weight Functions $w(l)$}
        \label{fig:true-wl}
    \end{subfigure}%
    \begin{subfigure}{0.45\textwidth}
        \centering
        \includegraphics[width=\linewidth]{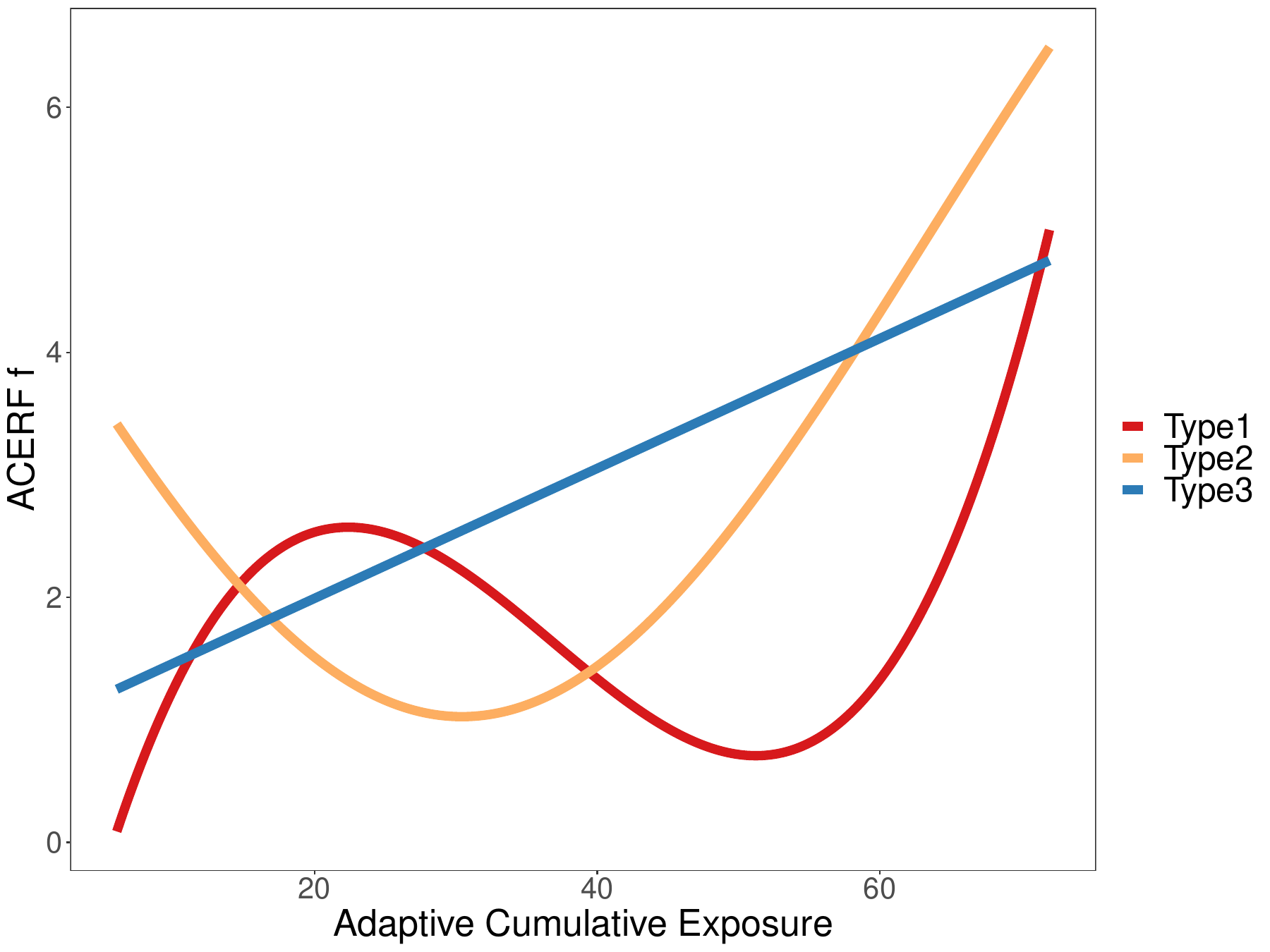}
        \caption{True ACERFs $f(E)$}
        \label{fig:true-fE}
    \end{subfigure}
    \begin{subfigure}{\textwidth}
        \centering
       \includegraphics[width=0.45\linewidth]{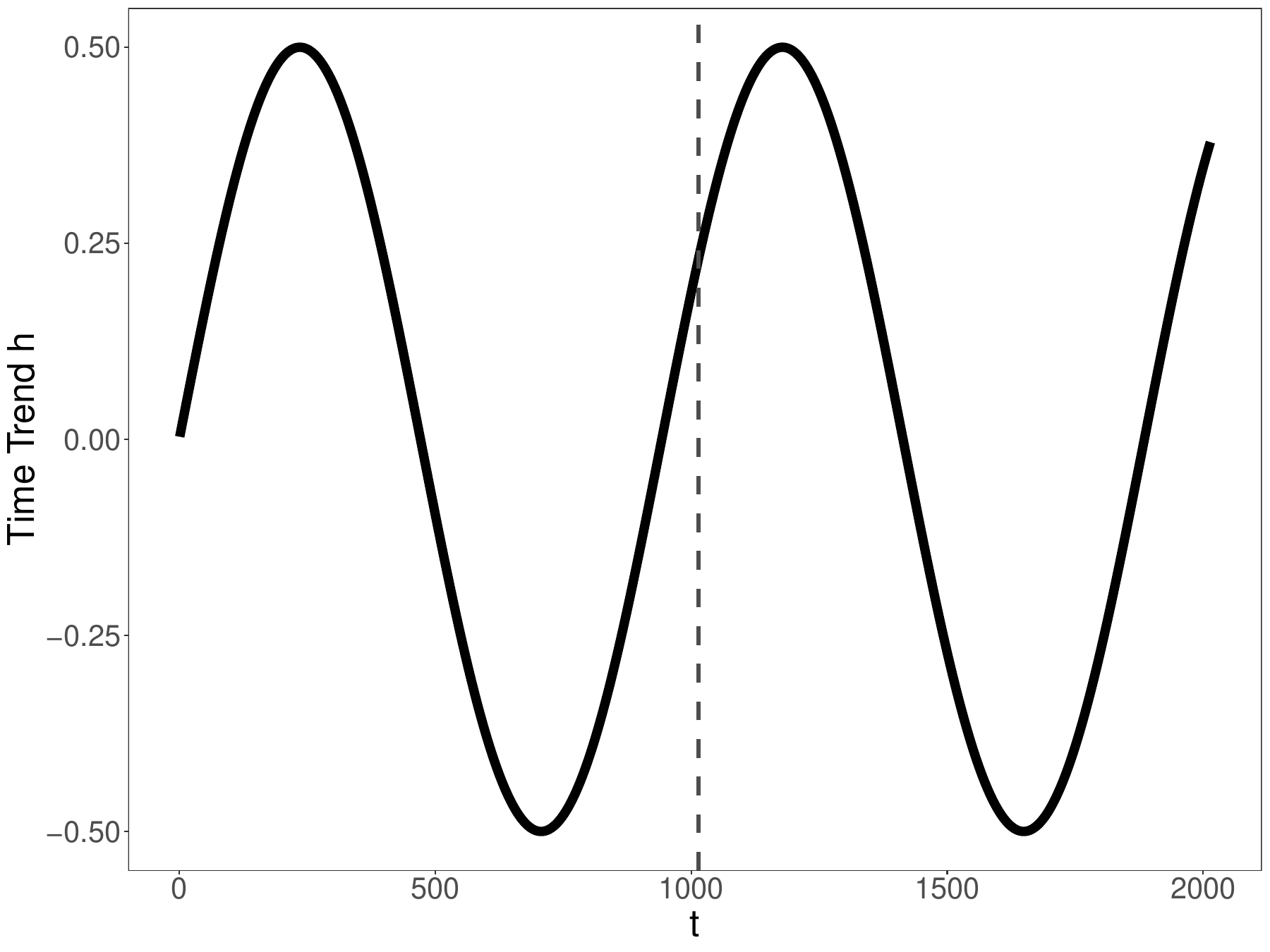}
        \caption{True Time Trend $h(t)$}
    \end{subfigure}
    \caption{Web Figure 1. True functions in Simulation A, including (a) three types of $w$, (b) three types of $f$, and (c) $h$ where the curve on the left of the dashed line is for the sample size of 1,000 and the full curve is for the sample size of 2,000.}
\end{figure}

We use the $\text{PM}_{2.5}$ concentrations to generate data in the simulation studies. In Web Figure 2a, we visualize the $\text{PM}_{2.5}$ concentrations from Jan. 1, 2001 to Oct. 11, 2003 (corresponding to a sample size of 1000).
To illustrate seasonality in the $\text{PM}_{2.5}$ concentrations, we fit a generalized additive model, regressing the log-transformed $\text{PM}_{2.5}$ concentrations on a smooth function of date for the time trend and a cyclic smooth function of month for seasonality (the residual diagnostics show no evidence of a lack of model fit). We plot the estimated function in Web Figure 2b, and observe a seasonal pattern. 

\begin{figure}[H]
    \centering
    \begin{subfigure}{\textwidth}
        \centering
        \includegraphics[width=0.7\linewidth]{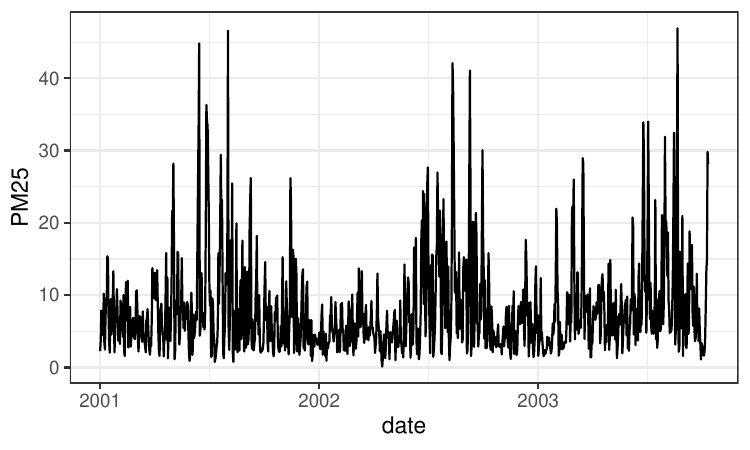}
        \caption{Visualization of the $\text{PM}_{2.5}$ concentrations.}
    \end{subfigure}
    \begin{subfigure}{\textwidth}
        \centering
        \includegraphics[width=\textwidth]{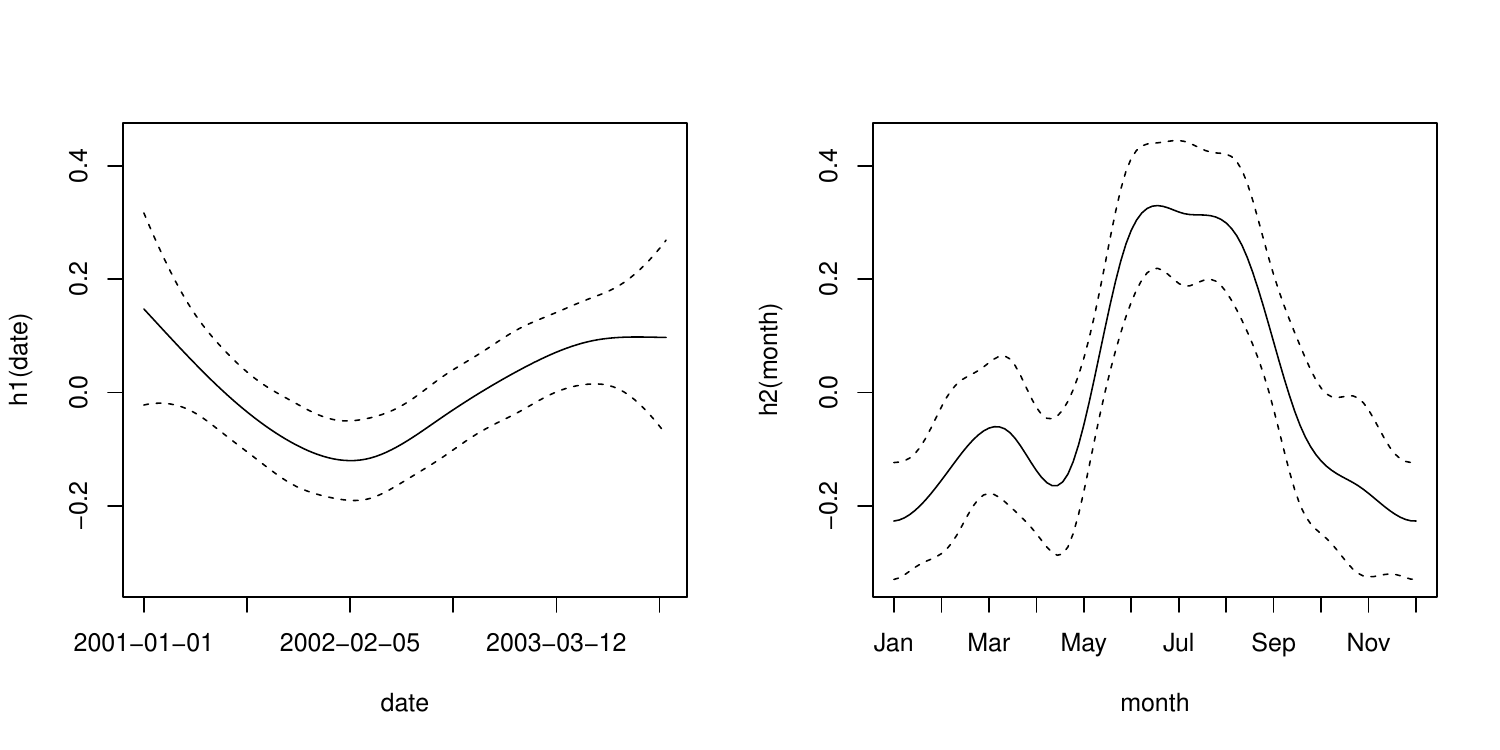}
        \caption{Estimated time trend and seasonality in $\text{PM}_{2.5}$ concentrations.}
    \end{subfigure}
    \caption{Web Figure 2. $\text{PM}_{2.5}$ concentrations in Waterloo from Jan. 1, 2001 to Oct. 11, 2003, corresponding to a sample size of 1000.}
\end{figure}

\subsection{Additional Simulation Results}

\subsubsection{Visualization of Simulation A Results}
\begin{figure}[H]
    \centering
    \begin{subfigure}{\textwidth}
        \centering
        \includegraphics[width=0.9\textwidth]{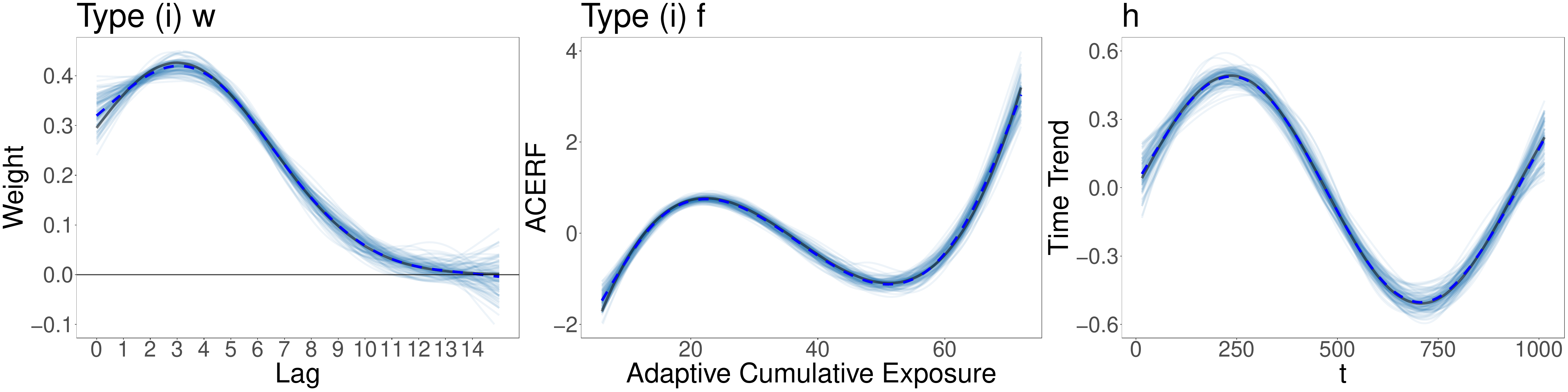}
        \caption{$N = 1000$}
    \end{subfigure}
    \begin{subfigure}{\textwidth}
        \centering
        \includegraphics[width=0.9\textwidth]{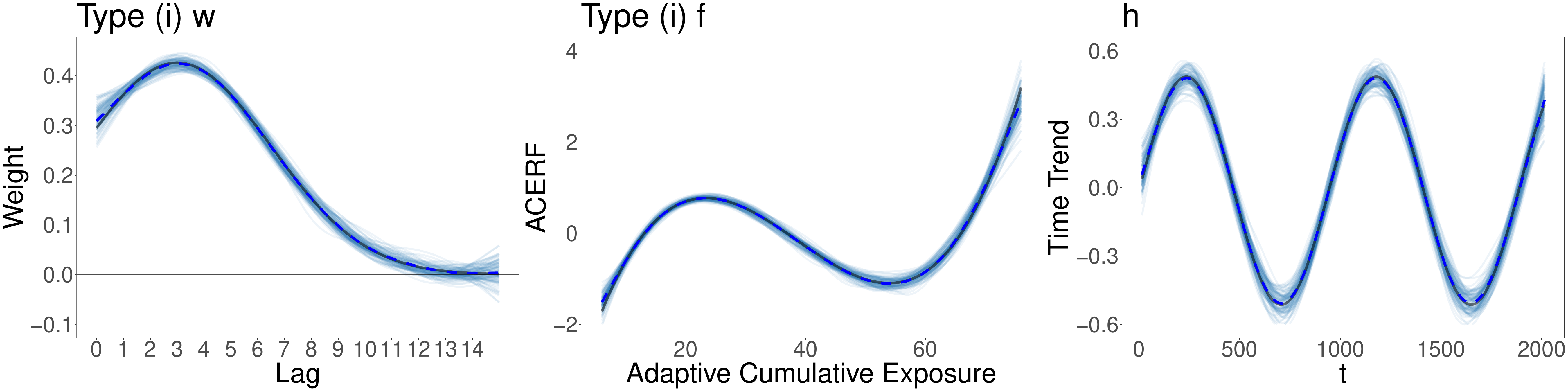}
        \caption{$N = 2000$}
    \end{subfigure}
    \caption{Web Figure 3. Visualization of Simulation A Results. The data are generated with type (i) weight function and type (i) ACERF. The sample sizes are 1,000 (a) and 2,000 (b). The estimates from the first 100 simulated data for the weight function $w(l)$ (first column), ACEFR $f(E)$ (second column) and the time trend $h(t)$ (third column) are labelled by blue curves. The dashed blue curves show the mean of the estimates over all of the 10,000 simulated datasets. The true functions are the solid black curves. }
    \label{fig:sim1}
\end{figure}

\begin{figure}[H]
    \centering
    \begin{subfigure}{\textwidth}
        \centering
        \includegraphics[width=0.9\textwidth]{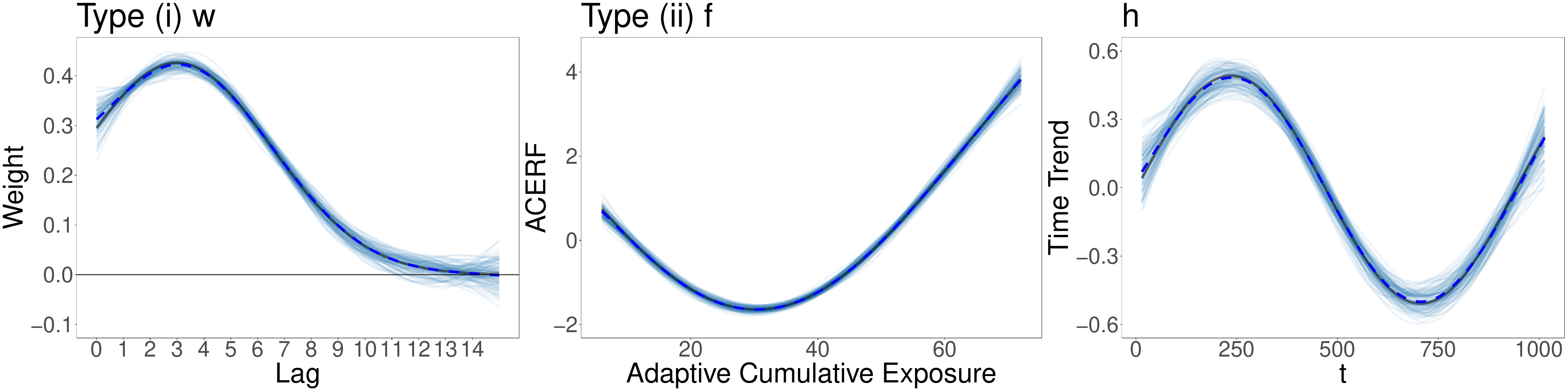}
        \caption{$N = 1000$}
    \end{subfigure}
    \begin{subfigure}{\textwidth}
        \centering
        \includegraphics[width=0.9\textwidth]{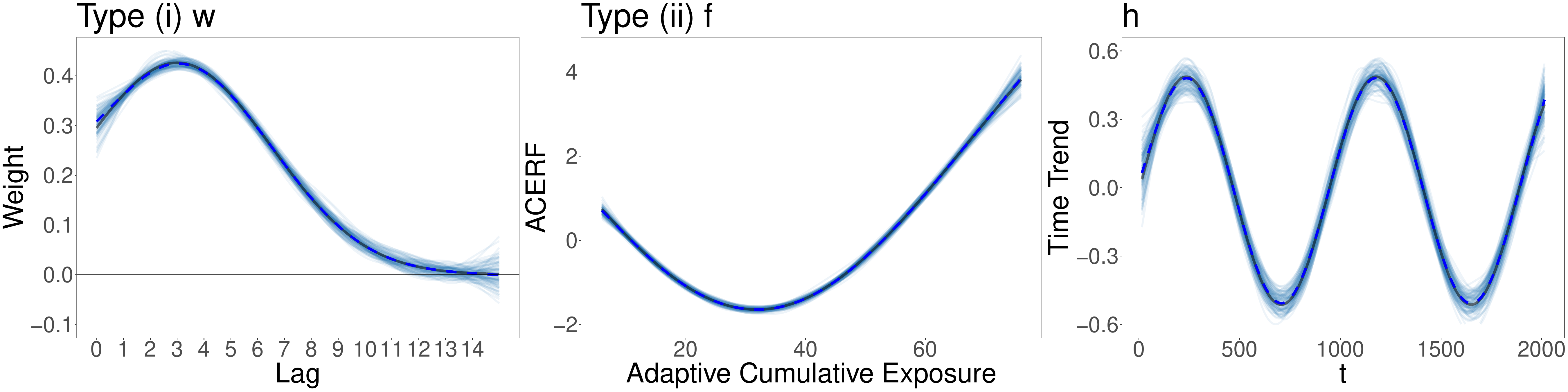}
        \caption{$N = 2000$}
    \end{subfigure}
    \caption{Web Figure 4. Visualization of Simulation A Results. The data are generated under type (i) weight function and type (ii) ACERF. The sample sizes are 1,000 (a) and 2,000 (b). The estimates from the first 100 simulated data for the weight function $w(l)$ (first column), ACEFR $f(E)$ (second column) and the time trend $h(t)$ (third column) are labelled by blue curves. The dashed blue curves show the mean of the estimates over all of the 10,000 simulated datasets. The true functions are the solid black curves. }
\end{figure}

\begin{figure}[H]
    \centering
    \begin{subfigure}{\textwidth}
        \centering
        \includegraphics[width=0.9\textwidth]{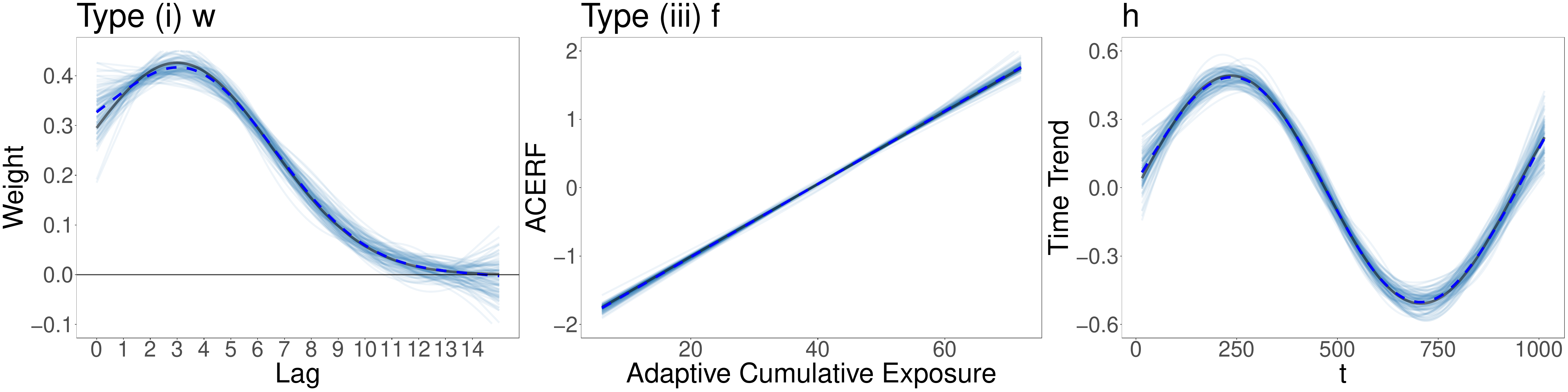}
        \caption{$N = 1000$}
    \end{subfigure}
    \begin{subfigure}{\textwidth}
        \centering
        \includegraphics[width=0.9\textwidth]{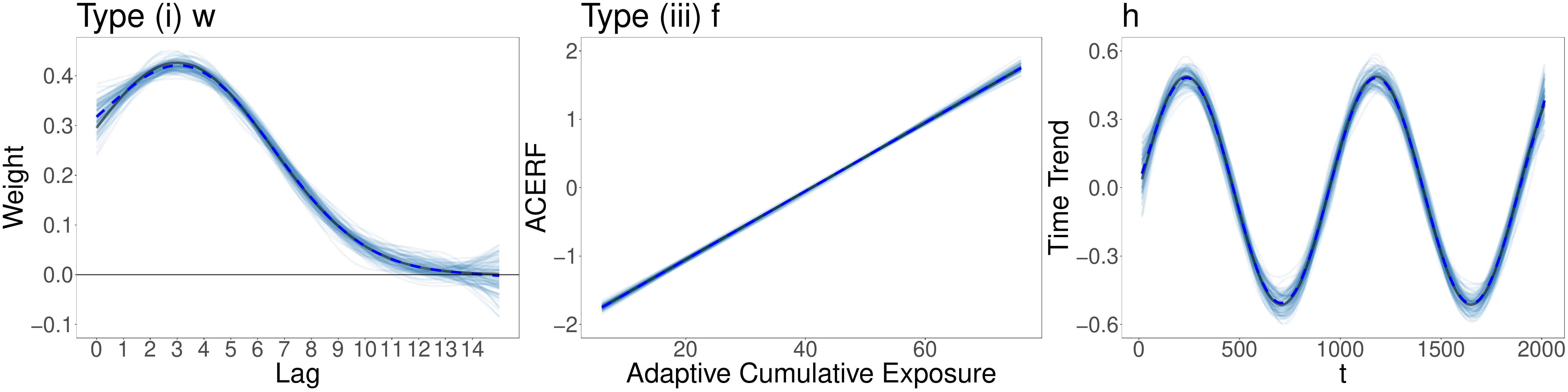}
        \caption{$N = 2000$}
    \end{subfigure}
    \caption{Web Figure 5. Visualization of Simulation A Results. The data are generated under type (i) weight function and type (iii) ACERF. The sample sizes are 1,000 (a) and 2,000 (b). The estimates from the first 100 simulated data for the weight function $w(l)$ (first column), ACEFR $f(E)$ (second column) and the time trend $h(t)$ (third column) are labelled by blue curves. The dashed blue curves show the mean of the estimates over all of the 10,000 simulated datasets. The true functions are the solid black curves. }
\end{figure}

\begin{figure}[H]
    \centering
    \begin{subfigure}{\textwidth}
        \centering
        \includegraphics[width=0.9\textwidth]{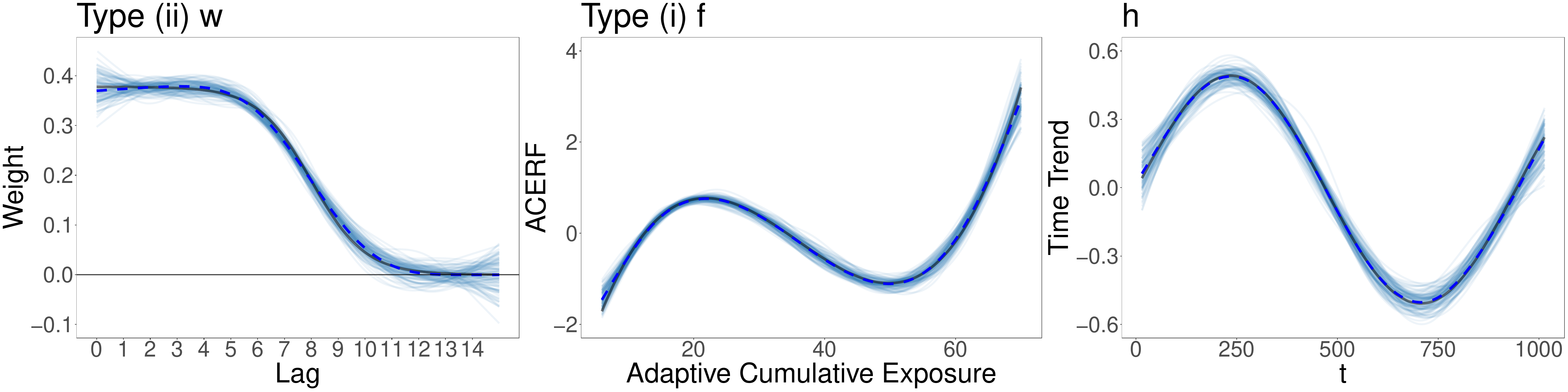}
        \caption{$N = 1000$}
    \end{subfigure}
    \begin{subfigure}{\textwidth}
        \centering
        \includegraphics[width=0.9\textwidth]{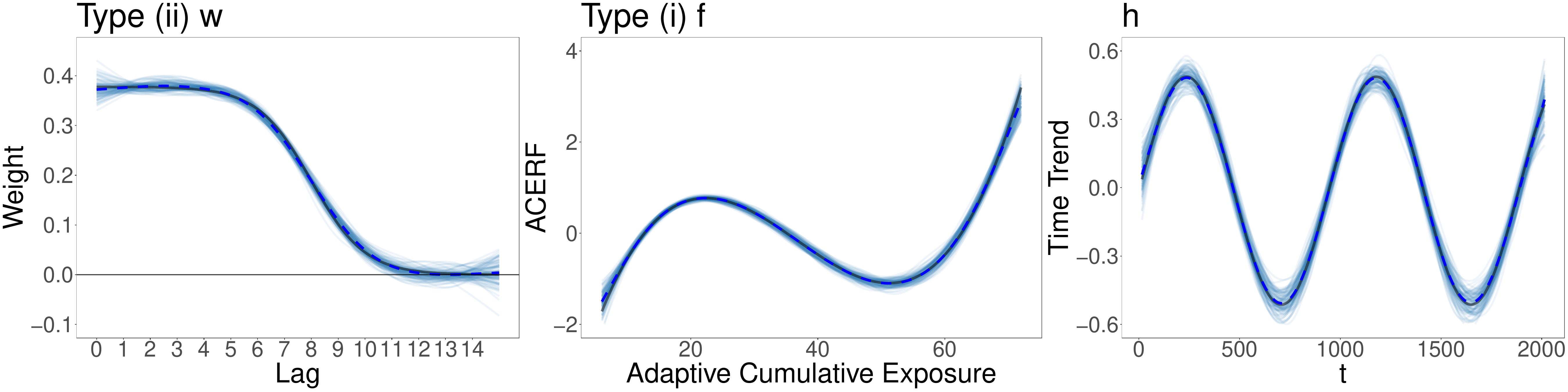}
        \caption{$N = 2000$}
    \end{subfigure}
    \caption{Web Figure 6. Visualization of Simulation A Results. The data are generated under type (ii) weight function and type (i) ACERF. The sample sizes are 1,000 (a) and 2,000 (b). The estimates from the first 100 simulated data for the weight function $w(l)$ (first column), ACEFR $f(E)$ (second column) and the time trend $h(t)$ (third column) are labelled by blue curves. The dashed blue curves show the mean of the estimates over all of the 10,000 simulated datasets. The true functions are the solid black curves. }
\end{figure}

\begin{figure}[H]
    \centering
    \begin{subfigure}{\textwidth}
        \centering
        \includegraphics[width=0.9\textwidth]{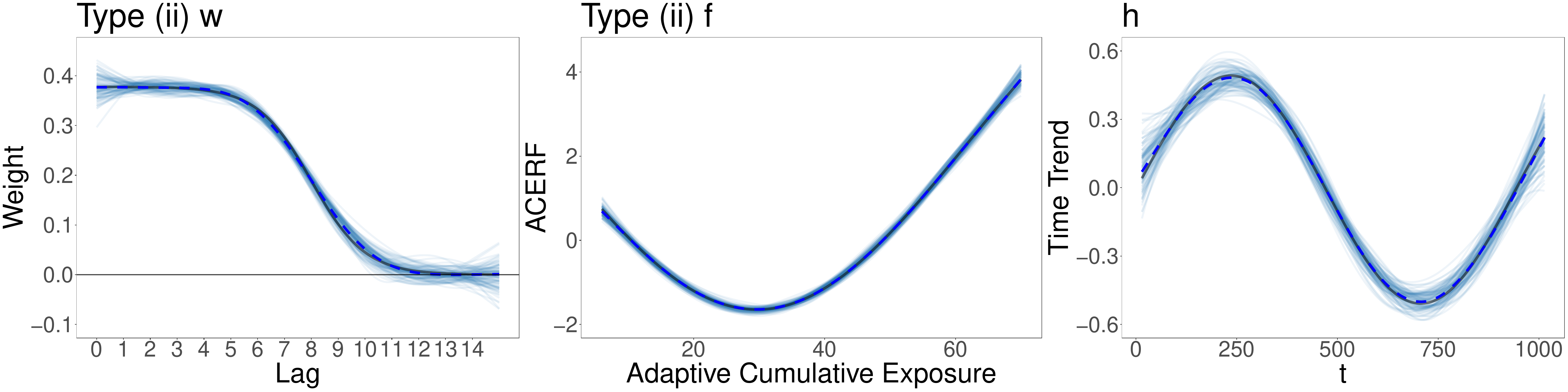}
        \caption{$N = 1000$}
    \end{subfigure}
    \begin{subfigure}{\textwidth}
        \centering
        \includegraphics[width=0.9\textwidth]{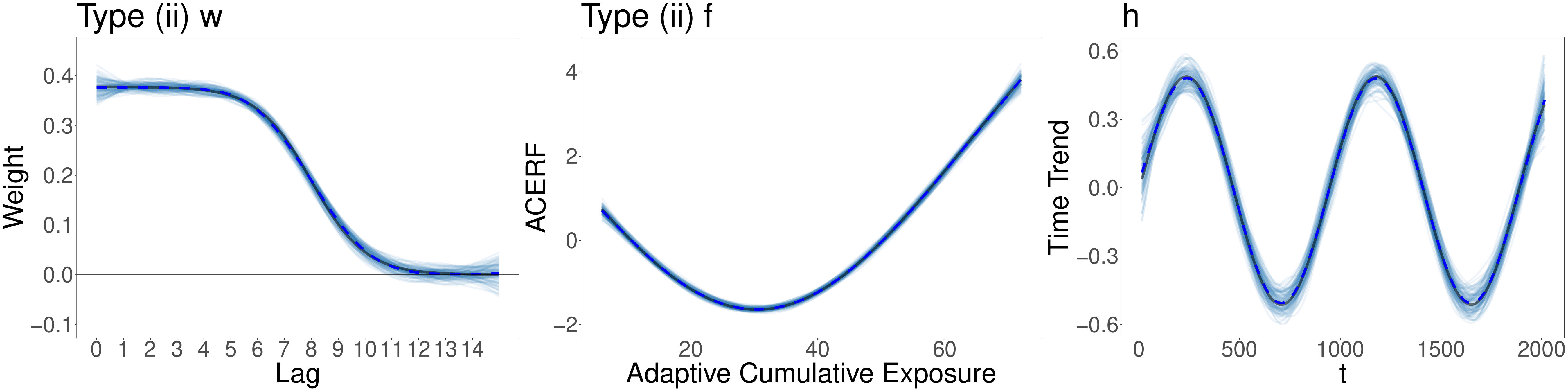}
        \caption{$N = 2000$}
    \end{subfigure}
    \caption{Web Figure 7. Visualization of Simulation A Results. The data are generated under type (ii) weight function and type (ii) ACERF. The sample sizes are 1,000 (a) and 2,000 (b). The estimates from the first 100 simulated data for the weight function $w(l)$ (first column), ACEFR $f(E)$ (second column) and the time trend $h(t)$ (third column) are labelled by blue curves. The dashed blue curves show the mean of the estimates over all of the 10,000 simulated datasets. The true functions are the solid black curves. }
\end{figure}

\begin{figure}[H]
    \centering
    \begin{subfigure}{\textwidth}
        \centering
        \includegraphics[width=0.9\textwidth]{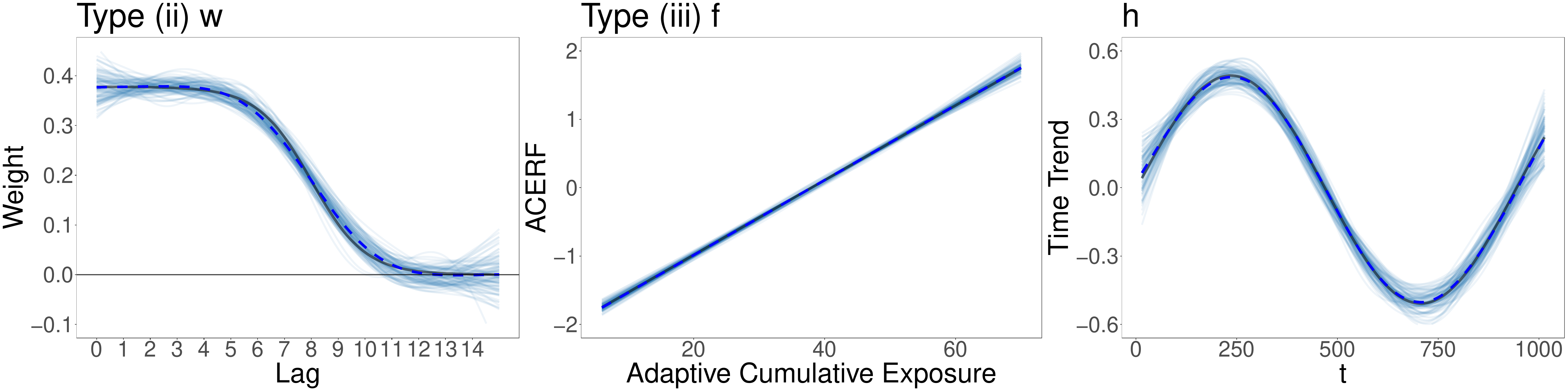}
        \caption{$N = 1000$}
    \end{subfigure}
    \begin{subfigure}{\textwidth}
        \centering
        \includegraphics[width=0.9\textwidth]{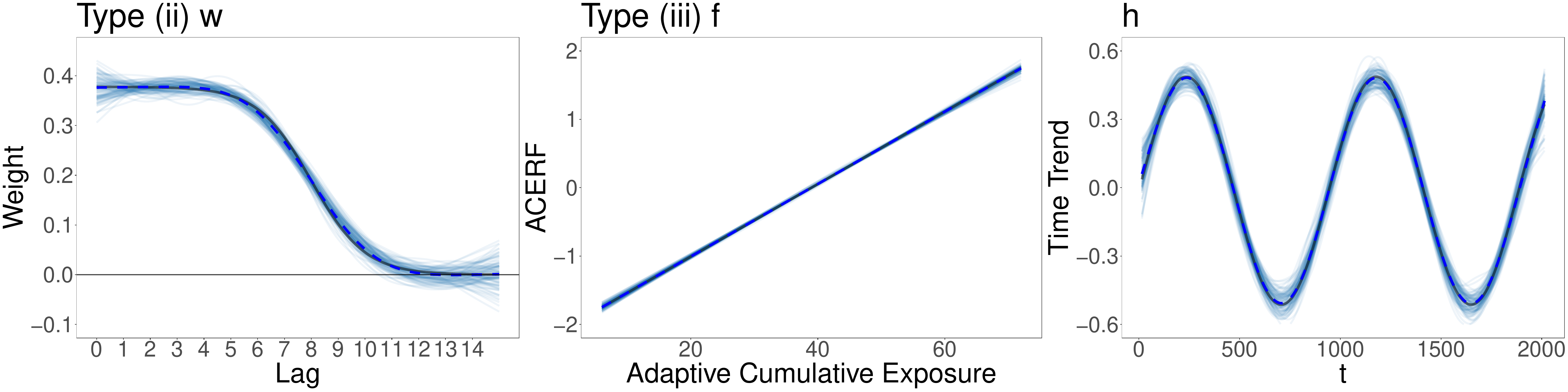}
        \caption{$N = 2000$}
    \end{subfigure}
    \caption{Web Figure 8. Visualization of Simulation A Results. The data are generated under type (ii) weight function and type (iii) ACERF. The sample sizes are 1,000 (a) and 2,000 (b). The estimates from the first 100 simulated data for the weight function $w(l)$ (first column), ACEFR $f(E)$ (second column) and the time trend $h(t)$ (third column) are labelled by blue curves. The dashed blue curves show the mean of the estimates over all of the 10,000 simulated datasets. The true functions are the solid black curves. }
\end{figure}

\begin{figure}[H]
    \centering
    \begin{subfigure}{\textwidth}
        \centering
        \includegraphics[width=0.9\textwidth]{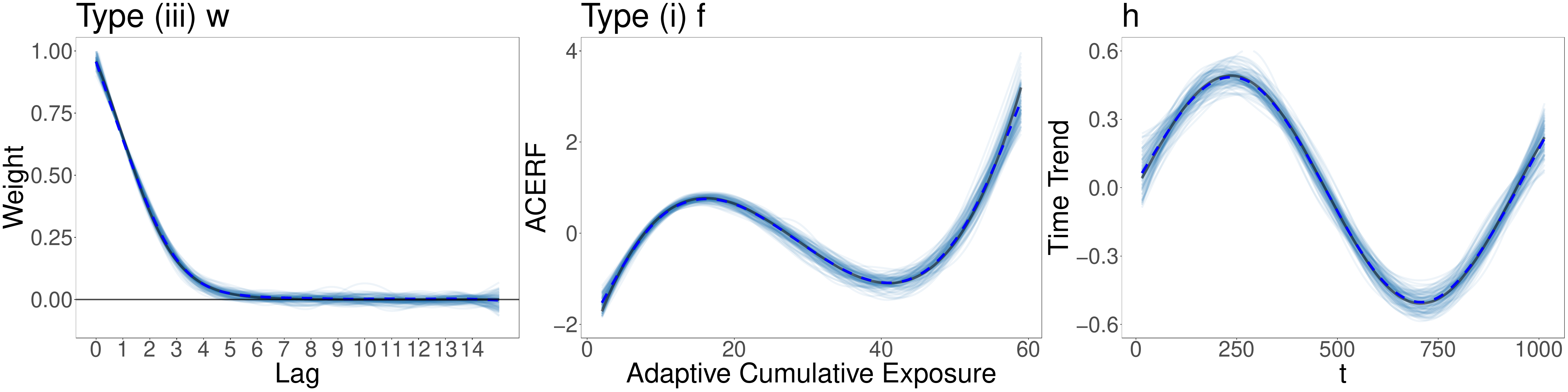}
        \caption{$N = 1000$}
    \end{subfigure}
    \begin{subfigure}{\textwidth}
        \centering
        \includegraphics[width=0.9\textwidth]{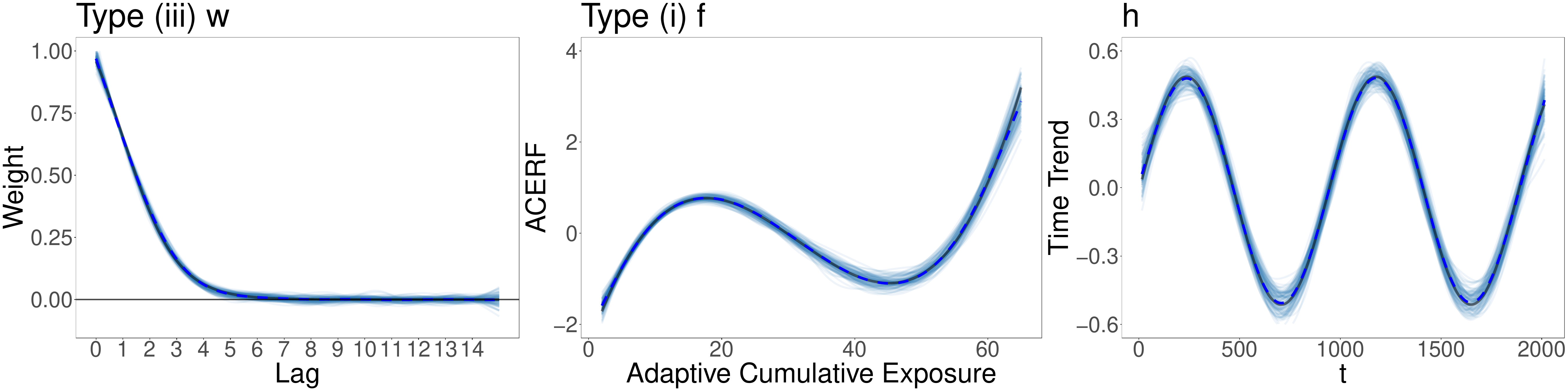}
        \caption{$N = 2000$}
    \end{subfigure}
    \caption{Web Figure 9. Visualization of Simulation A Results. The data are generated under type (iii) weight function and type (i) ACERF. The sample sizes are 1,000 (a) and 2,000 (b). The estimates from the first 100 simulated data for the weight function $w(l)$ (first column), ACEFR $f(E)$ (second column) and the time trend $h(t)$ (third column) are labelled by blue curves. The dashed blue curves show the mean of the estimates over all of the 10,000 simulated datasets. The true functions are the solid black curves. }
\end{figure}

\begin{figure}[H]
    \centering
    \begin{subfigure}{\textwidth}
        \centering
        \includegraphics[width=0.9\textwidth]{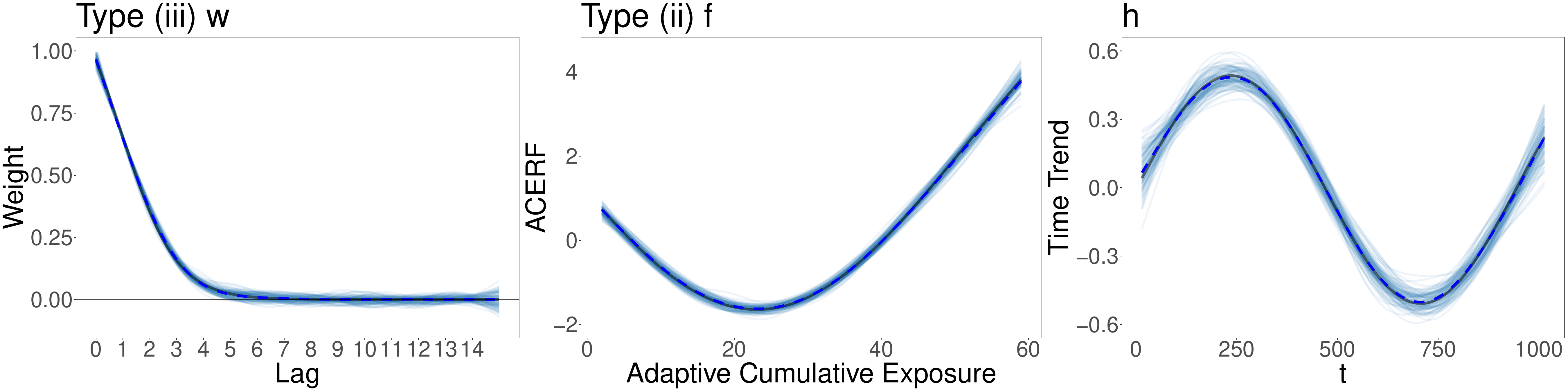}
        \caption{$N = 1000$}
    \end{subfigure}
    \begin{subfigure}{\textwidth}
        \centering
        \includegraphics[width=0.9\textwidth]{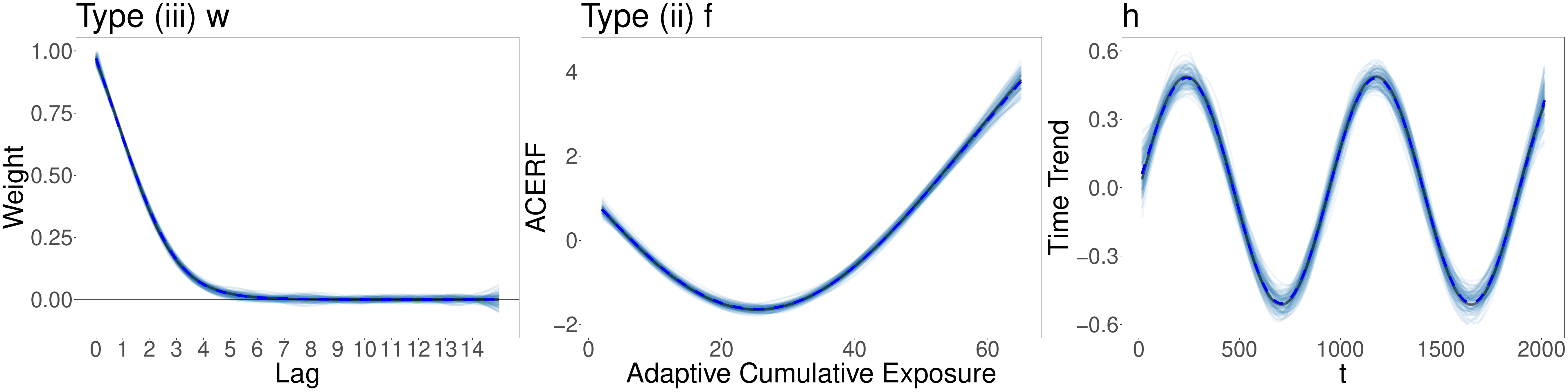}
        \caption{$N = 2000$}
    \end{subfigure}
    \caption{Web Figure 10. Visualization of Simulation A Results. The data are generated under type (iii) weight function and type (ii) ACERF. The sample sizes are 1,000 (a) and 2,000 (b). The estimates from the first 100 simulated data for the weight function $w(l)$ (first column), ACEFR $f(E)$ (second column) and the time trend $h(t)$ (third column) are labelled by blue curves. The dashed blue curves show the mean of the estimates over all of the 10,000 simulated datasets. The true functions are the solid black curves. }
\end{figure}

\begin{figure}[H]
    \centering
    \begin{subfigure}{\textwidth}
        \centering
        \includegraphics[width=0.9\textwidth]{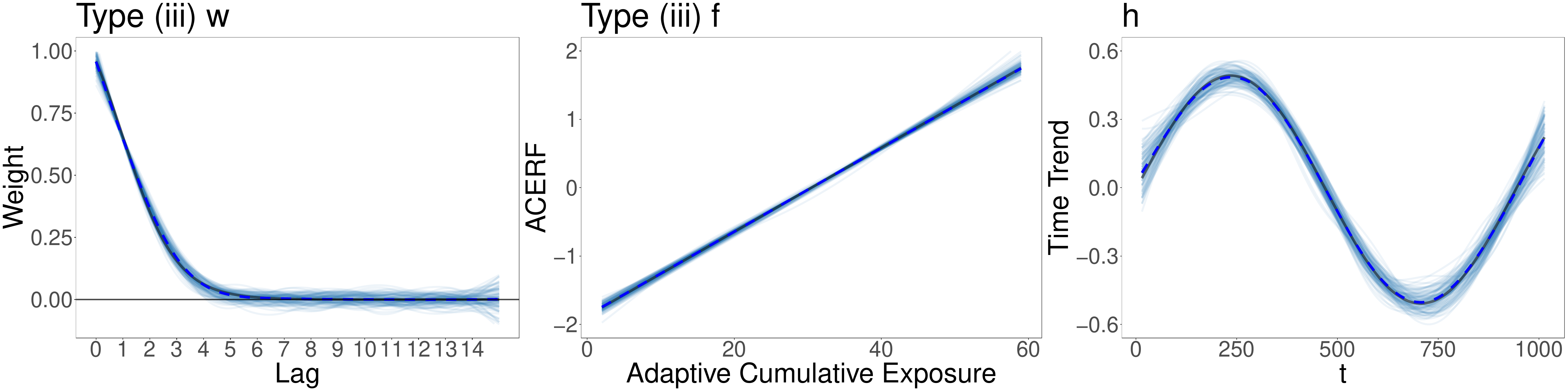}
        \caption{$N = 1000$}
    \end{subfigure}
    \begin{subfigure}{\textwidth}
        \centering
        \includegraphics[width=0.9\textwidth]{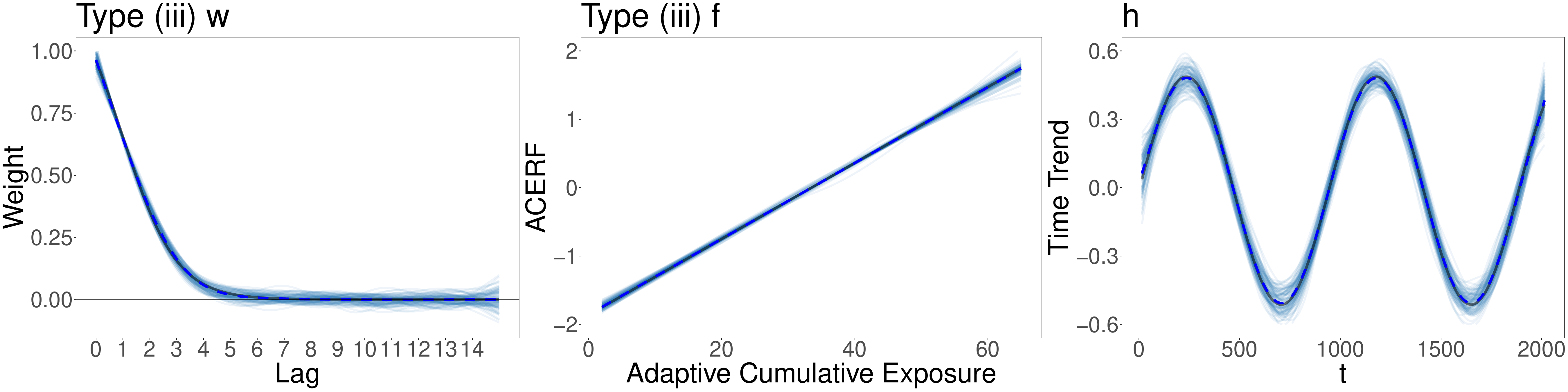}
        \caption{$N = 2000$}
    \end{subfigure}
    \caption{Web Figure 11. Visualization of Simulation A Results. The data are generated under type (iii) weight function and type (iii) ACERF. The sample sizes are 1,000 (a) and 2,000 (b). The estimates from the first 100 simulated data for the weight function $w(l)$ (first column), ACEFR $f(E)$ (second column) and the time trend $h(t)$ (third column) are labelled by blue curves. The dashed blue curves show the mean of the estimates over all of the 10,000 simulated datasets. The true functions are the solid black curves. }
\end{figure}

\subsubsection{Monte Carlo Standard Errors in Simulation A}

\begin{table}[H]
    \centering
    \caption{Web Table 3. Monte Carlo Standard Errors in Simulation A under different weight functions $w$ and ACERFs $f$. The Monte Carlo Standard Error for RMSE, 95\% confidence intervals coverage (Cvg) and average widths (Width) for $w$ and $f$ with the sample size $N = 1000$ and $2000$ are reported.}
    \label{tab:sim1}
    \begin{tabular}{@{}cccccccc@{}}
\toprule
      & \multicolumn{3}{c}{$w$}                 &  & \multicolumn{3}{c}{$f$}                 \\ \cmidrule(lr){2-4} \cmidrule(l){6-8} 
$N$ & RMSE        & Cvg        & Width        &  & RMSE        & Cvg        & Width        \\ \midrule
      & \multicolumn{3}{c}{\textbf{Type (i) $w$}} &  & \multicolumn{3}{c}{\textbf{Type (i) $f$}} \\
1000  & 0.000       & 0.001      & 0.000      &  & 0.001       & 0.001      & 0.000      \\
2000  & 0.000       & 0.000      & 0.000      &  & 0.000       & 0.001      & 0.000      \\
      & \multicolumn{3}{c}{}                    &  & \multicolumn{3}{c}{\textbf{Type (ii) $f$}} \\
1000  & 0.000       & 0.001      & 0.000      &  & 0.002       & 0.001      & 0.003      \\
2000  & 0.000       & 0.000      & 0.000      &  & 0.000       & 0.001      & 0.000      \\
      & \multicolumn{3}{c}{}                    &  & \multicolumn{3}{c}{\textbf{Type (iii) $f$}} \\
1000  & 0.000       & 0.001      & 0.000      &  & 0.000       & 0.002      & 0.000      \\
2000  & 0.000       & 0.001      & 0.000      &  & 0.000       & 0.002      & 0.000      \\
      &             &            &              &  &             &            &              \\
      & \multicolumn{3}{c}{\textbf{Type (ii) $w$}} &  & \multicolumn{3}{c}{\textbf{Type (i) $f$}} \\
1000  & 0.000       & 0.001      & 0.000      &  & 0.001       & 0.001      & 0.000      \\
2000  & 0.000       & 0.001      & 0.000      &  & 0.000       & 0.001      & 0.000      \\
      & \multicolumn{3}{c}{}                    &  & \multicolumn{3}{c}{\textbf{Type (ii) $f$}} \\
1000  & 0.000       & 0.001      & 0.000      &  & 0.001       & 0.001      & 0.001      \\
2000  & 0.000       & 0.000      & 0.000      &  & 0.000       & 0.001      & 0.000      \\
      & \multicolumn{3}{c}{}                    &  & \multicolumn{3}{c}{\textbf{Type (iii) $f$}} \\
1000  & 0.000       & 0.001      & 0.000      &  & 0.000       & 0.002      & 0.000      \\
2000  & 0.000       & 0.001      & 0.000      &  & 0.000       & 0.002      & 0.000      \\
      &             &            &              &  &             &            &              \\
      & \multicolumn{3}{c}{\textbf{Type (iii) $w$}} &  & \multicolumn{3}{c}{\textbf{Type (i) $f$}} \\
1000  & 0.000       & 0.001      & 0.000      &  & 0.001       & 0.001      & 0.001     \\
2000  & 0.000       & 0.000      & 0.000      &  & 0.000       & 0.001      & 0.000      \\
      & \multicolumn{3}{c}{}                    &  & \multicolumn{3}{c}{\textbf{Type (ii) $f$}} \\
1000  & 0.000       & 0.001      & 0.000      &  & 0.004       & 0.001      & 0.005      \\
2000  & 0.000       & 0.000      & 0.000      &  & 0.000       & 0.001      & 0.000      \\
      & \multicolumn{3}{c}{}                    &  & \multicolumn{3}{c}{\textbf{Type (iii) $f$}} \\
1000  & 0.000       & 0.001      & 0.000      &  & 0.000       & 0.002      & 0.000      \\
2000  & 0.000       & 0.001      & 0.000      &  & 0.000       & 0.002      & 0.000      \\ \bottomrule
\end{tabular}
\end{table}

\subsubsection{Results for Other Terms in Simulation A}

\begin{table}[H]
    \centering
    \caption{Web Table 4. Results for other terms in Simulation A under different weight functions and ACERFs. The RMSE, 95\% confidence intervals coverage (Cvg) and average widths (Width) for the time trend $h(t)$, and the Bias and RMSE for dispersion parameter $\theta$ with $N = 1000$ and $2000$ are reported.}
    \begin{tabular}{@{}ccccccc@{}}
    \toprule
          & \multicolumn{3}{c}{$h$} &  & \multicolumn{2}{c}{$\theta$} \\ \cmidrule(lr){2-4} \cmidrule(l){6-7} 
    $N$ & RMSE  & Cvg   & Width   &  & Bias         & RMSE         \\ \midrule
          & \multicolumn{6}{c}{\textbf{Type (i) $w$; Type (i) $f$}}      \\
    1000  & 0.037 & 0.960 & (0.155) &  & 0.104 &  0.744                  \\
    2000  & 0.037 & 0.966 & (0.159) & & 0.060 &  0.512                   \\
          & \multicolumn{6}{c}{\textbf{Type (i) $w$; Type (ii) $f$}}      \\
    1000  & 0.040 & 0.960 & (0.169) & & 0.163 &  0.902               \\
    2000  & 0.040 & 0.967 & (0.171) & & 0.090 &  0.598                 \\
          & \multicolumn{6}{c}{\textbf{Type (i) $w$; Type (iii) $f$}}      \\
    1000  & 0.036 & 0.961 & (0.153) &  & 0.090     &  0.697            \\
    2000  & 0.036 & 0.968 & (0.157) &  & 0.057     &  0.485            \\
          &       &       &         &  &              &              \\
          & \multicolumn{6}{c}{\textbf{Type (ii) $w$; Type (i) $f$}}      \\
    1000  & 0.037 & 0.960 & (0.156) &  & 0.091         & 0.736        \\
    2000  & 0.037 & 0.966 & (0.159) &  & 0.047        &  0.501        \\
          & \multicolumn{6}{c}{\textbf{Type (ii) $w$; Type (ii) $f$}}      \\
    1000  & 0.040 & 0.960 & (0.170) &  & 0.150        & 0.886       \\
    2000  & 0.040 & 0.967 & (0.172) &  & 0.090        & 0.594         \\
          & \multicolumn{6}{c}{\textbf{Type (ii) $w$; Type (iii) $f$}}      \\
    1000  & 0.036 & 0.960 & (0.152) &  & 0.092        & 0.686        \\
    2000  & 0.036 & 0.967 & (0.155) &  & 0.047        & 0.476        \\
          &       &       &         &  &              &              \\
          & \multicolumn{6}{c}{\textbf{Type (iii) $w$; Type (i) $f$}}      \\
    1000  & 0.038 & 0.958 & (0.157) &  & 0.028       & 0.961      \\
    2000  & 0.038 & 0.966 & (0.162) &  & 0.074       & 0.541        \\
          & \multicolumn{6}{c}{\textbf{Type (iii) $w$; Type (ii) $f$}}      \\
    1000  & 0.038 & 0.959 & (0.160) &  & 0.126        & 0.835         \\
    2000  & 0.038 & 0.967 & (0.161) &  & 0.071        & 0.516       \\
          & \multicolumn{6}{c}{\textbf{Type (iii) $w$; Type (iii) $f$}}      \\
    1000  & 0.038 & 0.961 & (0.160) &  & 0.115        & 0.792       \\
    2000  & 0.038 & 0.967 & (0.165) &  & 0.066        & 0.559       \\ \bottomrule
    \end{tabular}
\end{table}

\clearpage
\subsubsection{Results for Delta Method in Simulation A}
The only difference between the delta method and the sampling method lies in the confidence intervals for $w$. We report the Cvg and Width of the 95\% confidence intervals using the sampling method and delta method in Web Table 5 for comparison. We can observe the wide confidence intervals and over-coverage from the delta method. Hence, the delta method is less recommended than the sampling method. 

\begin{table}[H]
\centering
\caption{Web Table 5. Additional Results in Simulation A. The coverage (Cvg) and average widths (Width) of 95\% confidence intervals for $w$ using the sampling method and delta method are reported.}
\label{tab:delta}
\begin{tabular}{@{}ccccc@{}}
\toprule
                       & \multicolumn{4}{c}{$w$}                                  \\ \midrule
\multirow{2}{*}{$N$} & \multicolumn{2}{c}{Sampling} & \multicolumn{2}{c}{Delta} \\ \cmidrule(l){2-5} 
                       & Cvg          & Width         & Cvg        & Width        \\ \midrule
                       & \multicolumn{4}{c}{\textbf{Type (i) $w$; Type (i) $f$}}               \\
1000                   & 0.970        & (0.064)       & 0.991      & (0.094)      \\
2000                   & 0.979        & (0.049)       & 0.992      & (0.068)      \\
                       & \multicolumn{4}{c}{\textbf{Type (i) $w$; Type (ii) $f$}}               \\
1000                   & 0.977        & (0.055)       & 0.992      & (0.086)      \\
2000                   & 0.980        & (0.042)       & 0.993      & (0.062)      \\
                       & \multicolumn{4}{c}{\textbf{Type (i) $w$; Type (iii) $f$}}               \\
1000                   & 0.961        & (0.075)       & 0.991      & (0.120)      \\
2000                   & 0.973        & (0.060)       & 0.992      & (0.090)      \\
                       &              &               &            &              \\
                       & \multicolumn{4}{c}{\textbf{Type (ii) $w$; Type (i) $f$}}               \\
1000                   & 0.965        & (0.060)       & 0.987      & (0.090)      \\
2000                   & 0.966        & (0.044)       & 0.988      & (0.061)      \\
                       & \multicolumn{4}{c}{\textbf{Type (ii) $w$; Type (ii) $f$}}               \\
1000                   & 0.968        & (0.051)       & 0.989      & (0.079)      \\
2000                   & 0.973        & (0.038)       & 0.990      & (0.055)      \\
                       & \multicolumn{4}{c}{\textbf{Type (ii) $w$; Type (iii) $f$}}               \\
1000                   & 0.954        & (0.071)       & 0.986      & (0.113)      \\
2000                   & 0.964        & (0.055)       & 0.988      & (0.083)      \\
                       &              &               &            &              \\
                       & \multicolumn{4}{c}{\textbf{Type (iii) $w$; Type (i) $f$}}               \\
1000                   & 0.964        & (0.063)       & 0.979      & (0.144)      \\
2000                   & 0.975        & (0.050)       & 0.990      & (0.100)      \\
                       & \multicolumn{4}{c}{\textbf{Type (iii) $w$; Type (ii) $f$}}               \\
1000                   & 0.975        & (0.060)       & 0.991      & (0.139)      \\
2000                   & 0.979        & (0.047)       & 0.991      & (0.098)      \\
                       & \multicolumn{4}{c}{\textbf{Type (iii) $w$; Type (iii) $f$}}               \\
1000                   & 0.969        & (0.093)       & 0.996      & (0.256)      \\
2000                   & 0.976        & (0.078)       & 0.992      & (0.203)      \\ \bottomrule
\end{tabular}
\end{table}

\subsection{Supplementary Simulation: Linear $f$}
In Simulation A, the data are generated from the model:
$$
\log(\mu_t) = f \left\{\int_0^{15} w(l) X(t-l) dl\right\} + h(t),
$$
When $f$ is linear (Type (iii)), the model reduces to a distributed lag (linear) model (DLM):
$$
\log(\mu_t) = \beta \int_0^{15} w(l) X(t-l) dl + h(t),
$$
which can be fitted using the existing methods such as \citet{gasparrini2013distributed}. 
In this Supplementary Simulation, we compare the ACE-DLNM and DLM under the settings Simulation A with Type (iii) linear $f$. 
Since the function $w$ and the coefficient $\beta$ are not mutually identifiable in the DLM, 
separate estimates for $w$ and the linear $f$ cannot be directly obtained. 
Therefore, we compare the DLM and DLNM using RMSE, 95\% confidence intervals coverage (Cvg) and average widths (Width) for 
$f(\int_0^{15} w(l) X(t-l) dl)$ evaluated over $t\in \mathcal{T}$ across 10,000 replicates. 

The results are reported in Web Table 6. 
The DLM correctly specifying the linear association yields slightly lower RMSE, narrower confidence intervals, as expected. 
The ACE-DLNM, despite assuming a flexible $f$, performs closely to the correctly specified DLM. 
In practice, if there is no strong evidence supporting a linear association, 
one would favour the ACE-DLNM which behaves well in both linear and non-linear settings. 

\begin{table}[H]
\centering
\caption{Web Table 6. Simulation results under Type (iii) linear $f$. The ACE-DLNM and DLM are compared based on 
the RMSE, 95\% confidence intervals coverage (Cvg) and average widths (Width) for $f(\int_0^{15} w(l) X(t-l) dl)$ evaluated over $t\in \mathcal{T}$ across 10,000 replicates with sample size $N = 1000$ and $2000$.}
\begin{tabular}{@{}cccccccccc@{}}
\toprule
  &   &  & \multicolumn{3}{c}{ACE-DLNM} &  & \multicolumn{3}{c}{DLM} \\ \cmidrule(lr){4-6} \cmidrule(l){8-10} 
$w$       & $N$ &  & RMSE     & Cvg    & Width    &  & RMSE   & Cvg   & Width  \\ \midrule
Type (i)  &1000&  &  0.038    & 0.947  & 0.143   &  & 0.036  & 0.953 & 0.140 \\
          &2000&  &  0.028   &  0.953 & 0.107   &  &  0.027 & 0.957 &  0.104 \\
Type (ii)  &1000&  &  0.038  & 0.945 &  0.142  &  & 0.037 & 0.950 & 0.138 \\
          &2000&  & 0.028   & 0.952 &   0.105  &  & 0.027  & 0.954 & 0.103  \\
Type (iii)  &1000&  & 0.044   & 0.953   & 0.167   &  & 0.043  & 0.957 & 0.163 \\
            &2000&  & 0.033    & 0.957  &  0.126   &  &0.032   & 0.958 & 0.122  \\ \bottomrule
\end{tabular}
\end{table}

\clearpage

\subsection{Supplementary Simulation: Misspecification}
We conduct two supplementary simulation studies to investigate the sensitivity of the proposed ACE-DLNM framework
to misspecification of the model assumption. 
The ACE-DLNM framework assumes that the association function applies to the ACE. 
Here we consider two misspecification scenarios:  
(A) the association function applies directly to the exposure and remains constant over time; and 
(B) the association function applies directly to the exposure and changes with time. 
Under both scenarios, the DRF-DLNM is correctly specified.

\subsubsection{Misspecification Scenario A}
We generate data from a model where the association function applies directly to the exposure and remains constant over time, specifically: 
$$
\log (\mu_t) = \sum_{l=0}^{14} w(l) f(x_{t-l}) + h(t),
$$
where $w(\cdot)$ is Type (i) in Simulation A (Section 4.1). 
Under this data-generating mechanism, the DRF-DLNM is correctly specified, while the ACE-DLNM is misspecified except when $f$ is linear. 
We consider the function $f$ from linear to non-linear as shown in Web Figure 12. 
We use the integrated squared second derivatives, $J(f) = \int [f''(x)]^2 dx$, to quantify the roughness of the function $f$ \citep{green1993nonparametric}. 
A higher value of $J(f)$ indicates a rougher $f$, and $J(f) = 0$ if $f$ is linear. 
We calculate the relative roughness by dividing $J(f)$ by the value of the roughest function in Web Figure 12 shown in darkest blue. 
We would expect the ACE-DLNM to perform well with linear $f$, and to perform worse as the roughness $J(f)$ increases.

\begin{figure}[H]
    \centering
    \includegraphics[width=0.55\linewidth]{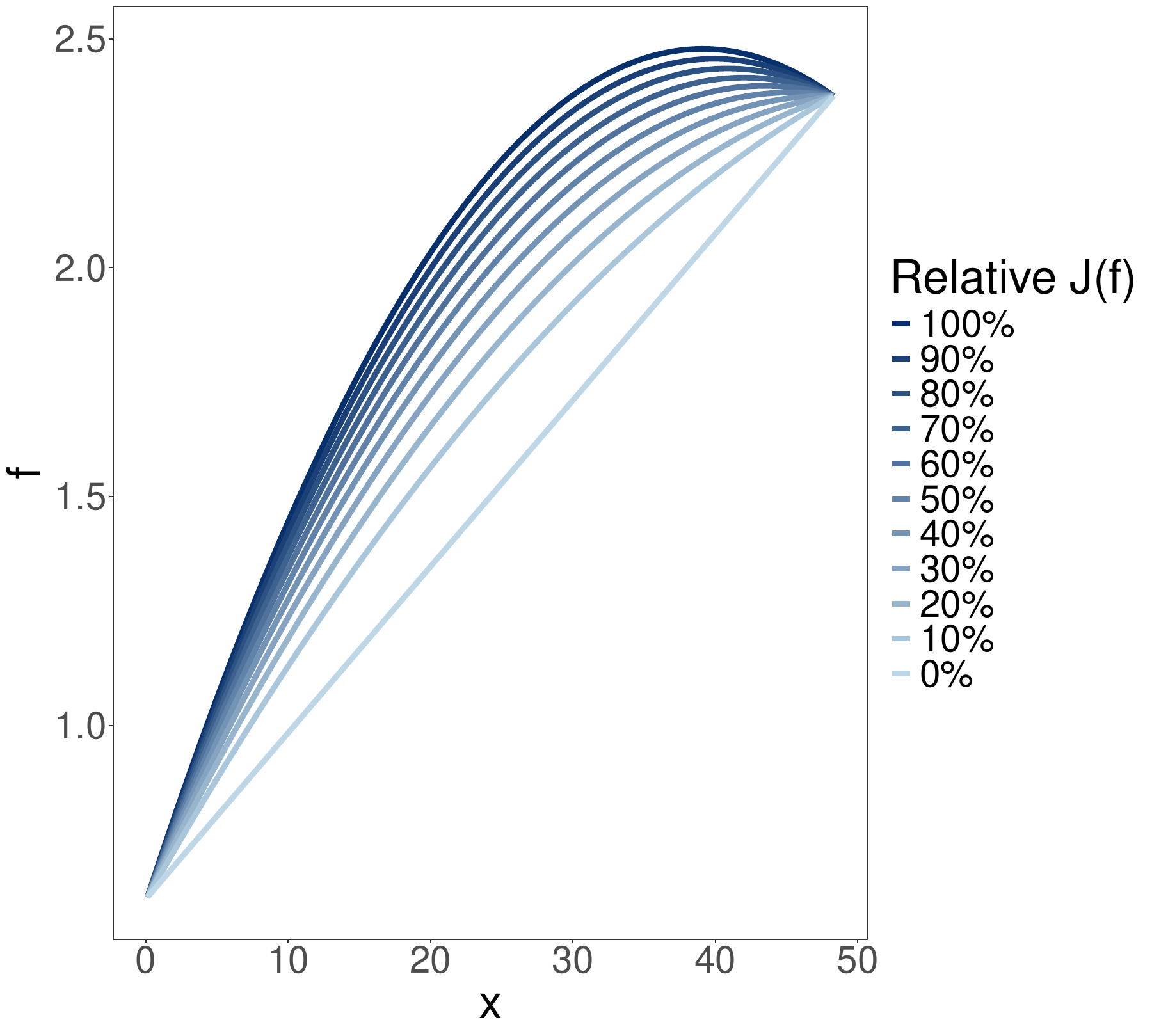}
    \caption{Web Figure 12. A set of functions $f$ used in Misspecification Scenario A. The relative roughness $J(f)$ is indicated by color, with the roughest function shown in the darkest blue.}
\end{figure}

We compare the models using RMSE and 95\% confidence intervals coverage (Cvg) for the distributed lag term $s_t = \sum_{l=0}^{14} w(l) f(x_{t-l})$ evaluated over $t\in \mathcal{T}$ across 10,000 replicates, reported in Web Table 7. 
The DRF-DLNM is correctly specified regardless of the shape of $f$, 
and performs better with low RMSE and near nominal coverage. 
When the $f$ is linear (relative $J(f)$ = 0), the ACE-DLNM is also correctly specified and achieves low RMSE and near nominal coverage, close to the DRF-DLNM. 
As $f$ becomes increasingly non-linear with higher roughness $J(f)$, the data-generating mechanism deviates from the model assumption of ACE-DLNM, leading to gradually increased RMSE and undercoverage. 
Overall, the ACE-DLNM maintains reasonable performance in this misspecification scenario, 
particularly when the association function is close to linear.

\begin{table}[H]
\centering
\caption{Web Table 7. Simulation results for Misspecification Scenario A. The ACE-DLNM and DRF-DLNM are compared based on the RMSE, 95\% confidence intervals coverage (Cvg) under different relative roughness of the association function $f$. }
\begin{tabular}{@{}ccccc@{}}
\toprule
     & \multicolumn{2}{c}{ACE-DLNM} & \multicolumn{2}{c}{DRF-DLNM} \\ \cmidrule(l){2-3} \cmidrule(l){4-5} 
Relative $J(f)$ & RMSE          & Cvg          & RMSE          & Cvg          \\ \midrule
0\% (Linear)  & 0.025 &   0.951      &   0.025       &   0.965      \\
10\%     &  0.039   &  0.869   &    0.032     &    0.966       \\
20\%     &  0.048    &  0.835   &   0.033       &    0.975     \\
30\%     &  0.055  & 0.806   &   0.034      &    0.977   \\
40\%     &  0.061  & 0.784  &    0.035     &     0.978    \\
50\%     &  0.067   & 0.765     &  0.036       &  0.979      \\
60\%     &  0.073   & 0.748  &  0.036      &    0.979     \\
70\%     &  0.078  & 0.732   &  0.037       &   0.979    \\
80\%     &  0.082   &  0.719   &   0.037       &    0.979     \\
90\%     &  0.087   &  0.707       &  0.037        &   0.979    \\
100\%    &  0.091    &  0.696      &  0.038       &     0.979     \\ \bottomrule
\end{tabular}
\end{table}

\subsubsection{Misspecification Scenario B}
In this scenario, we generate data from a model in which the association function applies directly to the exposure but changes with time. 
This corresponds to the framework of the DRF-DLNM, as considered in, for example, \citet{gasparrini2017penalized}. 
Specifically, we follow \citet{gasparrini2017penalized} to generate data from the model: 
$$
\log (\mu_t) = \sum_{l=0}^{40} \psi (x_{t-l}, l),
$$
where the exposure-lag-response function $\psi (x_{t-l}, l)$ is taken from Scenario 2 in \citet{gasparrini2017penalized}, as shown in 
Web Figure 13. The sample size is 5,114 and the exposure is the daily temperature in Chicago from 1987 to 2000; see \citet{gasparrini2017penalized} for details. 
We compare the ACE-DLNM and DRF-DLNM under this setting. 
As the two models are reparameterized differently, 
we compare the models using RMSE and 95\% confidence intervals coverage (Cvg) for the distributed lag term $s_t =  \sum_{l=0}^{40} \psi (x_{t-l}, l)$ evaluated over $t \in \mathcal{T}$ across 10,000 replicates, reported in Web Table 8. 
Because the data-generating mechanism deviates substantially from the model assumption of ACE-DLNM, 
it is difficult for the ACE-DLNM to capture the underlying model structure, yielding high RMSE and undercoverage. 
Similarly, in Simulation B in the main manuscript, the DRF-DLNM is misspecified and performs poorly. 

\begin{figure}[H]
    \centering
    \includegraphics[width=0.3\linewidth]{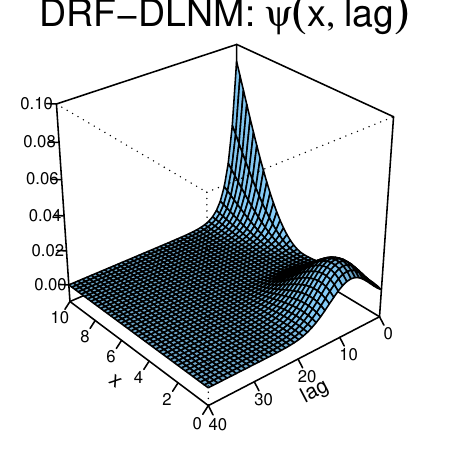}
    \caption{Web Figure 13. The exposure-lag-response function used in Misspecification Scenario B, following the simulation study in \citet{gasparrini2017penalized}. }
    \label{fig:fcurve}
\end{figure}

\begin{table}[H]
    \centering
    \caption{Web Table 8. Simulation results for Misspecification Scenario B. The ACE-DLNM and DRF-DLNM are compared based on the RMSE, 95\% confidence intervals coverage (Cvg).}
    \begin{tabular}{ccc}
    \toprule
             & RMSE & Cvg \\ \midrule
    ACE-DLNM & 0.017 & 0.424 \\
    DRF-DLNM & 0.007 & 0.900 \\ \bottomrule
    \end{tabular}
\end{table}

Overall, neither ACE-DLNM nor DRF-DLNM performs better in all settings --- they are based on different model assumptions 
and the relative performance depends on how well their assumptions align with the data-generating mechanism. 
In practice, one may select between the two models based on AIC or interpretability considerations, 
as we illustrated in Section 5.

\clearpage

\section{Canadian Air Pollution Data}

\subsection{Data Summary}
\begin{table}[H]
\centering
\caption{Web Table 9. Data Summary. We report the population in 2016, the daily average $\text{PM}_{2.5}$ concentration ($\mu g / m^3$), and daily average respiratory morbidity count, from January 1, 2001 to December 31, 2018, along with sample variance in brackets.}
\label{tab:summary}
\begin{tabular}{@{}lccccc@{}}
\toprule
                                       & Waterloo & Peel      & Hamilton & Calgary   & Vancouver \\ \midrule
Population (2016)                      & 535,154  & 1,381,739 & 536,917  & 1,498,778 & 2,463,431 \\  \addlinespace[0.3cm]
$\text{PM}_{2.5}$ ($\mu g / m^3$)              & 7.6     & 7.5      & 8.6     & 8.0      & 5.3      \\  
                                   & (33.4)   & (31.8)    & (38.0)   & (43.7)    & (15.4)    \\ \addlinespace[0.3cm]
Respiratory Morbidity            & 7.4     & 16.4     & 9.6     & 16.6     & 32.7     \\
                                   & (12.5)   & (37.2)    & (16.0)   & (38.8)    & (113.6)   \\ \bottomrule
\end{tabular}
\end{table}

\subsection{Additional Application Results}

\begin{figure}[H]
    \centering
    \includegraphics[width=0.7\linewidth]{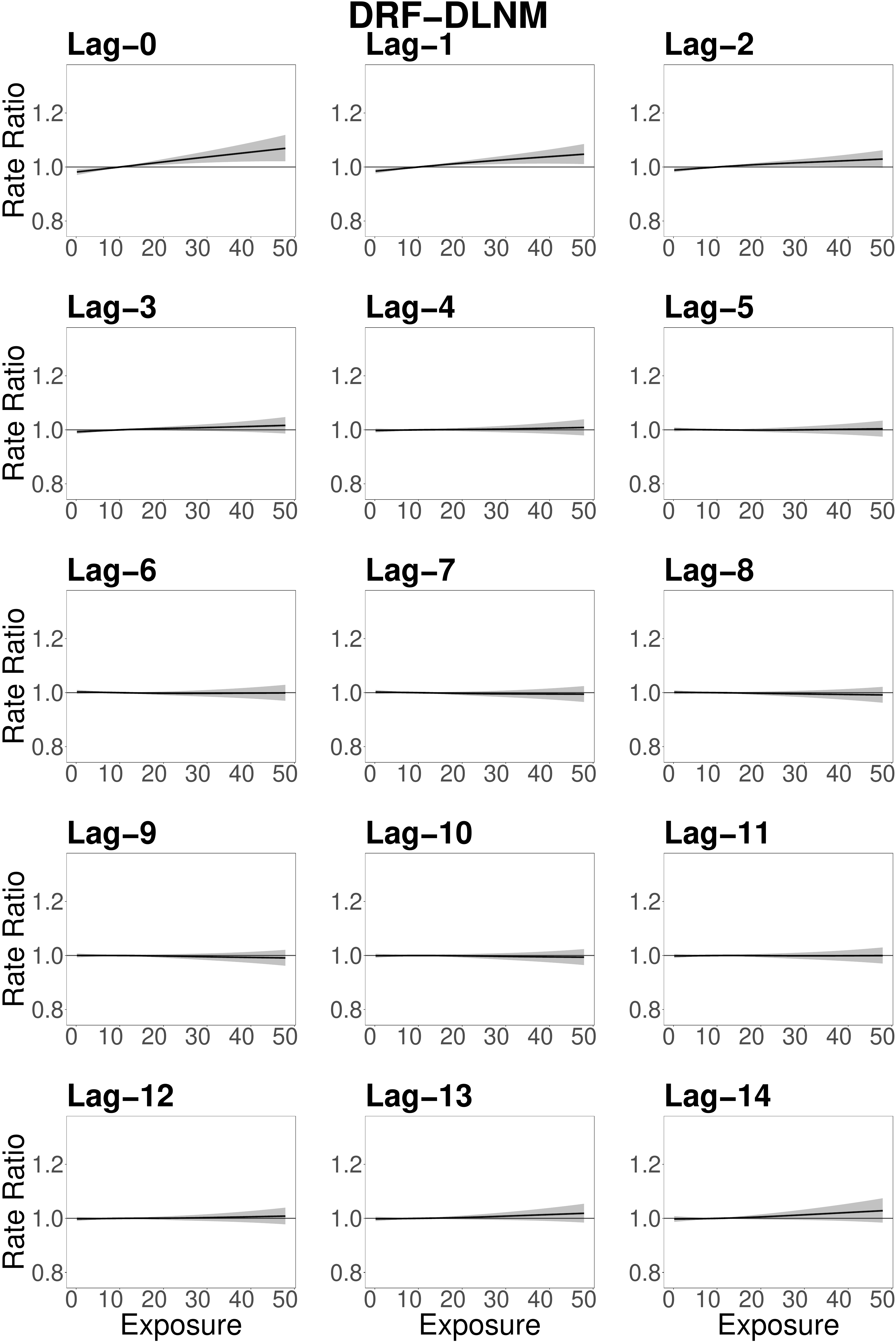}
    \caption{Web Figure 14. The estimated curves at each lag from the DRF-DLNM for the Hamilton dataset. }
\end{figure}

\begin{figure}[H]
    \centering
     \begin{subfigure}[t]{0.5\textwidth}
        \centering
        \includegraphics[width=\linewidth]{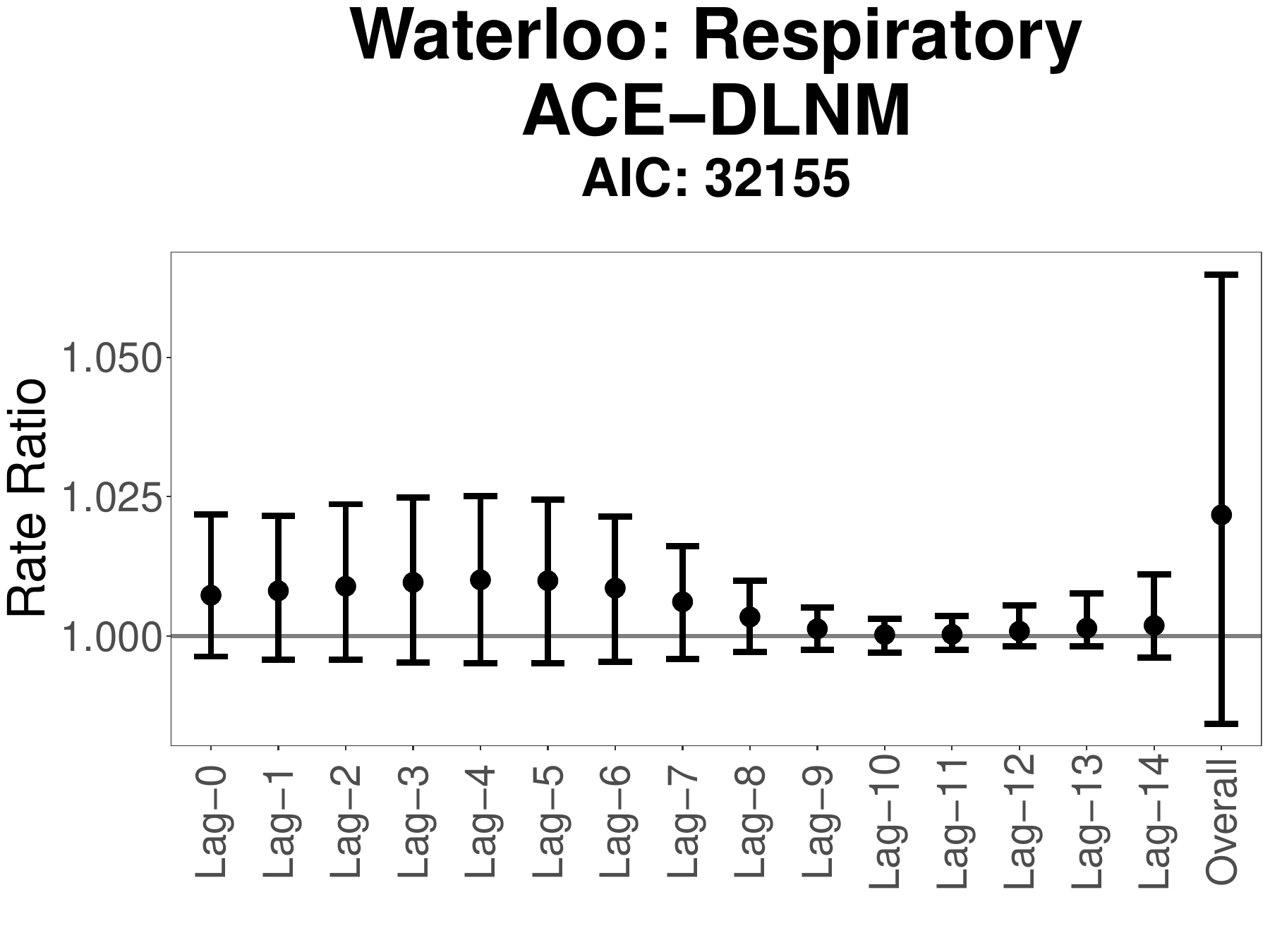}
        \caption{Rate ratios from ACE-DLNM}
    \end{subfigure}%
    \hfill
    \begin{subfigure}[t]{0.5\textwidth}
        \centering
        \includegraphics[width=\linewidth]{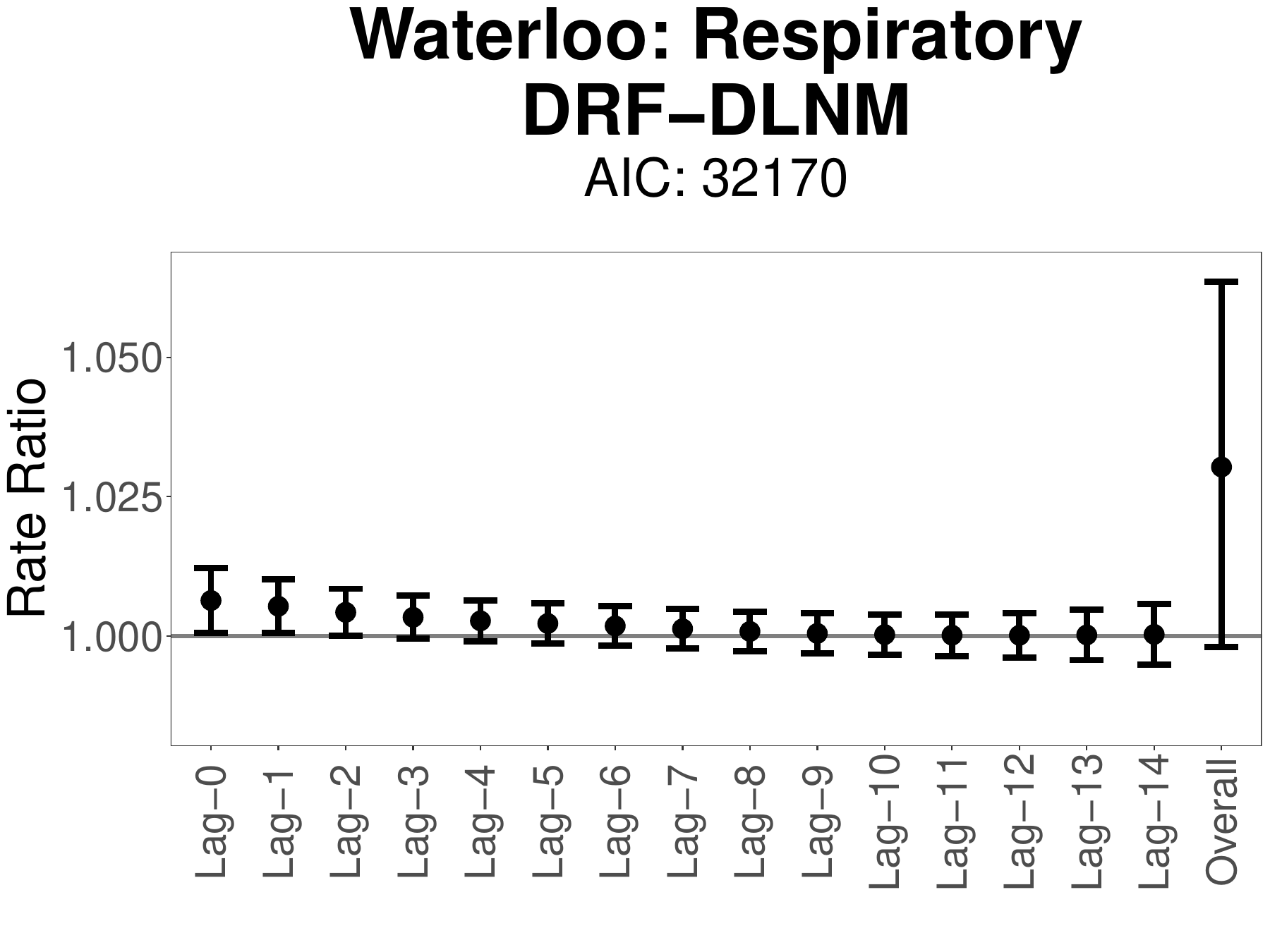}
        \caption{Rate ratios from DRF-DLNM}
    \end{subfigure}

    \vspace{0.5cm}
    \centering
    
    \begin{subfigure}[c]{0.48\textwidth}
    \centering
    \includegraphics[width=0.9\linewidth]{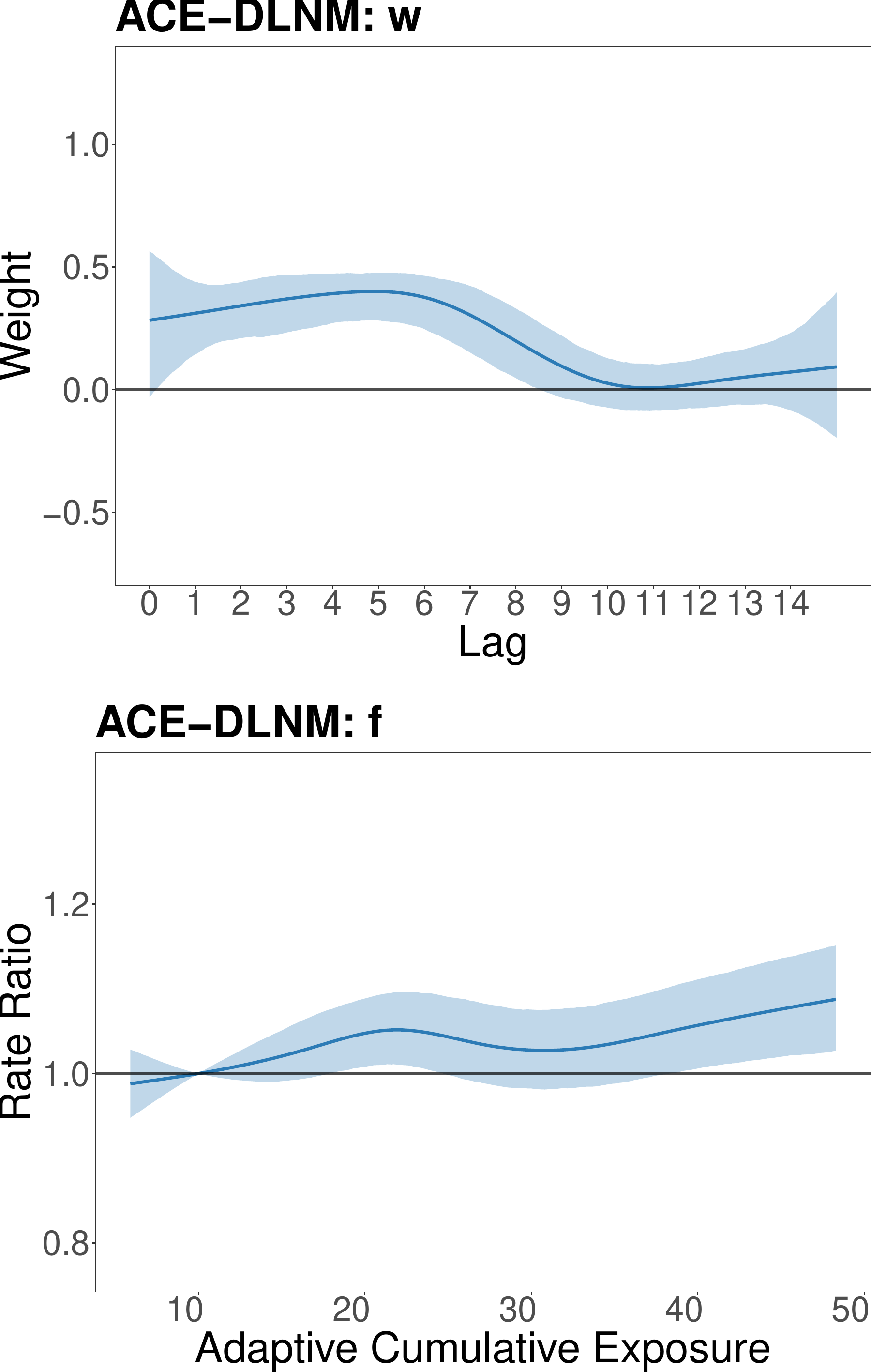}
    \caption{Estimated curves from ACE-DLNM}
    \end{subfigure}
    \hfill
    \begin{subfigure}[c]{0.48\textwidth}
    \centering
    \begin{subfigure}[b]{\textwidth}
      \centering
      \includegraphics[width=\linewidth]{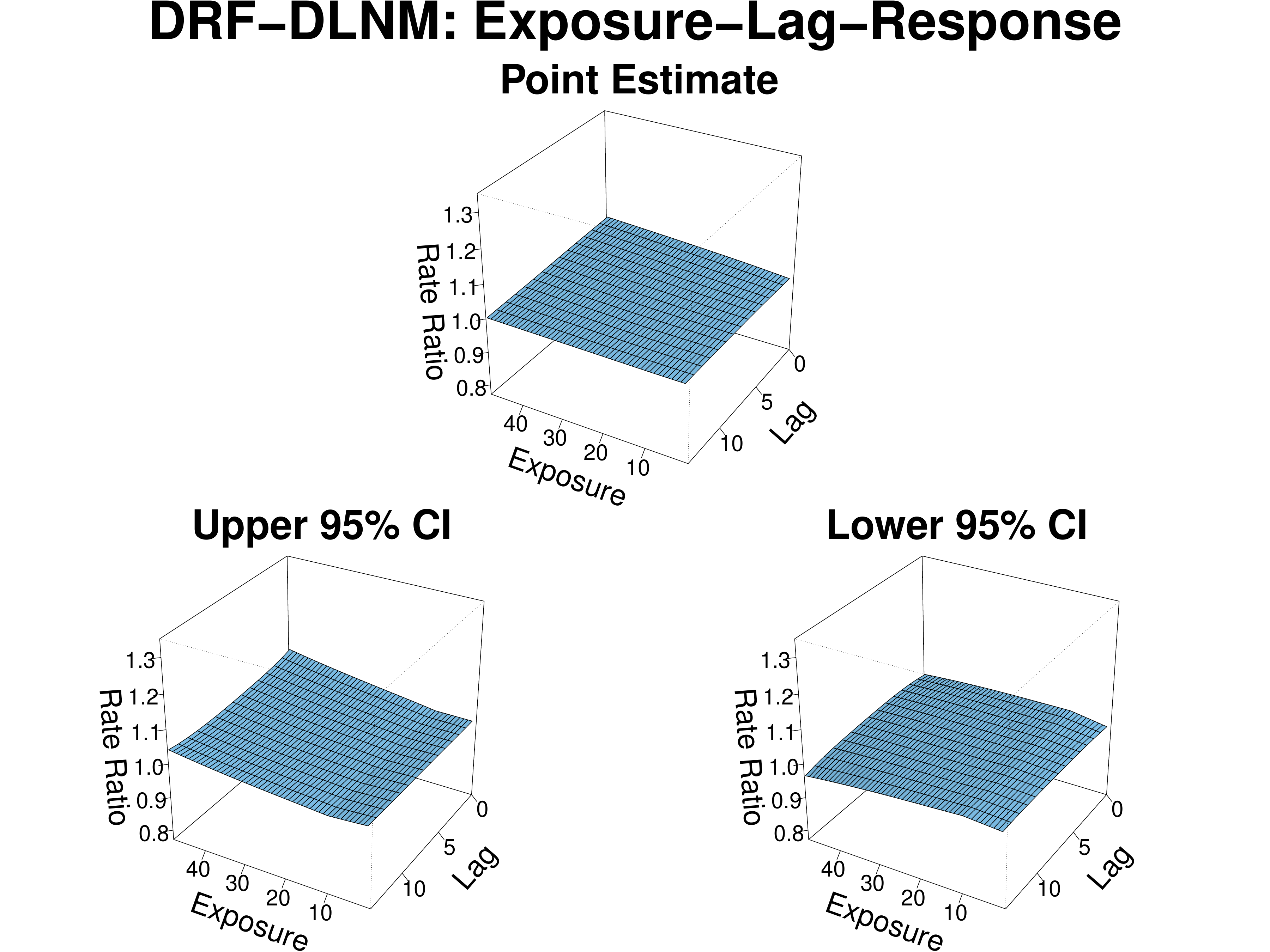}
      \caption{Estimated surfaces from DRF-DLNM}
    \end{subfigure}
    \begin{subfigure}[b]{\textwidth}
      \centering
      \includegraphics[width=0.9\linewidth]{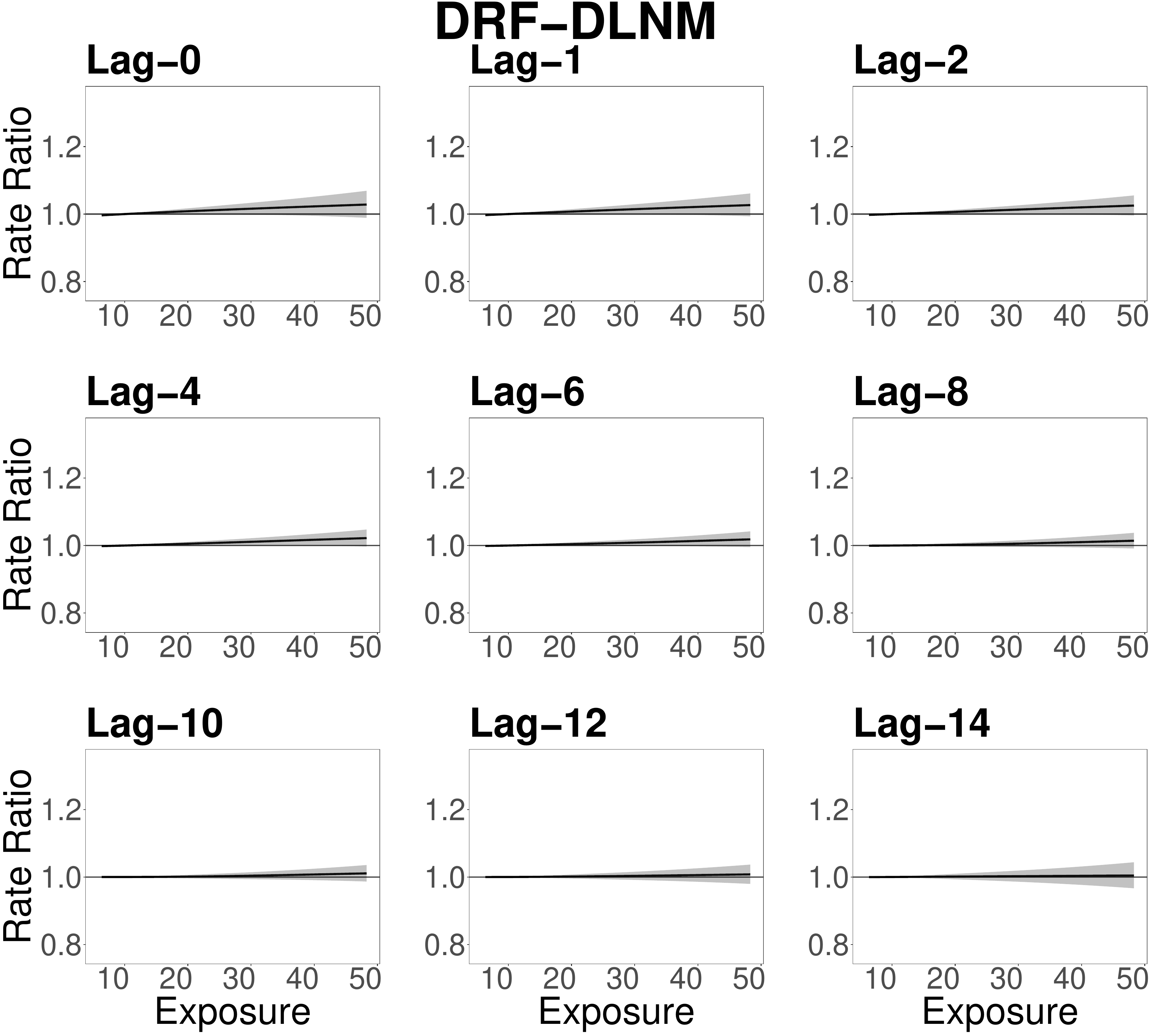}
      \caption{Estimated curves from DRF-DLNM}
    \end{subfigure}
  \end{subfigure}
    \caption{Web Figure 15. Comparisons between the ACE-DLNM and DRF-DLNM in the Waterloo dataset. The lag-specific and overall rate ratios from the ACE-DLNM (a) and DRF-DLNM (b) are plotted. Interpretation of the ACE-DLNM is based on the two curves shown in (c). Interpretation of the DRF-DLNM relies on the bivariate surfaces in (d), and we plot the estimated curves at selected lags in (e) (the full results are provided in Web Figure 16). The AICs for the two fitted models are reported. }
\end{figure}

\begin{figure}[H]
    \centering
    \includegraphics[width=0.7\linewidth]{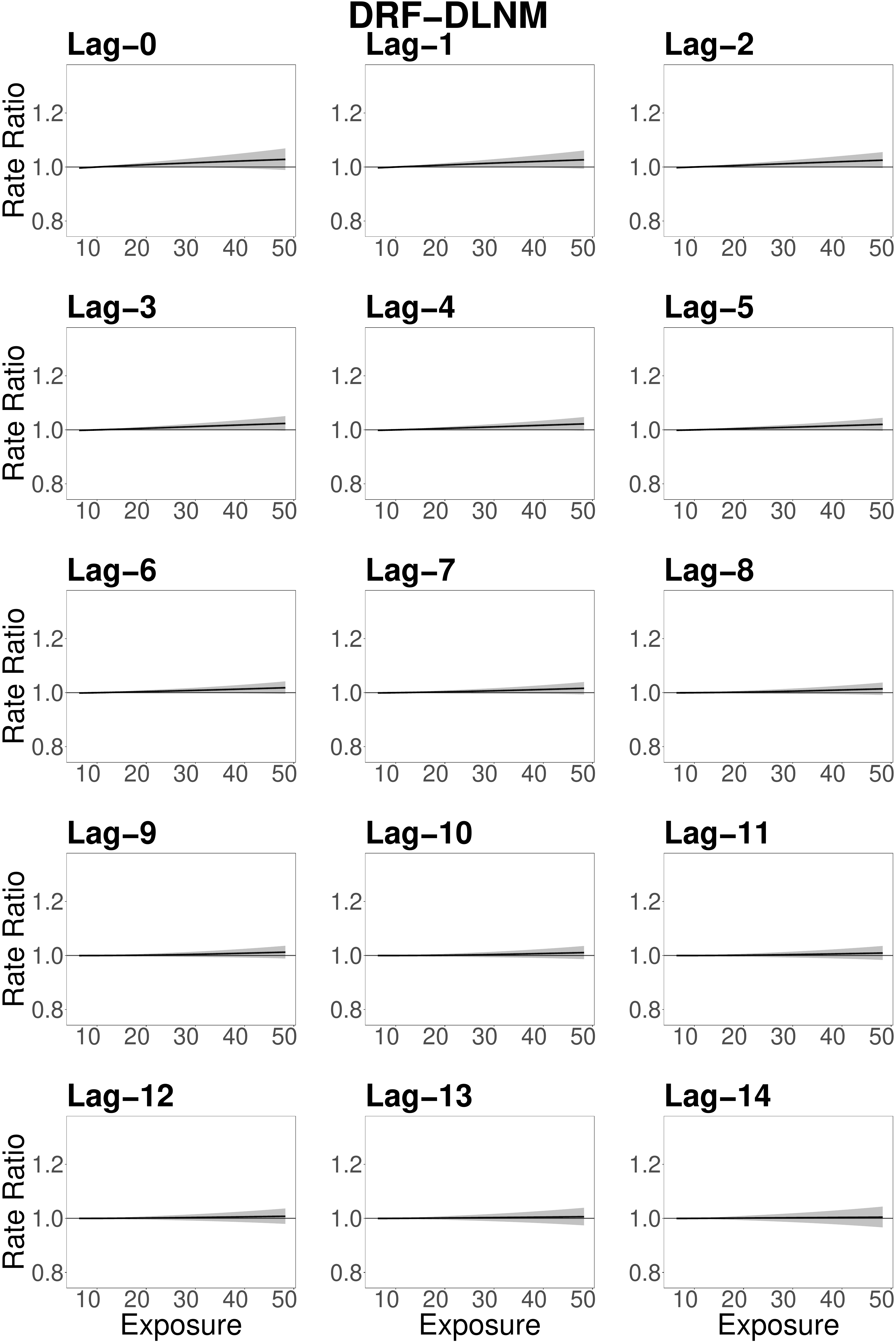}
    \caption{Web Figure 16. The estimated curves at each lag from the DRF-DLNM for the Waterloo dataset. }
    \label{fig:fcurve}
\end{figure}

\begin{figure}[H]
    \centering
     \begin{subfigure}[t]{0.5\textwidth}
        \centering
        \includegraphics[width=\linewidth]{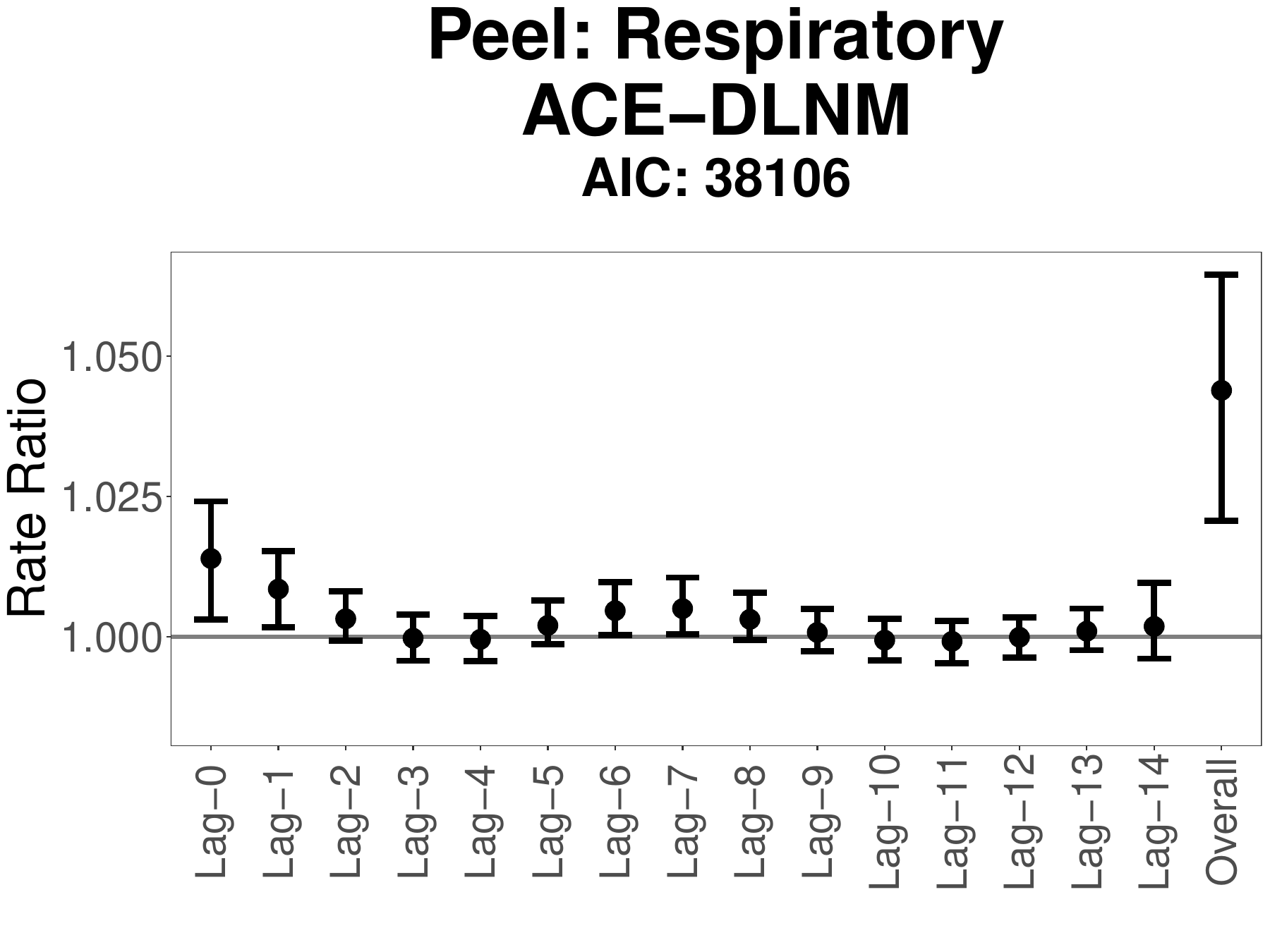}
        \caption{Rate ratios from ACE-DLNM}
    \end{subfigure}%
    \hfill
    \begin{subfigure}[t]{0.5\textwidth}
        \centering
        \includegraphics[width=\linewidth]{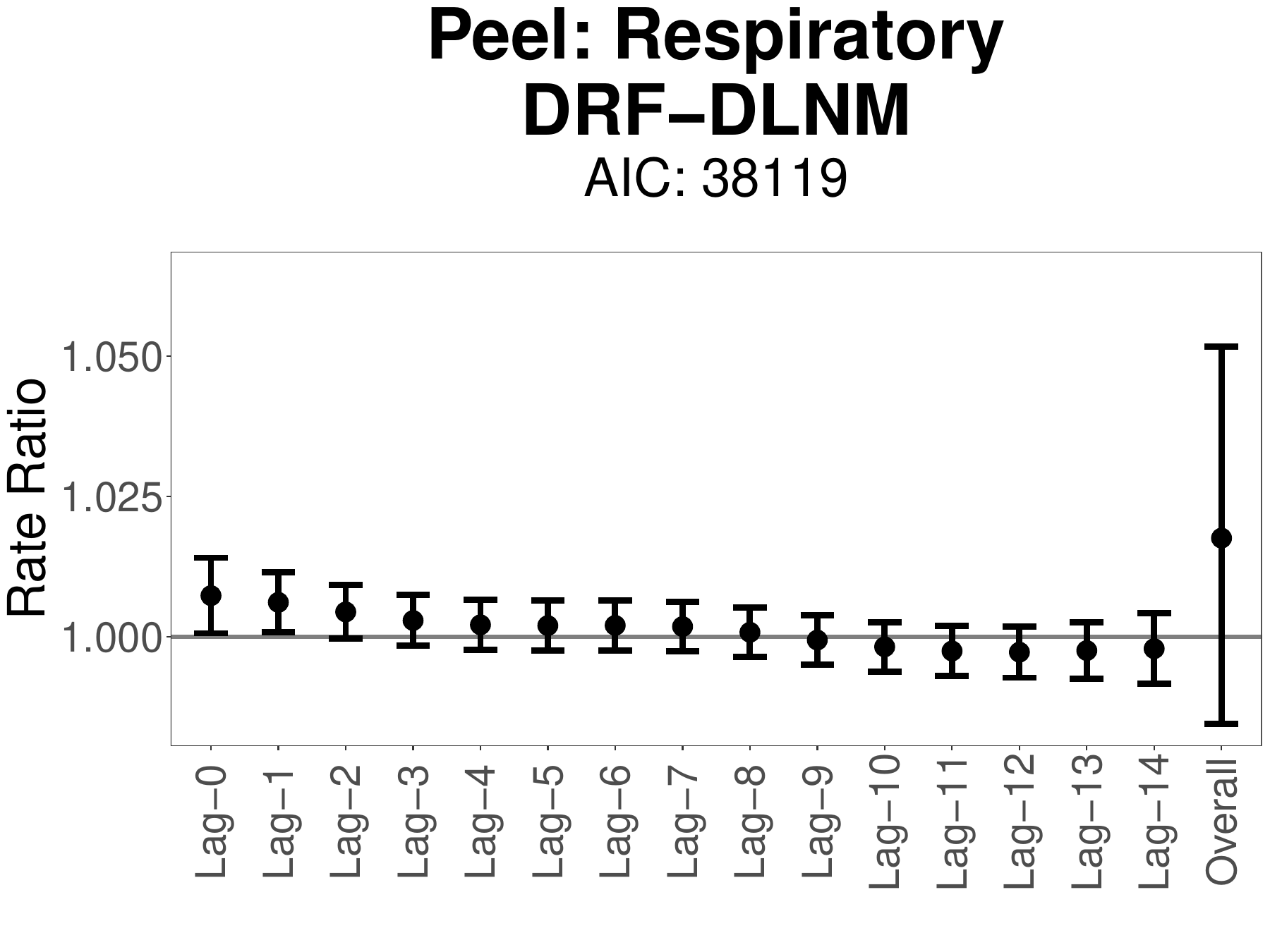}
        \caption{Rate ratios from DRF-DLNM}
    \end{subfigure}

    \vspace{0.5cm}
    \centering
    
    \begin{subfigure}[c]{0.48\textwidth}
    \centering
    \includegraphics[width=0.9\linewidth]{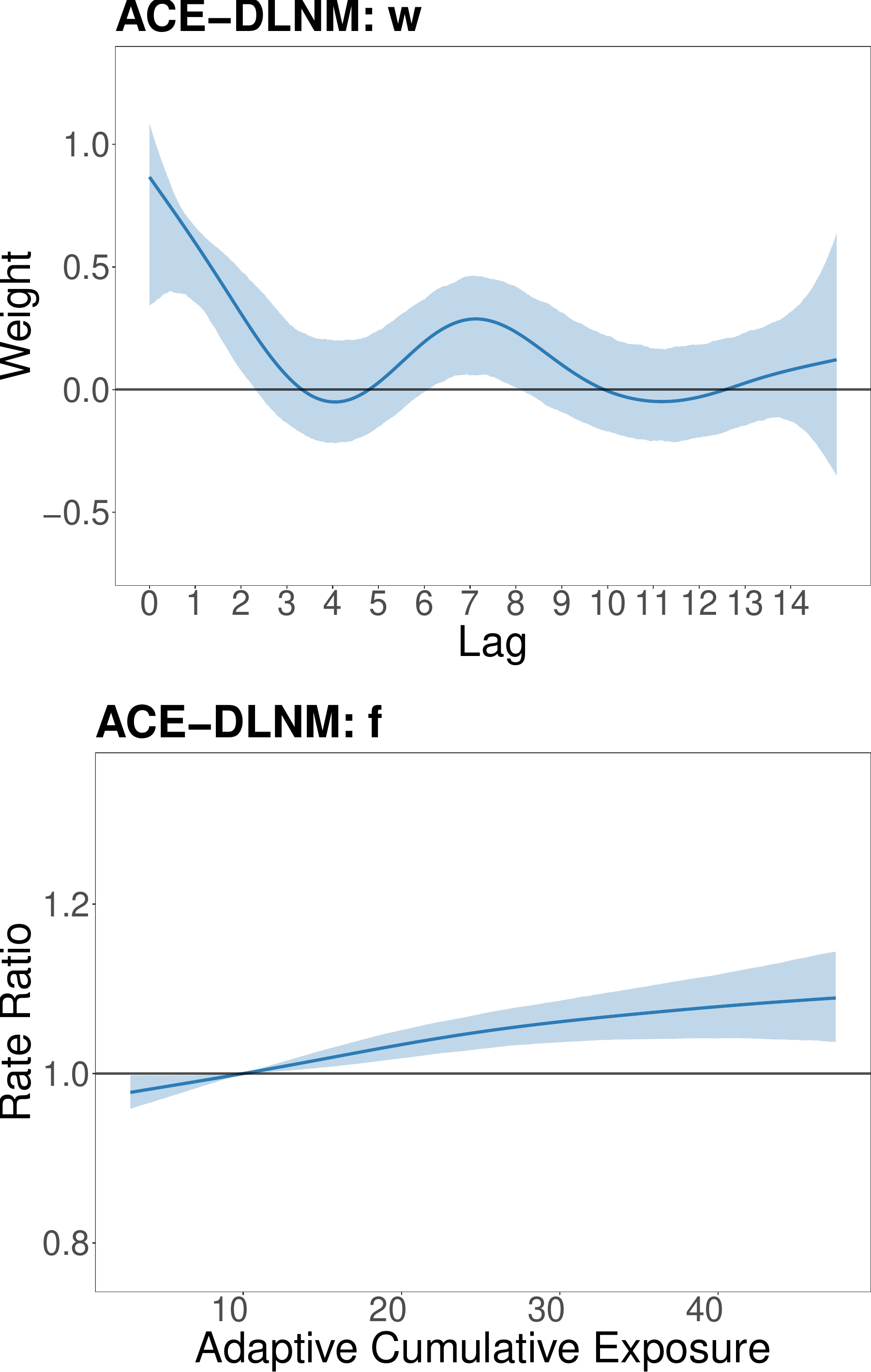}
    \caption{Estimated curves from ACE-DLNM}
    \end{subfigure}
    \hfill
    \begin{subfigure}[c]{0.48\textwidth}
    \centering
    \begin{subfigure}[b]{\textwidth}
      \centering
      \includegraphics[width=\linewidth]{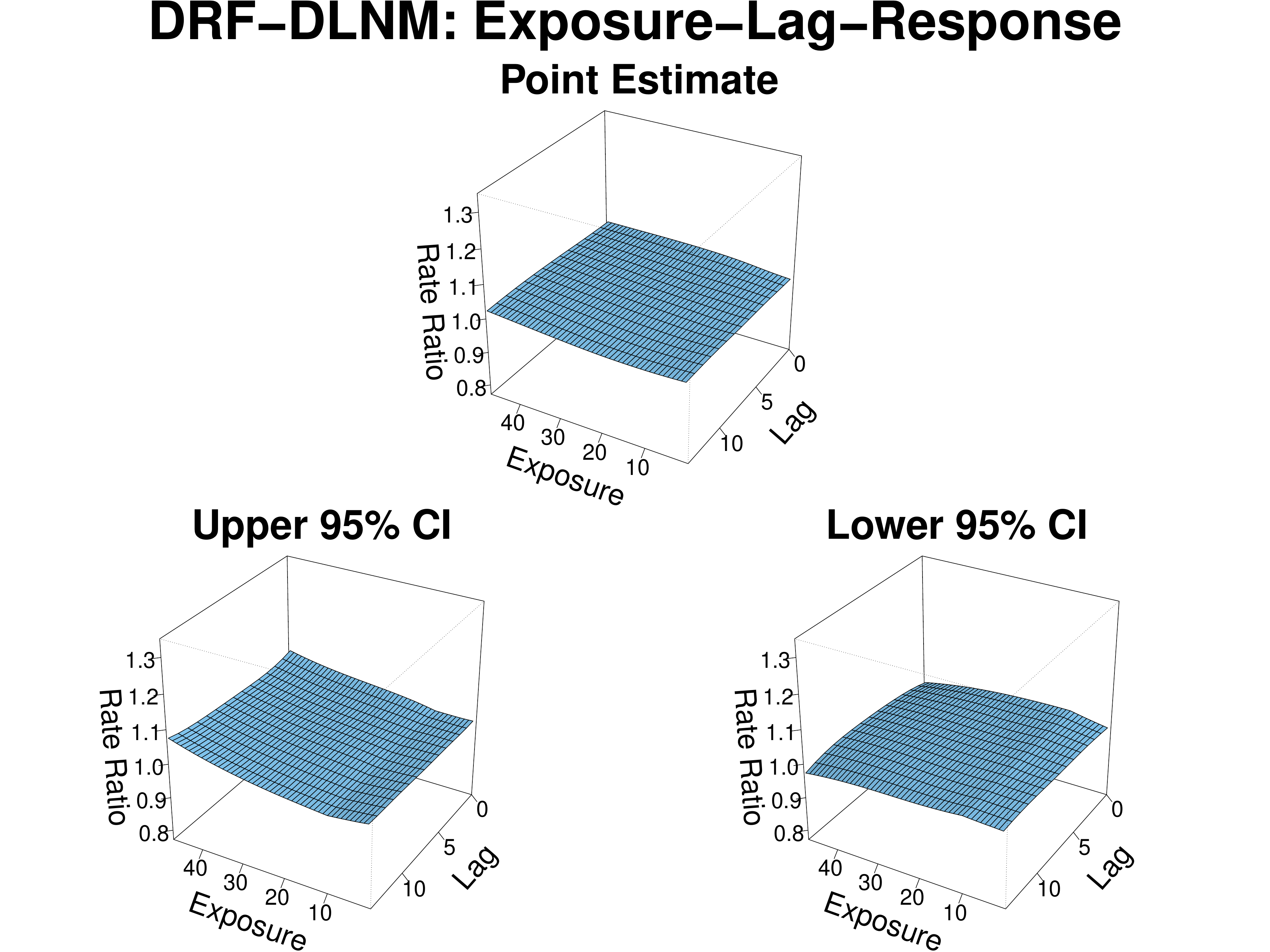}
      \caption{Estimated surfaces from DRF-DLNM}
    \end{subfigure}
    \begin{subfigure}[b]{\textwidth}
      \centering
      \includegraphics[width=0.9\linewidth]{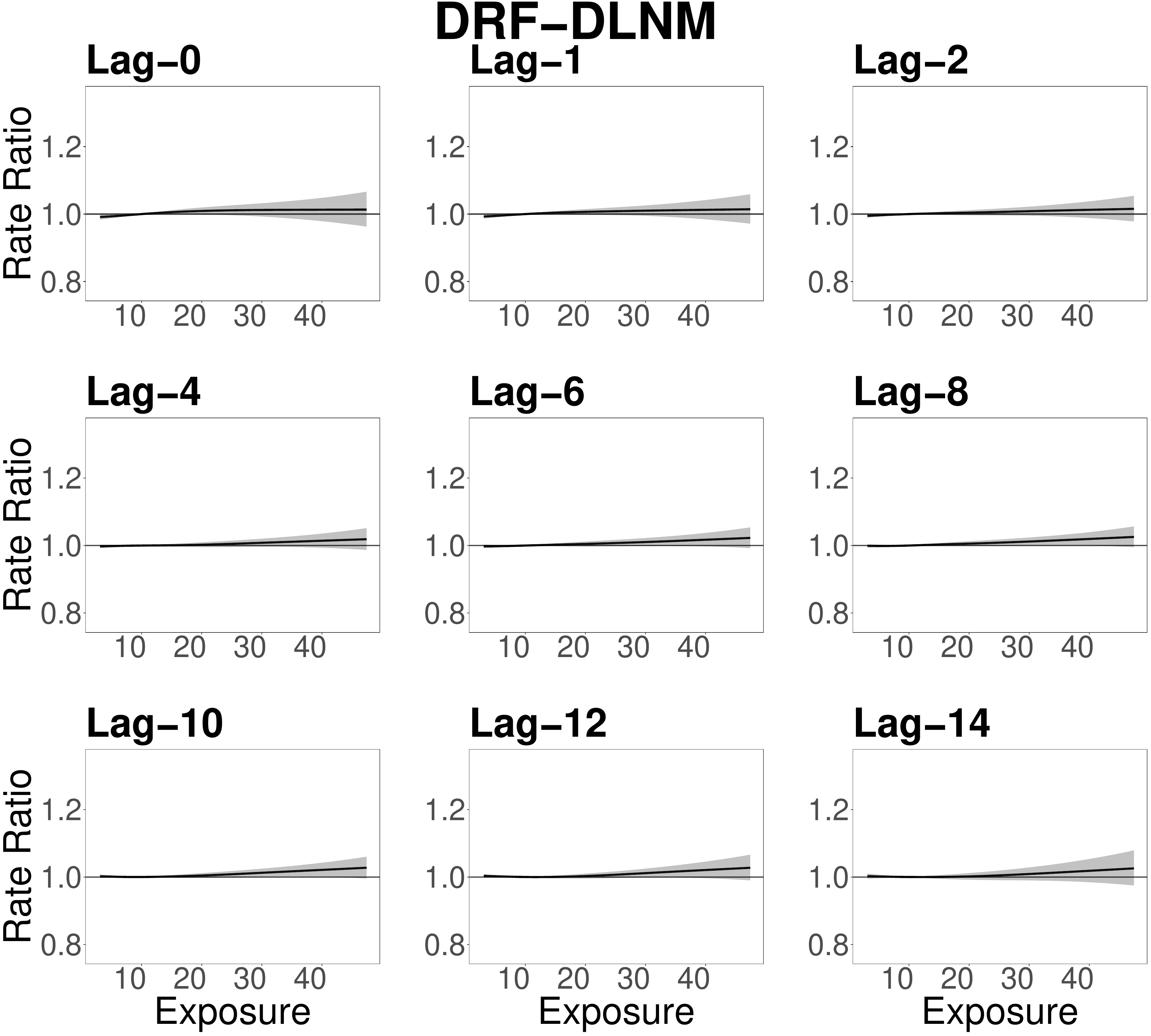}
      \caption{Estimated curves from DRF-DLNM}
    \end{subfigure}
  \end{subfigure}
    \caption{Web Figure 17. Comparisons between the ACE-DLNM and DRF-DLNM in the Peel dataset. The lag-specific and overall rate ratios from the ACE-DLNM (a) and DRF-DLNM (b) are plotted. Interpretation of the ACE-DLNM is based on the two curves shown in (c). Interpretation of the DRF-DLNM relies on the bivariate surfaces in (d), and we plot the estimated curves at selected lags in (e) (the full results are provided in Web Figure 18). The AICs for the two fitted models are reported. }
\end{figure}

\begin{figure}[H]
    \centering
    \includegraphics[width=0.7\linewidth]{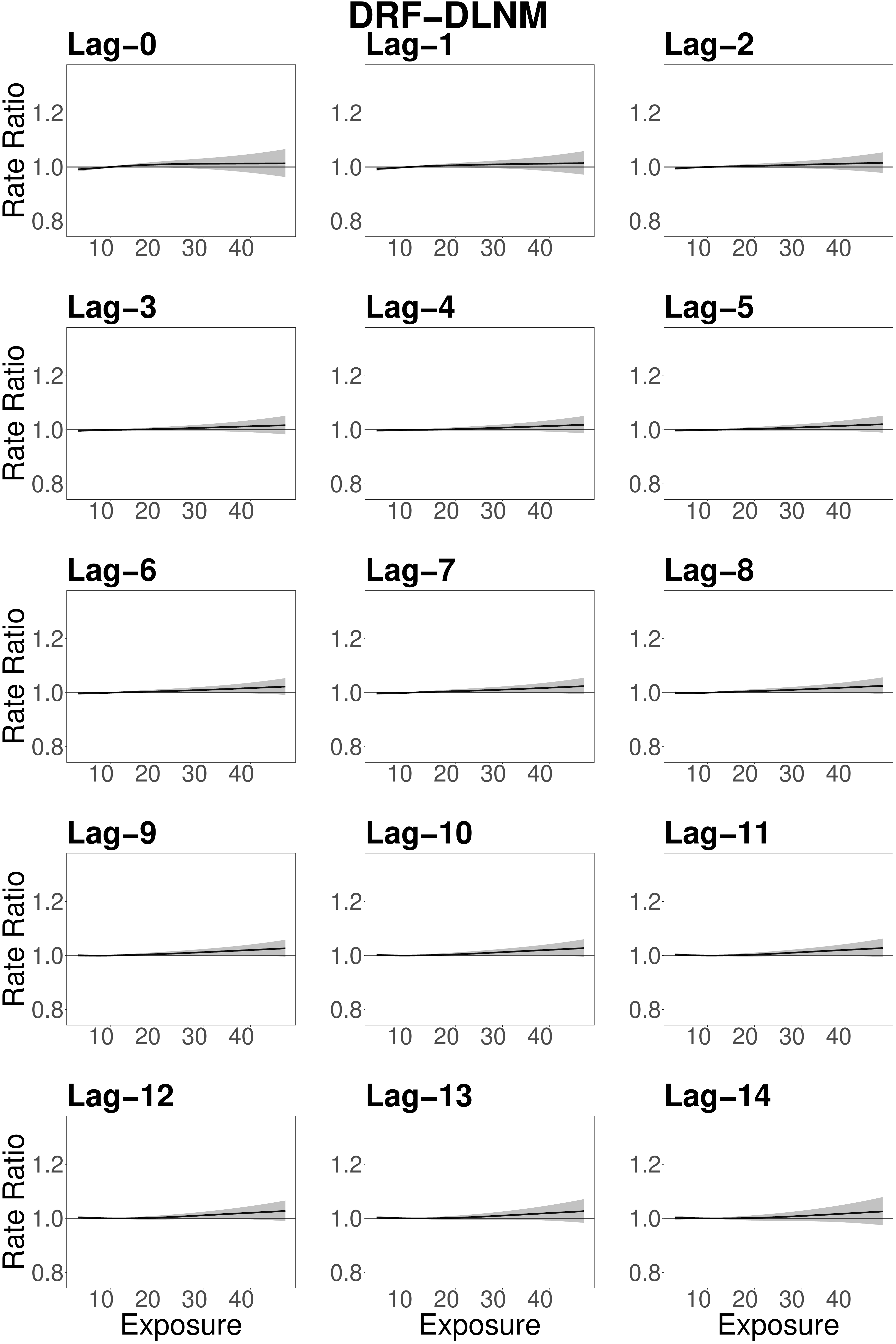}
    \caption{Web Figure 18. The estimated curves at each lag from the DRF-DLNM for the Peel dataset. }
    \label{fig:fcurve}
\end{figure}

\begin{figure}[H]
    \centering
     \begin{subfigure}[t]{0.5\textwidth}
        \centering
        \includegraphics[width=\linewidth]{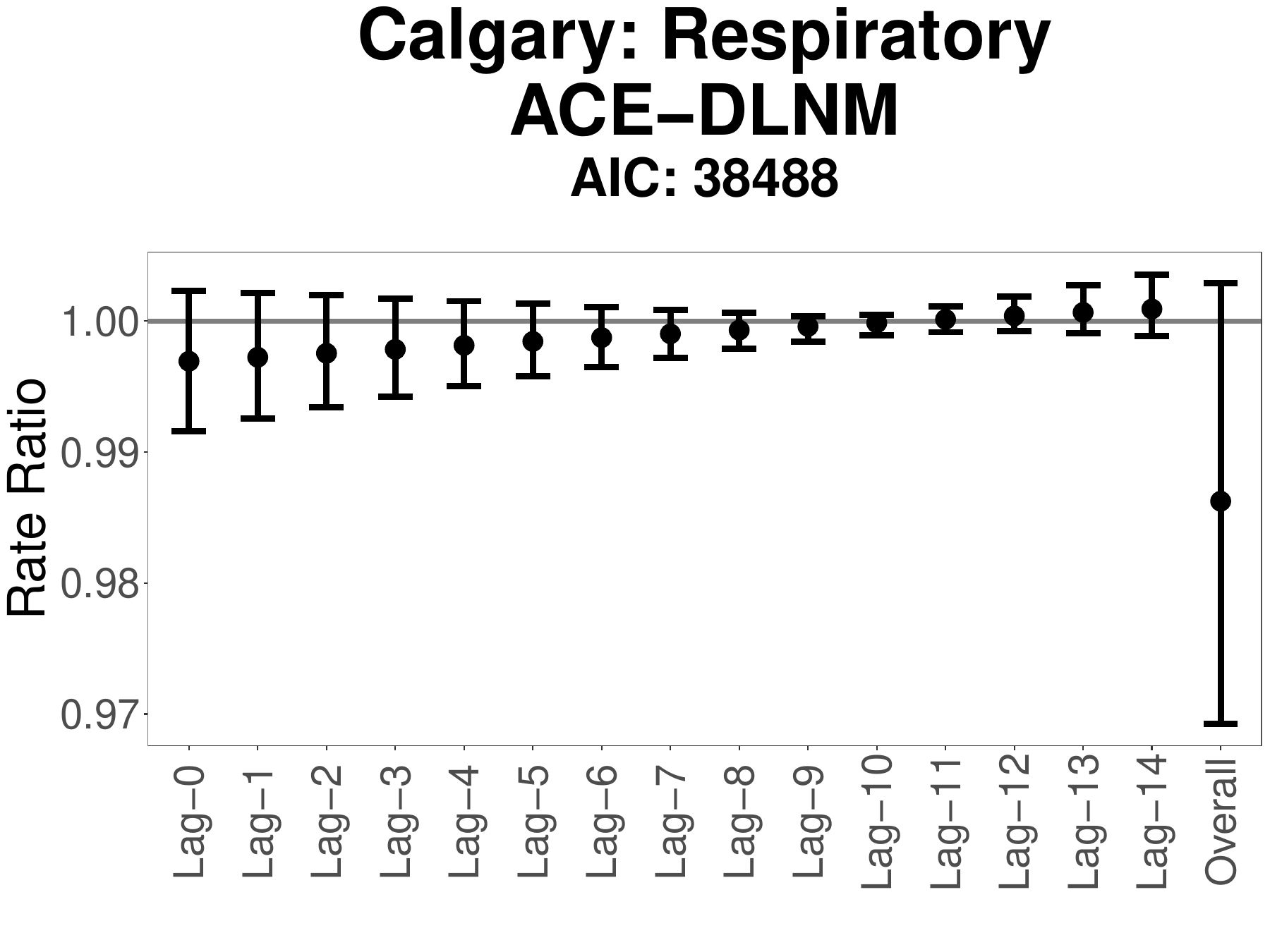}
        \caption{Rate ratios from ACE-DLNM}
    \end{subfigure}%
    \hfill
    \begin{subfigure}[t]{0.5\textwidth}
        \centering
        \includegraphics[width=\linewidth]{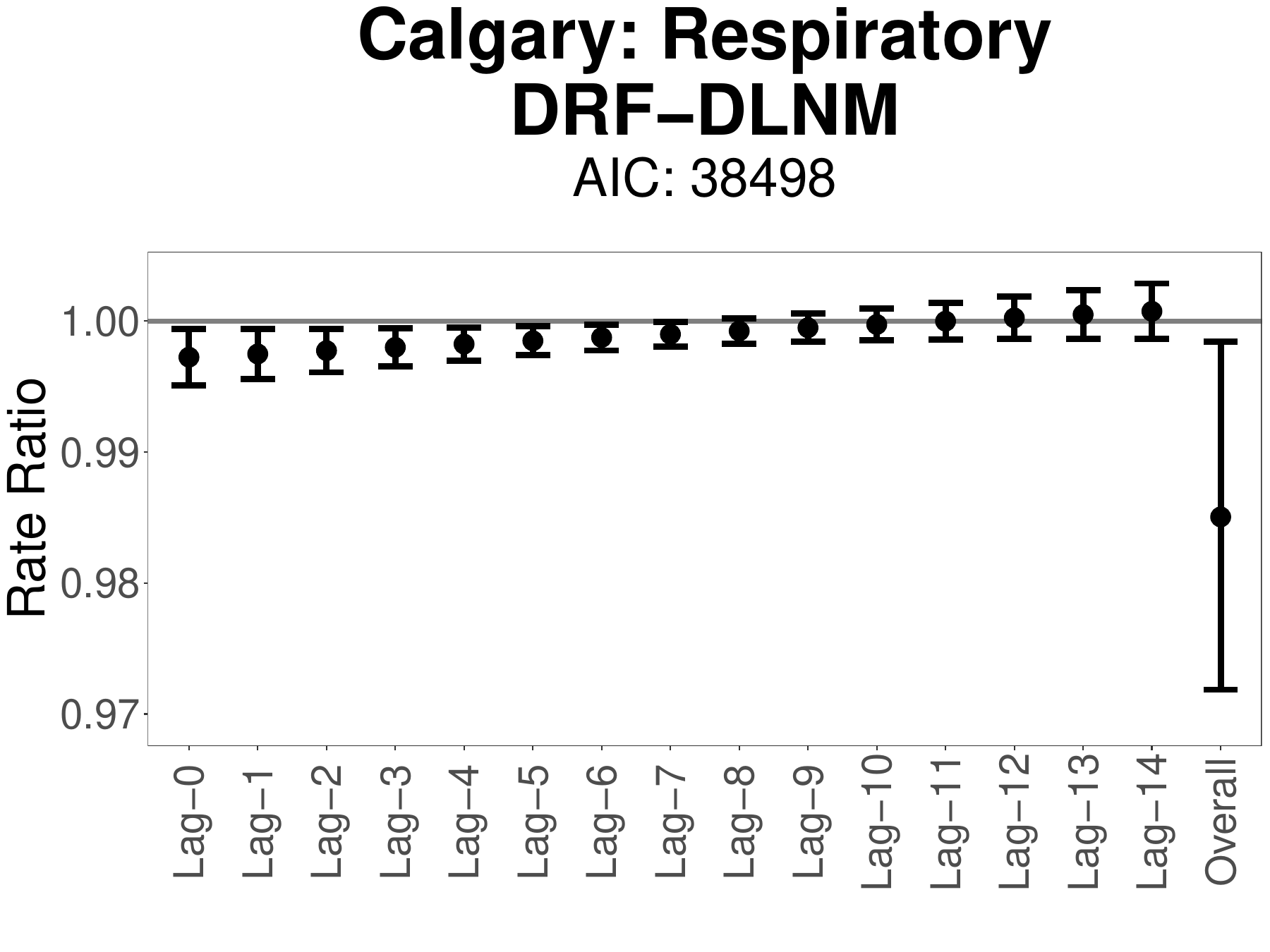}
        \caption{Rate ratios from DRF-DLNM}
    \end{subfigure}

    \vspace{0.5cm}
    \centering
    
    \begin{subfigure}[c]{0.48\textwidth}
    \centering
    \includegraphics[width=0.9\linewidth]{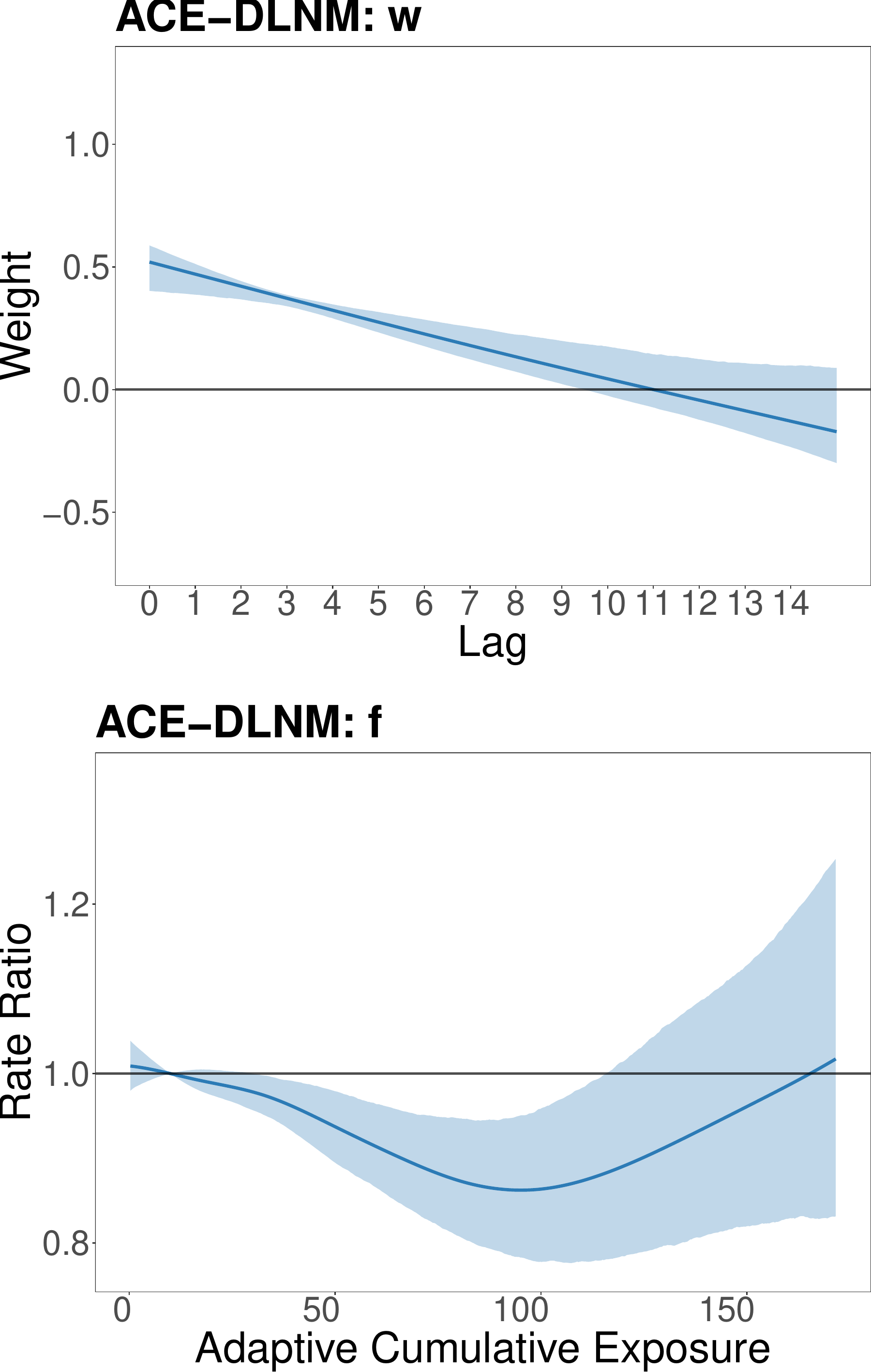}
    \caption{Estimated curves from ACE-DLNM}
    \end{subfigure}
    \hfill
    \begin{subfigure}[c]{0.48\textwidth}
    \centering
    \begin{subfigure}[b]{\textwidth}
      \centering
      \includegraphics[width=\linewidth]{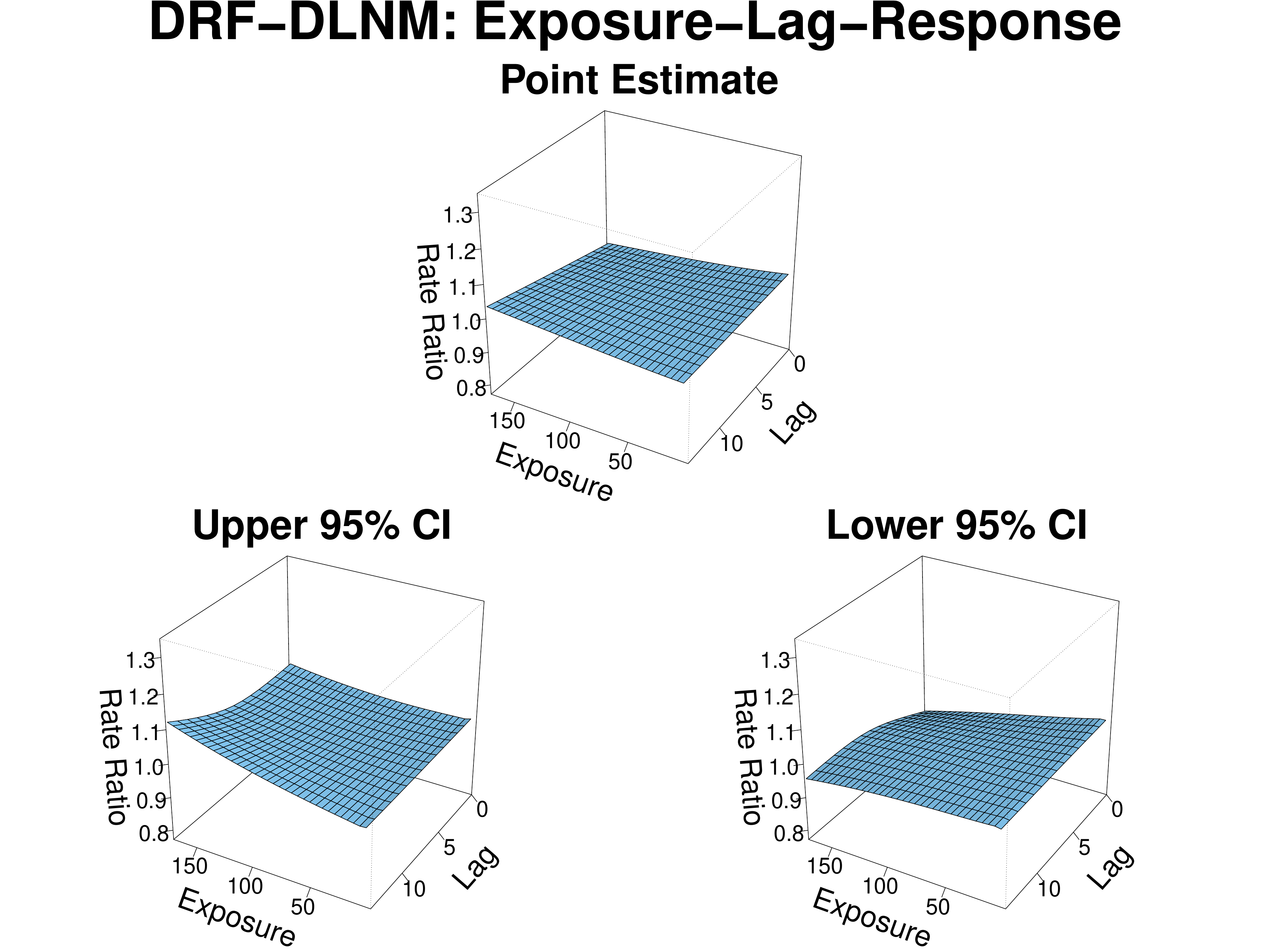}
      \caption{Estimated surfaces from DRF-DLNM}
    \end{subfigure}
    \begin{subfigure}[b]{\textwidth}
      \centering
      \includegraphics[width=0.9\linewidth]{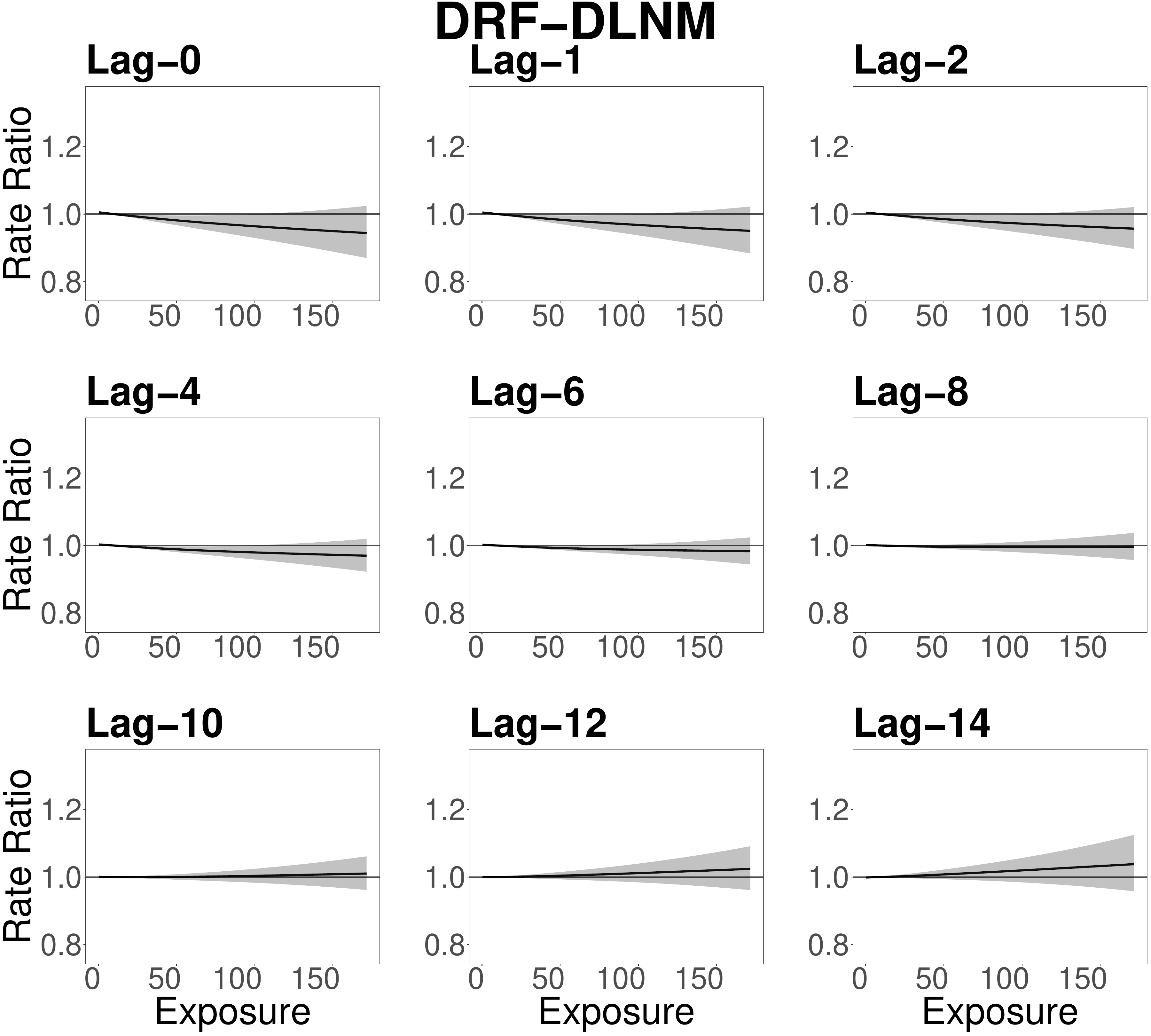}
      \caption{Estimated curves from DRF-DLNM}
    \end{subfigure}
  \end{subfigure}
    \caption{Web Figure 19. Comparisons between the ACE-DLNM and DRF-DLNM in the Calgary dataset. The lag-specific and overall rate ratios from the ACE-DLNM (a) and DRF-DLNM (b) are plotted. Interpretation of the ACE-DLNM is based on the two curves shown in (c). Interpretation of the DRF-DLNM relies on the bivariate surfaces in (d), and we plot the estimated curves at selected lags in (e) (the full results are provided in Web Figure 20). The AICs for the two fitted models are reported. }
\end{figure}

\begin{figure}[H]
    \centering
    \includegraphics[width=0.7\linewidth]{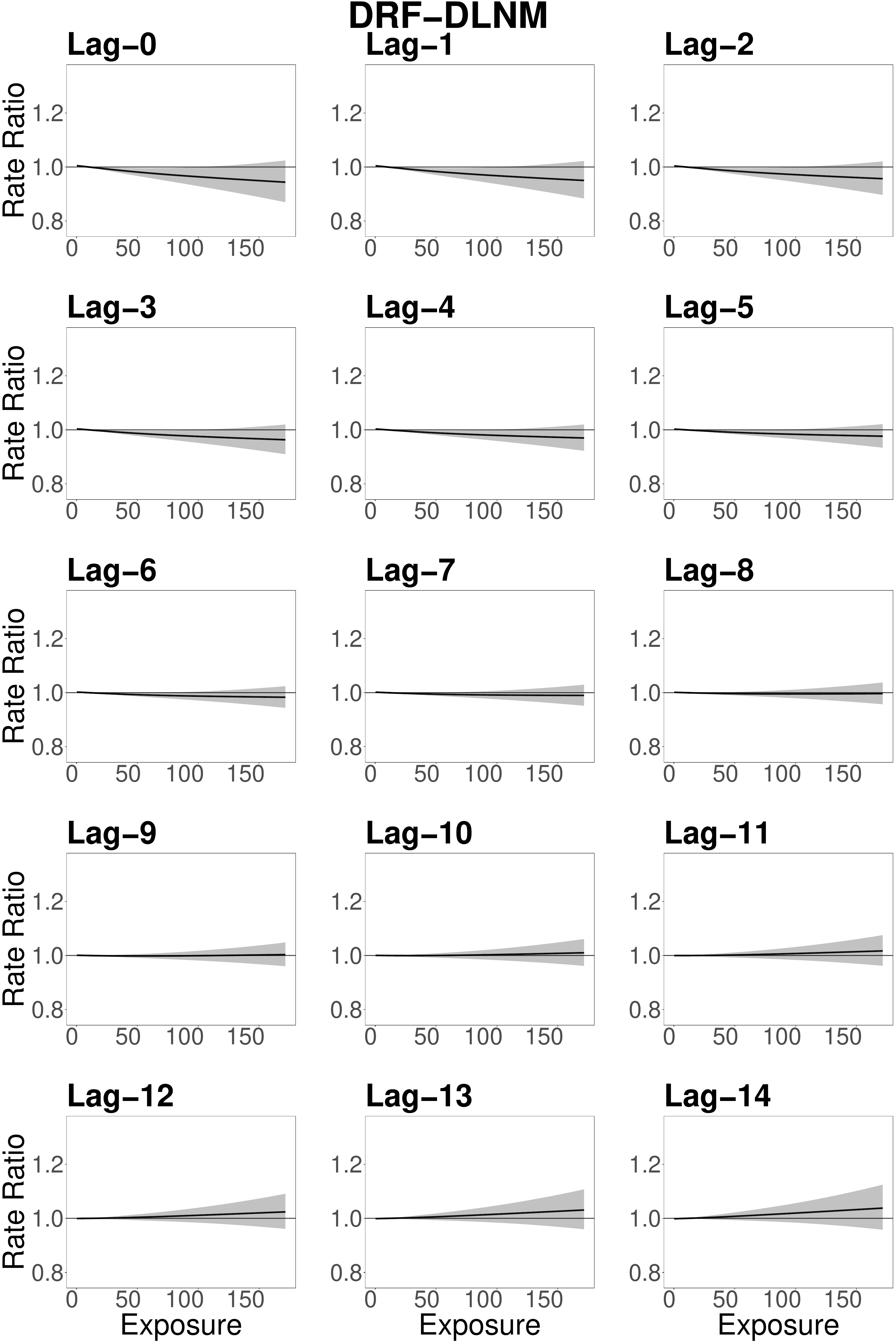}
    \caption{Web Figure 20. The estimated curves at each lag from the DRF-DLNM for the Calgary dataset. }
    \label{fig:fcurve}
\end{figure}

\begin{figure}[H]
    \centering
     \begin{subfigure}[t]{0.5\textwidth}
        \centering
        \includegraphics[width=\linewidth]{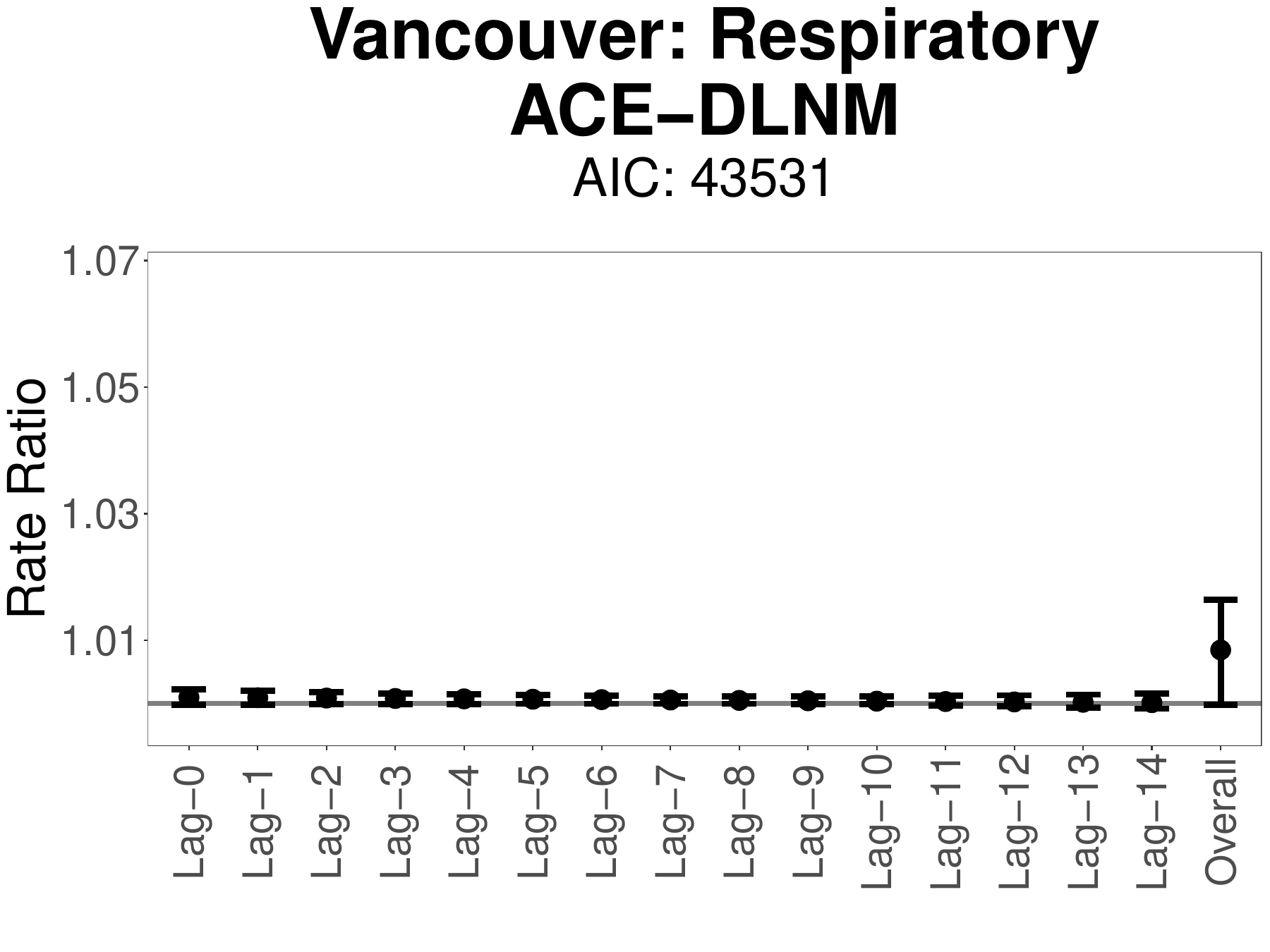}
        \caption{Rate ratios from ACE-DLNM}
    \end{subfigure}%
    \hfill
    \begin{subfigure}[t]{0.5\textwidth}
        \centering
        \includegraphics[width=\linewidth]{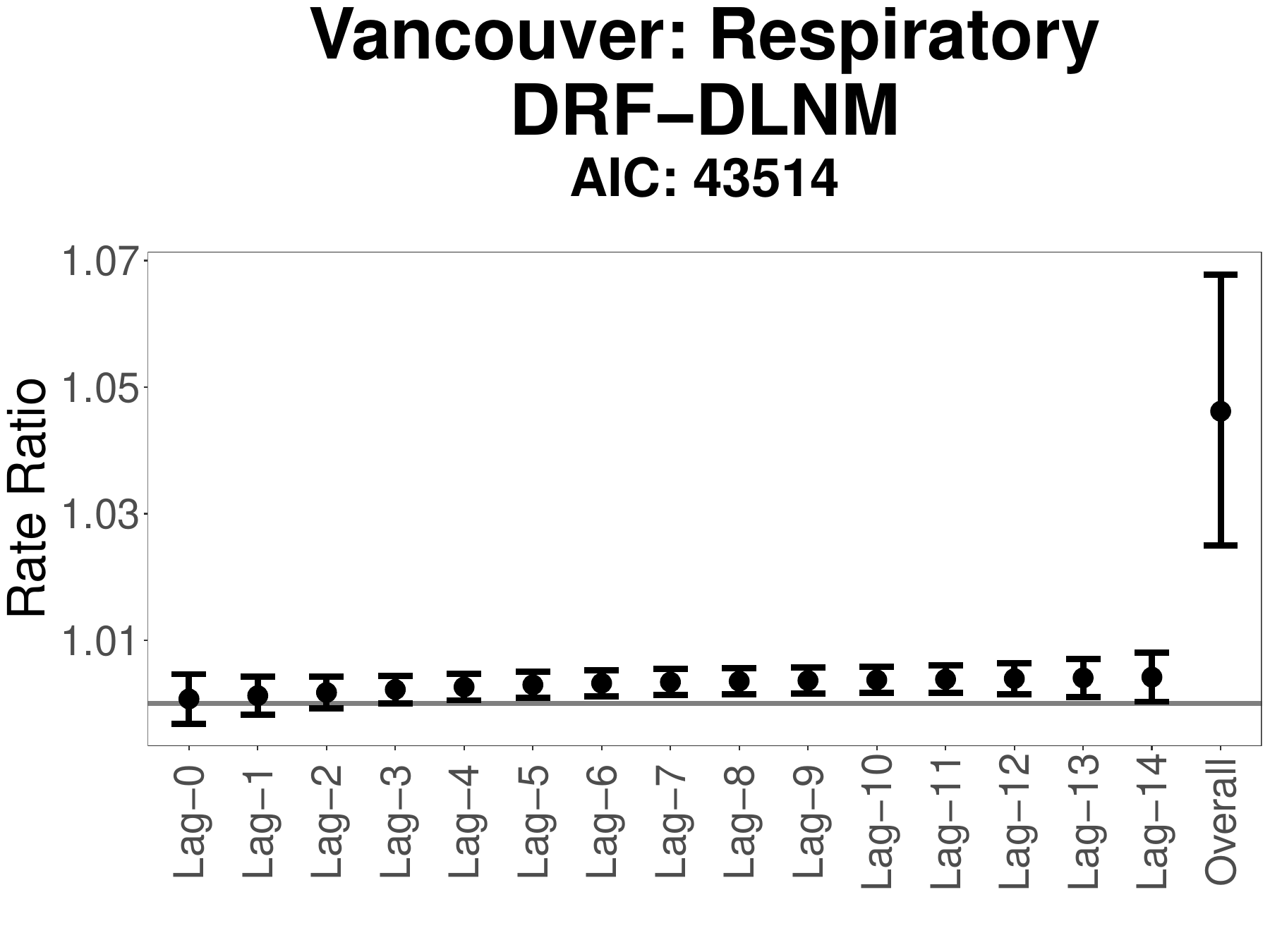}
        \caption{Rate ratios from DRF-DLNM}
    \end{subfigure}

    \vspace{0.5cm}
    \centering
    
    \begin{subfigure}[c]{0.48\textwidth}
    \centering
    \includegraphics[width=0.9\linewidth]{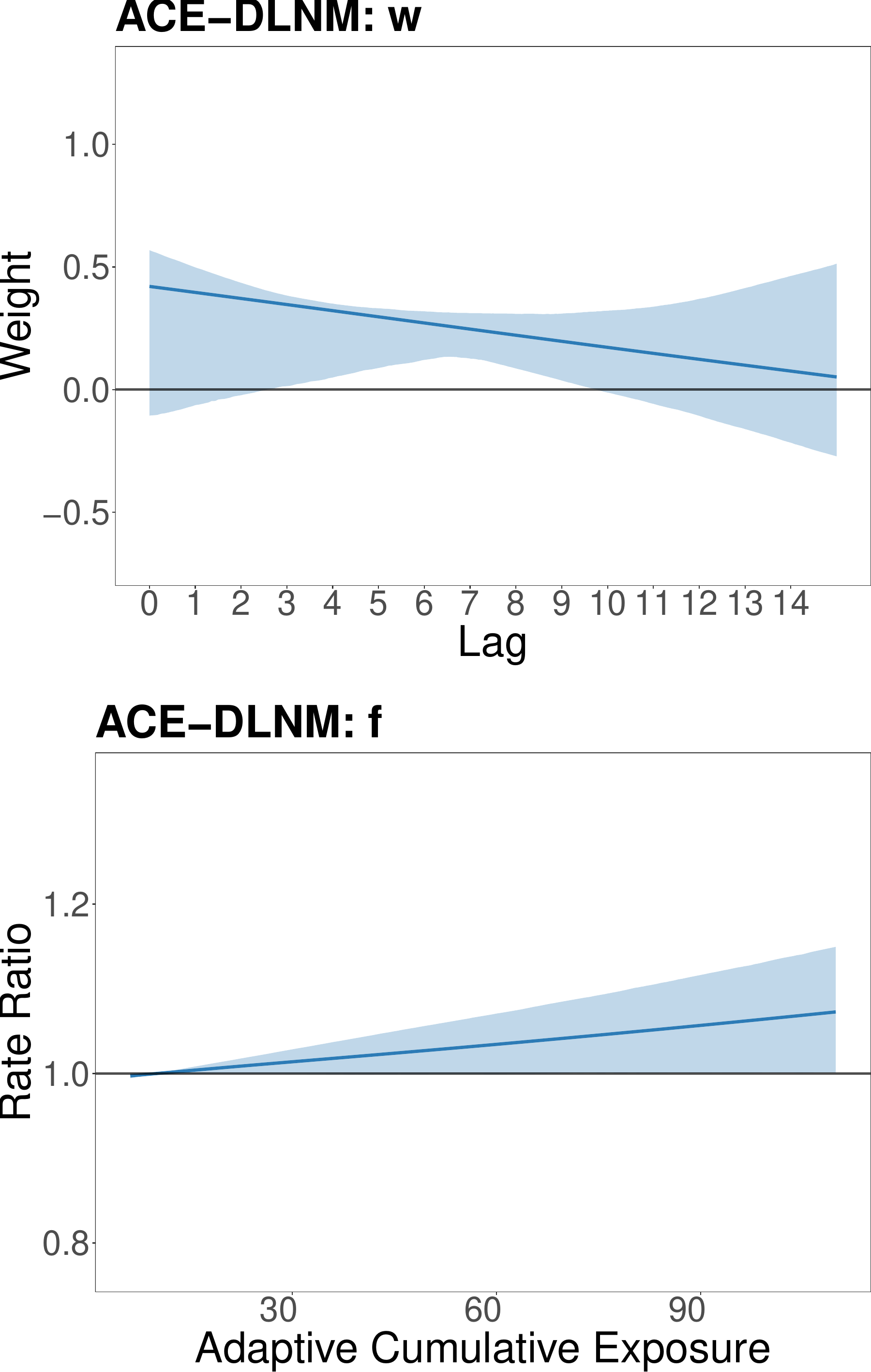}
    \caption{Estimated curves from ACE-DLNM}
    \end{subfigure}
    \hfill
    \begin{subfigure}[c]{0.48\textwidth}
    \centering
    \begin{subfigure}[b]{\textwidth}
      \centering
      \includegraphics[width=\linewidth]{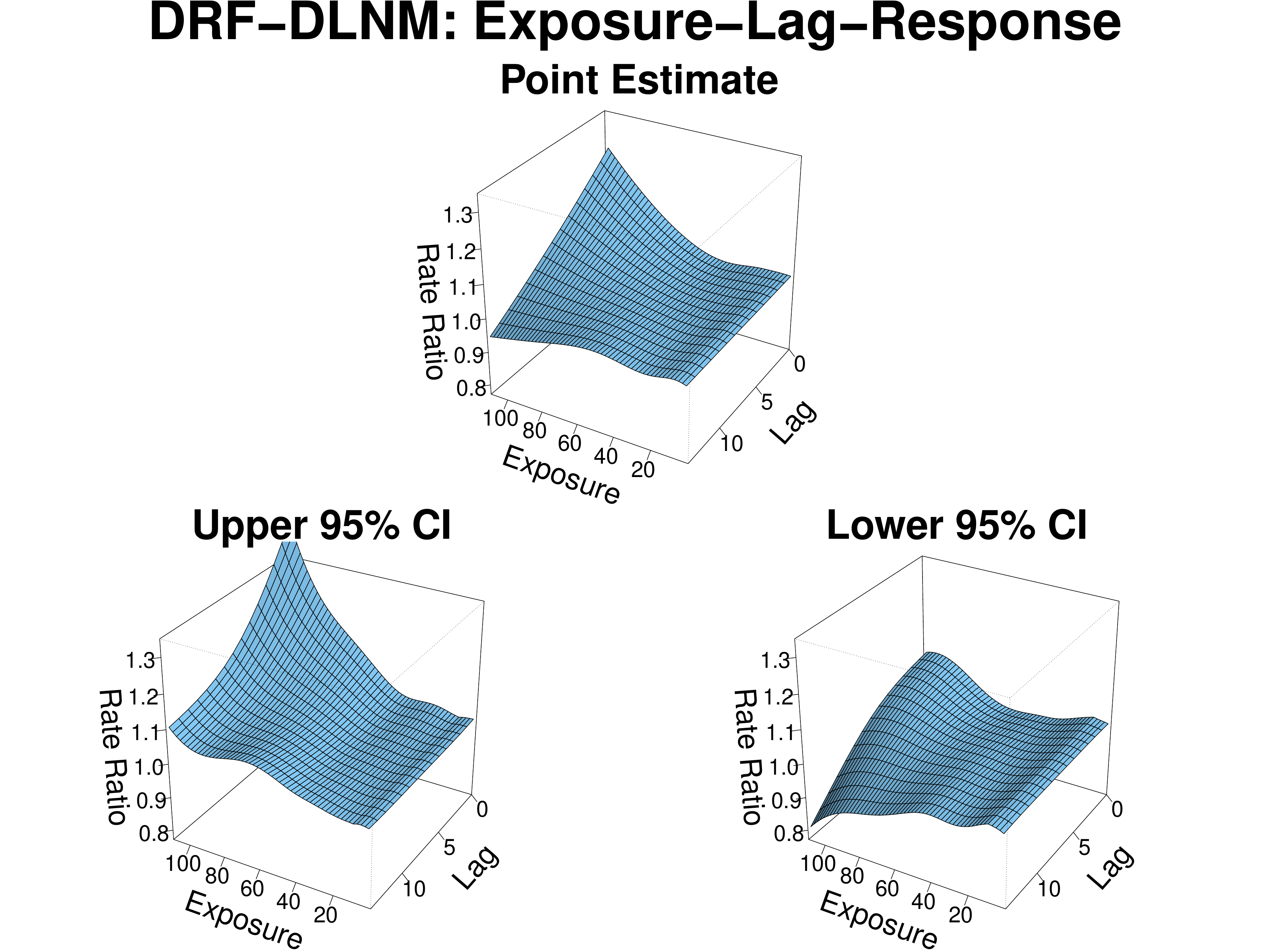}
      \caption{Estimated surfaces from DRF-DLNM}
    \end{subfigure}
    \begin{subfigure}[b]{\textwidth}
      \centering
      \includegraphics[width=0.9\linewidth]{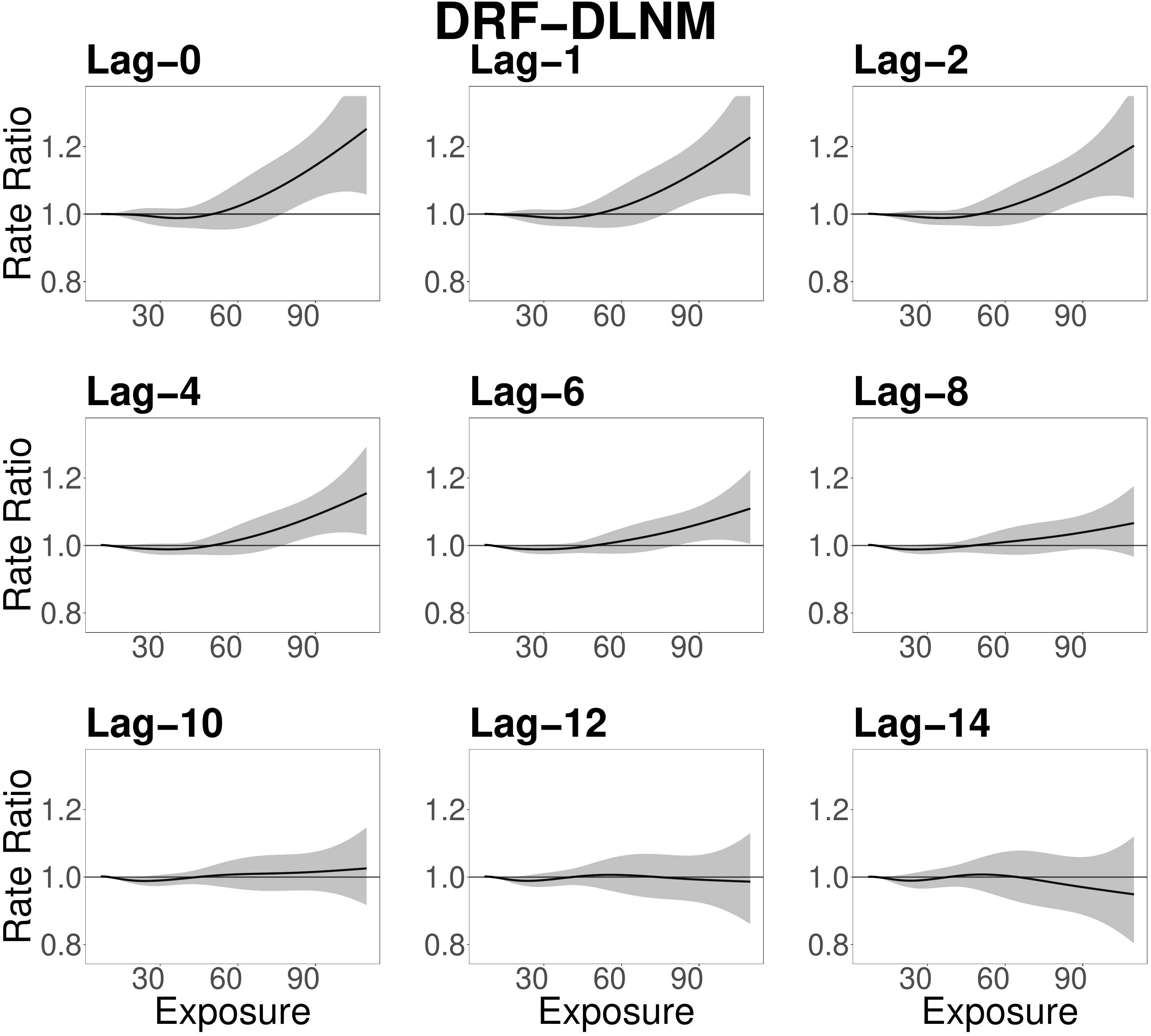}
      \caption{Estimated curves from DRF-DLNM}
    \end{subfigure}
  \end{subfigure}
    \caption{Web Figure 21. Comparisons between the ACE-DLNM and DRF-DLNM in the Vancouver dataset. The lag-specific and overall rate ratios from the ACE-DLNM (a) and DRF-DLNM (b) are plotted. Interpretation of the ACE-DLNM is based on the two curves shown in (c). Interpretation of the DRF-DLNM relies on the bivariate surfaces in (d), and we plot the estimated curves at selected lags in (e) (the full results are provided in Web Figure 22). The AICs for the two fitted models are reported. }
\end{figure}

\begin{figure}[H]
    \centering
    \includegraphics[width=0.7\linewidth]{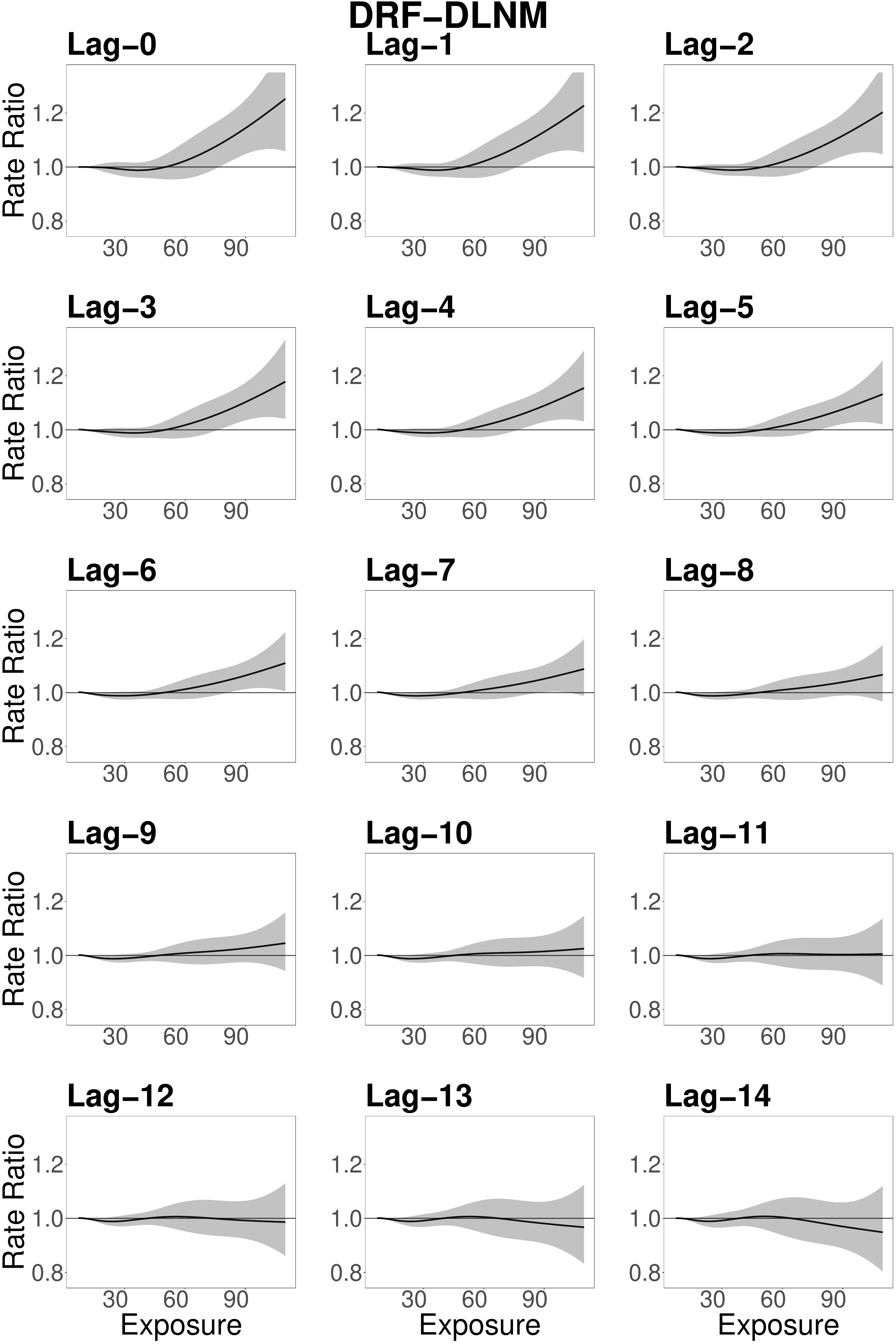}
    \caption{Web Figure 22. The estimated curves at each lag from the DRF-DLNM for the Vancouver dataset. }
    \label{fig:fcurve}
\end{figure}

\begin{figure}[H]
    \centering
     \begin{subfigure}{0.68\textwidth}
        \centering
        \includegraphics[width=\linewidth]{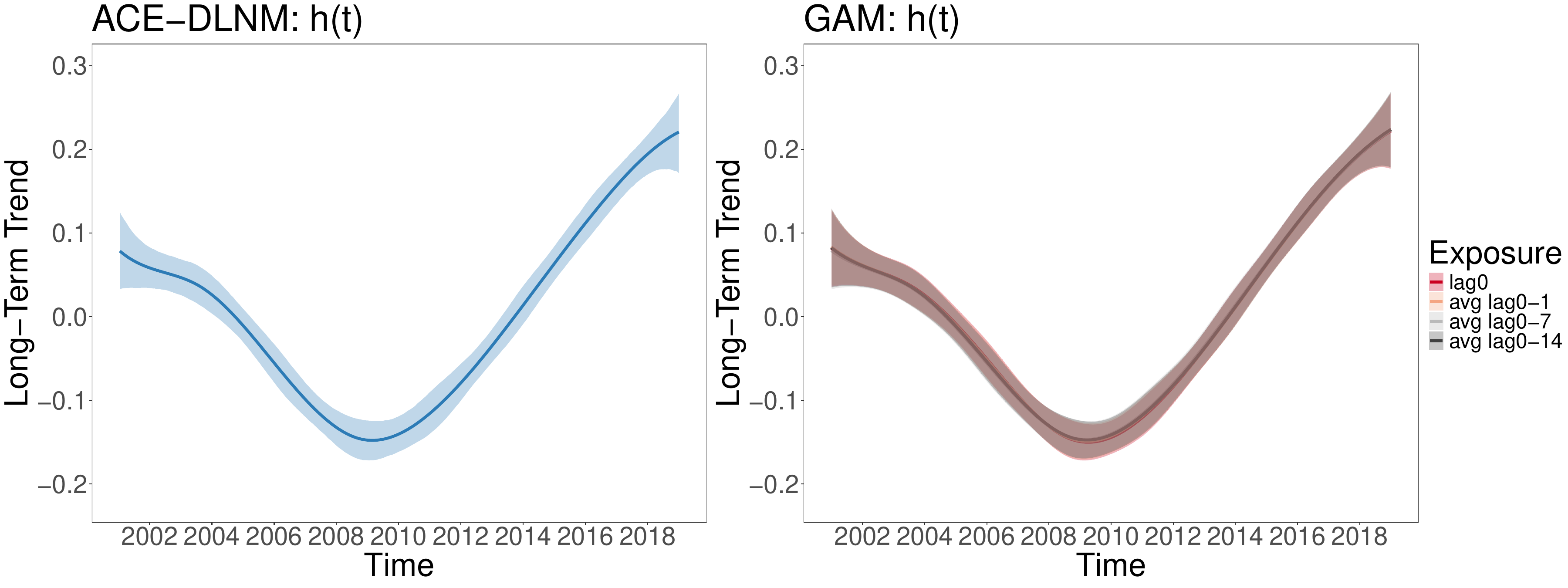}
    \end{subfigure}%

    \vspace{0.1cm}
    \begin{subfigure}{0.68\textwidth}
        \centering
        \includegraphics[width=\linewidth]{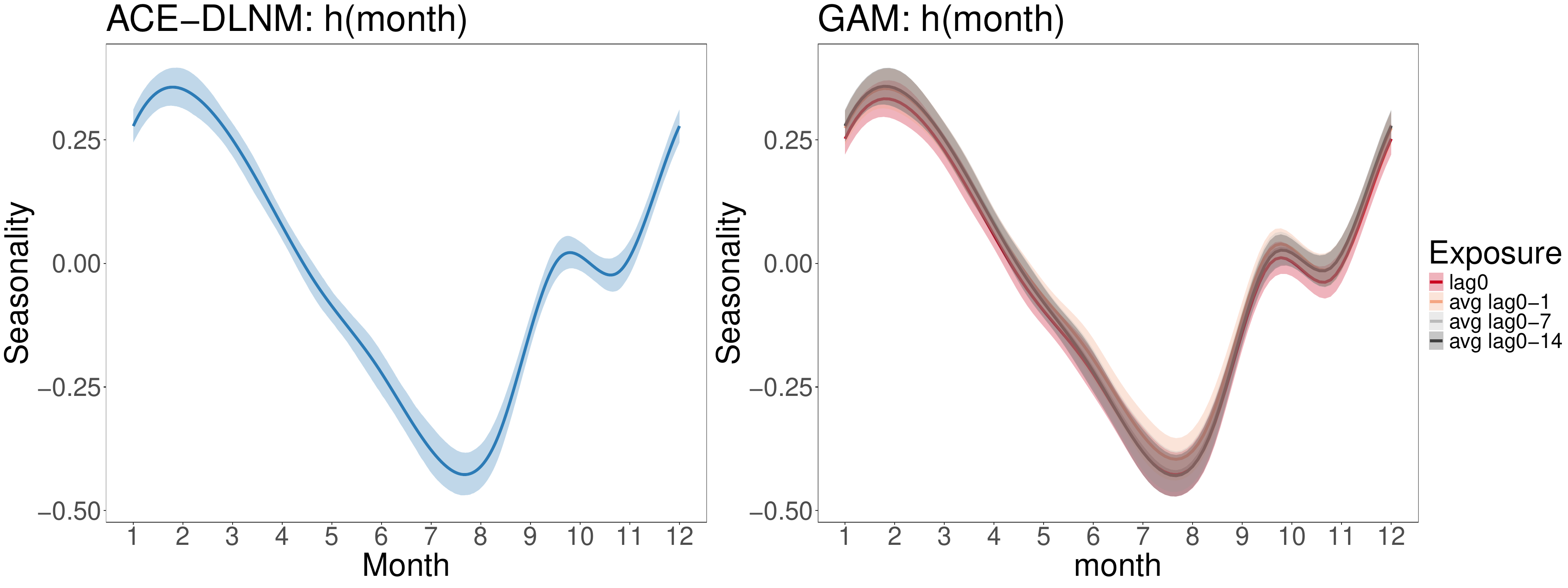}
    \end{subfigure}%

    \vspace{0.1cm}
   \begin{subfigure}{0.68\textwidth}
        \centering
        \includegraphics[width=\linewidth]{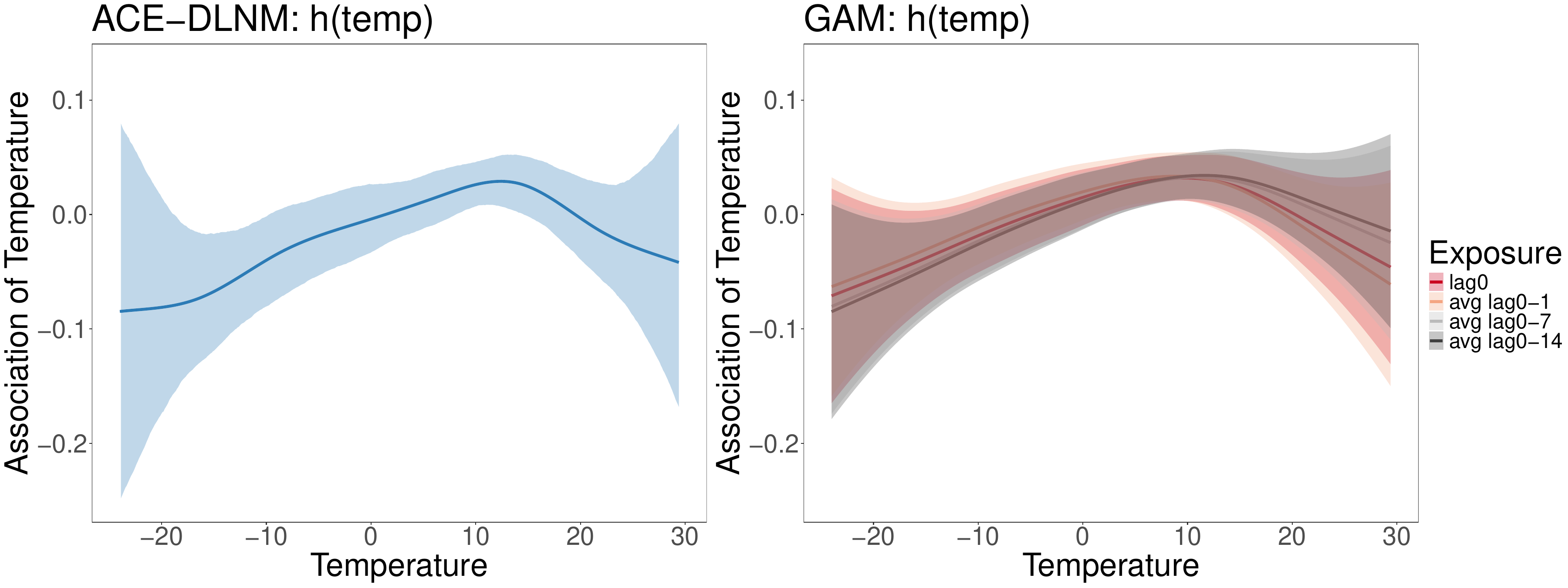}
    \end{subfigure}%

     \vspace{0.1cm}
   \begin{subfigure}{0.68\textwidth}
        \centering
        \includegraphics[width=\linewidth]{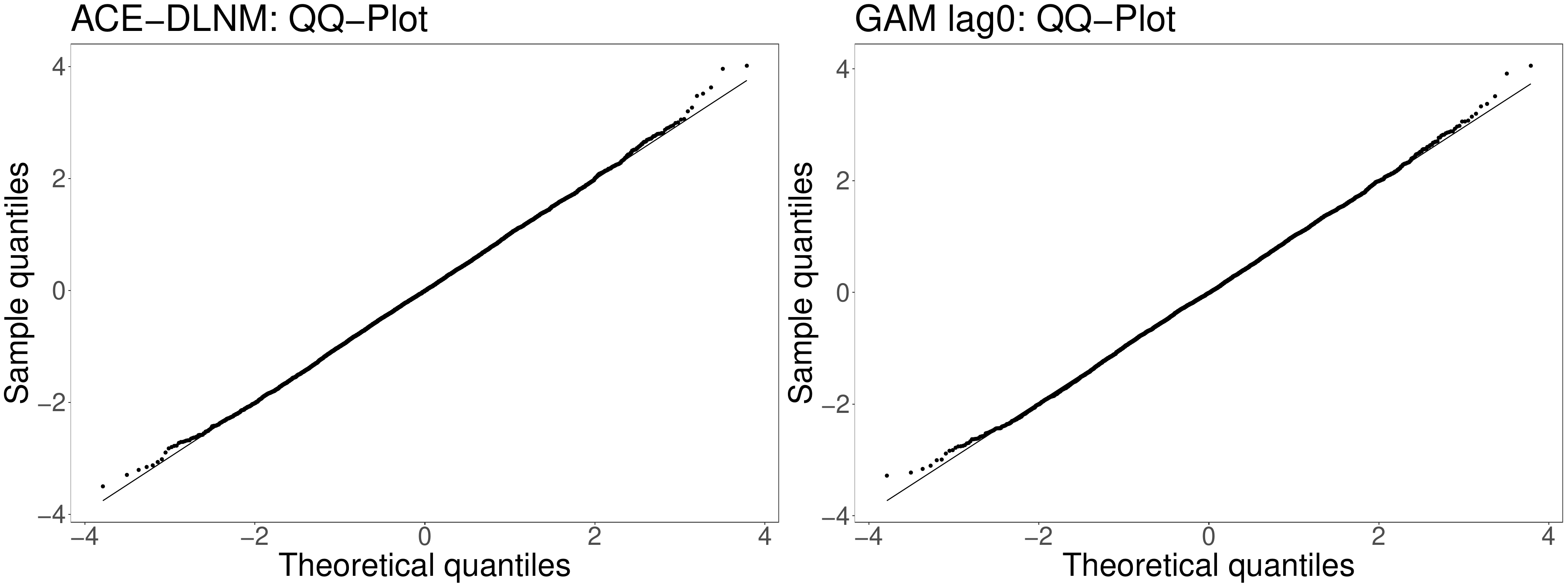}
    \end{subfigure}%

    \vspace{0.1cm}
   \begin{subfigure}{0.68\textwidth}
        \centering
        \includegraphics[width=\linewidth]{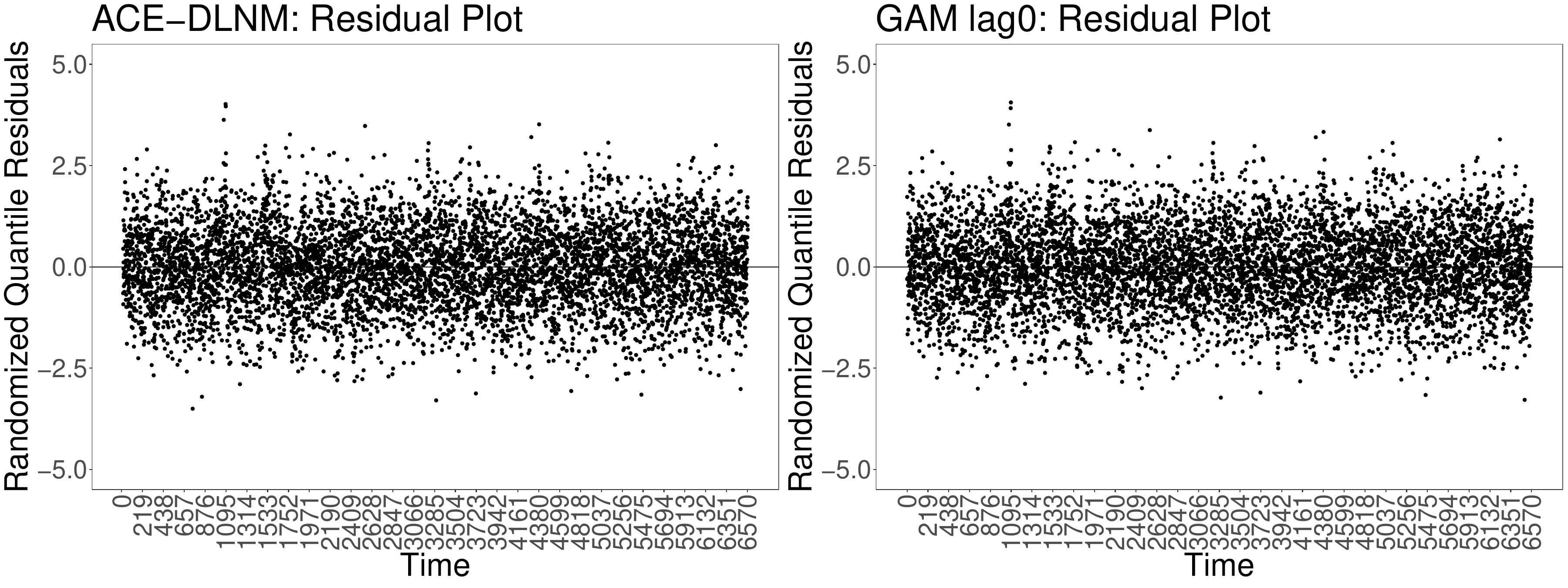}
    \end{subfigure}%
    \caption{Web Figure 23. Additional Results for Respiratory Morbidity in Waterloo. }
\end{figure}

\begin{figure}[H]
    \centering
     \begin{subfigure}{0.68\textwidth}
        \centering
        \includegraphics[width=\linewidth]{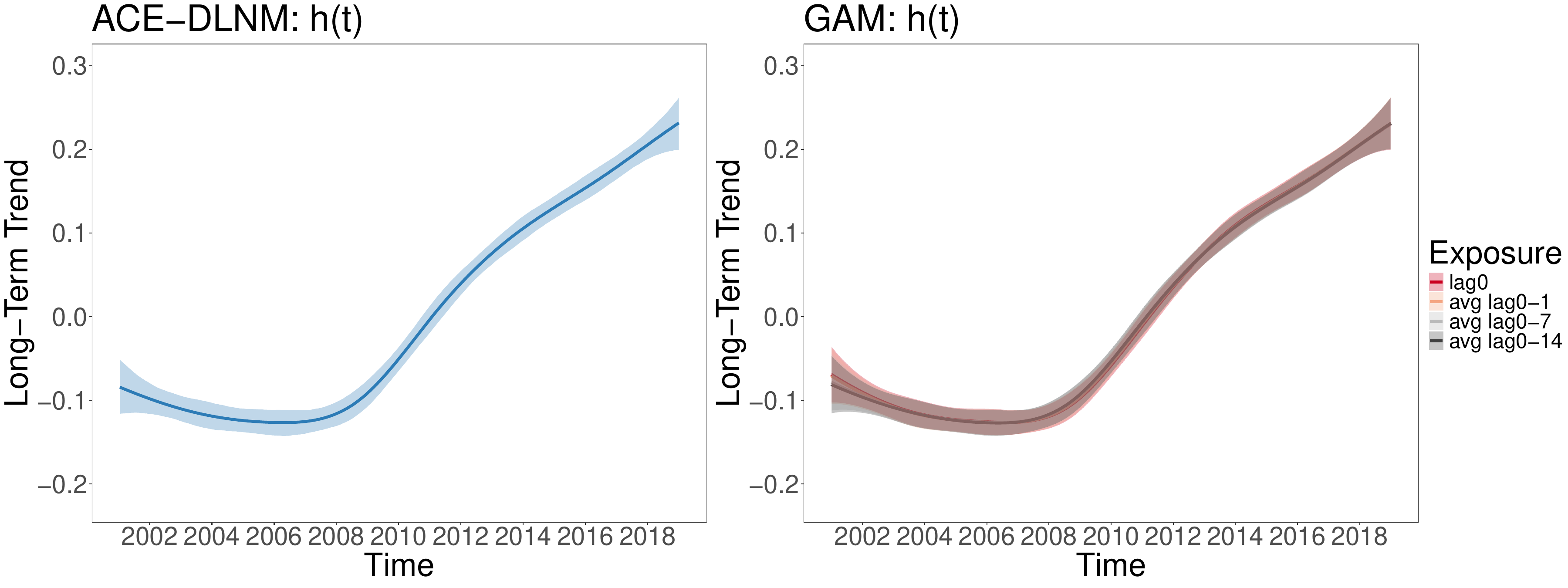}
    \end{subfigure}%

    \vspace{0.1cm}
    \begin{subfigure}{0.68\textwidth}
        \centering
        \includegraphics[width=\linewidth]{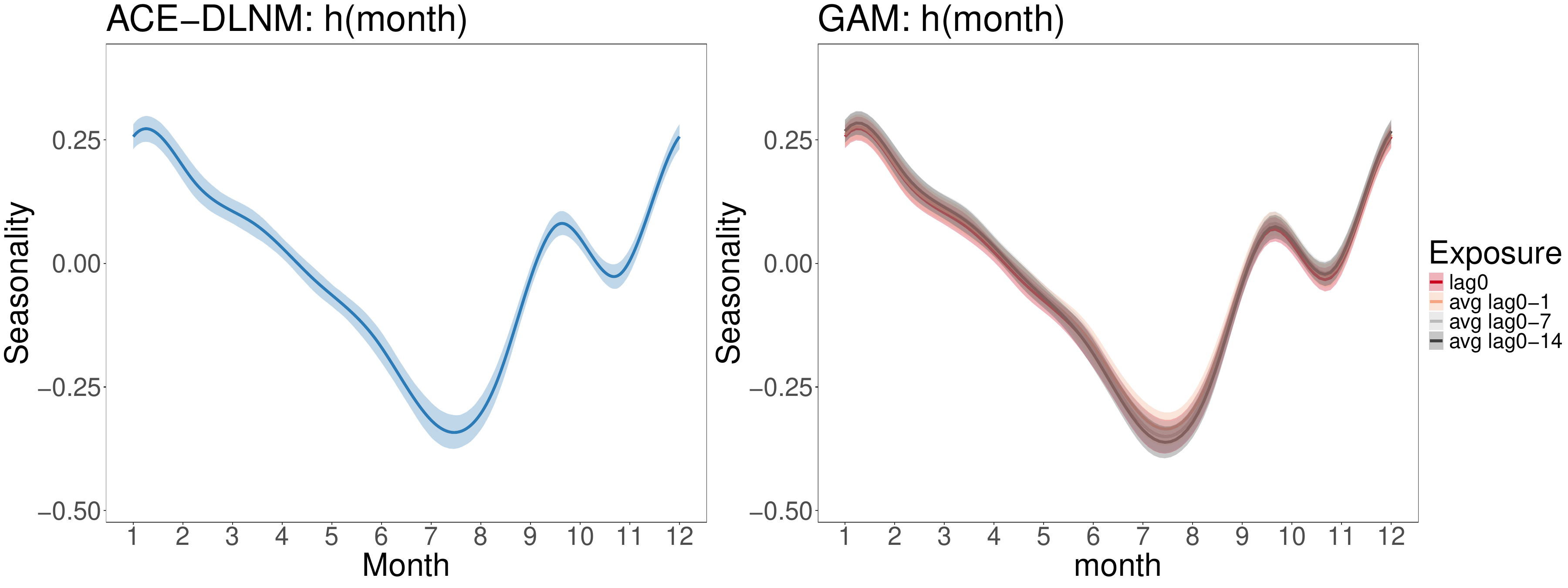}
    \end{subfigure}%

    \vspace{0.1cm}
   \begin{subfigure}{0.68\textwidth}
        \centering
        \includegraphics[width=\linewidth]{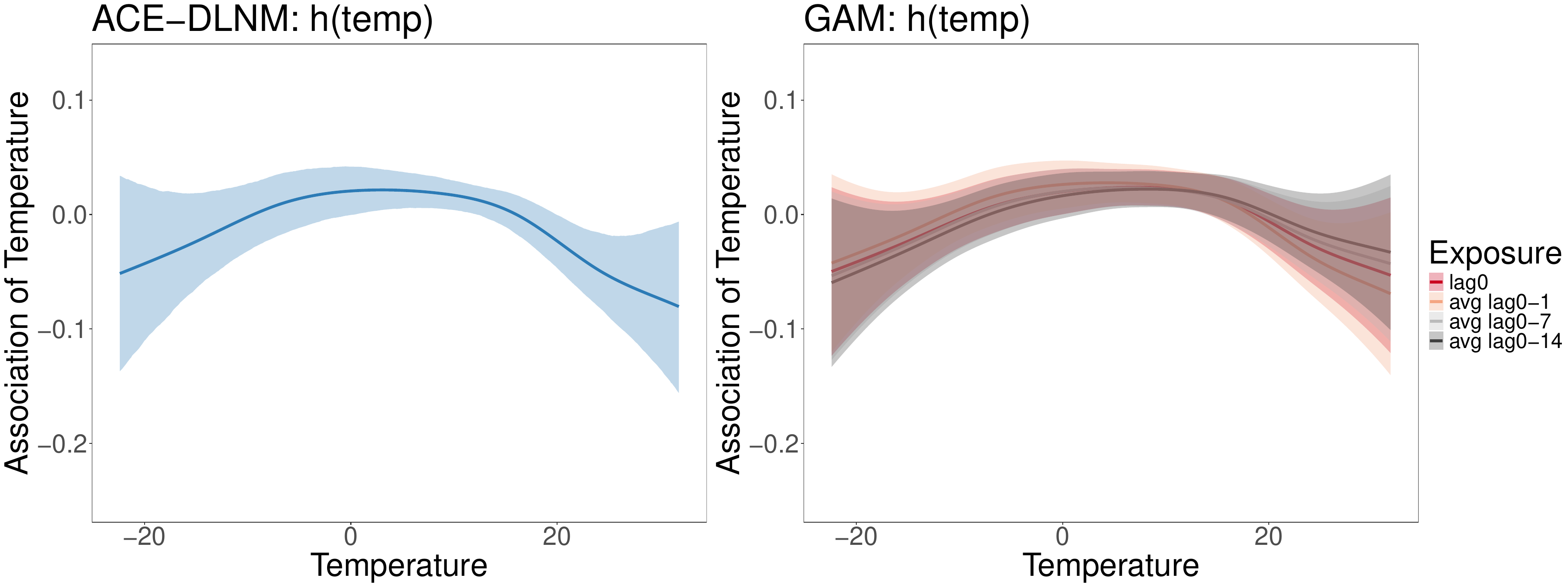}
    \end{subfigure}%

     \vspace{0.1cm}
   \begin{subfigure}{0.68\textwidth}
        \centering
        \includegraphics[width=\linewidth]{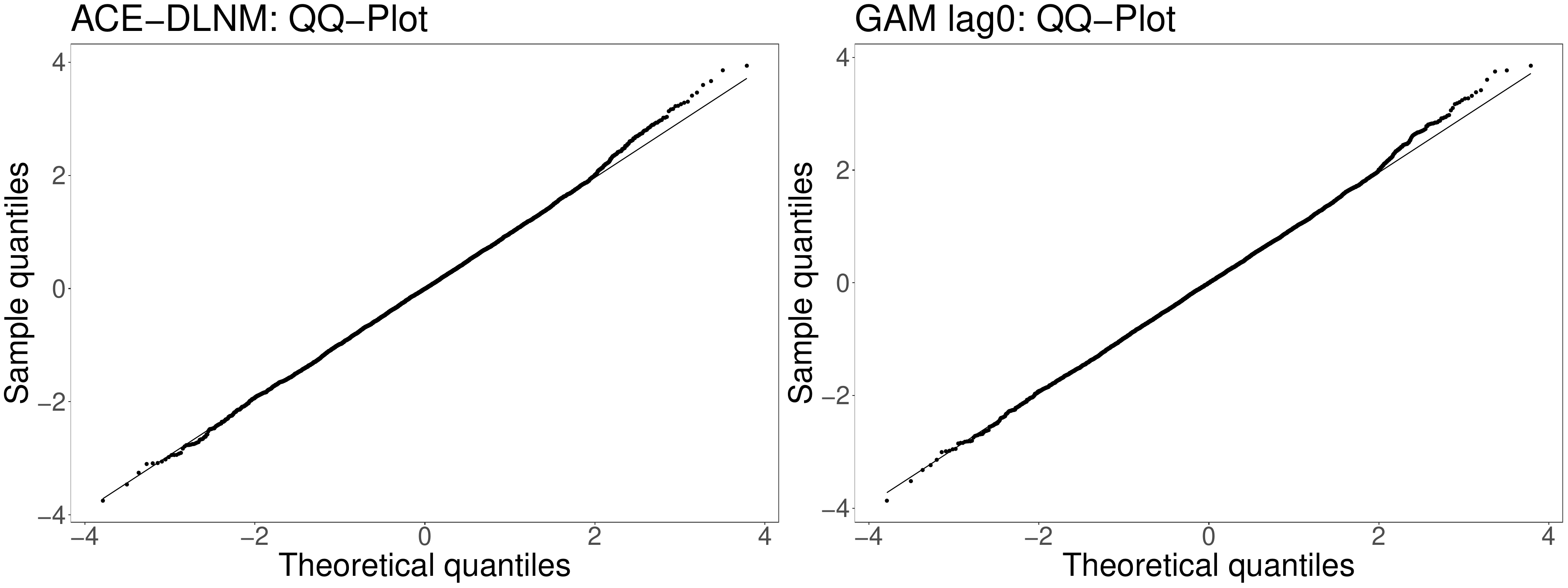}
    \end{subfigure}%

    \vspace{0.1cm}
   \begin{subfigure}{0.68\textwidth}
        \centering
        \includegraphics[width=\linewidth]{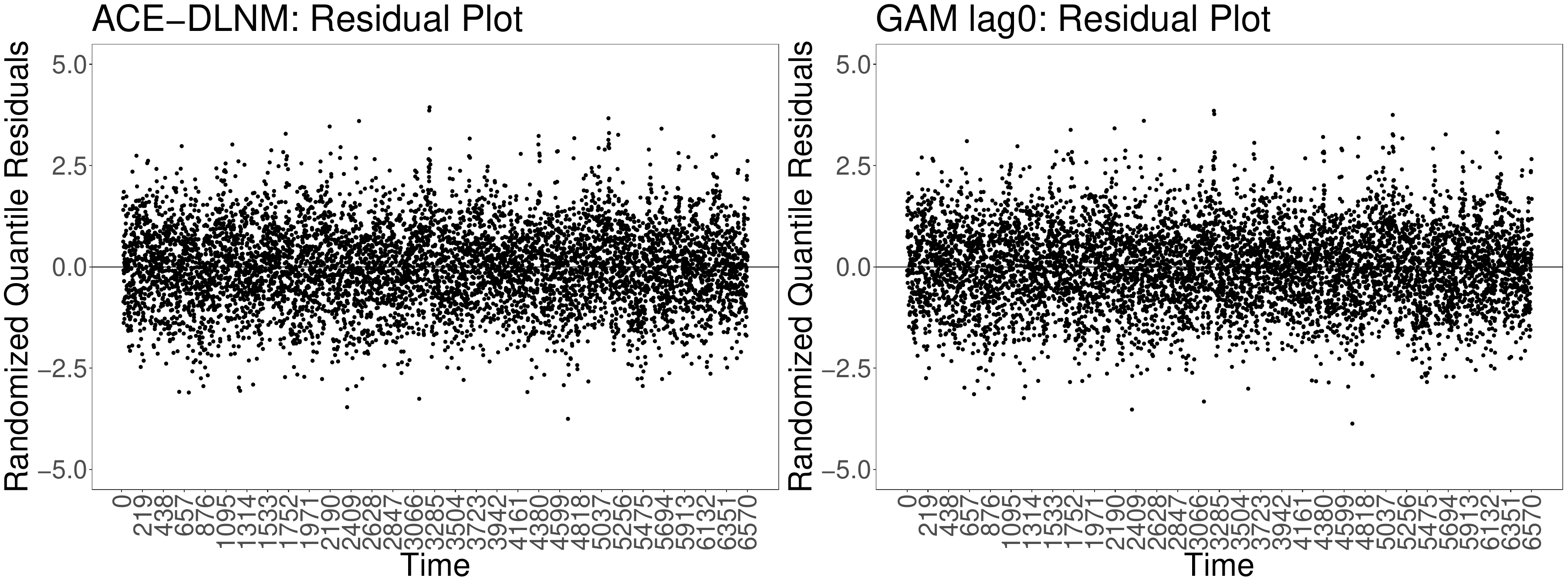}
    \end{subfigure}%
    \caption{Web Figure 24. Additional Results for Respiratory Morbidity in Peel.}
\end{figure}

\begin{figure}[H]
    \centering
     \begin{subfigure}{0.68\textwidth}
        \centering
        \includegraphics[width=\linewidth]{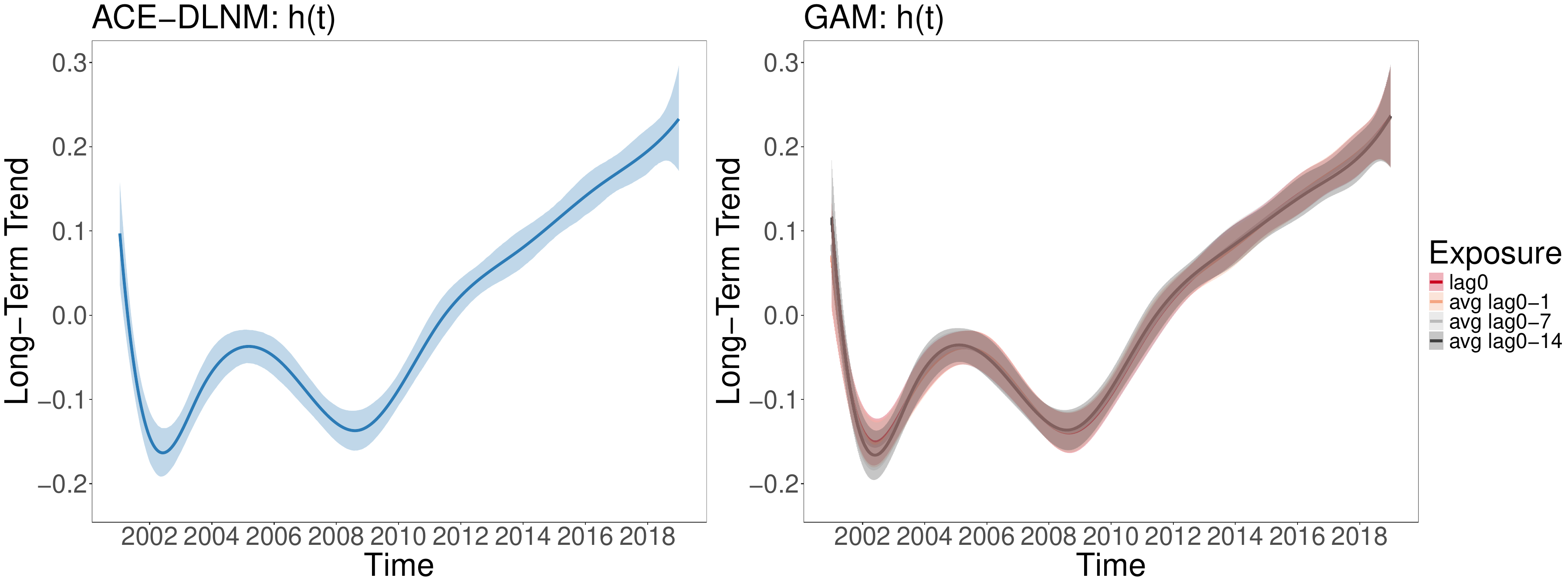}
    \end{subfigure}%

    \vspace{0.1cm}
    \begin{subfigure}{0.68\textwidth}
        \centering
        \includegraphics[width=\linewidth]{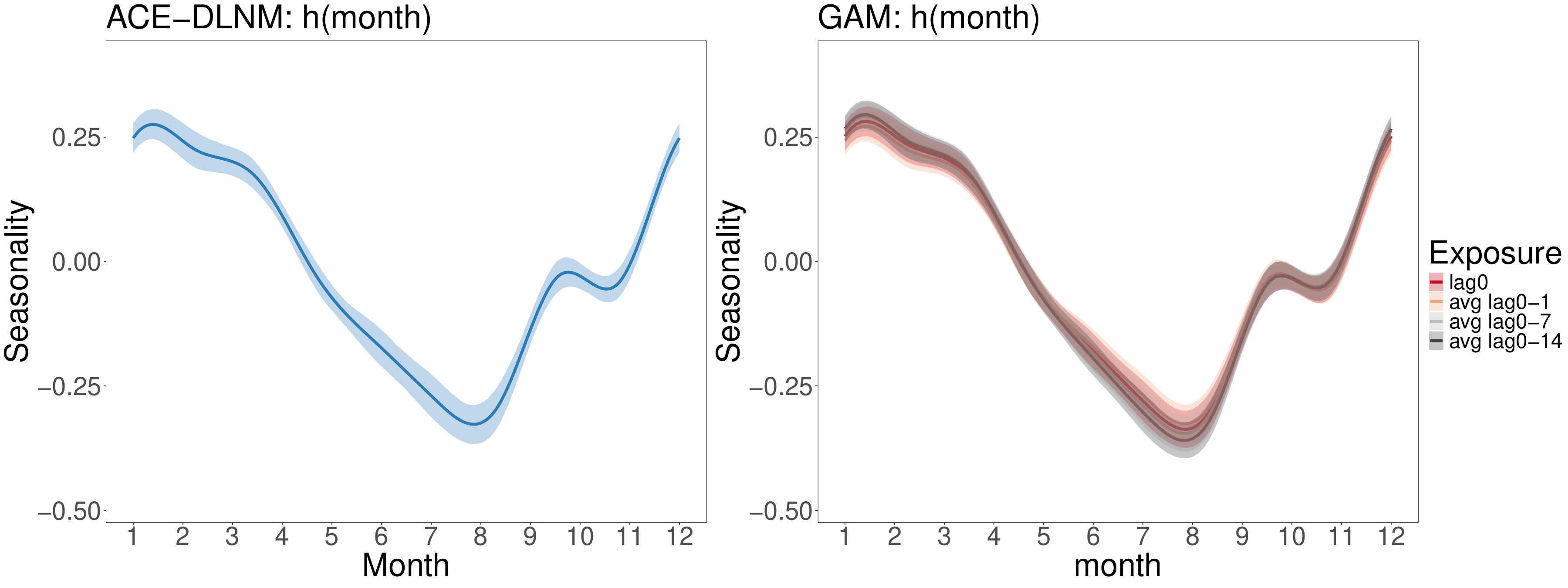}
    \end{subfigure}%

    \vspace{0.1cm}
   \begin{subfigure}{0.68\textwidth}
        \centering
        \includegraphics[width=\linewidth]{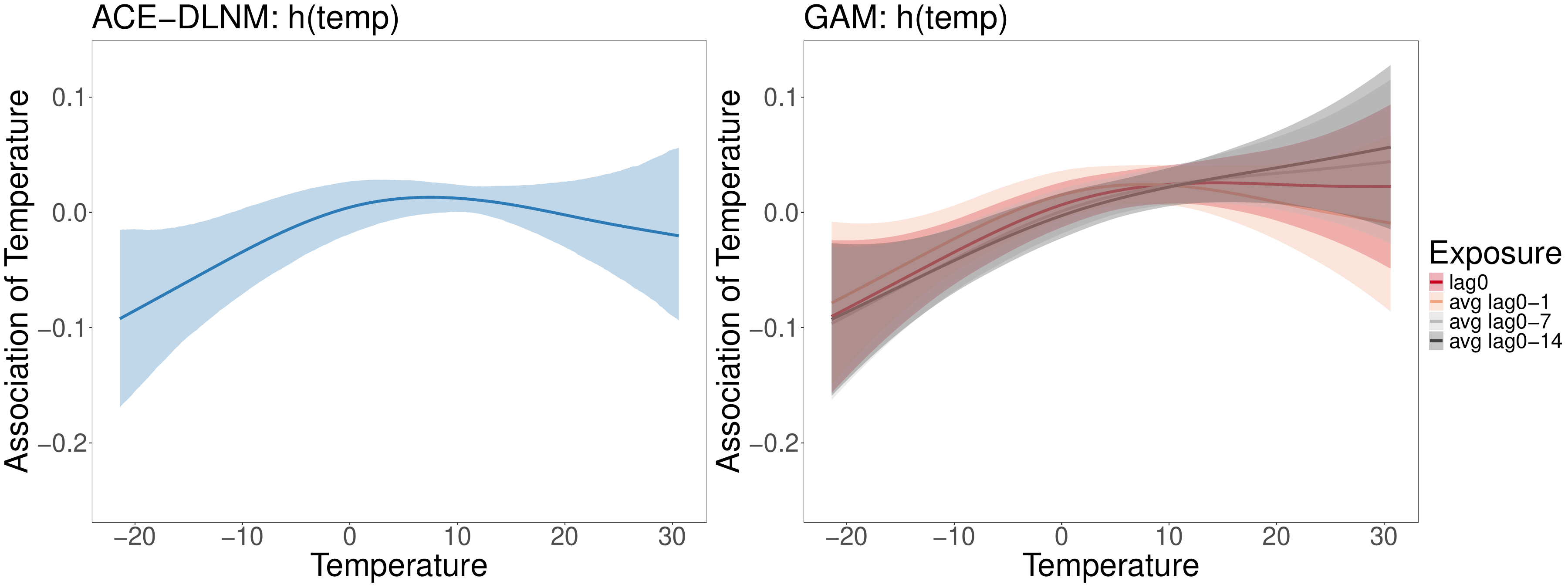}
    \end{subfigure}%

     \vspace{0.1cm}
   \begin{subfigure}{0.68\textwidth}
        \centering
        \includegraphics[width=\linewidth]{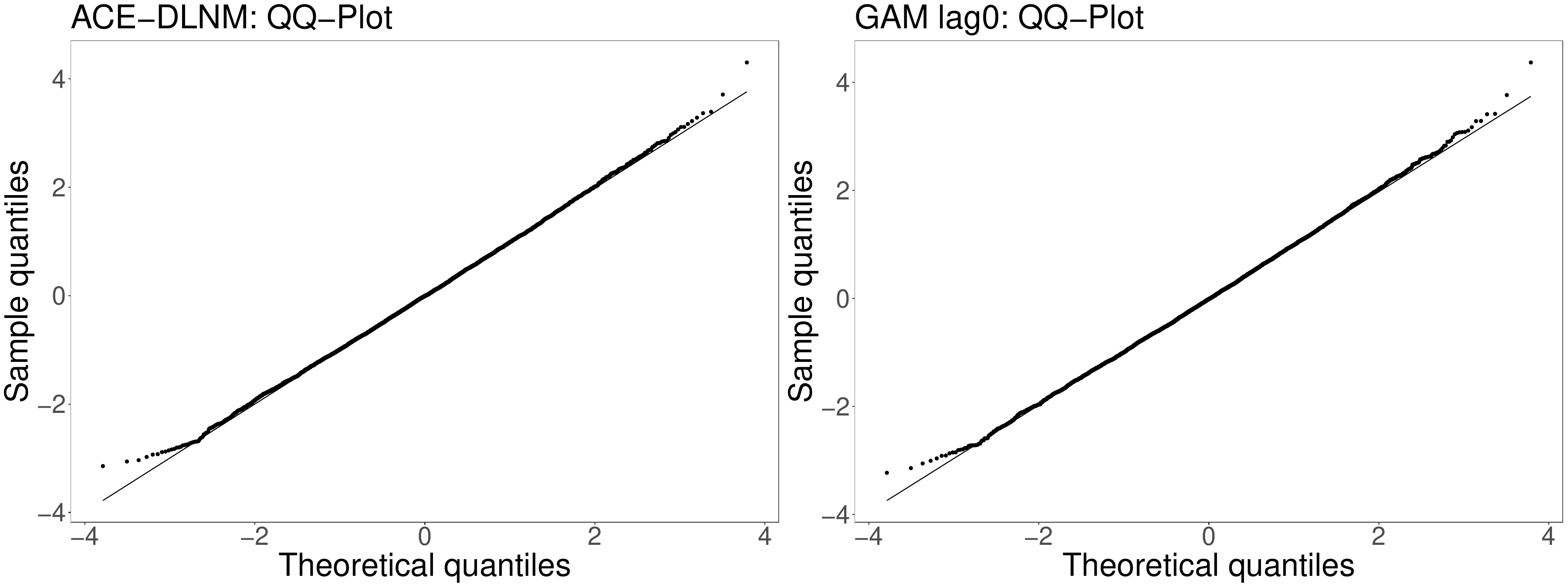}
    \end{subfigure}%

    \vspace{0.1cm}
   \begin{subfigure}{0.68\textwidth}
        \centering
        \includegraphics[width=\linewidth]{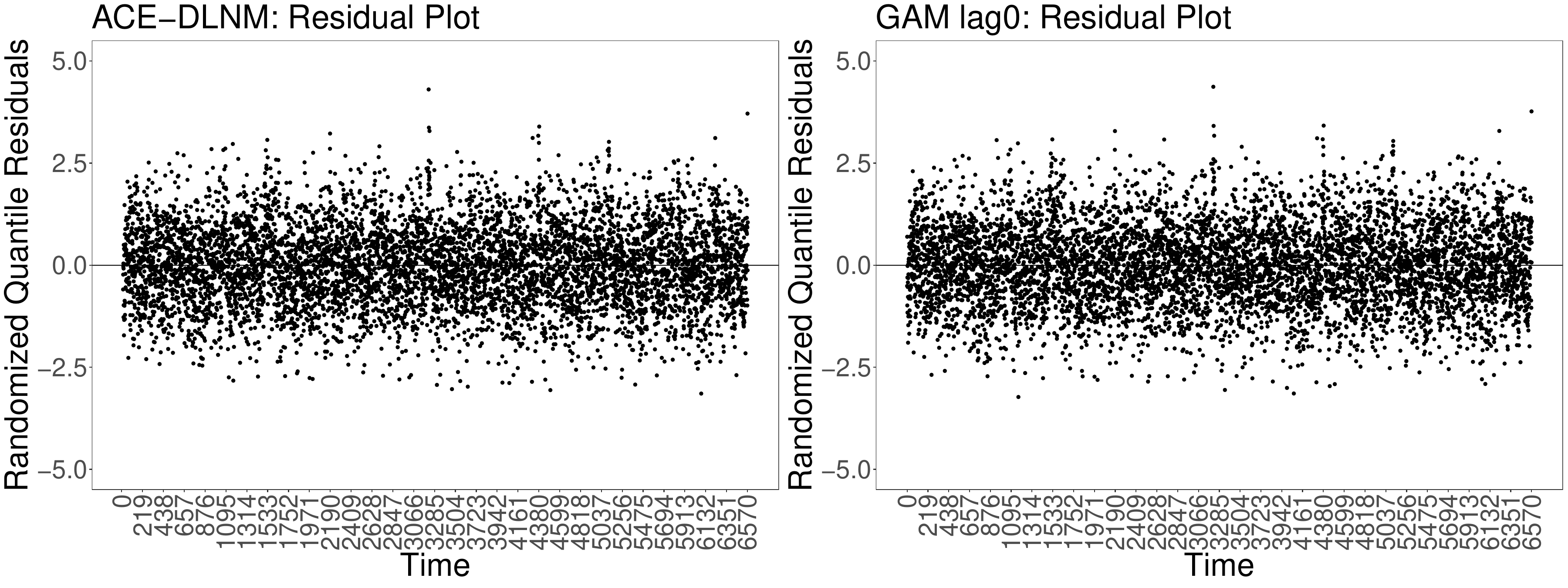}
    \end{subfigure}%
    \caption{Web Figure 25. Additional Results for Respiratory Morbidity in Hamilton.}
\end{figure}

\begin{figure}[H]
    \centering
     \begin{subfigure}{0.68\textwidth}
        \centering
        \includegraphics[width=\linewidth]{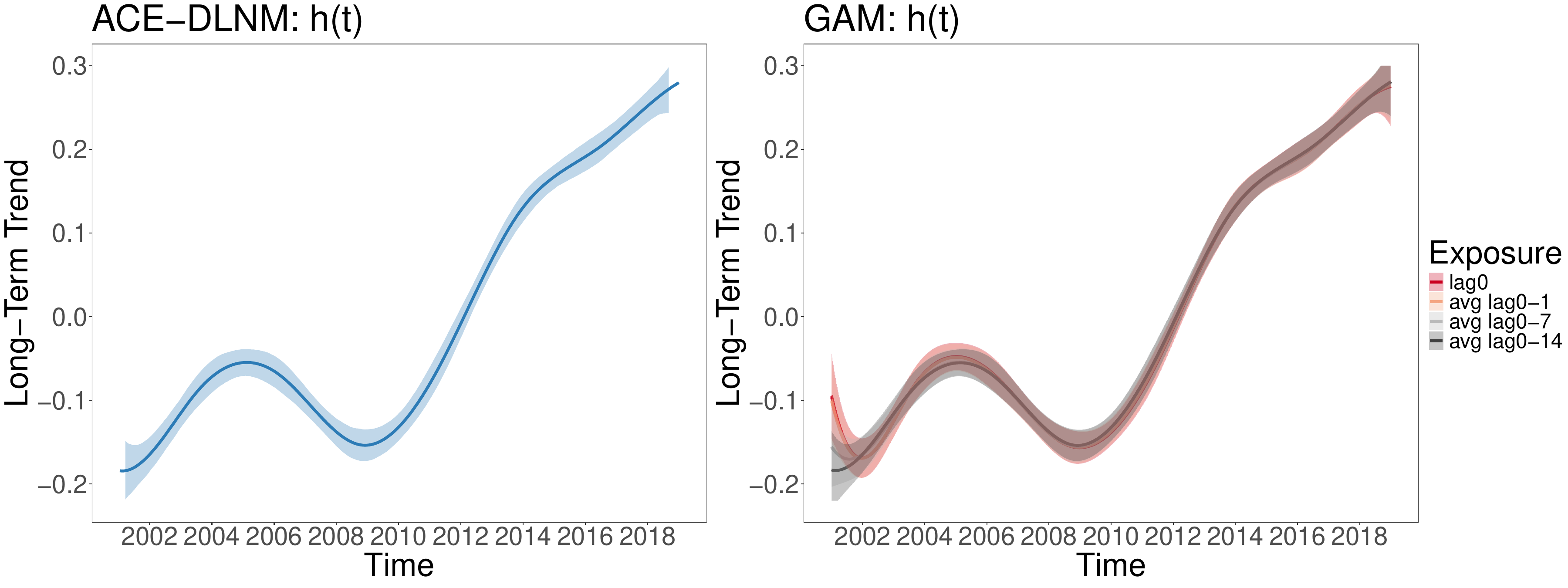}
    \end{subfigure}%

    \vspace{0.1cm}
    \begin{subfigure}{0.68\textwidth}
        \centering
        \includegraphics[width=\linewidth]{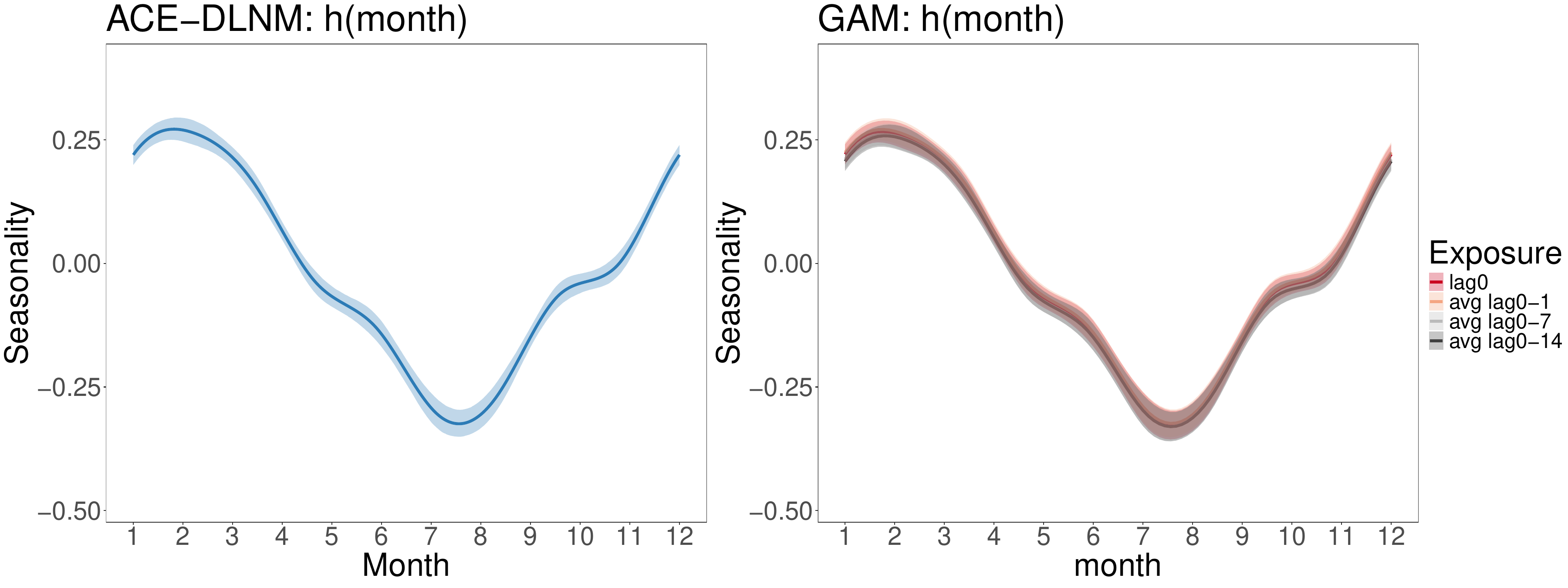}
    \end{subfigure}%

    \vspace{0.1cm}
   \begin{subfigure}{0.68\textwidth}
        \centering
        \includegraphics[width=\linewidth]{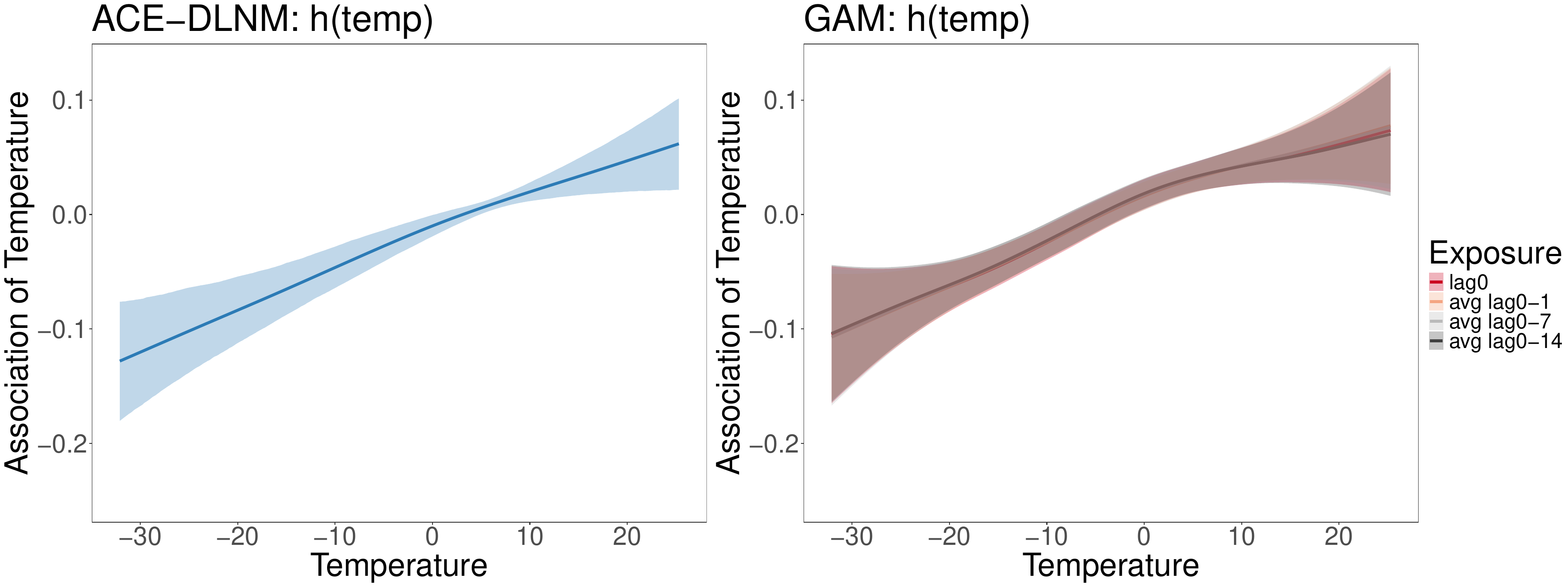}
    \end{subfigure}%

     \vspace{0.1cm}
   \begin{subfigure}{0.68\textwidth}
        \centering
        \includegraphics[width=\linewidth]{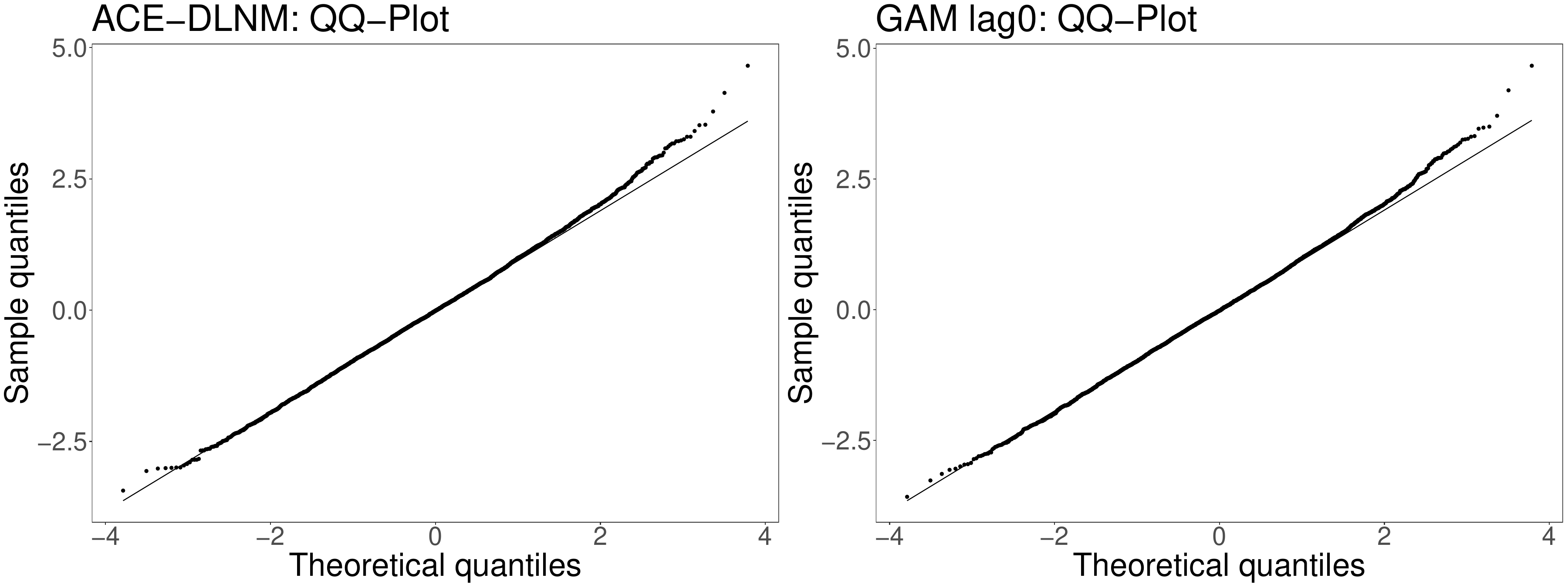}
    \end{subfigure}%

    \vspace{0.1cm}
   \begin{subfigure}{0.68\textwidth}
        \centering
        \includegraphics[width=\linewidth]{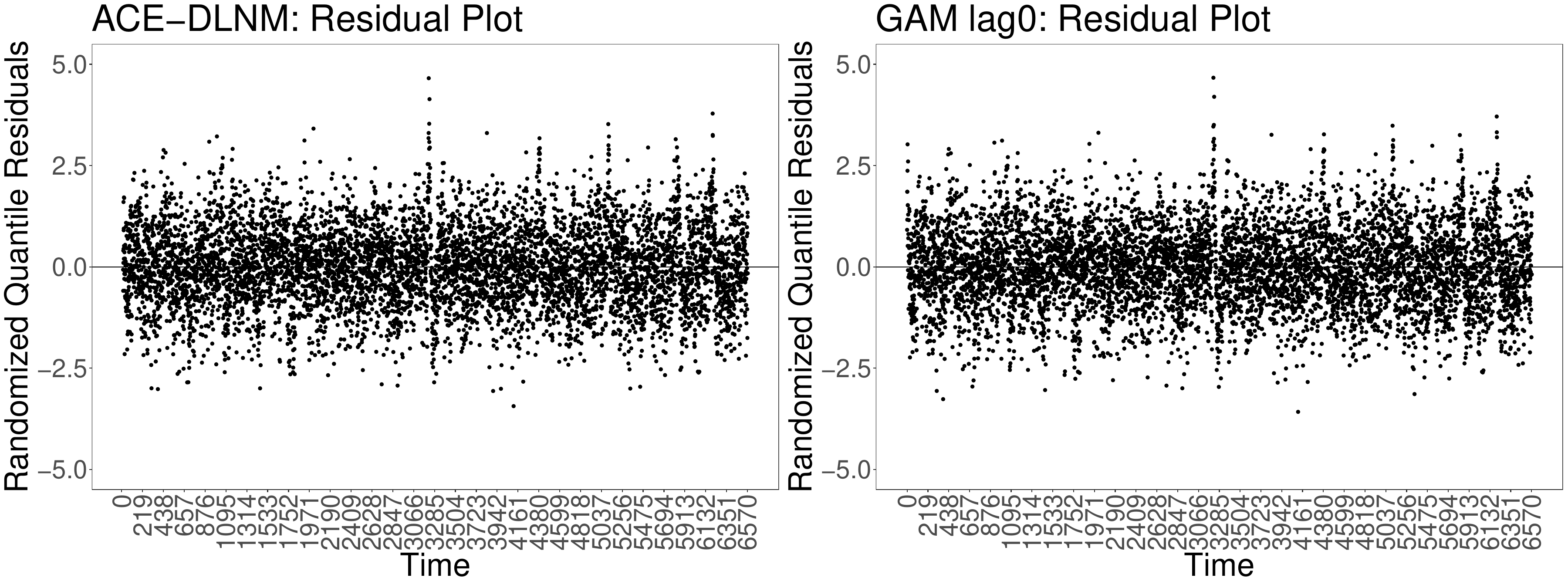}
    \end{subfigure}%
    \caption{Web Figure 26. Additional Results for Respiratory Morbidity in Calgary.}
\end{figure}

\begin{figure}[H]
    \centering
     \begin{subfigure}{0.68\textwidth}
        \centering
        \includegraphics[width=\linewidth]{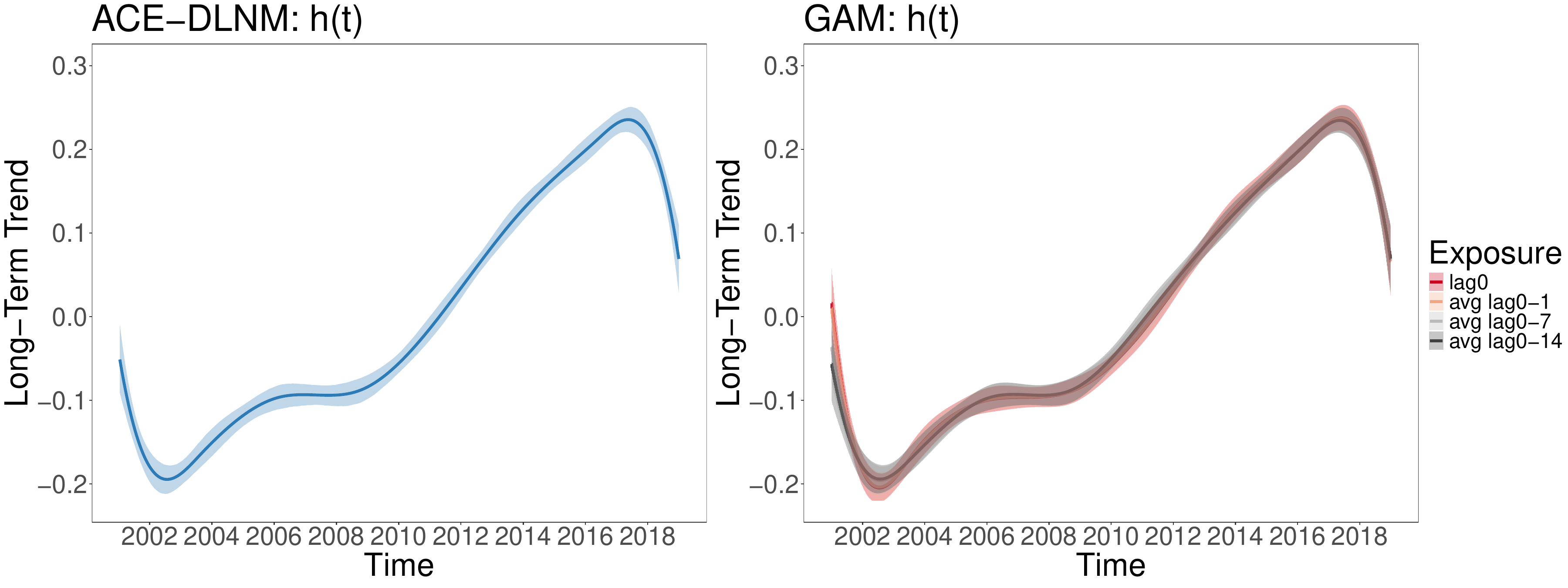}
    \end{subfigure}%

    \vspace{0.1cm}
    \begin{subfigure}{0.68\textwidth}
        \centering
        \includegraphics[width=\linewidth]{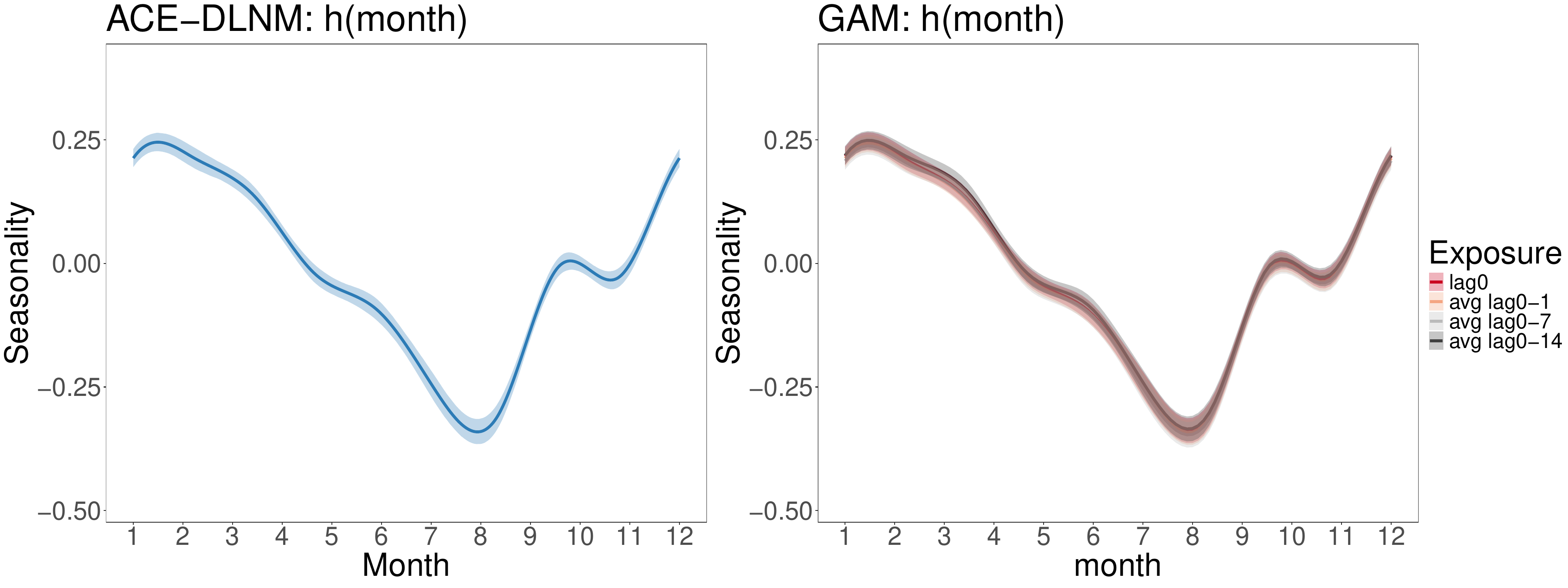}
    \end{subfigure}%

    \vspace{0.1cm}
   \begin{subfigure}{0.68\textwidth}
        \centering
        \includegraphics[width=\linewidth]{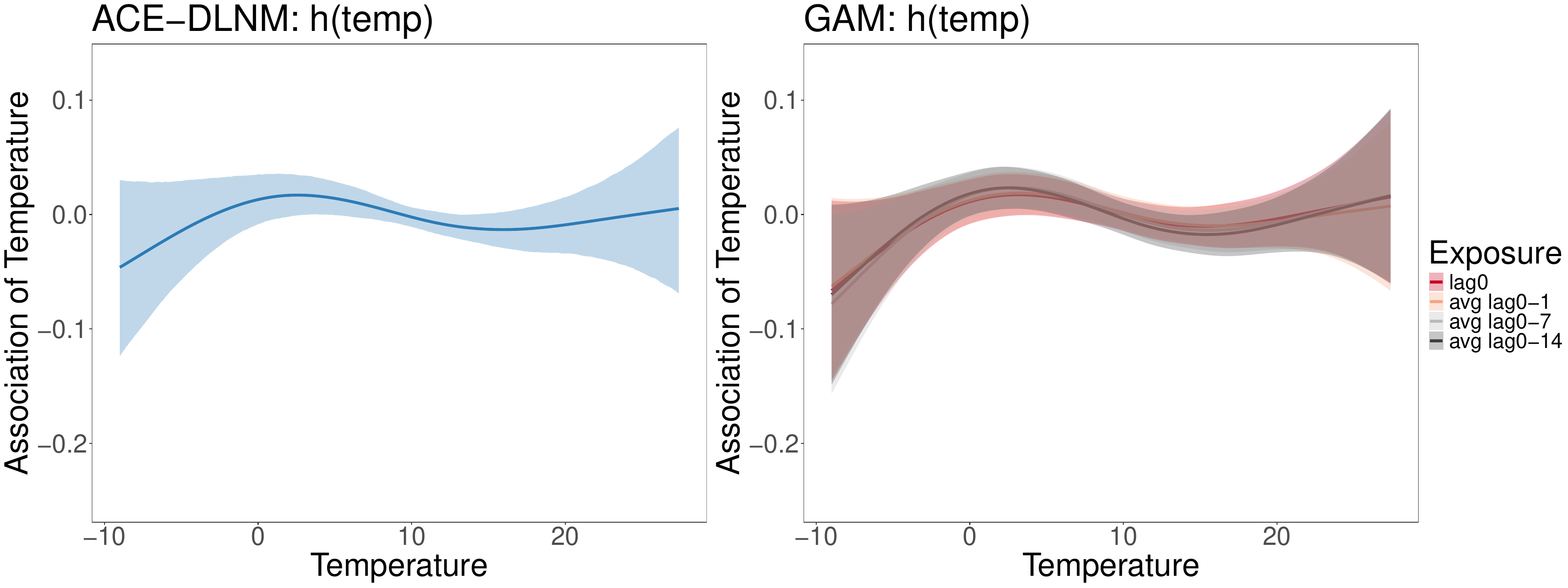}
    \end{subfigure}%

     \vspace{0.1cm}
   \begin{subfigure}{0.68\textwidth}
        \centering
        \includegraphics[width=\linewidth]{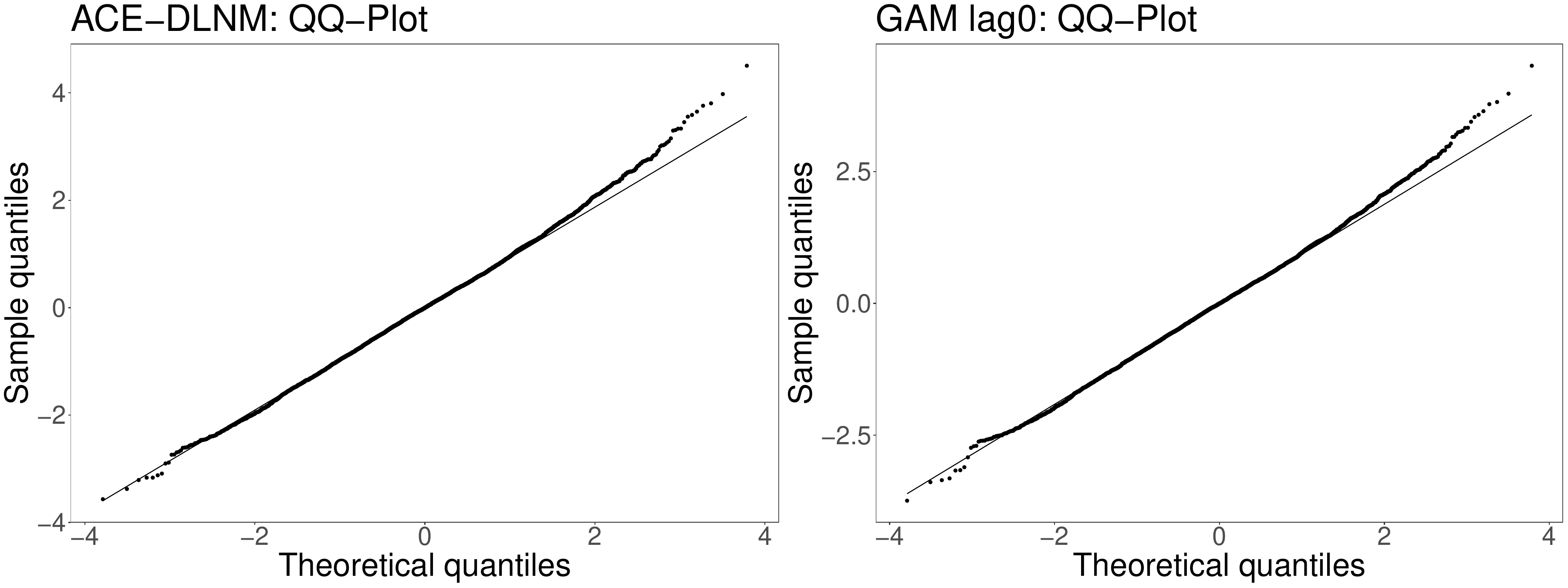}
    \end{subfigure}%

    \vspace{0.1cm}
   \begin{subfigure}{0.68\textwidth}
        \centering
        \includegraphics[width=\linewidth]{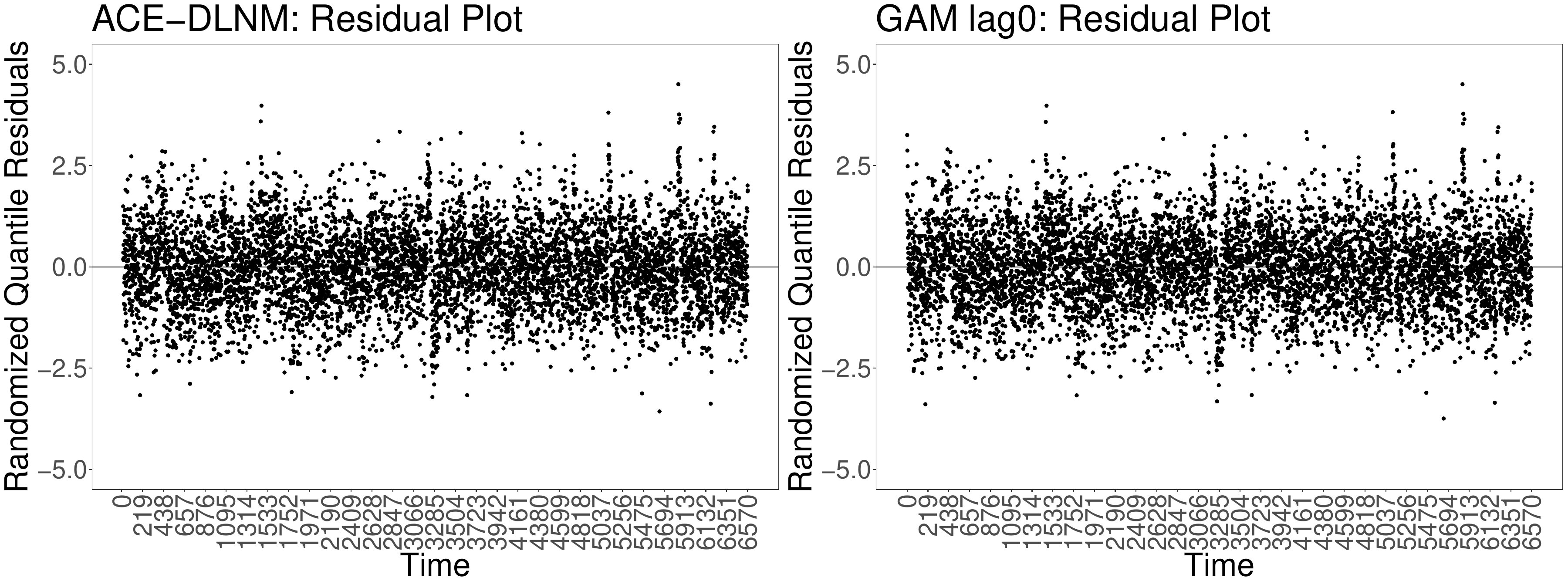}
    \end{subfigure}%
    \caption{Web Figure 27. Additional Results for Respiratory Morbidity in Vancouver.}
\end{figure}

\begin{table}[H]
\centering
\caption{Web Table 10. Estimates for the linear components and the dispersion parameter for respiratory morbidity in Waterloo.}
\begin{tabular}{@{}lcccc@{}}
\toprule
          & \multicolumn{2}{c}{ACE-DLNM}        & \multicolumn{2}{c}{GAM Lag0} \\ \cmidrule(l){2-3} \cmidrule(l){4-5} 
          & Point Est. & CI Est.          & Point Est. & CI Est.          \\ \midrule
Intercept     & 1.967      & (1.884, 2.050)            & 1.898      & (1.867, 1.929)   \\
$\beta$ Mon  & 0.192      & (0.157, 0.227)   & 0.192      & (0.157, 0.228)   \\
$\beta$ Tue  & 0.155      & (0.119, 0.191)   & 0.156      & (0.120, 0.192)   \\
$\beta$ Wed  & 0.096      & (0.060, 0.133)  & 0.097       & (0.060, 0.133)   \\
$\beta$ Thu  & 0.113      & (0.077, 0.149)   & 0.114      & (0.078, 0.151)   \\
$\beta$ Fri  & 0.072      & (0.037, 0.108)   & 0.075      & (0.038, 0.111)   \\
$\beta$ Sat  & -0.049     & (-0.086, -0.010)  & -0.048     & (-0.086, -0.011)  \\
$\log \theta$     & 4.046      &                  & 3.903     &                  \\ \bottomrule
\end{tabular}
\end{table}

\begin{table}[H]
\centering
\caption{Web Table 11. Estimates for the linear components and the dispersion parameter for respiratory morbidity in Peel.}
\begin{tabular}{@{}lcccc@{}}
\toprule
          & \multicolumn{2}{c}{ACE-DLNM}        & \multicolumn{2}{c}{GAM Lag0} \\ \cmidrule(l){2-3} \cmidrule(l){4-5} 
          & Point Est. & CI Est.          & Point Est. & CI Est.          \\ \midrule
Intercept     & 2.712      & (2.651, 2.773)            & 2.679      & (2.657, 2.702)   \\
$\beta$ Mon  & 0.183      & (0.158, 0.209)   & 0.184      & (0.158, 0.209)   \\
$\beta$ Tue  & 0.203      & (0.178, 0.229)   & 0.204      & (0.179, 0.229)   \\
$\beta$ Wed  & 0.134      & (0.109, 0.161)  & 0.136       & (0.110, 0.161)   \\
$\beta$ Thu  & 0.135      & (0.109, 0.160)   & 0.135      & (0.109, 0.161)   \\
$\beta$ Fri  & 0.084      & (0.058, 0.109)   & 0.084      & (0.059, 0.110)   \\
$\beta$ Sat  & -0.038     & (-0.065, -0.011)  & -0.037     & (-0.064, -0.011)  \\
$\log \theta$     & 4.116      &                  & 4.128     &                  \\ \bottomrule
\end{tabular}
\end{table}

\begin{table}[H]
\centering
\caption{Web Table 12. Estimates for the linear components and the dispersion parameter for respiratory morbidity in Hamilton.}
\begin{tabular}{@{}lcccc@{}}
\toprule
          & \multicolumn{2}{c}{ACE-DLNM}        & \multicolumn{2}{c}{GAM Lag0} \\ \cmidrule(l){2-3} \cmidrule(l){4-5} 
          & Point Est. & CI Est.          & Point Est. & CI Est.          \\ \midrule
Intercept     & 2.113      & (2.060, 2.164)            & 2.171      & (2.143, 2.199)   \\
$\beta$ Mon  & 0.126      & (0.096, 0.156)   & 0.126      & (0.094, 0.157)   \\
$\beta$ Tue  & 0.148      & (0.117, 0.179)   & 0.149      & (0.118, 0.180)   \\
$\beta$ Wed  & 0.093      & (0.062, 0.124)  & 0.093       & (0.061, 0.124)   \\
$\beta$ Thu  & 0.104      & (0.073, 0.136)   & 0.105      & (0.073, 0.136)   \\
$\beta$ Fri  & 0.100      & (0.070, 0.132)   & 0.101      & (0.069, 0.132)   \\
$\beta$ Sat  & -0.041     & (-0.073, -0.009)  & -0.040     & (-0.073, -0.008)  \\
$\log \theta$     & 4.265      &                  & 4.283     &                  \\ \bottomrule
\end{tabular}
\end{table}

\begin{table}[H]
\centering
\caption{Web Table 13. Estimates for the linear components and the dispersion parameter for respiratory morbidity in Calgary.}
\begin{tabular}{@{}lcccc@{}}
\toprule
          & \multicolumn{2}{c}{ACE-DLNM}        & \multicolumn{2}{c}{GAM Lag0} \\ \cmidrule(l){2-3} \cmidrule(l){4-5} 
          & Point Est. & CI Est.          & Point Est. & CI Est.          \\ \midrule
Intercept     & 2.735      & (2.512, 2.951)            & 2.643      & (2.614, 2.671)   \\
$\beta$ Mon  & 0.176      & (0.150, 0.202)   & 0.177      & (0.151, 0.203)   \\
$\beta$ Tue  & 0.154      & (0.128, 0.181)   & 0.156      & (0.130, 0.182)   \\
$\beta$ Wed  & 0.142      & (0.116, 0.168)  & 0.143       & (0.117, 0.169)   \\
$\beta$ Thu  & 0.146      & (0.120, 0.173)   & 0.149      & (0.123, 0.175)   \\
$\beta$ Fri  & 0.103      & (0.077, 0.128)   & 0.104      & (0.078, 0.131)   \\
$\beta$ Sat  & -0.011     & (-0.037, 0.017)  & -0.010     & (-0.036, 0.017)  \\
$\log \theta$     & 3.929      &                  & 3.924     &                  \\ \bottomrule
\end{tabular}
\end{table}

\begin{table}[H]
\centering
\caption{Web Table 14. Estimates for the linear components and the dispersion parameter for respiratory morbidity in Vancouver.}
\begin{tabular}{@{}lcccc@{}}
\toprule
          & \multicolumn{2}{c}{ACE-DLNM}        & \multicolumn{2}{c}{GAM Lag0} \\ \cmidrule(l){2-3} \cmidrule(l){4-5} 
          & Point Est. & CI Est.          & Point Est. & CI Est.          \\ \midrule
Intercept     & 3.405      & (3.348, 3.461)            & 3.352      & (3.332, 3.372)   \\
$\beta$ Mon  & 0.166      & (0.147, 0.186)   & 0.166      & (0.147, 0.186)   \\
$\beta$ Tue  & 0.181      & (0.162, 0.201)   & 0.181      & (0.162, 0.201)   \\
$\beta$ Wed  & 0.164      & (0.145, 0.184)  & 0.164       & (0.145, 0.184)   \\
$\beta$ Thu  & 0.117      & (0.098, 0.137)   & 0.117      & (0.098, 0.137)   \\
$\beta$ Fri  & 0.146      & (0.128, 0.165)   & 0.146      & (0.126, 0.165)   \\
$\beta$ Sat  & -0.034     & (-0.054, -0.013)  & -0.034     & (-0.054, -0.014)  \\
$\log \theta$     & 4.268      &                  & 4.267     &                  \\ \bottomrule
\end{tabular}
\end{table}

\clearpage
\subsection{Sensitivity to the Bounds of ACE}

We use the Hamilton data as an example to investigate the sensitivity to bounds of the ACE. 
As introduced in Section 3.2, we use the pre-determined bounds for ACE through the Cauchy-Schwarz inequality, and then define the basis function and compute the penalty matrix accordingly. 
In the Hamilton dataset, the predetermined bounds are $[-94,94]$ (rounded to integers in the implementation), 
and the range of the estimated ACE is $[-4.264, 69.800]$. 
We first re-fit the model using narrower bounds $[-70, 70]$, but the inner optimization algorithm cannot converge. 
We re-fit the model by setting the bounds as $[-120, 120]$, and find that the results remain the same; see Web Figure 28. 

\begin{figure}[H]
    \centering
     \begin{subfigure}[t]{0.5\textwidth}
        \centering
        \includegraphics[width=\linewidth]{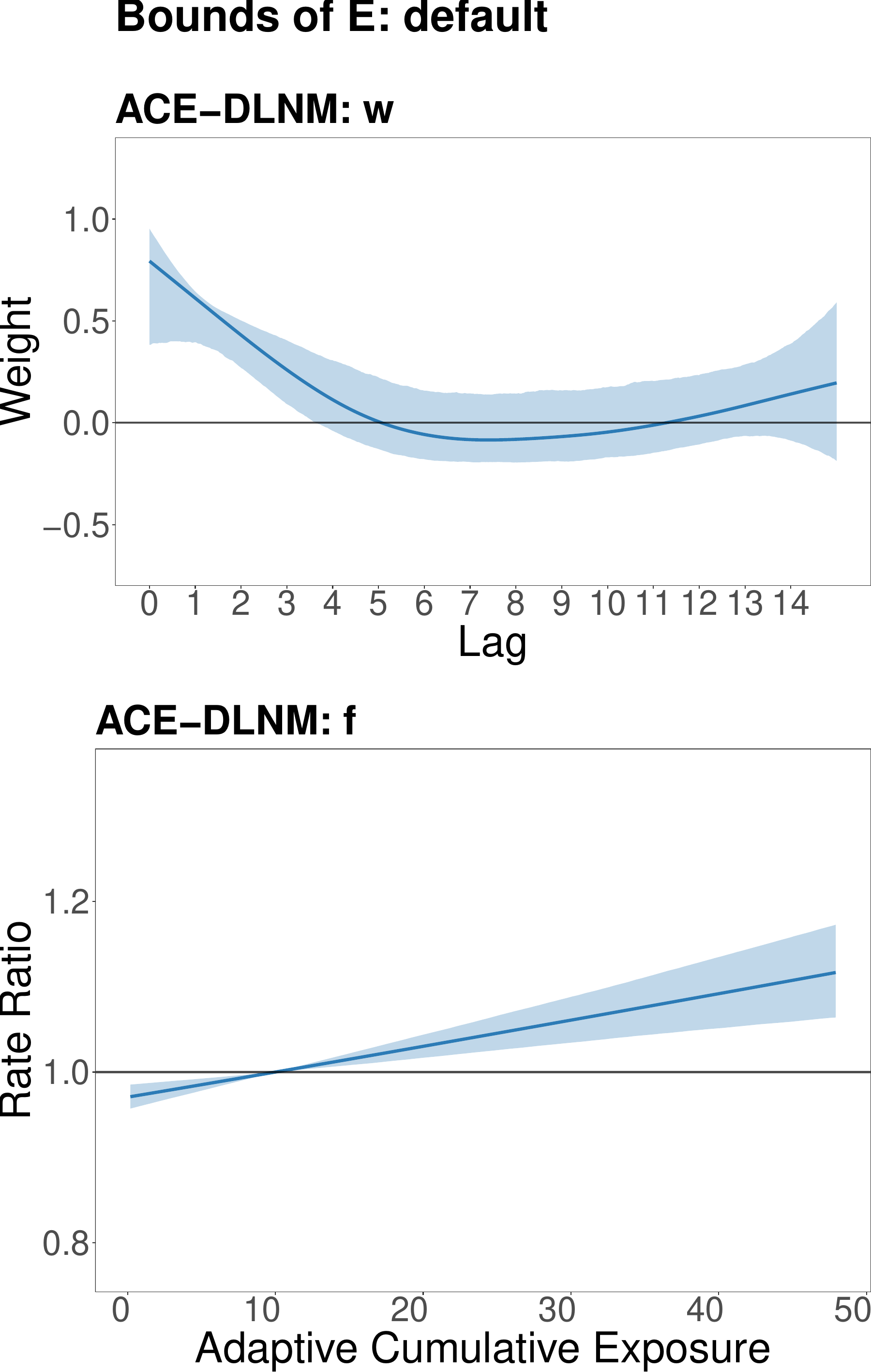}
        \caption{Bounds of $E$: default}
    \end{subfigure}%
    \hfill
    \begin{subfigure}[t]{0.5\textwidth}
        \centering
        \includegraphics[width=\linewidth]{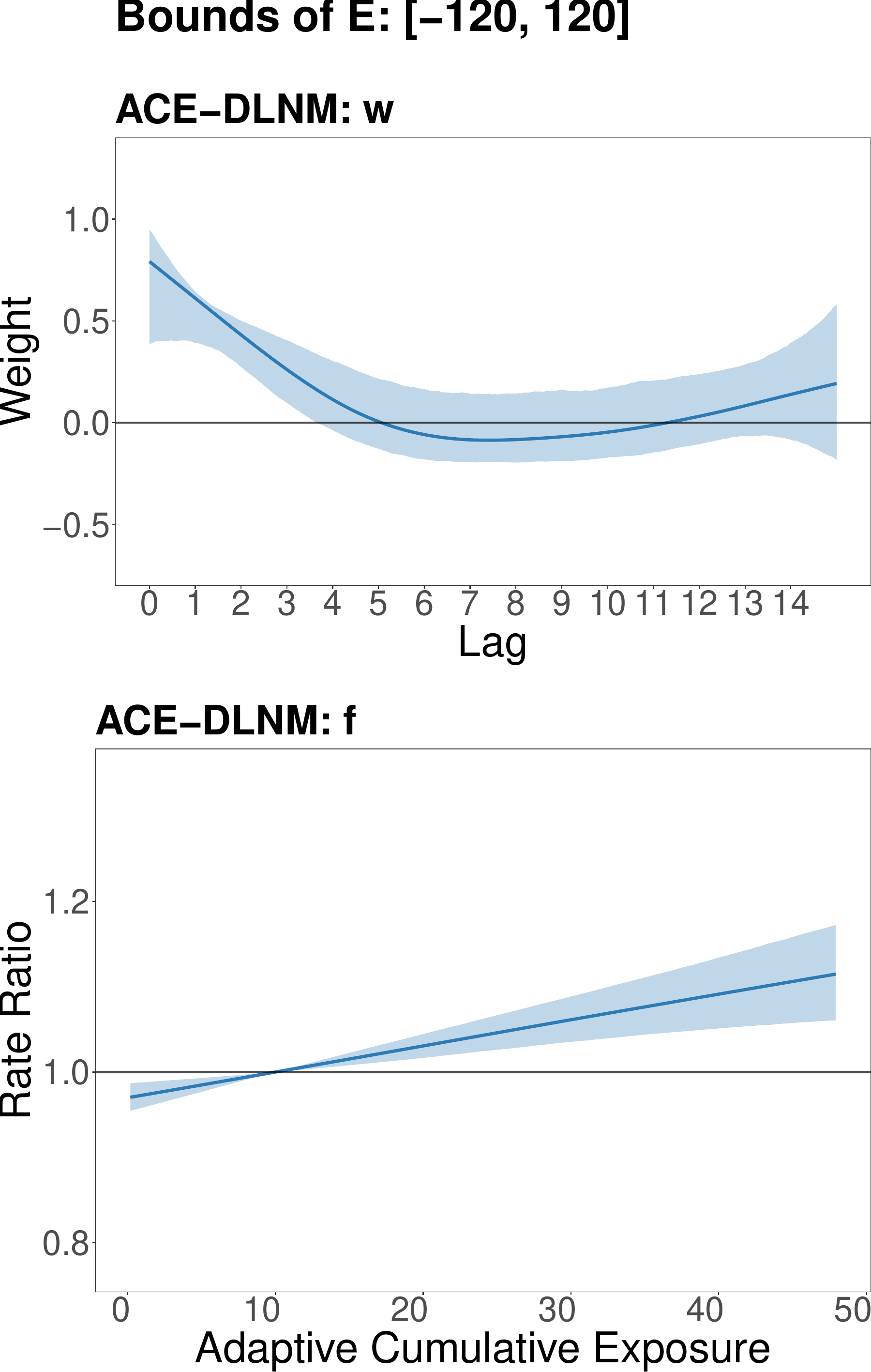}
        \caption{Bounds of $E$: $[-120,120]$}
    \end{subfigure}%
    \caption{Web Figure 28. Estimates of $w$ and $f$ for the Hamilton dataset using (a) the default bounds of E ($[-94,94]$ in this example) and (b) $[-120,120]$ as the bounds of E.}
\end{figure}

\clearpage
\subsection{Sensitivity to the Initial Values}

We use the Hamilton data as an example to investigate the sensitivity to initial values of the outer optimization. 
The default and recommended choice of initial values is introduced in \ref{s:opt}. 
We re-fit the model by setting the initial values as 0, and find that the results remain the same; see Web Figure 29. 
However, when using 0 as the initial values, the BFGS algorithm converges in 48 iterations, 
while the BFGS algorithm with the recommended initial values converges in 37 iterations. 

\begin{figure}[H]
    \centering
     \begin{subfigure}[t]{0.5\textwidth}
        \centering
        \includegraphics[width=\linewidth]{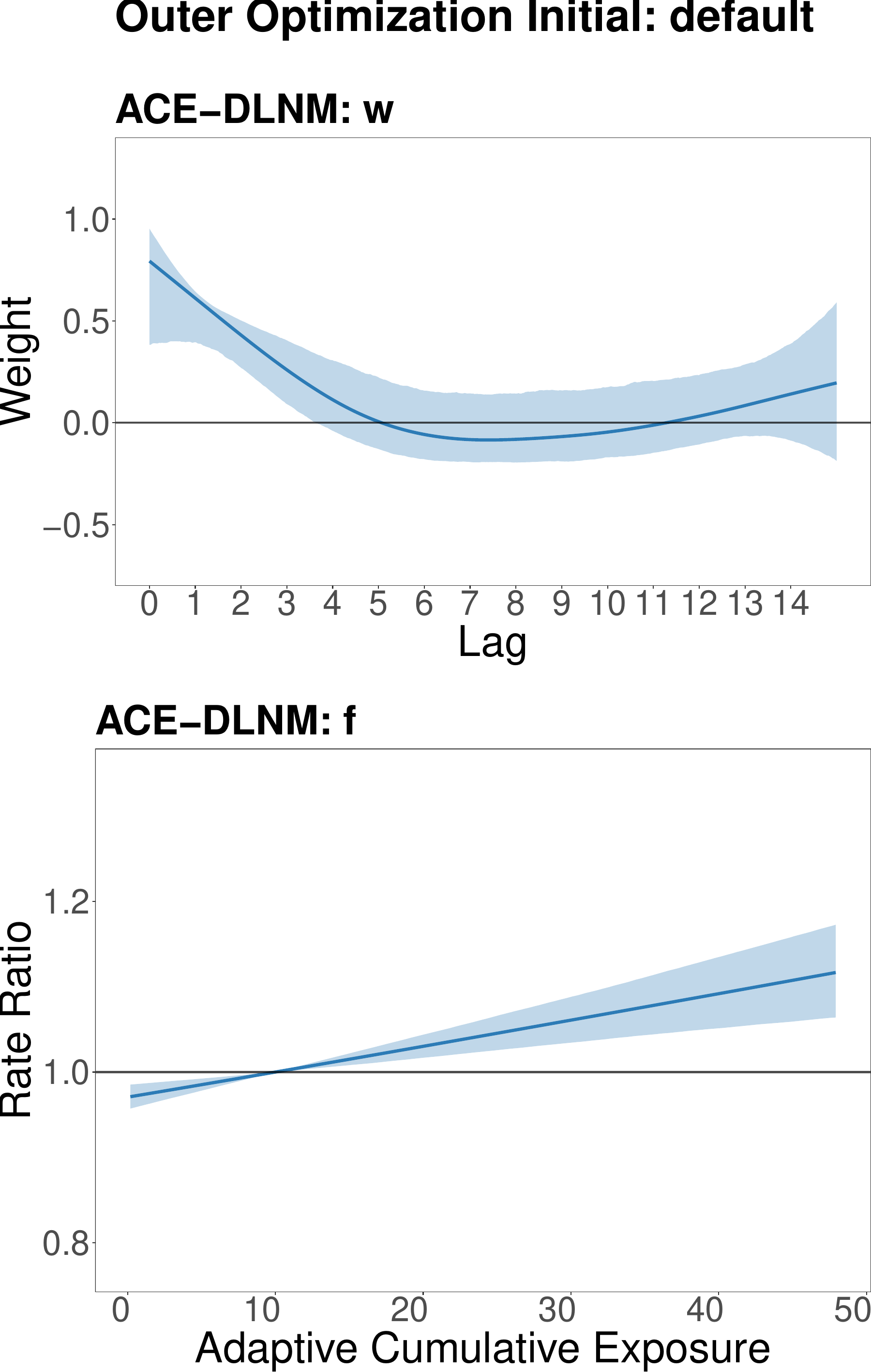}
        \caption{initial values: default}
    \end{subfigure}%
    \hfill
    \begin{subfigure}[t]{0.5\textwidth}
        \centering
        \includegraphics[width=\linewidth]{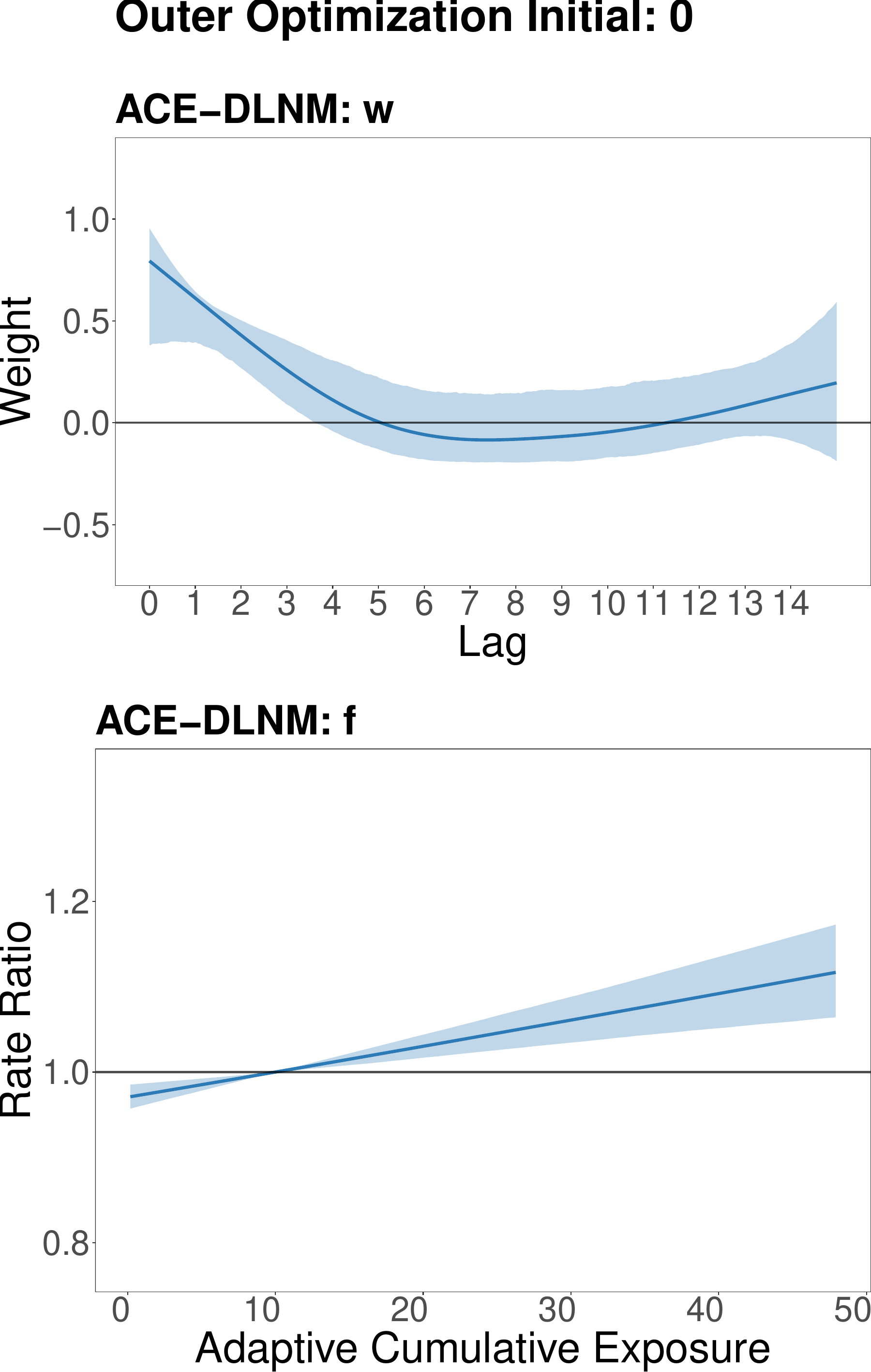}
        \caption{initial values: 0}
    \end{subfigure}%
    \caption{Web Figure 29. Estimates of $w$ and $f$ for the Hamilton dataset using (a) the default initial values and (b) 0 as the initial values.}
\end{figure}

\clearpage
\subsection{Negative Binomial and Poisson Distribution}
We compare the ACE-DLNMs under negative binomial response assumption and Poisson response assumption in terms of AIC. 
The model comparison results are reported in Web Table 15. 
For all cities, the negative binomial distribution yields smaller AICs. 
The results support our choice of the negative binomial response, which is the default for the proposed framework implemented in the R package {\tt aceDLNM}. 

\begin{table}[H]
\centering
\caption{Web Table 15. AICs for ACE-DLNMs with negative binomial response (default in {\tt aceDLNM}) and Poisson response. }
\label{tab:nb_poisson}
\begin{tabular}{@{}lcc@{}}
\toprule
          & Negative Binomial & Poisson  \\ \midrule
Waterloo         & 32155          & 32232 \\ %
Peel         & 38106          & 38314 \\ %
Hamilton         &  33869          & 33923 \\
Calgary         & 38488          & 38786 \\ %
Vancouver         & 43531          & 44038 \\ \bottomrule
\end{tabular}
\end{table}

\end{document}